\RequirePackage{fix-cm}
\documentclass{svjour3}                     
\smartqed  
\usepackage{graphicx}
\usepackage{mathptmx}      
\usepackage{amssymb}
\usepackage{amsmath}
\usepackage{color}
\usepackage{xspace}
\usepackage{natbib}
\usepackage{mwe,tikz}
\usepackage[percent]{overpic}

\newcommand{\msun}{\ensuremath{M_{\odot}}}
\newcommand{\lsun}{\ensuremath{L_{\odot}}}
\newcommand{\rsun}{\ensuremath{R_{\odot}}}

\newcommand{\Teff}{\ensuremath{T_{\rm eff}}}
\newcommand{\vinf}{\ensuremath{\upsilon_{\infty}}}
\newcommand{\mdot}{\ensuremath{\dot{M}}}

\newcommand{\msunyr}{\ensuremath{M_{\odot} {\rm yr}^{-1}}}

\newcommand{\beq}{\begin{equation}}
\newcommand{\eeq}{\end{equation}}
\newcommand{\beqa}{\begin{eqnarray}}
\newcommand{\eeqa}{\end{eqnarray}}

\newcommand{\nbeq}{\begin{equation*}}
\newcommand{\neeq}{\end{equation*}}

\newcommand{\kms}{\ensuremath{{\rm km}\,{\rm s}^{-1}}}

\newcommand{\rarrow}{\rightarrow}
\newcommand{\dd}{{\rm d}}

\newcommand{\HeII}{He\,{\sc ii}}

\newcommand\NIII{N\,{\sc iii}}
\newcommand\NIV{N\,{\sc iv}}
\newcommand\NV{N\,{\sc v}}

\newcommand{\FeI}{Fe\,{\sc ii}}

\newcommand{\FeIII}{Fe\,{\sc iii}}
\newcommand{\FeIV}{Fe\,{\sc iv}}

\newcommand{\FeVIII}{Fe\,{\sc viii}}
\newcommand{\FeX}{Fe\,{\sc x}}

\newcommand{\Hd} {H$_{\rm \delta}$}
\newcommand{\Hg} {H$_{\rm \gamma}$}

\newcommand{\Ha} {H$_{\alpha}$\xspace}

\newcommand{\MV}{\ensuremath{M_V}}

\newcommand{\Rstar}{\ensuremath{R_{\ast}}}

\newcommand{\Dmom}{\ensuremath{D_{\rm mom}}}

\newcommand{\logg}{\ensuremath{\log g}}
\newcommand{\YHe}{\ensuremath{Y_{\rm He}}}

\newcommand{\vesc}{\ensuremath{\upsilon_{\rm esc}}}
\newcommand{\vth}{\ensuremath{\upsilon_{\rm th}}}

\newcommand{\vrot}{\ensuremath{\upsilon_{\rm rot}}}

\newcommand{\vsini}{\ensuremath{\upsilon{\thinspace}\sin{\thinspace}i}}

\newcommand{\dvdr}{\ensuremath{\dd \upsilon/\dd r}}

\newcommand{\taur}{\ensuremath{\tau_{\rm Ross}}}

\newcommand{\taus}{\ensuremath{\tau_{\rm Sob}}}

\newcommand{\nivem}{\ensuremath{\lambda4058}}
\newcommand{\nivab}{\ensuremath{\lambda6380}}
\newcommand{\niiir}{\ensuremath{\lambda4640}}

\newcommand{\Lx}{\mbox{$L_{\rm X}$}}
\newcommand{\myr}{\mbox{$M_\odot\,{\rm yr}^{-1}$}}
\newcommand{\lsim}{\raisebox{-.4ex}{$\stackrel{<}{\scriptstyle \sim}$}}
\newcommand{\msim}{\raisebox{-.4ex}{$\stackrel{>}{\scriptstyle \sim}$}}

\newcommand{\nh}{\ensuremath{N_\mathrm{H}}\xspace}
\newcommand{\feka}{\ensuremath{\mathrm{Fe}~\mathrm{K}\alpha}\xspace}
\newcommand{\fekb}{\ensuremath{\mathrm{Fe}~\mathrm{K}\beta}\xspace}
\newcommand{\Mcl}{\ensuremath{M_\mathrm{cl}}\xspace}
\newcommand{\Macc}{\ensuremath{M_\mathrm{acc}}\xspace}
\newcommand{\Rcl}{\ensuremath{R_\mathrm{cl}}\xspace}
\newcommand{\Racc}{\ensuremath{R_\mathrm{acc}}\xspace}
\newcommand{\tfl}{\ensuremath{t_\mathrm{fl}}\xspace}
\newcommand{\vcl}{\ensuremath{\upsilon_\mathrm{fl}}\xspace}

\journalname{Space Science Reviews}

\begin{document}

\title{Towards a unified view of inhomogeneous stellar winds in isolated supergiant stars and supergiant high mass X-ray binaries}
\titlerunning{Unified view of of inhomogeneous stellar winds in supergiant stars and HXMB}

\author{Silvia Mart\'inez-N\'u\~nez \and
        Peter Kretschmar \and
        Enrico Bozzo \and
        Lidia M.\ Oskinova \and
        Joachim Puls \and
        Lara Sidoli \and
        Jon Olof Sundqvist \and
        Pere Blay \and
        Maurizio Falanga \and
        Felix F\"urst \and
        Angel G\'{\i}menez-Garc\'{\i}a \and
        Ingo Kreykenbohm \and
        Matthias K\"{u}hnel \and
        Andreas Sander \and
        Jos\'e Miguel Torrej\'on \and
        J\"orn Wilms
}

\authorrunning{S. Mart\'inez-N\'u\~nez et al.}

\institute{S.~Mart\'inez-N\'u\~nez \at
Instituto de F\'isica de Cantabria (CSIC-Universidad de Cantabria), E-39005, Santander, Spain
\email{silvia.martinez.nunez@gmail.com}   
\and
P.~Kretschmar \at
European Space Astronomy Centre (ESA/ESAC), Science Operations Department
P.O. Box 78, E-28691, Villanueva de la Ca\~{n}ada, Madrid, Spain
\email{Peter.Kretschmar@esa.int}
\and
E.~Bozzo \at
ISDC, University of Geneva, Chemin d’Ecogia 16, Versoix, 1290, Switzerland
\email{Enrico.Bozzo@unige.ch}
\and
L.M.~Oskinova \at
Institut f\"ur Physik und Astronomie, Universit\"at Potsdam,
Karl-Liebknecht-Str. 24/25, D-14476 Potsdam, Germany
\and
J.~Puls \at
Universit\"atssternwarte der Ludwig-Maximilians-Universit\"at M\"unchen,
Scheinerstrasse 1, 81679, M\"unchen, Germany
\and
L.~Sidoli \at
INAF, Istituto di Astrofisica Spaziale e Fisica Cosmica - Milano, via E. Bassini 15, I-20133 Milano, Italy
\and
J.O.~Sundqvist\at
Centro de Astrobiolog{\'i}a, CSIC-INTA,\ Ctra. Torrej{\'o}n a Ajalvir km.4, 28850 Madrid, Spain
\&
Instituut voor Sterrenkunde, KU Leuven, Celestijnenlaan 200D, 3001 Leuven, Belgium
 \and
P.~Blay\at
Nordic Optical Telescope - IAC, P.O.Box 474, E-38700, Santa Cruz de La Palma
Santa Cruz de Tenerife, Spain
\and
M.~Falanga \at
International Space Science Institute (ISSI), Hallerstrasse 6, CH-3012 Bern, Switzerland
\&
International Space Science Institute in Beijing, No. 1 Nan
Er Tiao, Zhong Guan Cun, Beijing 100190, China
\and
F.~F\"urst \at
Cahill Center for Astronomy and Astrophysics, California Institute of
Technology, Pasadena, CA 91125, USA
\and
A~G\'{\i}menez-Garc\'{\i}a \at
Instituto Universitario de F\'isica Aplicada a las Ciencias y las Tecnolog\'ias, University of Alicante, 
P.O. Box 99, E03080 Alicante, Spain
\and
I.~Kreykenbohm \at
Dr. Karl Remeis-Observatory \& ECAP, Universit{\"a}t Erlangen-N\"urnberg, Sternwartstr. 7, D-96049 Bamberg, Germany
\and
M.~K\"{u}hnel \at
Dr. Karl Remeis-Observatory \& ECAP, Universit{\"a}t Erlangen-N\"urnberg, Sternwartstr. 7, D-96049 Bamberg, Germany
\and
A.~Sander \at
Institut f\"ur Physik und Astronomie, Universit\"at Potsdam,
Karl-Liebknecht-Str. 24/25, D-14476 Potsdam, Germany
\and
J.~M.~Torrej\'on \at
Instituto Universitario de F\'isica Aplicada a las Ciencias y las Tecnolog\'ias, University of Alicante, 
P.O. Box 99, E03080 Alicante, Spain
\and
J. Wilms \at
Dr. Karl Remeis-Observatory \& ECAP, Universit{\"a}t Erlangen-N\"urnberg, Sternwartstr. 7, D-96049 Bamberg, Germany
}

\date{Published at Journal of Space Science Reviews, Springer online 07 March 2017}

\maketitle

\begin{abstract}

Massive stars, at least $\sim$ 10 times more massive than the Sun,
have two key properties that make them the main drivers of evolution of star
clusters, galaxies, and the Universe as a whole. On the one hand, the outer layers of massive stars
are so hot that they produce most of the ionizing ultraviolet radiation of galaxies; in fact, the 
first massive stars helped to re-ionize the Universe after its Dark Ages. Another important property of
massive stars are the strong stellar winds and outflows they produce. This mass loss,
and finally the explosion of a massive star as a supernova or a gamma-ray burst, provide a significant input 
of mechanical and radiative energy into the interstellar space. These two properties together make massive stars 
one of the most important cosmic engines: they trigger the star formation and enrich 
the interstellar medium with heavy elements, that ultimately leads to formation of Earth-like rocky 
planets and the development of complex life. The study of massive star winds is thus a 
truly multidisciplinary field and has a wide impact on different areas of astronomy.  

In recent years observational and theoretical evidences have been growing that these winds are not smooth and 
homogeneous as previously assumed, but rather populated by dense ``clumps''. The presence of 
these structures dramatically affects the mass loss rates derived from the study of stellar winds. 
Clump properties in isolated stars are nowadays inferred mostly through indirect methods 
(i.e., spectroscopic observations of line profiles in various
wavelength regimes, and their analysis based on tailored,
inhomogeneous wind models). The limited characterization 
of the clump physical properties (mass, size) obtained so far have led to large uncertainties 
in the mass loss rates from massive stars. Such uncertainties limit our 
understanding of the role of massive star winds in galactic 
and cosmic evolution. 

Supergiant high mass X-ray binaries (SgXBs) are among the brightest X-ray 
sources in the sky. A large number of them consist of a neutron star accreting  
from the wind of a massive companion and producing a powerful X-ray source.
The characteristics of the stellar wind together with the complex interactions
between the compact object and the donor star determine the observed
X-ray output from all these systems. Consequently, the use of SgXBs 
for studies of massive stars is only possible when the physics of the stellar winds, the 
compact objects, and accretion mechanisms are combined 
together and confronted with observations.

This detailed review summarises the current knowledge on the theory and observations of winds 
from massive stars, as well as on observations and accretion processes in wind-fed high mass X-ray binaries. 
The aim is to combine in the near future all available theoretical diagnostics and observational 
measurements to achieve a unified picture of massive star winds in isolated objects and in binary systems. 


\end{abstract}

\section{Introduction}
\label{sec:intro}

Massive stars ($M_{\rm initial} \msim 10\,M_\odot$) play an important role 
in the evolution of star clusters and galaxies. Massive stars generate 
ionizing ultraviolet radiation, and heat the dust.  The winds of massive 
stars,  and  their final explosions as supernovae or gamma-ray bursts
provide a significant input of energy and chemically enriched matter into the interstellar 
medium \citep{kud2002}. Massive stars are among the most important
drivers of cosmic evolution, they regulate star formation and, 
together with low-mass stars, enrich the interstellar medium 
with heavy elements. Among the bright X-ray sources in the sky a significant number 
consists of a compact object accreting from the wind of such massive stars. These winds 
are fast (with typical terminal velocities up to 2500\,km\,s$^{-1}$), dense (with mass-loss 
rates up to $\dot{M}~\msim~10^{-5}-10^{-7}$\,\myr), and driven by line scattering of the star's intense 
continuum radiation field. Examples of a system comprising a massive star and a compact object 
are Cyg X-1/HDE 226868, the first detected 
stellar-mass black hole, and Vela~X-1, the prototype of wind accreting neutron star X-ray binaries.
Both in isolated massive stars and in binary systems with accreting compact objects, the basic picture of 
the wind formation and wind accretion process has been established for decades. However, 
new findings concerning inhomogeneities in the massive star winds and the unexpectedly pronounced 
X-ray variability in some wind-fed binaries questioned our previous understanding of these systems.  

The first quantitative description of line-driven stellar winds 
was provided in the seminal paper by \citet{CAK}, which assumed a stationary, 
homogeneous, and spherically symmetric outflow. Later works
\citep[e.g.,][]{Owocki88,Feldmeier97a,Feldmeier97b,Dessart05}
showed that the line-driven hot star winds are in fact unstable 
\footnote{Already \citet{LS70} pointed out that radiative 
line-driving is subject to a strong instability.} to velocity
perturbations (the so-called "line-driven instability", hereafter LDI), 
leading to high-speed rarefactions that steepen into strong shocks,
whereby most material is compressed into spatially narrow  'clumps' (or
shells in 1-D simulations) separated by large regions of much lower
densities. The presence of clumps in the winds of massive stars is supported 
by numerous observational evidences in many different 
wavebands \citep[see][for comprehensive overviews]{ham2008,sund2011}.
In numerical simulations, the LDI is observed to generate strong wind shocks, which 
provide a possible explanation \citep{Feldmeier97a,Feldmeier97b} for the soft X-ray emission 
observed from "normal" (putatively single, non-magnetic) OB-stars, as well as for their  
lack of significant time-variability \citep[see][for a recent review]{Naze2013}. 
Clumps affect several stellar wind diagnostics in a non-trivial way, and discussions are 
on-going to infer the physical properties of these structures from the
results of the most recent observational campaigns. 

Additional independent observational evidence of clumped stellar winds comes  
from supergiant high mass X-ray binaries (SgXBs), i.e. those systems in which a compact object 
(a black hole or a neutron star) orbits a supergiant O-B star. \citet{sako2003} was the first 
to review spectroscopic results obtained by X-ray observatories for several wind-fed SgXBs. They 
concluded that the observed spectra and time variability of these objects could be best explained 
by assuming that accretion onto the compact object is taking place from a highly structured 
stellar wind where cool dense clumps are embedded in a rarefied
photoionized gas. Similar studies were later carried out on a number of bright SgXBs, 
including 4U\,1700$-$37 \citep{vdm2005},
Vela\,X-1 \citep{krey2008,Furst2010,martinez-nunez2014a}, Cyg~X-1 \citep{misko2011a}, and 
GX~301$-$2 \citep{Fuerst2011}.  
Although the presence of structured clumped winds in SgXBs seems thus well established, 
there is a still considerable uncertainty in the physical
properties of those clumps and the mechanisms by which the structured wind is able to feed the
compact object. Particularly puzzling is the pronounced X-ray variability (a factor of $\sim$100-1000 higher 
than in classical SgXBs) of the supergiant fast X-ray transients (SFXTs) sources. This variability is 
unlikely to be only due to the presence of massive structures 
in the wind of the supergiant stars and requires {\it ad-hoc} assumptions on the on-going accretion processes. 

The layout of the review is as follows: Section~\ref{basicphysics} first introduces
the basic physics of line-driven winds in detail, from the pioneering 'CAK model'
to modern simulations including small and large scale structures in the wind.
The section continues with the theory of accretion of these winds onto compact
objects and especially neutron stars, treating also different accretion regimes
and the inhibition of accretion. Section~\ref{stellarwindpars} discusses the
determination of stellar and wind parameters by quantitative
spectroscopy in the optical and UV regime, including the effects of
wind clumping on the mass loss diagnostics. Measurements and diagnostics
in the X-ray regime are discussed in Section~\ref{xraypars}, together
with caveats when applying these. This section also summarises
the current knowledge on both the "classical" SgXBs and the SFXTs. 
Finally, Section~\ref{conclusions} summarises the
main currently open questions on stellar winds and wind properties 
of massive stars.

\section{Basic physics}
\label{basicphysics}

We summarise in the next two sections the basic physics of the line-driven winds in massive stars and accretion processes in wind-fed binaries. 

\subsection{Basics of line-driven winds}
\label{basicsldw}

A decisive property of hot, massive stars is their stellar wind, with 
typical mass-loss rates (for solar metallicity), $\mdot \approx
10^{-7} {\ldots} 10^{-5}$ \msun/yr, and terminal velocities, \vinf,
ranging from 200 {\ldots} 3,500 \kms. The origin of these winds is
attributed to {\it radiative line-driving}, i.e., stellar continuum
photons are scattered in a multitude of spectral lines and transfer
their momentum to the wind. Since this process requires a large number
of photons (i.e., a high luminosity), such winds occur in the hottest
stars, like O-type stars of all luminosity classes, but also in cooler
BA-supergiants, because of their larger radii. Efficient line-driving
further requires a large number of spectral lines close to the
flux-maximum and a high interaction probability (i.e., a significant
optical depth). Since most spectral lines originate from various
metals, a strong dependence of \mdot\ on metallicity is thus to be
expected, and such line-driven winds should only play a minor role (if
at all) in the early Universe.\footnote{Contrasted to the almost
metallicity-independent, porosity-moderated continuum-driven winds
hypothesized by \citet{OGS04}.} The theory of line-driven winds has
been pioneered by \citet{LS70} and particularly by \citet[][henceforth
'CAK']{CAK}, with essential improvements regarding a
quantitative description and application provided by \citet{FA86} and
\citet{PPK}. Line-driven winds have been reviewed by \citet{KP00} and
more recently by \citet{Puls08}. In the following, we will briefly
consider some relevant aspects, mostly in terms of the 'standard
model' and the theory developed by CAK.

\subsubsection{The CAK model and beyond}\label{basicsldw:CAK}
From studying the temporal variability of typical wind-features (UV
P-Cygni profiles, \Ha, \HeII4686, see Sect.~\ref{stellarwindpars})
and from analysing these lines, it turned out that the global
quantities describing the outflow (\mdot, \vinf) typically show only
little variations. This and other evidence motivates the definition of
a stationary, spherically symmetric, and homogeneous {\it standard
model}. Effects from rotation and magnetic fields are briefly outlined
at the end of this section, and deviations from a homogeneous
structure are discussed in Sects.~\ref{sec:smallscale} and
\ref{sec:largescale}.

In such a standard model, the mass-loss rate $\mdot = 4\pi r^2 \rho(r)
\upsilon(r)$ remains constant over the wind, and the equation of motion is
governed by pressure terms and external forces, in our case the
inward gravitational pull and an outward directed radiative acceleration.
For simplicity, the Thomson-acceleration due to electron scattering
will be included as a correction to gravity\footnote{Both
accelerations depend on $r^{-2}$, at least in a homogeneous medium.} in
terms of the conventional Eddington-Gamma, $\Gamma_{\rm Edd} = g_{\rm
Thomson}/g_{\rm grav} \propto L/M$, and the remaining {\it continuum}
acceleration can be neglected in most hot star winds.

Thus, the `only' difficulty regards calculating the radiative line
force. Basically, this force can be derived from the momentum transfer
occurring during the absorption and (re-) emission of (mostly) stellar
photons, where on average the emission process cancels out because of
its fore-aft-symmetry. Since most photons are absorbed in {\it metal}
lines, the momentum needs to be redistributed to the bulk plasma (H
and He), by means of Coulomb collisions \citep{SP92}\footnote{A
significant drift between metallic ions and the bulk plasma or even a
complete decoupling of certain ions might become possible in winds of
low metallicity and/or low density, e.g., \citet{Babel95, Krticka03,
Krticka06, OP02}.}. 

In certain frequency intervals, the line-density can be so high that
photons on their way out of the wind are not only scattered in one
line before they escape, but also in a second one, a third one, etc.,
until they ultimately find their way out. This process is called
multi-line-scattering, and leads to a certain complexity in {\it
analytical} calculations of the line force.\footnote{For details, see,
e.g., \citet{FC83, Puls87, LA93, Gayley95}.} For simplicity, we assume
instead that each line can be treated separately, i.e., that stellar
photons can interact with only one line and then leave the wind,
irrespective of line-density. As shown by \citet{Puls87}, this is not
too bad an approximation for OBA-stars.\footnote{In the dense winds of
Wolf-Rayet stars (see below), multi-line scattering needs to be
accounted for.} In this case, the total line force can be calculated
by summing up the individual contributions from all participating
lines, expressed in terms of illuminating intensity and line opacity.
In rapidly expanding atmospheres, this expression can be simplified by
means of the so-called Sobolev approximation \citep{Sobo60}. If at
first we only consider radially streaming photons (relaxed later on),
the radiative line-acceleration for line $i$ at rest
transition-frequency $\nu_{0,i}$ results in
\beq
\label{gradi}
g_{{\rm rad},i}=\frac{L_{\nu_{i}} \nu_{0,i}}{c^2} 
\frac{\dvdr}{4\pi r^2 \rho}\Bigl[1-\exp{\bigl(-\frac{k_{{\rm L},i} s_{\rm e}
\rho \vth}{\dvdr}\bigr)}\Bigr] =
\frac{L_{\nu_{i}} \nu_{0,i}}{c^2} 
\frac{\dvdr}{4\pi r^2 \rho}\Bigl[1-{\rm e}^{-\taus}\Bigr]
\eeq
with $L_{\nu_{i}}$ the spectral luminosity, $s_{\rm e}$ the
mass-absorption coefficient for Thomson scattering, and $\vth$ the 
thermal velocity for a representative ion. $k_{{\rm L},i}$ is the so-called
line-strength, corresponding to the ratio between frequency-integrated
line-opacity $\bar \chi_i$ and Thomson-scattering opacity $s_{\rm e}$ over a typical
line-width $\Delta \nu_{{\rm Dop},i}$,
\beq
k_{{\rm L},i}=\frac{\bar \chi_i}{s_{\rm e} \rho \Delta \nu_{{\rm Dop},i}}.
\eeq
For the dominating resonance lines from major ions, 
$k_{{\rm L},i}$ is roughly constant over the wind. A
line-strength of unity thus refers to a weak line of 
continuum electron-scattering strength, whereas strong lines 
can have $k_{{\rm L},i} \approx 10^6$ or even more. 

The most intriguing quantity appearing in Eq.~\ref{gradi} is the
radial velocity gradient, which results from the Doppler-effect
experienced by the absorbing matter in an expanding medium. As obvious
from Eq.~\ref{gradi}, the radiative acceleration from optically thin
lines (with line optical depth in Sobolev approximation $\taus < 1$)
is proportional to $k_{{\rm L},i}$ and does not depend on velocity and
density, whilst for optically thick lines ($\taus > 1$) $g_{{\rm
rad},i}$ becomes independent of line-strength (saturation), but now
depends on $(\dvdr)/\rho$.

The basic trick of CAK was to write the total line acceleration,
i.e., the sum over all contributing lines $i$, as an integral over a
line-strength distribution 
\beq 
\label{gradtot1}
g_{\rm rad}^{\rm tot} = \sum_i g_{{\rm rad},i} \rarrow \int \!\! \int g_{\rm rad}(k_{\rm L}, \nu)
\, \dd N(k_{\rm L}, \nu)
\eeq
where this distribution depends on line-strength and frequency. From
some preliminary empirical arguments which have been confirmed
meanwhile (e.g., \citealt{Puls00} and Fig.~\ref{dNdk}), CAK assumed a power-law
distribution w.r.t. $k_{\rm L}$ and a frequential distribution $\propto
1/\nu$,
\beq
\label{lsdf}
\dd N(k_{\rm L}, \nu) = -N_0 k_{\rm L}^{\alpha-2} \dd k_{\rm L} \dd\nu/\nu. 
\eeq
In this case, integrals in Eq.~\ref{gradtot1} can be solved 
analytically, and one obtains
\beq
\label{gradtot2}
g_{\rm rad}^{\rm tot}=\frac{L}{4\pi r^2 c^2} \frac{s_{\rm e}\vth N_0
\Gamma(\alpha)}{1-\alpha} \Bigl(\frac{\dvdr}{s_{\rm e}\vth
\rho}\Bigr)^\alpha:=\frac{s_{\rm e}L}{4\pi r^2 c}k_{\rm CAK}k_1^\alpha=
g_{\rm grav}(r)\Gamma_{\rm Edd} k_{\rm CAK}k_1^\alpha,
\eeq
where $\Gamma(\alpha)$ is the Gamma-function,
\beq
k_{\rm CAK} = \frac{\vth}{c}\frac{N_0\Gamma(\alpha)}{1-\alpha}, \qquad
k_1=\frac{\dvdr}{s_{\rm e} \rho \vth}.
\eeq
$k_{\rm CAK}$ (on the order of 0.1 for O-stars and early B-stars) is one of the
so-called force-multiplier parameters, and $k_1$ the line-strength
where the exponent in Eq.~\ref{gradi}, the optical depth in Sobolev
approximation, becomes unity.\footnote{$k_1$ corresponds to $t^{-1}$ in
the notation of CAK.} $\alpha$ is the 2nd force-multiplier parameter
($\approx 0.6 {\ldots} 0.7$ for O-star winds, see Fig.~\ref{dNdk},
left panel), either corresponding to
the slope of the line-strength distribution function, Eq.~\ref{lsdf},
or alternatively interpreted as the ratio of line-acceleration from
optically thick lines to the total line acceleration. 

After accounting for non-radial photons and ionization effects
(not discussed here), we can insert the total line acceleration into
the time-independent equation of motion. The 
resulting non-linear differential equation
can be solved either numerically (e.g., \citealt{PPK, FA86}) or,
applying certain simplifications, also analytically (e.g.,
\citealt{Kud89, OGS04}), and the mass-loss rate results as the
eigenvalue of the problem. Overall, we obtain the following scaling
relations:
\beqa
\label{scalerel}
\mdot &\propto& \Bigl(k_{\rm CAK}\,\frac{L}{\lsun}\Bigr)^{1/\alpha'}
                \Bigl(\frac{M}{\msun}(1-\Gamma_{\rm
		Edd})\Bigr)^{1-1/\alpha'} \propto \, 
		k_{\rm CAK}^{1/\alpha'}\,\frac{L}{\lsun}\,\Bigl(\frac{\Gamma_{\rm Edd}}
                {1-\Gamma_{\rm Edd}}\Bigr)^{1/\alpha'-1} \nonumber \\
\upsilon(r) & = & \vinf\Bigl(1-\frac{\Rstar}{r}\Bigr)^\beta	 \nonumber \\	    
\vinf &\approx& \frac{2.25 \alpha}{1-\alpha} \Bigr(\frac{2 G M 
                (1-\Gamma_{\rm Edd})}{\Rstar}\Bigl)^{\frac{1}{2}} =
                \frac{2.25 \alpha}{1-\alpha}\vesc
\eeqa
Here, $\alpha' = \alpha - \delta$, where $\delta$ ($\approx 0.1$ for
O-stars) is \citeauthor{Abbott82}'s (\citeyear{Abbott82}) ionization
parameter, and \vesc\, is the photospheric escape velocity, corrected
for the radiative acceleration by Thomson scattering. The exponent
$\beta$ is on the order of 0.8 for O-dwarfs, and on the order of 1.3
{\ldots} 2 for BA-supergiants. 
To overcome some inherent problems with this initial CAK
formulation (e.g., the artificial dependence on a fiducial thermal
speed, see above), \citet{Gayley95} used a somewhat different
definition for the line-strength, as well as a line-statistics with an
exponential cut-off, and reformulated the standard CAK approach. While
the two formulations give identical results, the new one provides a
somewhat modified expression for the mass-loss rate, which is
frequently used nowadays:
\begin{figure}[t]
\begin{minipage}{0.52\textwidth}
\resizebox{\hsize}{!}
  {\includegraphics{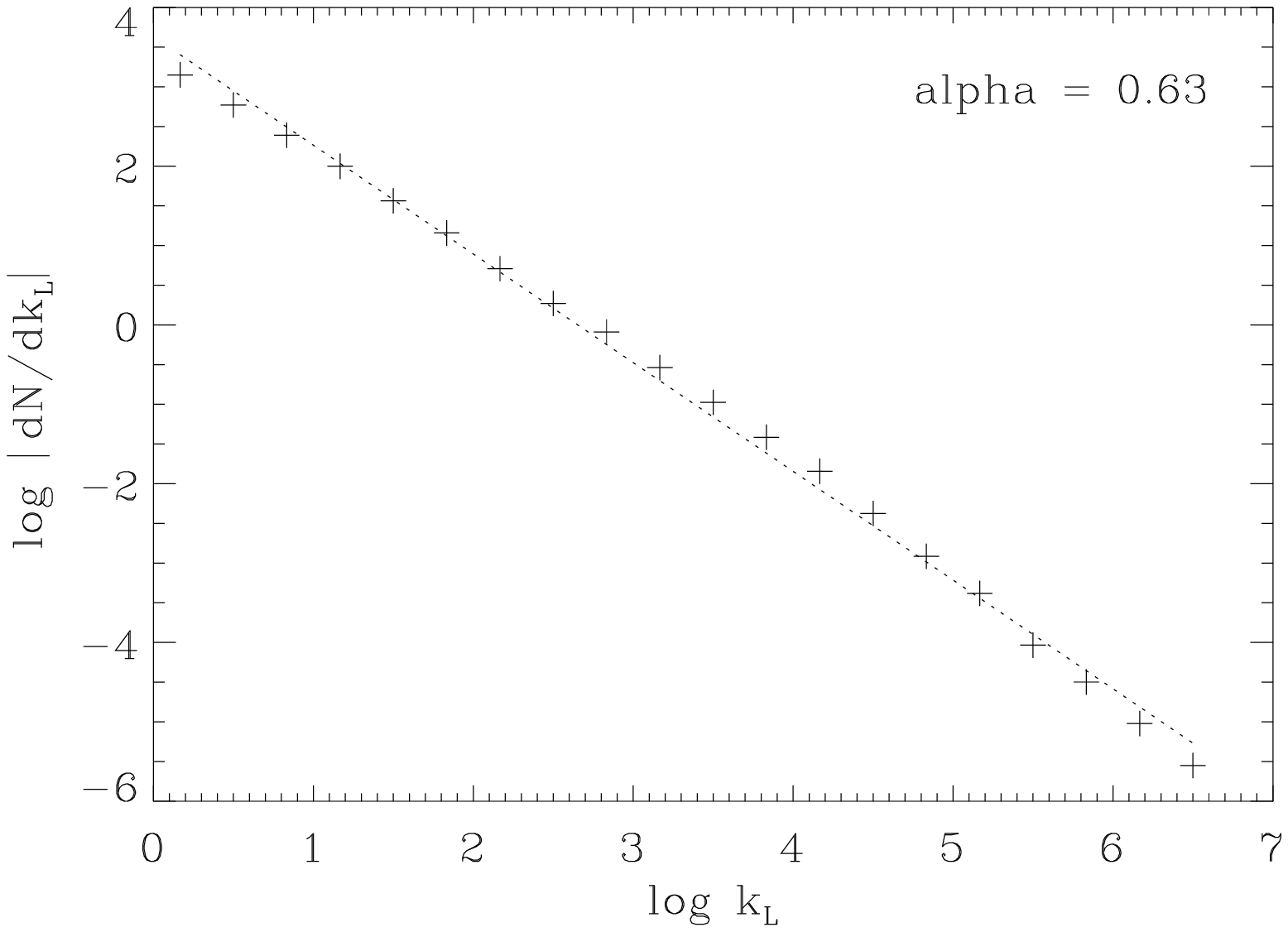}} 
\end{minipage}
\hspace{-0.04\textwidth}
\begin{minipage}{0.52\textwidth}
\resizebox{\hsize}{!}
  {\includegraphics[angle=90]{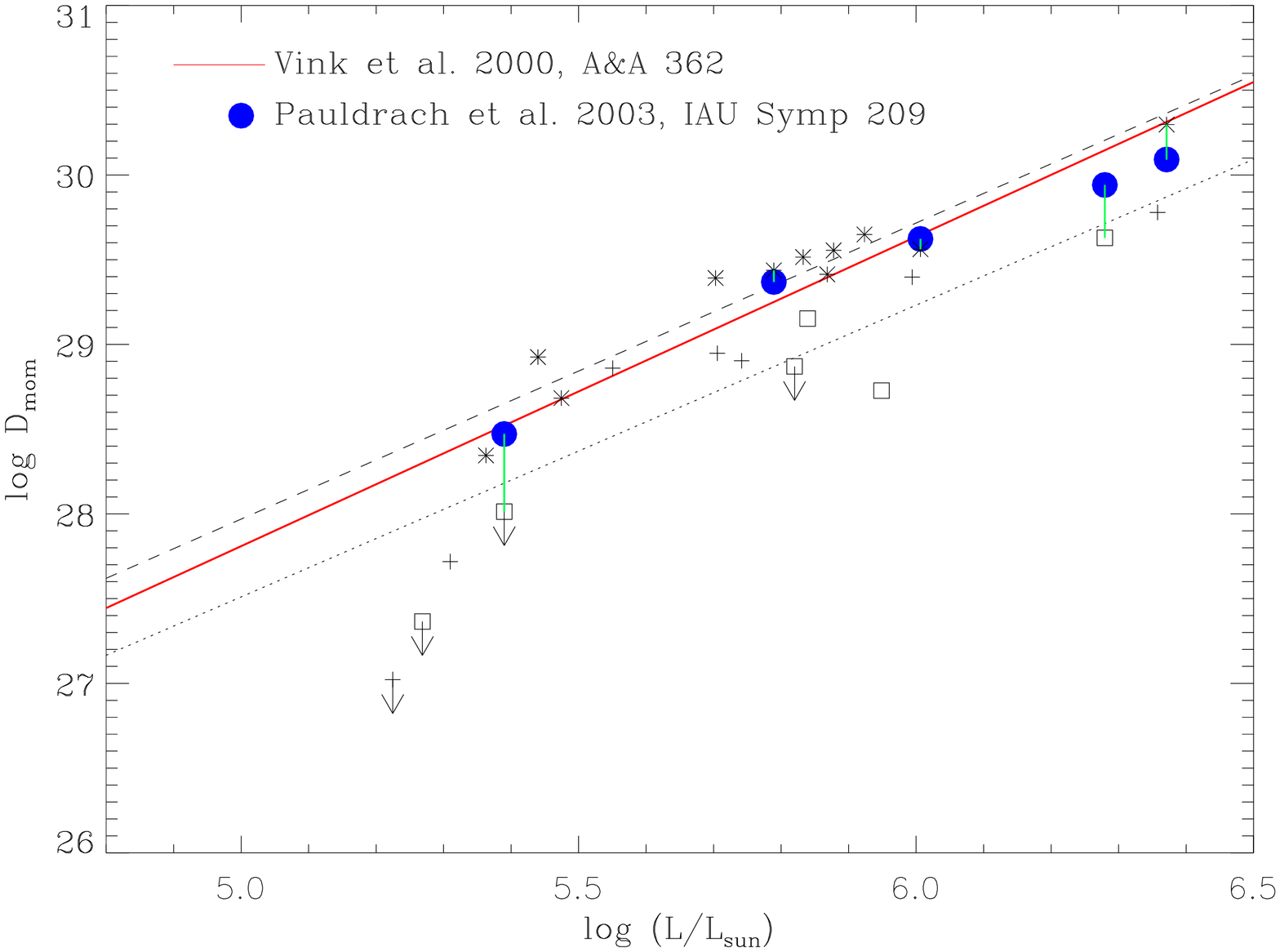}} 
\end{minipage}
\caption{{\bf Left:} Frequency integrated line-strength distribution
function for an O-type wind (\Teff = 40 kK, solar abundance), and
corresponding power-law fit (slope = $\alpha-2$). {\bf Right:}
Observed WLR for Galactic O-stars (asterisks, plus-signs and
rectangles for luminosity classes I, III and V objects, respectively).
Data from \citet{Puls96}. Dashed: linear regression to l.c. I objects;
dotted: linear regression to luminosity class. III and V objects. Analytical
considerations (see text) and theoretical models (from
\citealt{Vink00}, solid red) do not predict a dependence on luminosity
class. The `observed' difference is presumably due to the neglect of
wind-inhomogeneities (clumping) in the mass-loss analysis (see also
\cite{Repolust04}). Note that
the observed wind-momenta deviate towards low values below $\log
L/\lsun < 5.2$ (`weak wind problem', cf. \citealt{Marcolino09} and
references therein, and \citealt{Huen2012} for a potential
explanation). Blue dots indicate theoretical
models from \citet{Pauldrach03} calculated for five stars from the
observed sample, resulting in similar wind-momenta as predicted by
\citet{Vink00}.}
\label{dNdk}
\end{figure}
\beq
\mdot 
 \propto\Bigl(\frac{L}{\lsun}\Bigr)^{1/\alpha'}  
              \Bigl(\frac{1}{\bar Q}\frac{M}{\msun}(1-\Gamma_{\rm Edd})\Bigr)^{1-1/\alpha'} 
	      \propto \frac{L}{\lsun}\,\Bigl(\frac{\bar Q \Gamma_{\rm Edd}}
              {1-\Gamma_{\rm Edd}}\Bigr)^{1/\alpha'-1} \nonumber 
\eeq
%
For $\delta = 0$, the relation between 
$\bar Q$ ($\approx 2000$ for O-stars) and  $k_{\rm CAK}$ is given by 
$\bar Q^{1/\alpha-1} = c/\vth \bigl((1-\alpha)k_{\rm CAK}\bigr)^{1/\alpha}$.  

Finally, by using the scaling relations for \mdot\, and \vinf\,
(Eqs.~\ref{scalerel}), and approximating $\alpha' \approx 2/3$, one
obtains the so-called wind-momentum luminosity relation - WLR - \citep{Kud95},
\beq
\label{wlr}
\log \Dmom = \log \Bigl(\mdot \vinf
\bigl(\frac{R}{\rsun}\bigr)^{\frac{1}{2}}\Bigr) \approx \frac{1}{\alpha'}
\log \Bigl(\frac{L}{\lsun}\Bigr) + \mbox{offset({\it Z}, spectral type)},
\eeq
which relates the {\it modified} wind-momentum rate \Dmom\, with the
stellar luminosity {\it alone}. The dependence on $M$ and $\Gamma_{\rm Edd}$
(difficult to `measure') vanishes since the product of
$(M(1-\Gamma_{\rm Edd}))^{1-1/\alpha'}$ and $(\vesc R^{1/2})$
becomes negligible as long as $\alpha'$ is close to 2/3. The offset in
Eq.~\ref{wlr} depends on metallicity and spectral type, mostly because
the effective line number and thus $k_{\rm CAK}$ (or $\bar Q$) 
depend on these quantities, via different opacities and contributing
ions.

Originally, it had been suggested to use a carefully calibrated WLR as
an independent tool to measure extragalactic distances, from the
spectroscopic analysis of extragalactic A-supergiants and their winds,
and by solving for the stellar radius via Eq.~\ref{wlr}. Meanwhile, 
however, the WLR is mostly used to test the validity of the line-driven
wind theory itself (e.g., Fig.~\ref{dNdk}, right panel).

Various theoretical models have been computed during
recent decades, based on a more or less exact calculations of
the line-force (i.e., discarding the statistical approach and
accounting for non Local Thermodynamic Equilibrium (hereafter, 
non-LTE) effects). Most prominent are the models by
\citet{Vink00, Vink01}, relying on a Monte Carlo approach, the models
by \citet{Pauldrach87} and \citet{Pauldrach94,Pauldrach01}
(`WM-Basic'), calculating the line-force in a Sobolev-approach, and
the models by \citet{KK00, KK01, KK04}, which include a more-component
description (metal ions plus H/He). All these models agree in their
quantitative predictions (e.g., Fig.~\ref{dNdk}, right panel), in
particular regarding the metallicity dependence of the mass-loss rate,
$\mdot \propto Z^{0.6{\ldots}0.7}$.\footnote{The metallicity
dependence of \vinf\ is rather weak, $\vinf \propto Z^{0.06{\ldots}0.13}$
\citep{Leitherer92, Krticka06}.}

The most impressive observational confirmation of the theoretical
concept of line-driven winds and their metallicity dependence has been
provided by \citet{Mokiem07}, compiling observed stellar- and
wind-parameters from Galactic, LMC and SMC O-stars, and analysing the
corresponding WLRs. Accounting for wind-inhomogeneities (see
Sect.~\ref{sec:smallscale}) in an approximate way, they derive $\mdot
\propto Z^{0.72 \pm 0.15}$, in very good agreement with theoretical
predictions.


\paragraph{The bi-stability jump} One of the still unsolved problems
regarding line-driven winds is the reality of the so-called
bi-stability jump\footnote{introduced by \citet{PauldrachPuls90} to
explain the bi-stable behaviour of wind models for P~Cygni.}, which
should affect the mass-loss rates of B-supergiants (important in the
context of SgXBs). As it turns out \citep{Puls00, Vink01}, the
mass-loss rates of radiation driven winds are mostly determined by
iron (group) lines. Below \Teff\ $\approx$ 23kK, the ionization of Fe
(in the lower wind) switches abruptly from \FeIV\ to \FeIII, which has
much more lines close to flux maximum. Consequently, the mass-loss
rate is predicted to increase for such cooler stars by roughly a
factor of five or more, whilst \vinf\ should decrease by a factor of
two \citep{Vink01}. Thus, the wind-momentum rates for B-supergiants
(like in Vela~X-1) are predicted to be larger than of O-stars 
of similar luminosity.

Though a gradual decrease in $\vinf$ (more precisely, in the ratio
$\vinf/\vesc$) over the range \Teff\ = 23kK to 18kK has been confirmed
in many studies (e.g., \citealt{Groenewegen89, Crowther06, MP08}),
this is not true for the predicted increase in \mdot. Detailed
investigations of B-supergiants by \citet{Crowther06} and \citet{MP08}
do {\it not} show such a behaviour, but rather indicate that their
mass-loss rates are lower (or similar) to those from O-stars at the
same luminosity. Since the theoretical predictions are quite robust,
whereas the formation of the prime mass-loss indicator, \Ha, is quite
complex in the B-supergiant range \citep{petrov2014}, further
investigations are required to solve this long-standing issue. One
might note, however, that all present evolutionary codes for massive
stars incorporate the theoretical mass-loss predictions, and that the
predicted bi-stability jump has a large effect on the evolution and
rotation of B-supergiants and beyond \citep{Vink10, Markova14}. If the
jump in \mdot\ were not present, significant changes in such
evolutionary phases are to be expected.

\paragraph{Mass loss from Wolf-Rayet stars} From early on, the
mass-loss properties of Wolf-Rayet stars posed a major problem for
theoretical explanations, since they are considerably larger compared
to O-stars of similar luminosity. Though \citet{LA93} showed that
line-overlap effects, coupled with a significantly stratified
ionization balance, can help a lot to increase the mass-loss, it were
\citet{GH05, GH06, GH07} who showed that there are two ingredients
that might produce the observed large mass-loss rates in parallel with
high terminal velocities. First, a high Eddington-$\Gamma$ is
necessary to provide a low effective gravity and to enable a deep
lying sonic point at high temperatures. Then, a high mass-loss rate
leading to an {\it optically thick} wind can be initiated either by
the `hot' Fe-opacity bump (around 160 kK, for the case of WCs and
WNEs) or the cooler one (around 40 to 70 kK, for the case of
WNLs).\footnote{The importance of these opacity bumps was pointed out
already by \citet{NL02}.} Alternative wind models have been
constructed by \citet{Vink11}, who argue that for $\Gamma_{\rm Edd} >
0.7$ the winds (more precisely, the pseudo-continuum) become optically
thick already at the sonic point, which should enable a high \mdot.
Nevertheless, there are still a number of details to be worked out
before these winds are completely understood.

\paragraph{Impact of (fast) rotation} When stars rotate
rapidly, their photospheres become oblate, the effective temperature
decreases from pole towards equator (`gravity darkening'), and the
wind is predicted to become {\it prolate} in most cases (because of
the larger illuminating polar fluxes), with a fast and dense polar
outflow, and a slow and thinner equatorial one
\citep{CO95}.\footnote{All these effects become significant if the
rotational speed exceeds roughly 70\% of the critical one.} Whilst
stellar oblateness and gravity darkening have been confirmed (at least
the basic effects) by means of interferometry \citep{deSouza03,
Monier07}, the predictions on the wind-structure of rapidly rotating
stars have {\it not} been verified by observations so far
(\citealt{Puls10} and references therein): first, only few stars in
such phases are known (but they exist, e.g., the most extremely
rotating massive star detected by \citealt{Dufton11} rotates very
close to critical), and second, the tools to analyse the atmospheres
and winds (multi-D models!) of such stars are rare, if they exist at
all. 
Note that even for moderate rotation the wind is
predicted to become asymmetric (though to a lesser extent), and the
formation of important mass-loss diagnostics such as \Ha becomes
affected (e.g., \citealt{PetrenzPuls96}).

\paragraph{Impact of magnetic fields} Recent spectropolarimetric
surveys (mostly performed by the international Magnetism in Massive
Stars, MiMeS, collaboration, e.g., \citealt{Wade12}, and work done by
S. Hubrig and collaborators, e.g., \citealt{Hubrig13}) have revealed
that roughly 10\% of all massive stars have a large-scale, organized 
magnetic field in their outer stellar layers (the incidence of
internal fields might be higher), on the order of a couple of hundred
to several thousand Gauss. The origin of these fields is still
unknown, though most evidence points to quite stable fossil fields
formed sometimes during early phases of stellar formation
\citep{Alecian13}. The interaction of these fields with the stellar
wind has been theoretically investigated by ud-Doula, Owocki and
co-workers in a series of publications (summarised in
\citealt{udDoula13}), and two different scenarios have been
identified, depending on rotational speed and field strength. For
not too fast rotation (when the Alv\'en radius is smaller than the
Keplerian co-rotation radius), a magnetically confined wind
is predicted, in which the gravitational pull on the trapped wind
plasma creates large regions of {\it infalling} material, whereas for
fast rotation and strong confinement\footnote{For example, a large ratio
between magnetic and wind energy.} one obtains a rigidly rotating
magnetosphere (Alv\'en radius larger than Keplerian radius (
these radii are defined in Sect.~\ref{sec:regimes}),
in which the centrifugal force prevents the trapped material from
falling back to the stellar surface. Both scenarios are consistent
with observational findings \citep{Petit13}, and are nowadays called
dynamical and centrifugal magnetospheres. These two populations can be
differentiated by their distinct \Ha emission: slowly rotating O-type
stars with narrow, strong emission consistent with a dynamical
magnetosphere, and more rapidly rotating B-type stars with broader,
often double-peaked, emission associated with a centrifugal
magnetosphere. First attempts \citep{Sundqvist12} to simulate the \Ha
emission from the dynamical magnetospheres of prototypical O-stars
(denoted by the spectral type qualifiers `f?p') have been quite
successful, thought these initial investigations certainly need to be
repeated within a multi-D NLTE approach.

\subsubsection{Small scale structures}
\label{sec:smallscale}

Although the standard theory of line-driven winds outlined in the
previous section assumes a stable, time-independent and homogeneous
wind, it is since long known that the radiation line-force in fact is
subjected to a very strong, intrinsic instability \citep{Milne26, LS70}.
Below we review the theoretical background for this fundamental
instability, whereas the corresponding observational background
regarding small-scale wind structure is given in Sect.~\ref{sec:mdotclump}.

\paragraph{Linear perturbation theory} 
Following \citet{Owocki84}, let us assume a small velocity
perturbation of the conventional sinusoidal form $\delta \upsilon = \delta \upsilon_0
e^{i(kx-wt)}$, where the wave number $k$ is the inverse of the
perturbation wavelength and the circular frequency $w$ may be complex
(to account for potential exponential growth or damping of the initial
perturbation). For a spherically symmetric wind, in a frame co-moving
with the underlying mean flow, and neglecting gas pressure terms, this
circular frequency is given by
\begin{equation} 
  w = i \ \delta g_{\rm rad}/\delta \upsilon, 
\end{equation} 
where $\delta g_{\rm rad} = g_{\rm rad} - g_{\rm rad,0}$ is the
response of the unperturbed line force $g_{\rm rad,0}$ to the velocity
perturbation. For a line that is optically thick in the mean flow,
\citet{Owocki84} showed the ratio $\delta g_{\rm rad}/\delta \upsilon$ can be
expressed by a ``bridging law''
\begin{equation} 
  \frac{\delta g_{\rm rad}}{\delta \upsilon} \approx \Omega i k \frac{L_{\rm
      Sob}}{1+i k L_{\rm Sob}}, 
\end{equation} 
where $\Omega \approx \upsilon_0/L_{\rm Sob}$ is the growth rate of the
perturbation, and $L_{\rm Sob} = \upsilon_{\rm th}/(d\upsilon_0/dr)$ the radial
Sobolev length\footnote{the radial extent of that zone where photons
can be absorbed by a specific line.} of the unperturbed flow, moving
with $\upsilon_0$ in the stellar frame. Thus, for large-scale perturbations
with wavelengths \textit{longer} than this Sobolev length ($k L_{\rm
  Sob} << 1$),
\begin{equation} 
  w \approx i^2 \Omega k L_{\rm Sob} \approx - k \ \upsilon_0. 
\end{equation} 
Inserting this into the perturbation $\delta \upsilon = \delta \upsilon_0
e^{i(kr-wt)}$ gives rise to radiative-acoustic waves of zero growth
and with phase speed $w/k \sim -\upsilon_0$, propagating backwards in the
co-moving frame. The properties of such "Abbott waves''
\citep{Abbott80} thus imply that a line-force computed within the
Sobolev approximation (like for all models discussed in the previous
section) is marginally stable.

However, for short wavelength perturbations near or below this Sobolev
length ($k L_{\rm Sob} > 1$), we obtain instead
\begin{equation} 
  w \approx i \Omega \approx i \frac{\upsilon_0}{L_{\rm Sob}},
\end{equation} 
which is complex with $\Im(w) > 0$, and so when inserted into the
perturbation-expression above results in an exponential growth of the
initial velocity perturbation. Since the growth rate $\Omega \sim
\upsilon_0/L_{\rm Sob} \sim \upsilon_0/\upsilon_{\rm th} (d\upsilon_0/dr)$ is a factor $\sim
\upsilon_0/\upsilon_{\rm th}$ larger than the wind expansion rate $d\upsilon_0/dr$, this
implies small-scale perturbations can be amplified by an enormous
amount $\upsilon_0/\upsilon_{\rm th} \sim 100$ e-folds within this linear theory for
a pure absorption-line-driven flow!

This strong \textit{line-deshadowing instability} (LDI)\footnote{or
  alternatively simply line-driven instability, also LDI.} can be
somewhat damped by asymmetries in the scattered, diffuse component of
the line-force \citep{Lucy84}. \citet{Owocki85} (see also
\citealt{Owocki96}) showed that accounting for the diffuse force in
the perturbation theory outlined above still gives a very un-stable
outer wind, but where the LDI growth rate can be strongly damped close
to stellar surface, and even become zero at the photospheric
boundary. Recently, \citet{Sundqvist13} showed that including simple
stellar limb-darkening breaks this cancellation of the LDI by the
diffuse damping at the stellar surface, and so leads to a net
instability growth rate and to an unstable wind also in near
photospheric layers.

\paragraph{Non-linear numerical simulations} The operation of this
fundamental and remarkably strong LDI has been confirmed by
time-dependent numerical hydrodynamical wind modelling using a
non-Sobolev radiation line-force \citep{Owocki88, Feldmeier95, OP99,
  Dessart05, Sundqvist13}. Such simulations show that the non-linear
growth of the LDI leads to high-speed rarefactions that steepen into
strong reverse shocks (resulting, e.g., in high-energy emission
observable in soft X-ray band-passes around $\sim$\,1\,keV, see also
below), whereby the wind plasma becomes compressed into spatially
narrow `clumps' separated by large regions of rarefied gas. This
characteristic structure is the theoretical basis for our current
understanding and interpretation of \textit{wind clumping} (see
Sect.~\ref{stellarwindpars}).

\begin{figure}
\begin{minipage}{0.5\textwidth}
\resizebox{\hsize}{!}
  {\includegraphics[angle=90]{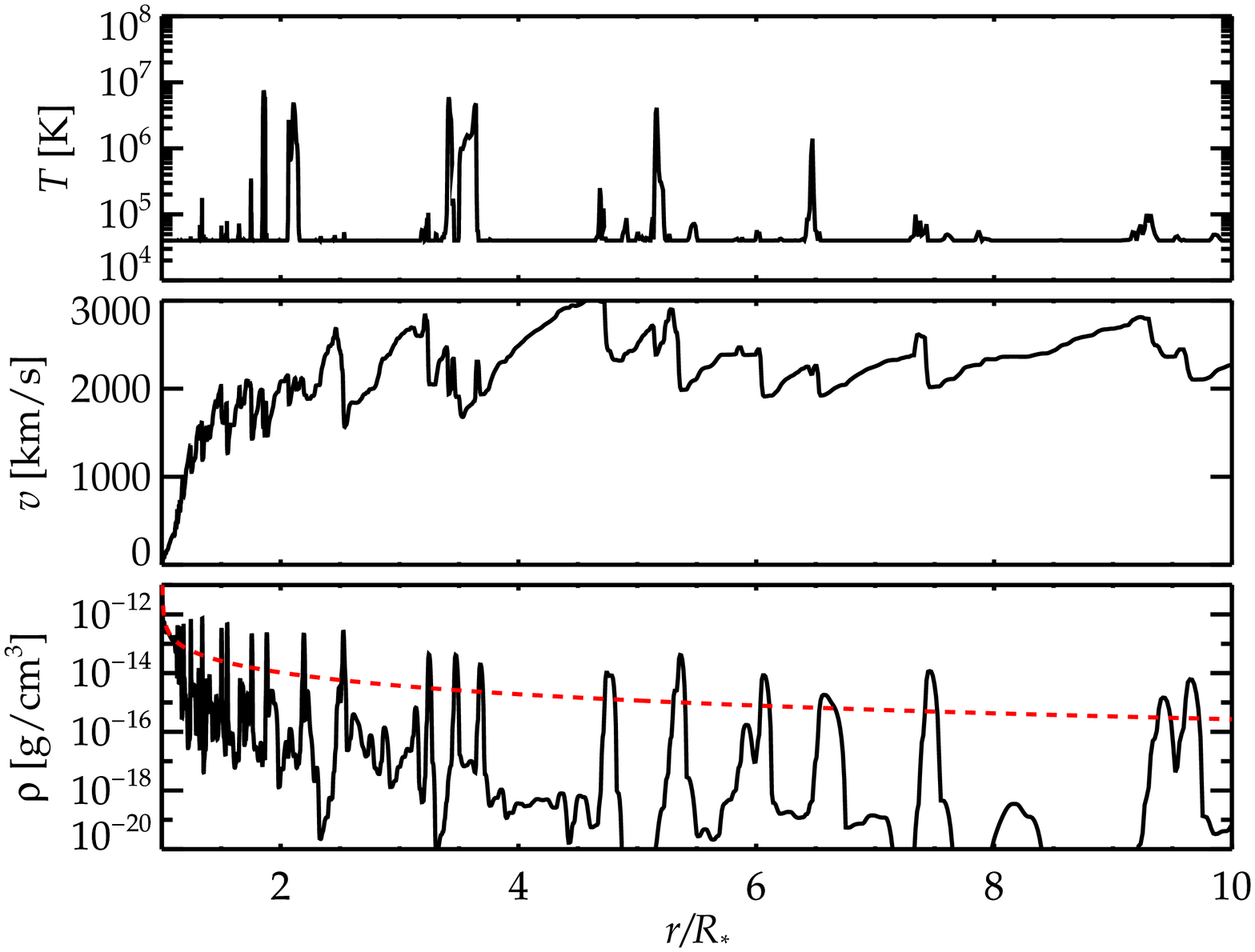}} 
\end{minipage}
\hspace{-0.02\textwidth}
\begin{minipage}{0.52\textwidth}
\resizebox{\hsize}{!}
  {\includegraphics[angle=90]{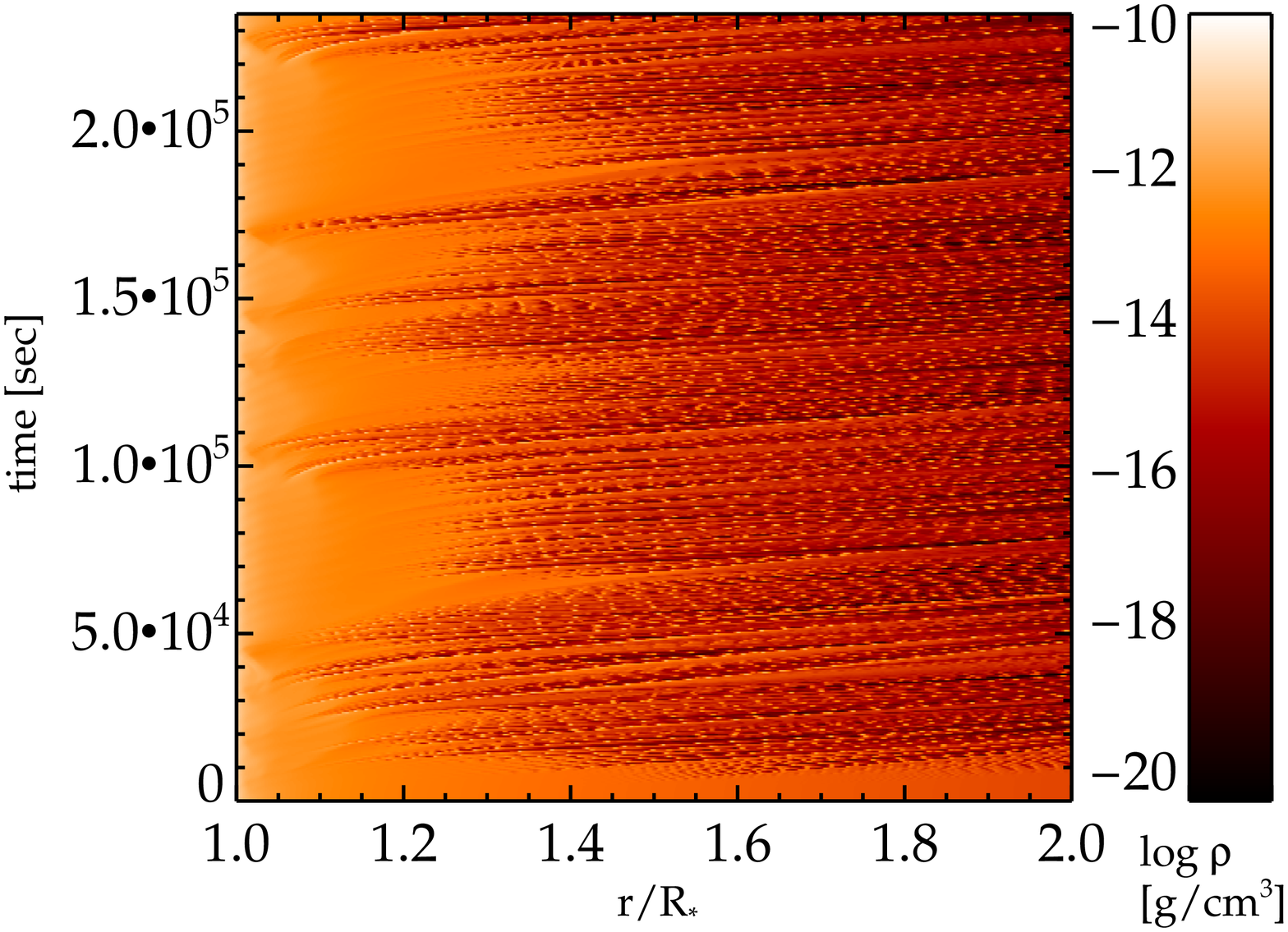}} 
\end{minipage}
\caption{{\bf Left:} Temperature, velocity, and density snapshots of
  an LDI simulation (see text) that has reached a well-developed
  phase. The dashed red line in the bottom panel shows the density at
  $t=0$ sec. {\bf Right:} Contour-map of the inner-wind time-evolution
  of the density between $t=0$ and $t=23$ ksec in the same simulation
  (the characteristic wind flow time of this simulation is $t \approx
  R_*/\upsilon_{\infty} \approx 20 R_{\odot}/2000 \ \rm km/s \approx 7
  \ ksec$).}
\label{Fig:ldi}
\end{figure}

The left panel in Fig.~\ref{Fig:ldi} illustrates this typical
structure by plotting density, velocity, and temperature snapshots of
a spherically symmetric LDI simulation computed from an initial steady
Sobolev-based model following \citet{Sundqvist13}, and the right panel
in the figure shows a contour-map of the density time evolution for
the same model. The line force in this simulation is calculated using
the ``Smooth Source function'' method \citep[SSF,][]{Owocki96}, which
allows one to follow the non-linear evolution of the strong, intrinsic
instability, while simultaneously accounting for the stabilizing
effect of the scattered, diffuse radiation field. The simulation here
further includes the effects of stellar limb darkening and
photospheric density perturbations (the latter in the form of simple
sound waves). The model displays clumps with very short characteristic
radial length scales, on the order of a Sobolev length $L_{\rm Sob}$
or below (as expected from the linear theory above), and shows typical
clump-densities an order of magnitude higher than the mean wind
density. As can be seen from the figure, the rarefied regions in
between the dense clumps are much more extended and with very low
densities, on order a percent of the mean density or even lower. The
inclusion of limb-darkening and photospheric perturbations further
leads to strong, variable wind structure also close to the
photosphere, at $r \approx 1.05-1.2 R_{\star}$ (in contrast to earlier
simulations which typically had this onset only at $r \approx 1.5
R_{\star}$, e.g. \citealt{Runacres02}). An X-ray study of the
B0I star QV Nor, the massive donor of the 4U1538-52 binary system,
established an upper limit of $\sim$ $1.25{R}_{*}$ on the radial onset of
clumping \citep{Torrejon2015}. We note though, that although
the basic properties and predictions of LDI-caused structure are
robust, quantitative details regarding, e.g., the onset of structure
and over-densities of clumps depend on various assumptions made in the
computations of the source function and the line-force, on the level
of photospheric perturbations, and on the dimensionality of the
problem (see discussion in \citealt{Sundqvist13}).

The presence of strong, embedded wind shocks in LDI simulations
further provides a generally accepted explanation for the X-rays
observed from single O-stars without very strong magnetic fields.  The
jump velocities are typically on order $\sim 300$\,km/s, giving
characteristic temperatures $T \sim 10^6$\,K and associated quite soft
X-ray emission at $\sim$\,1\,keV and below. In 1-D simulations like
those above, \citet{Feldmeier97b} demonstrated that the velocity
dispersion of individual shells ('clumps') results in collisional
merging of such shells, which creates regions of relatively dense, hot
gas and X-ray emission at levels quite comparable to those typically
observed for single O-stars. More realistically though, thin-shell
instabilities and associated effects can be expected to break up the
spherical shell structure into a more complex, multi-dimensional
structure. Because of the computational expense of calculating the
line-force at each time step, however, radiation-hydrodynamical LDI
simulations have generally been limited to 1-D. But first attempts to
construct 2-D simulations have been carried out by \citet{Dessart03,
  Dessart05}. Such models typically show that the LDI first manifests
itself as strong density compressions mimicking corresponding 1-D
simulations. But as these initial shell structures are accelerated
outwards, they become disrupted by velocity shearing and
Rayleigh-Taylor and/or thin-shell instabilities, resulting in lateral
structure all the way down to the grid-scale (see
\citealt{Dessart03}).
However, it may well be that these initial 2-D models exaggerate
somewhat the level of lateral disruption, as they do not yet properly
treat the lateral component of the diffuse line-force. Presuming this
could damp azimuthal velocity perturbations on scales below the
lateral Sobolev length $L_{\phi} = \upsilon_{\rm th} r/\upsilon_{r}$
\citep{Rybicki90}, lateral breakup may be prohibited below scales of
order $\phi \approx L_{\phi}/r \approx \upsilon_{\rm th}/\upsilon_{\rm r} \approx
0.5$\,degrees. Further work is required to examine this important
issue by an adequate incorporation of the lateral line-force into
multi-D LDI simulations.

\paragraph{Characteristic clump mass} 
With the discussed caveats in mind, we may nonetheless use the basic
predictions of the ab-initio hydrodynamical wind simulations above to
make a simple order of magnitude estimate of a characteristic clump
mass: at a typical radius $r \approx 2 R_*$, the radial and lateral
Sobolev lengths are on the same order, i.e. $d\upsilon_r/dr \approx \upsilon_r/r$,
and so we estimate a clump volume by $V_{\rm cl} \sim L_{Sob}^3 \sim
(\upsilon_{\rm th}/(\upsilon_r/r))^3$. We further assume the clump density is set by
$\rho_{\rm cl} \approx \langle \rho \rangle f_{\rm cl}$ with a
clumping factor (see further Sect. 3.2) $f_{\rm cl} \approx 10$ and
mean density $\langle \rho \rangle = \dot{M}/(4 \pi r^2 \upsilon_r)$ for an
average stellar mass-loss rate $\dot{M}$. Thus a characteristic
clump-mass is $M_{\rm cl} \approx \rho_{\rm cl} V_{\rm cl} \approx
\dot{M} f_{\rm cl} \upsilon_{\rm th}^3 r/(4 \pi \upsilon_r^4)$. Inserting typical
numbers for a hot supergiant, $\dot{M} \approx 10^{-6} M_{\odot}/yr$,
$\upsilon_{\rm th} \approx 10 \ \rm km/s$, $R_\star \approx 20 R_{\odot}$,
and $\upsilon_r \approx 1000 \ \rm km/s$, then yields $M_{\rm cl} \approx
10^{18}$\,g. Although this indeed should be viewed only as a rough
order of magnitude estimate, we note that since the LDI predicts small
clumps, where the characteristic length scale is the Sobolev length,
it is generally quite difficult to reconcile the small-scale
structures predicted by basic line-driven wind theory with
clump-masses that are much (i.e. orders of magnitude) higher than this
estimate. This is further discussed in Sect.~\ref{xraypars}, in the context of
empirical constraints obtained from X-ray observations of SgXBs.

\subsubsection{Large scale structures}
\label{sec:largescale} \noindent Besides the small scale structures
discussed in the previous section, the winds from massive stars also
harbour structures of larger scale, inferred predominantly from the
so-called discrete absorption components (DACs).

\smallskip \noindent 
{\it Discrete absorption components} are optical depth enhancements in
the absorption troughs of unsaturated UV P~Cygni profiles, observed in
most O- and early B-star winds \citep{HP89}, and also in late
B-supergiants\footnote{Absorption components
have been also found in at least one WN7 star \citep{PrinjaSmith92},
and in the Balmer lines of the LBV P~Cygni \citep{Markova86}.} \citep{BatesGilheany90}. 
Typically, these DACs accelerate towards the blue wing of the profile
on time scales of a few days, becoming narrower as they approach an
asymptotic velocity.\footnote{This asymptotic value is frequently used
to estimate $\vinf$ (e.g., \citet{HP89} and
\citet{Prinjaetal90}).} Their acceleration and recurrence time
scales are correlated with the star rotational period \citep{Prinja88,
Henrichs88, Kaper99}. Since the acceleration of most strong DACs is
much slower than the mean wind acceleration \citep{Prinjaetal92,
Prinja94}, DACs might arise from a slowly evolving perturbation
through which the wind material flows (higher density or lower
velocity gradient/velocity plateau: \citet{Lamersetal82}, or both:
\citet{FO92}). Although a spherically-symmetric disturbance most
likely can be ruled out \citep{Prinja92, Howarth92}, the structure
must cover a substantial fraction of the stellar disk in order to
produce the observed strong absorption features.

\paragraph{Dynamical models - Co-rotating Interaction Regions} 
Summarizing the above findings and arguments, DACs should originate
from \textit{coherent} structures of significant lateral extent, with an
increased optical line depth due to density and/or velocity field
effects, and a clock coupled to stellar rotation. As a potential
candidate mechanism which is compatible with all these constraints,
\cite{Mullan84, Mullan86} suggested co-rotating interaction regions
(CIRs), which are well-studied in the solar wind.\footnote{Indeed,
this suggestion was made already before all constraints were known.}
By means of 2-D time-dependent hydrodynamic simulations, \citet{CO96} 
investigated this scenario in detail for the case of a rotating O-star
wind. Their most promising model comprises photospheric disturbances
due to a bright stellar spot in the equatorial plane.\footnote{to be
regarded as a representative for other potential disturbances, e.g.,
localized magnetic fields or non-radial pulsations.} Because of this disturbance, 
the line acceleration becomes locally enhanced (more flux!). Consequently, 
the local mass loss increases, and thus the density already near the star. Further out 
in the wind, the effect of the disturbance weakens (since the opening angle of the
disturbance decreases, and the radiation from undisturbed
regions of the stellar surface begins to dominate), and the radiative
force cannot accelerate the higher density material as strongly as the
unperturbed wind. In summary, a stream of higher density and lower
velocity, compared to the undisturbed mean flow, is generated. Where the faster mean flow
collides supersonically with the slow material at its trailing edge, a
CIR of enhanced density is formed, and a (non-linear) signal is sent
{\it back} towards the star which forms a sharp, propagating discontinuity
in the radial velocity gradient (see also \citealt{FeldmeierShlosman00,
FeldmeierShlosman02}).
In the {\it stellar} frame, this feature travels slowly outwards, and a
velocity plateau-like structure is formed between the trailing
discontinuity (`kink') and the CIR compression.

By calculating the corresponding line optical depths and
time-dependent synthetic profiles, \citet{CO96} showed that these
slowly moving kinks, together with their low velocity gradients ($\tau
\propto (\dvdr)^{-1}$, cf. Eq.~\ref{gradi}), are the likely origin of
the observed DACs rather than the CIRs themselves.
Moreover, the {\it apparent} acceleration of the synthetic DACs was
found to be slower than the acceleration of the mean wind (as
required), and also their recurrence time scale is strictly correlated
with $\vrot$ (because of the link of the CIRs and the rotating
photosphere). On the other hand, no correlation between
acceleration time scale and rotational speed appeared in the
synthetic profiles, contrasted to the observational indications.  

By means of an instructive kinematic investigation, \citet{Hamann01}
showed that the low drift rate of the DACs is not a consequence of the
CIRs themselves, but a consequence of the difference between mean flow
and the velocity field of the pattern in which the features form. An
upstream propagating pattern (as the kink in the Cranmer-Owocki model)
inevitably results in a wavelength drift with a slower apparent
acceleration than displayed by features formed in the mean flow {\it or
within the CIR itself.} Any of these drifts, however, were shown to be
independent on the rotation rate, leaving the observed
correlation (if actually present) still unexplained, at least if the
CIRs are induced by disturbances {\it locked} to the stellar surface,
as assumed by both \citet{Hamann01} and \citet{CO96}.

In a detailed study, \citet{LobelBlomme08} relaxed this assumption in
order to allow for a {\it quantitative} analysis of the temporal evolution of
DACs for the fast-rotating B0.5 Ib supergiant HD\,64760, one of the
targets of the {\it IUE MEGA Campaign} (see below). A large grid
of 3-D models and dynamic spectra for different spot parameters
(brightness, opening angle and velocity) has been computed, producing
a best fit for a model with two spots of unequal brightness and size
on opposite sides of the equator, with spot velocities being five
times slower than \vrot.\footnote{Motivated by a suggestion from
\citet{Kaufer06} who detected non-radial pulsations in the
photospheric lines of HD\,64760. To explain the observed \Ha
variability, they invoked the beat period between two of these periods
(lower than the rotational one!) to be responsible for the CIRs in the
wind.} All basic conclusions of \citet{CO96} could
be confirmed, particularly the importance of velocity plateaus in
between kinks and CIRs. Moreover, models with non-locked spots
displayed a correlation of DAC acceleration time scale and \vrot, since the
internal clock is now set by the rotational speed of the spots,
whilst a change in the stellar rotational speed modifies the
underlying bulk flow (densities and velocities) via a different
centrifugal acceleration such that the time-scales are no longer
conserved when \vrot\ is changed. 

One may note that the above hydro-simulations have been performed by
calculating the line-force in Sobolev approximation, because of
computational feasibility. An improved approach regarding this aspect
has been provided by \citet{Owocki99}, who calculated CIR models based
on a {\it non-local} line force calculated from a three-ray
treatment. Also here, somewhat "rounded" kinks followed by a velocity
plateau are present! Unfortunately, however, the line-driven
instability (LDI, see Sect.~\ref{sec:smallscale}) effectively destroys all
macro-structure, though this might be related to a still inadequate
three-ray treatment and/or insufficient resolution, leading to too low
a lateral damping. Certainly, future work is required to clarify these
problems.

\paragraph{Rotational Modulations (RMs)} The observed correlations of the DAC
properties with rotational speed inspired the {\it IUE MEGA Campaign}
\citep{Massaetal95}, during which three prototypical massive stars
(including HD~64760 from above) were monitored almost continuously
over 16 days. Most importantly, a new type of variability was
detected, namely {\it periodic modulations} in the UV wind lines,
derived from the dynamic spectra of HD\,64760 \citep{Prinjaetal95}. 
Similar features (though not as pronounced) have been detected in only
two other stars, $\zeta$~Pup \citep{Massaetal95, Howarth95} and
$\xi$~Per (O7.5 III (n)((f)), \citealt{Kaper99}). Note that these
features occur in parallel with the conventional DACs.

The peculiar phase properties of the modulation features indicate the 
presence of absorbing material with the same phase at two different
projected velocities. By means of a kinematical model and corresponding
synthetic spectra, \citet{OCF95} and \citet{Fullerton97} suggested that
these features are formed in azimuthally extended, co-rotating and
spirally-shaped wind structures linked to a surface density which is
modulated by non-radial pulsations, 
and they are observational "evidence for co-rotating wind streams rooted
in surface variations" \citep{OCF95}. Meanwhile, \citet{Lobel13}
performed 3-D hydro-simulations to investigate the rotational
modulation features. To fit the observations, the RMs required quite
a regular `spoke-like' radial density pattern that was simulated as a 
result from the action of pressure waves at the lower wind boundary.
These structures are then maintained in the wind by an increased 
line acceleration in front of the (rotating) pressure wave.

\paragraph{Density contrast in CIRs and RMs} With respect to one of
our central topics, the interaction between (inhomogeneous) winds of
supergiants and compact objects in SgXBs, the typical density contrast
of CIRs and RMs is of prime importance. Though this value depends on
the specific situation (size of spots, amplitude of sound-wave,
rotation rate, etc.), the above hydro-simulations indicate a rather
mild density contrast, namely over-densities of factors $\sim 1.3$
\citep{LobelBlomme08} to $\sim 3$ \citep{CO96} for the CIRs and a
maximum over-density of a factor of 1.17 within 10 \Rstar\ for the RM
region of HD\,64760 \citep{Lobel13}. Together with the rather small
filling factor, these inhomogeneities thus have only a small impact on
density squared \mdot\ diagnostics (see Sect.~\ref{stellarwindpars}), on the
order of 2 to 3\% in \mdot, to be compared with typical {\it factors}
of $\sqrt{f_{\rm cl}} \sim \sqrt{10} \sim 3$ resulting from the
small-scale structures discussed in the previous section. We note
though, that since these large-scale structures must cover a
substantial fraction of the stellar disc (see above), their individual
masses can be very high despite their modest
density-enhancements. Following the previous section, we may estimate
$M_{\rm cir} \sim R_*^2 \delta R \rho_{\rm cir} \sim R_*^2 \delta R
\dot{M} f_{\rm cir}/(4 \pi r^2 \upsilon_r) \sim 10^{21-22} g$, where we have
assumed a geometrical CIR-height $\delta R \approx 0.1-1 R_\star$, and
used the same numbers as before for a typical hot supergiant but now
assuming a characteristic CIR over-density set by $f_{\rm cir}
=2$. These type of large-scale wind structures can thus have much
higher individual masses than the small-scale structures normally
associated with wind clumping (see also Sect.~\ref{xraypars}).


%
%
%
\subsection{Wind accretion theory}
\label{sec:windtheory} 

In SgXBs, an O-B supergiant is gravitationally bound to a compact object and the latter  
can accrete a significant fraction of the stellar wind. In most of these systems, and for all 
the systems that are of interest for this review, the massive star does not fill its Roche lobe 
and thus an accretion disk is not expected to form around the compact object \citep[see, e.g.,][]{frank2002}. 
In the remaining part of this review we thus restrict our discussion to purely wind-fed systems 
and consider only the case in which the compact object is a neutron star (by far the majority of  
SgXBs are proven to host such kind of compact objects, the only known exceptions being 
Cyg~X-1, GRS~1915+105 and LMC X-1). 
We provide first a description of the simplest theoretical wind-accretion treatment, i.e. the so-called 
Bondi-Hoyle-Lyttleton (BHL) accretion theory, and then summarise the most recent 
developments. In this second part we identify different possible accretion regimes,  depending on  
the neutron star properties (magnetic field, spin period), as well as on the physical parameters 
of the supergiant wind (mass outflow rate, velocity, density).  
We also provide a short section summarizing the basic physics of the interaction between the 
X-rays produced as a consequence of the accretion onto the compact object and the stellar wind. 
Finally, the obtained results using a simplified analytical treatment are compared to the achievements 
of recent numerical simulations. 

\subsubsection{Bondi-Hoyle-Lyttleton accretion}
\label{sec:bondi}

In the following, we briefly summarise elements of the BHL theory. 
For a detailed review, we refer the reader to \citet{Edgar2004}.

The original work by \citet{HoyleLyttleton1939} considered accretion by a point mass (star) passing
through a nebulous cloud as a mechanism for terrestrial climate change. This work was later updated 
by \citet{BondiHoyle1944}, who included in the analysis the presence of an accretion wake behind the 
point mass. All these results were first applied in the context of accreting X-ray
binaries by \citet{DavidsonOstriker73}. Below, we follow their approach \citep[see][for details]{osk2012,
Bozzo2016}.

Let us consider a neutron star of mass $M_{\rm NS}$, travelling at a relative speed $\upsilon_{\rm rel}$ 
through a gas density $\rho$. If $\upsilon_{\rm rel}$ is supersonic, then 
the gas flows in as described by \citet{BondiHoyle1944}. The mass
accretion rate onto the compact object is given by: 
\begin{equation}
\dot{M}_{\rm acc}=\pi \zeta R_{\rm acc}^2 \upsilon_{\rm rel} \rho(a,t)
\label{eq:acrrate}
\end{equation} 
where $\rho(a,t)$ is the wind density as a function of the orbital separation $a$ and time $t$.  
$\zeta$ is a numerical factor introduced to correct for the radiation pressure
and the finite cooling time of the gas. In moderately luminous X-ray sources $\zeta\lsim 1$, 
and it is usually assumed $\zeta \equiv 1 $\citep{DavidsonOstriker73}.
The Bondi-Hoyle radius, or ``accretion radius''  
\begin{equation}
\Racc=\frac{2GM_{\rm NS}}{\upsilon_{\rm rel}^2}
\label{eq:raccr}
\end{equation}
represents a measurement of the distance at which the neutron star is able to 
gravitationally focus and capture the surrounding material. 
At this radius, the supersonically inflowing gas from the supergiant companion 
passes through a bow-shock and begins to free-fall toward the compact object. 
The relative speed $\upsilon_{\rm rel}$ can be expressed as 
\begin{equation}
\upsilon^2_{\rm rel}=\upsilon^2_{\rm NS}(a) + \upsilon^2_{\rm w}(a,t), 
\label{eq:vrel}
\end{equation}
where $\upsilon_{\rm NS}$ is the velocity of the neutron star 
at the orbital separation $a$
and $\upsilon_{\rm w}$ is the stellar wind velocity.

In non-stationary stellar winds, the wind density
$\rho(a,t)$ and the wind velocity $\upsilon_{\rm w}(a,t)$ depend on time (see
Section\,\ref{basicsldw}).
Combining equations (\ref{eq:acrrate}) and (\ref{eq:raccr}), the 
accretion rate onto the neutron star can be written as:  
\begin{equation}
\dot{M}_{\rm acc}=4\pi \zeta \frac{(GM_{\rm NS})^2}{\upsilon^3_{\rm rel}}\rho(a,t).
\label{eq:sac}  
\end{equation} 
The accretion of matter onto the neutron star produces X-rays. 
Assuming a direct conversion of the kinetic energy gained by the
infalling matter to radiation at the neutron star surface, 
the X-ray luminosity of the system can be expressed as: 
\begin{equation}
  L_{\rm X} = \frac{G M_{\rm NS} \dot{M}_{\rm acc}}{R_{\rm NS}}
  \label{eq:lx_base}
\end{equation}
where $R_{\rm NS}$ is the neutron star radius. 
In the following, we consider commonly used values 
for the neutron star parameters, i.e., 
$R_{\rm 6}$ for the neutron star radius in units of 10$^6$~cm and
$M_{\rm 1.4}$ the neutron star mass in units of 1.4~$M_{\odot}$. 
A reasonable scale for the mass accretion rate obtained from
Eq.~\ref{eq:sac} is $\dot{M}_{\rm 16}$, i.e., $\dot{M}_{\rm acc}$ in units of
10$^{16}$~g~s$^{-1}$. We can thus write: 
\begin{equation}
L_{\rm X} \simeq 2 \times 10^{36} \frac{M_{1.4} \dot{M}_{16}}{R_{6}}~{\rm erg~s^{-1}},  
\label{eq:lx_q}
\end{equation}

Using the fact that in the cases of interest the neutron star velocity $\upsilon_{\rm NS}$ is usually 
much lower than the wind velocity $\upsilon_{\rm w}$, Eq.~\ref{eq:vrel} reduces to  
$\upsilon_{\rm rel}\simeq \upsilon_{\rm w}$. We thus obtain: 
\begin{equation}
R_{\rm acc} \approx \frac{3.7\times10^{10}}{\upsilon_{8}^{2}}~{\rm cm.} 
\label{eq:racc_q}
\end{equation}
where $\upsilon_{8}$ is the wind velocity in units of 10$^8$~cm~s$^{-1}$.

The predicted X-ray luminosity (Eq.~\ref{eq:lx_q}) can be rewritten in terms 
of measurable properties of the massive star wind. If we approximate the wind from 
the supergiant star as smooth and spherically symmetric, then we have at the neutron 
star location (orbital separation $a$):
\begin{equation}
\rho_{\rm w} (a) = \frac{\dot{M}_{\rm w}}{4 \pi a^2 \upsilon_{\rm w}(a)}.
\label{eq:rho}
\end{equation}
Including this in Eq.~\ref{eq:acrrate} (with $\zeta \equiv 1$) we get:
\begin{equation}
\frac{\dot{M}_{\rm acc}}{\dot{M}_{\rm w}} = \frac{1}{4} \frac{R_{\rm acc}^2}{a^2}. 
\label{eq:dotmratio}
\end{equation}
Using Eq.~\ref{eq:dotmratio} and Eq.~\ref{eq:racc_q} in Eq.~\ref{eq:lx_q}, we obtain: 
\begin{equation}
L_{\rm X} = 2 \times 10^{35} {\dot{M}_{-6} \over a_{10d}^{2}\, \upsilon_8^{4}}~{\rm erg~s^{-1}},  
\label{eq:lx}
\end{equation}
where $\dot{M}_{-6}$ is the mass-loss rate of the primary star in units 
of 10$^{-6}$~M$_{\odot}$~yr$^{-1}$,  
and  $a_{\rm 10d}$~cm, is the orbital separation for a massive binary in which $M_{\rm NS}+M_{\ast}$=30$M_{\odot}$ 
and the NS is on a circular orbit with a period of 10~days: 
\begin{equation}
a = 4.2 \times 10^{12} P_{\rm 10d}^{2/3} M_{30}^{1/3} = 4.2 \times 10^{12} a_{\rm 10d}. 
\label{eq:orbsep}
\end{equation}
In the remaining part of this section, we omit from all equations the symbols $R_{6}$ and $M_{\rm 1.4}$, 
as the range spanned by these parameters are marginally affecting the results compared to the larger 
uncertainties on more relevant quantities (e.g., the wind velocity, mass loss rate, the neutron star 
spin period and its magnetic field). 

The regime in which accretion at the rate defined by Eq.~\ref{eq:dotmratio} is allowed, is 
usually termed ``direct accretion'' and corresponds to the maximum achievable X-ray luminosity 
in a wind-fed neutron star SgXB. It is interesting to note from Eq.~\ref{eq:dotmratio} that, even 
in the direct accretion regime, wind-fed binaries are relatively inefficient accreting systems 
because in all cases of interest $R_{\rm acc}$ $\ll$ $a$ and thus only a tiny fraction of the 
total mass lost by the primary star can reach the surface of the compact object. 
Another interesting property of these systems can be deduced from Eq.~\ref{eq:lx_q}. 
As in wind-fed SgXBs no accretion disk is present to mediate the transport of matter, the relevant time 
scale for changes in the mass accretion rate onto the compact object is roughly comparable to the 
free-fall time at $R_{\rm acc}$, i.e., a few to hundred seconds \citep{frank2002}. This means 
that any change occurring in either the mass inflow rate from the supergiant star, or in the velocity/density of the 
wind around the neutron star, would produce a virtually immediate 
variation in the source X-ray luminosity. For this reason, the pronounced 
X-ray variability of SgXBs is usually ascribed to the presence of structures (clumps) 
in the winds of their supergiant companions. However there are a number of complications in the 
accretion process (see Sect.~\ref{sec:regimes}) that makes the evaluation of the stellar wind properties 
from the variations of the X-ray luminosity in these systems challenging. We will discuss more these aspects in Sect.~\ref{xraypars}.

\subsubsection{Different accretion regimes}
\label{sec:regimes}

The simplified scenario depicted in Sect.~\ref{sec:bondi} provides only a very rough 
estimate of the mass accretion rate that a neutron star can experience due to the capture of 
wind from its massive companion. In particular, the direct accretion regime cannot apply in all 
circumstances as the neutron star rotation and magnetic field can dramatically affect the plasma 
entry through the compact object magnetosphere and thus the resulting mass accretion rate. In this 
section we describe all the accretion regimes that a neutron star can experience depending on 
the different key system parameters (i.e. the mass-loss rate from the primary star, the stellar 
wind velocity, the neutron star spin period, the magnetic field and the orbital separation). 

We also generalize the treatment by relaxing the assumption of Eq.~\ref{eq:rho} and assuming that 
the wind velocity and density at the location of the neutron star can be more generally a non-trivial 
function of time and of the separation between the primary star and the compact object, i.e. 
$\rho_{\rm w}$=$\rho_{\rm w}(a,t)$ and $\upsilon_{\rm w}$=$\upsilon_{\rm w}(a,t)$.  

Neutron stars in massive binaries are known to be strongly magnetized,  
as their relatively young age (a few 10$^6$~yr) would not allow for a significant decay of the 
dipolar magnetic field with which they were endowed at birth (B$\gtrsim$10$^{12}$~G). The 
magnetosphere around the neutron star has a non-negligible effect on the development of the 
accretion flow and begins to completely dominate its dynamics once the material gets close 
to the so-called magnetospheric radius, $R_{\rm M}$. For wind accreting systems, $R_{\rm M}$ can 
be roughly estimated by equating the ram pressure of the 
material flowing toward the neutron star with the local magnetic pressure. 
Scaling to typical parameters, we obtain: 
\begin{equation}
R_{\rm M}=1.3\times10^{9} \rho_{-12}^{-1/6} \upsilon_{8}^{-1/3} \mu_{30}^{1/3}\,{\rm cm}.  
\label{eq:rm}
\end{equation}
where $\rho_{-12} $=$ \rho_{\rm w}$/10$^{-12}$~g~cm$^{-3}$, $\mu$ = $B_{\rm NS}$ $R_{\rm NS}^{3}$ is the 
magnetic moment of the NS and $\mu_{30}$=$\mu$/(10$^{30}$)~G~cm$^{3}$ (i.e., 
considering $B_{\rm NS}$=10$^{12}$~G). The direct accretion regime described in Sect.~\ref{sec:bondi} 
requires that $R_{\rm M}$$<$$R_{\rm acc}$, in such a way that the accretion flow has enough time 
to settle in a free-fall spherical symmetric motion before encountering the neutron star magnetosphere.  
Combining Eq.~\ref{eq:rm} and \ref{eq:racc_q}, we can see that the condition $R_{\rm M}$$<$$R_{\rm acc}$ is violated if 
\begin{equation}
\mu_{30} \gtrsim 2.3 \times 10^{4} \rho_{-12}^{1/2} \upsilon_8^{-5}. 
\end{equation}
Assuming reasonable values for the supergiant wind velocity ($\upsilon_8$ $\simeq$ 0.5-2),  
it turns out that the magnetospheric radius can extend beyond the accretion radius only for very strongly magnetized neutron stars with 
$B_{\rm NS}$ $\gtrsim $10$^{14}$~G. As discussed by \citet{Bozzo2008}, such magnetic field would not 
be completely unreasonable if it is assumed  that ``magnetars'' are hosted in SgXBs 
\citep[see, e.g.,][for a recent review]{rea2011}. The regime in which 
$R_{\rm M}$ $\gtrsim$ $R_{\rm acc}$ is called the "magnetic inhibition of accretion" regime, as 
the magnetosphere of the neutron star in this case inhibits the gravitational focusing of the wind material and 
decreases the effective accretion rate onto the compact object. It is noteworthy that a similar 
scenario was also envisaged by a number of authors to predict the X-ray luminosity of isolated magnetars 
accreting through the interstellar medium \citep{harding1992,mori2003,toropina2006}. 
Before entering the details of the X-ray luminosity released during the magnetic inhibition of accretion, it is necessary to introduce 
also the concept of the corotation radius. 

It is known since the early 70s that accretion onto a 
magnetized neutron star cannot occur unperturbed if the rotation of the compact object is slow 
enough to allow $R_{\rm M}$ to reside within the corotation radius: 
\begin{equation}
R_{\rm co}=3.7\times10^{9} P_{s2}^{2/3}~cm  
\label{eq:rco}
\end{equation}
(here $P_{s2}$ is the neutron star spin period in units of 100~s). $R_{\rm co}$ represents the 
distance from the neutron star at which material attached  to the magnetic field lines of the 
compact object (corotating with it) reaches a velocity comparable with 
the local Keplerian velocity. The condition $R_{\rm M}$$<$$R_{\rm co}$ thus ensures that the 
centrifugal force at $R_{\rm M}$ is low enough for the incoming material to be accreted onto 
the neutron star rather than ejected by its fast rotation. When $R_{\rm M}$$>$$R_{\rm co}$, the 
neutron star centrifugal gate closes and the ``propeller'' regime sets-in, halting the incoming flow and 
centrifugally inhibiting accretion \citep{IllarionovSunyaev75}. By combining Eq.~\ref{eq:rco} 
and \ref{eq:rm}, we can see that this case occurs if
\begin{equation}
P_{\rm s2}\lesssim 0.2 \upsilon_{8}^{-1/2} \rho_{-12}^{-1/4} \mu_{30}^{1/2}
\end{equation}
We thus have to distinguish different possibilities. 

\begin{itemize}
\item {\it The super-keplerian magnetic inhibition regime}. This case applies when both conditions 
$R_{\rm M}$$\gtrsim$$R_{\rm acc}$ and $R_{\rm M}$$\gtrsim$$R_{\rm co}$ are satisfied. As argued before, 
the magnetic barrier in this case halts the mass inflowing from the companion star and inhibits  
the gravitational focusing of this material within the accretion radius. Furthermore, the wind material 
that approaches the neutron star magnetosphere is pushed away from the compact object due to the 
centrifugal gate. In this regime, no accretion is possible and the lowest X-ray luminosity 
is achieved. \citet{Bozzo2008} argued that the main contribution to the X-ray luminosity in 
this case is due to shocks close to the magnetospheric radius and to the dissipation of the neutron star 
rotational energy. The two contributions can be estimated as  
\begin{equation}
L_{\rm shock}\simeq\frac{\pi}{2} R_{\rm M}^2 \rho_{\rm w} \upsilon_{\rm w}^3 = 
2.7\times10^{30} \mu_{30}^{2/3} \rho_{-12}^{2/3} \upsilon_{8}^{7/3} ~{\rm erg ~s^{-1}}  
\label{eq:lshock}
\end{equation}
and 
\begin{equation}
L_{\rm sd1} \simeq \pi R_{\rm M}^2 \rho_{\rm w} \upsilon_{\rm w} (R_{\rm M}\Omega)^2\simeq  
3.5\times10^{30} \mu_{30}^{4/3} \rho_{-12}^{1/3} \upsilon_{8}^{-1/3} P_{\rm s2}^{-2}~{\rm erg ~s^{-1}},  
\label{eq:lsd1}
\end{equation}
respectively (here $\Omega$=2$\pi$/$P_{\rm s}$).  

\item {\it The sub-keplerian magnetic inhibition regime}. This case applies when 
$R_{\rm M}$$\gtrsim$$R_{\rm acc}$ but $R_{\rm M}$$<$$R_{\rm co}$. At odds with the previous case, 
despite the lack of gravitational focusing within $R_{\rm acc}$, the effective gravity at the 
neutron star magnetosphere points toward the compact object. The inflowing material is thus 
not pushed away by the rotating star but tends to be accreted at a rate that is regulated by 
the efficiency of magnetohydrodynamics instabilities at $R_{\rm M}$. As the stellar wind material 
is passing by the neutron star with high velocities ($\upsilon_8$$\simeq$1), the most relevant 
instability allowing matter to penetrate the compact object magnetosphere 
is the Kelvin-Helmholtz instability \citep[hereafter KHI;][]{harding1992}. The mass 
inflow rate through the magnetosphere resulting from the KHI is: 
\begin{equation}
\dot{M}_{\rm KHI} = 2 \pi R_{\rm M}^2 \rho_{\rm ps} \upsilon_{\rm conv} = 2 \pi R_{\rm M}^2 \rho_{\rm ps} \upsilon_{\rm sh} \eta_{\rm KHI} (\rho_i/\rho_e)^{1/2}(1+\rho_i/\rho_e)^{-1}  
\end{equation}
where $\upsilon_{\rm conv}$ is the typical velocity at which matter crosses the magnetospheric boundary, 
$\rho_{\rm ps}$ is the density of the wind material being shocked and passing through the shock in 
front of the neutron star magnetosphere, $\upsilon_{\rm sh}$ is the shear velocity between the wind and the 
neutron star magnetosphere, $\eta_{\rm KHI}$$\simeq$0.1 is an efficiency parameter, 
and $\rho_i$ ($\rho_e$) is the density immediately inside (outside) $R_{\rm M}$.  
\citet{Bozzo2008} argued that $\upsilon_{\rm sh}$=$max(\upsilon_{\rm ps}, \upsilon_{\rm rot})$, depending on the 
position along the neutron star magnetosphere. Here $\upsilon_{\rm ps}$ is the matter post-shock 
velocity dominating close to the stagnation point, while  
$\upsilon_{\rm rot}$=2$\pi$$R_{\rm M}$$P_{\rm s}^{-1}$ is the magnetosphere rotational velocity 
dominating the shear away from the stagnation point. The ratio $\rho_i$/$\rho_e$ can be estimated 
numerically from the mass conservation equation 
$R_{\rm M}^2$ $\rho_e$ $\upsilon_{\rm conv}$ $\simeq$ $R_{\rm M}$$h_t$$\rho_i$ $\upsilon_{\rm ff}(R_{\rm M})$. Here we 
considered that matter crossing the KH unstable layer is rapidly brought at corotation with the 
neutron star and free-falls toward its surface before being accreted. The height of the KHI layer 
is assumed to be $h_t$ $\simeq$ $R_{\rm M}$. \citet{Bozzo2008} discussed that this is a reasonable 
assumption as long as $\rho_i$ $\lesssim$ $\rho_e$, even though a full detailed study of the KHI 
in these conditions is still lacking. Under these assumptions, the X-ray luminosity in 
the sub-keplerian magnetic inhibition regime can be estimated as 
$max$($L_{\rm KH1}$,$L_{\rm KH2}$)\,,where 
\begin{eqnarray}
L_{\rm KH1} = 2.0\times10^{35} \eta_{\rm KH} \mu_{30}^{2/3}\rho_{-12}^{2/3} \upsilon_{8}^{1/3}
(\rho_{\rm i}/\rho_{\rm e})^{1/2} (1+\rho_{\rm i}/\rho_{\rm e})^{-1}~{\rm erg~s^{-1}}, 
\end{eqnarray}
and 
\begin{eqnarray}
L_{\rm KH2} = 6.5\times10^{35} \eta_{\rm KH} P_{\rm s2}^{-1} \upsilon_8^{-1} \rho_{-12}^{1/2} \mu_{30} 
(\rho_{\rm i}/\rho_{\rm e})^{1/2} (1+\rho_{\rm i}/\rho_{\rm e})^{-1} ~{\rm erg~s^{-1}}.  
\label{eq:lx_kh2}
\end{eqnarray}
In this regime, we thus expect to observe relatively faint X-ray sources. 

\item {\it The supersonic propeller regime.} When $R_{\rm M}$$<$$R_{\rm acc}$, the wind material 
is gravitationally focused within the accretion radius and fills the region between 
$R_{\rm acc}$ and $R_{\rm M}$.  The magnetic gate is open and the 
properties of this envelope surrounding the neutron star are determined by the physical 
processes regulating the interaction between matter and magnetic field 
at $R_{\rm M}$ \citep{davies1981}. If radiative losses are negligible 
within the envelope, then the latter is in hydrostatic equilibrium and stationary on 
dynamical time-scales \citep[this requires a sufficiently low mass-loss rate from the primary star, 
see Eq.~B2 in][]{Bozzo2008}. In this case the pressure and density within the envelope can be written as: 
\begin{eqnarray}
p(R)=\rho_{\rm ps} \upsilon_{\rm ps}^2 \left[1+(1/(1+n))8R_{\rm a}/R\right]^{n+1} \\
\rho(R)=\rho_{\rm ps}\left[1+(1/(1+n))8R_{\rm a}/R\right]^{n}. ~~~~~~ 
\label{eq:hydro} 
\end{eqnarray} 
If the condition $R_{\rm M}$ $\gtrsim$ $R_{\rm co}$ applies, accretion is, however, still inhibited 
by the centrifugal barrier and the system enters the supersonic propeller regime. As the rotation 
of the magnetosphere is supersonic, dissipations at $R_{\rm M}$ induce turbulent motions that 
convect the dissipated neutron star rotational energy up through the outer boundary of the 
envelope, where this energy is dissipated. In this case, $n$=1/2 and the correct 
location of $R_{\rm M}$ should be computed by equating the neutron star magnetic pressure 
with the atmosphere pressure given above. We obtain: 
 \begin{equation}
R_{\rm M}=3.2\times10^{8} \upsilon_8^{2/9} \mu_{30}^{4/9} \rho_{-12}^{-2/9} ~{\rm cm}
\label{eq:rmsuper} 
\end{equation}
(where it was assumed that $R_{\rm M}$ $\ll$ $R_{\rm acc}$). In the supersonic propeller regime, 
the material around the neutron star is pushed away from the rotating magnetosphere and accretion cannot occur. 
Dissipations at $R_{\rm M}$ are expected to provide the largest contribution to the X-ray luminosity: 
\begin{equation}
L_{\rm X} = 2\pi R_{\rm M}^2 \rho(R_{\rm M}) c_{\rm s}^3
(R_{\rm M})\simeq8.2\times10^{34} \upsilon_8^{-1} \rho_{-12} ~{\rm erg ~s}^{-1}   
\end{equation}
($c_{\rm s}$ is the sound velocity). 

\item {\it The subsonic propeller regime.} When both conditions $R_{\rm M}$$<$$R_{\rm acc}$ and 
$R_{\rm M}$$<$$R_{\rm co}$ are satisfied, it is usually believed that the envelope around the 
neutron star still cannot accrete at the full regime indicated by Eq.~\ref{eq:acrrate} until material 
is cooled down below a certain critical temperature. If cooling is not efficient, the system enters 
the subsonic propeller regime. This is one of the most discussed regimes of wind accretion and 
different authors have made different assumptions regarding physical processes occurring 
in this case.  

The subsonic propeller regime was initially investigated by \citet{davies79} 
and \citet{davies1981}, using theoretical findings proposed by \citet{elsner1977} and 
\citet{arons1976}. The latter authors argued that, in a wind-fed system, the material accumulating 
around the neutron star magnetosphere cannot penetrate it until its temperature is low enough for the 
magnetohydrodynamic instabilities to be efficient at $R_{\rm M}$. 
Both \citet{elsner1977} and \citet{arons1976} suggested the Rayleigh-Taylor instability (hereafter RTI) 
to be the most efficient instability in transporting material across the neutron star magnetic 
field lines. A criterion to trigger the RTI was derived from first principles \citep{bernstein57}, 
assuming a situation of hydromagnetic equilibrium in which the fluid velocity vanishes at any point. 
Under these conditions, the boundary between the plasma and a vacuum magnetic field is unstable for 
the RTI if the temperature of the plasma is \citep[see also,][]{ikhsanov96,ikhsanov05}:  
\begin{equation}
T < T_{\rm crit}(R_{\rm M}) \simeq 0.3 T_{\rm ff}(R_{\rm M})=0.3 GM_{\rm NS} m_{\rm p}/kR_{\rm M}.   
\label{eq:temp}  
\end{equation}
Here, $m_{\rm p}$ is the proton mass and $k$ is the Boltzmann constant. 
In a more detailed treatment, it can be seen that $T_{\rm crit}$ has also a non-negligible dependence on 
the longitude of the neutron star magnetosphere, with the equatorial region being the least 
stable zone. If the condition of Eq.~\ref{eq:temp} is not satisfied, accretion is largely inhibited. 
In this situation, the neutron star spins at subsonic velocities within the surrounding shell  
and loses its rotational energy due to friction. As long as the dissipated rotational energy 
of the neutron star keeps the material in the spherical shell hot, the subsonic 
propeller regime can be sustained and the structure of the shell can be approximated 
as adiabatic. The magnetospheric radius can thus be estimated by using $n=3/2$ in Eq.~\ref{eq:hydro}:  
\begin{equation}
R_{\rm M} = 3.8\times10^{9} \upsilon_8^{6/7} \mu_{33}^{4/7} \rho_{-12}^{-2/7}~{\rm cm}. 
\label{eq:rmsubsonic}
\end{equation}
In the subsonic propeller regime the luminosity produced by the dissipation of the neutron star 
rotational energy can be written as:    
\begin{eqnarray}
L_{\rm sd3} =  2\pi R_{\rm M}^5 \rho(R_{\rm M}) \Omega^3 
= 6.2\times10^{35}  P_{\rm s2}^{-3}\cdot \\
\cdot R_{\rm M10}^5 \rho_{-12} (1+16 R_{\rm acc10}/(5 R_{\rm M10}))^{3/2} ~{\rm erg ~s}^{-1} \nonumber ,   
\end{eqnarray}
where $R_{\rm M10}$=$R_{\rm M}$/$10^{10}$~cm is given by Eq.~\ref{eq:rmsubsonic}, and 
$R_{\rm acc10}$=$R_{\rm acc}$/$10^{10}$~cm. A number of different prescriptions for the 
X-ray luminosity produced in the subsonic regime were proposed by different 
authors, and we refer the reader to the most relevant publications for further details 
\citep[see, e.g.,][and references therein]{ikhsanov01a,ikhsanov01b,ikhsanov01c}. It is noteworthy 
that in all cases the assumed key characteristics of this regime are a strong spin-down of 
the neutron star  and a moderate X-ray luminosity 
\citep[$\lesssim$10$^{33}$-10$^{34}$~erg~s$^{-1}$][]{ikhsanov01c,ikhsanov07}. 

The approach to the subsonic propeller regime discussed above was criticised by \citet{burnard83}. 
These authors argued that,  when the neutron star rotational velocity is taken into account, the 
basic assumptions behind Eq.~\ref{eq:temp} are no longer valid and different conclusions can be 
reached. In particular, the presence of a shear velocity between the neutron star magnetosphere 
and the material in the shell would excite the KHI. Even in those situations in which 
the RTI is not efficiently working due to the high temperature of the plasma, material could 
still be pushed through the compact object magnetic field lines by the KHI and be finally accreted 
onto its surface. \citet{burnard83} presented all calculations needed to estimate the 
amount of material that can be accreted through the KHI depending on the NS magnetic field and 
spin period (as well as from the physical conditions of matter outside the NS magnetosphere). 
Following the simplified treatment of \citet{Bozzo2008}, the mass luminosity in the subsonic propeller regime permitted by the KHI can be written as: 
\begin{eqnarray}
& & L_{\rm KH3} \simeq G M_{\rm NS} \dot{M}_{\rm KH}/R_{\rm NS}= 
1.5\times10^{39} \eta_{\rm KH} P_{\rm s2}^{-1} R_{\rm M10}^3 \rho_{-12} \cdot \nonumber \\
& & \cdot (1+16 R_{\rm acc10}/(5 R_{\rm M10}))^{3/2} 
(\rho_{\rm i}/\rho_{\rm e})^{1/2}
(1+\rho_{\rm i}/\rho_{\rm e})^{-1} ~{\rm erg ~s}^{-1} 
\end{eqnarray}
It is thus found that, even in the subsonic propeller regime, a significant amount of material 
can still be accreted onto the NS at odds with what was considered before. The conditions under
which the RTI becomes inefficient and the KHI begins to control matter penetration through 
the neutron star magnetosphere are still highly debated and difficult to be assessed. The problem 
is that these conditions depend strongly on the physical properties of matter and magnetic field 
at the magnetospheric boundary, which is an extremely complex region to be treated analytically. 
The conditions for the development of magnetohydrodynamic instabilities in the boundary are also 
affected by the X-ray photon field produced by the central accreting neutron star. In most of the 
cases of interest for this review, the photon flux changes significantly on rapid time scales 
(few to hundreds of seconds), thus making any analysis of the magnetospheric 
region even more complicated. 

\item {\it The subsonic settling accretion regime.} As mentioned before, it is usually assumed 
that when Eq.~\ref{eq:temp} is satisfied, the subsonic propeller is not applicable and all 
material captured by the neutron star at $R_{\rm acc}$ can be accreted at the Bondi-Hoyle rate 
(Eq.~\ref{eq:dotmratio}). \citet{Shakura2012} studied this process in a more detailed 
way. These authors solved analytically the structure of the spherical shell around a 
wind-accreting neutron star, assuming that its magnetospheric boundary is RTI unstable 
(i.e. assuming that Eq.~\ref{eq:temp} is valid and satisfied). They found that the Bondi-Hoyle rate 
can only be achieved when the X-ray luminosity from the central 
neutron star is $\gtrsim$4$\times$10$^{36}$~erg~s~$^{-1}$ (for typical values of the systems of 
interest for this review) and the inflowing material from the companion can be efficiently 
Compton cooled down already at $R_{\rm acc}$. At lower luminosities, a subsonic settling accretion 
regime is achieved. In this regime, the mean radial velocity,  $\upsilon_{\rm r}$, of the plasma filling 
the region between $R_{\rm M}$ and $R_{\rm acc}$ is lower 
than the free-fall velocity $\upsilon_\mathrm{ff}$ ($\upsilon_{\rm r}=f(\upsilon)\upsilon_\mathrm{ff}$, $f(\upsilon)<1$) and depends 
on the cooling time, $t_\mathrm{cool}$, near the magnetosphere, so that 
$f(\upsilon)\sim [t_\mathrm{ff}(R_\mathrm{acc})/t_\mathrm{cool}(R_\mathrm{acc})]^{1/3}$.
Two different sub-regimes are identified, depending on the dominant cooling mechanism at the 
magnetospheric radius. The authors showed that if Compton cooling at the magnetospheric 
radius dominates over radiative losses, then 
\begin{equation}
f(\upsilon)\sim 0.3 (L_{\rm X}/10^{36}~{\rm erg~s^{-1}})^{-2/11} \mu_{30}^{6/11}, 
\label{eq:sh03} 
\end{equation}
and thus the mass accretion rate onto the neutron star is reduced by roughly a factor of 3 
compared to the Bondi-Hoyle rate. While if radiative cooling dominates, then 
\begin{equation}
f(\upsilon)\sim 0.1 \zeta^{14/27} (L_{\rm X}/10^{36}~{\rm erg~s^{-1}})^{6/27} \mu_{30}^{2/27}, 
 \label{eq:sh01}
\end{equation}
and the mass accretion rate onto the NS is reduced by a factor of $\sim$10 compared to the 
Bondi-Hoyle rate (for typical parameters). In the above equation, $\zeta$$<$1 is a numerical 
parameter describing the extension of the transition zone inside and outside the 
neutron star magnetosphere \citep{Shakura2013}. The transition between the Compton and the 
radiative cooling regimes is mainly regulated by the neutron star X-ray luminosity. This value 
of the luminosity can be determined by comparing the cooling time scale in the two cases. As 
detailed in \citet{Shakura2013}, the time scale for the Compton cooling is 
\begin{equation}
t_{\rm Compton} \simeq 10 (R_{\rm M}/10^9~{\rm cm})^2(L_{\rm X}/10^{36}~{\rm erg~s^{-1}})^{-1}\,{\rm s}
\end{equation}
while for the radiative cooling is 
\begin{equation}
t_{\rm radiative} \simeq 1000 (R_{\rm M}/10^9~{\rm cm})(L_{\rm X}/10^{36}~{\rm erg~s^{-1}})^{-1}(f(\upsilon)/0.3)\,{\rm s}
\end{equation}
For typical parameters, it can be seen that Compton cooling dominates for\linebreak[4] 
$L_{\rm X}$ $\gtrsim$3$\times$10$^{35}$~erg~s$^{-1}$. 

\item {The direct accretion regime.} At sufficiently high mass inflow rate from the companion star, 
the material outside the neutron star magnetospheric boundary reaches a critical density above 
which it is cooled down efficiently for the largest accretion rate to take place (i.e. the 
Bondi-Hoyle rate in Sect.~\ref{sec:bondi}). According to the calculations in \citet{davies1981} 
and \citet{Bozzo2008}, the critical density above which such transition occurs can be written as: 
\begin{equation}
\rho_{\rm lim_{-12}} = 8.3\times10^2 P_{\rm s2}^{-3} R_{\rm M10}^{5/2} (1+16 R_{\rm acc10}/(5 R_{\rm M10}))^{-3/2}. 
 \label{eq:rholim}
\end{equation} 
The accretion of material with the above density would roughly release an X-ray luminosity of  
$\gtrsim$4$\times$10$^{36}$~erg~s~$^{-1}$, that is the transition luminosity estimated by 
\citet{Shakura2012}. In the direct accretion regime, all material captured at $R_{\rm acc}$ 
can be readily accreted onto the NS, and the X-ray luminosity released from the system is: 
\begin{equation}
L_{\rm X} = L_{\rm acc}= G M_{\rm NS} \dot{M}_{\rm capt} / R_{\rm NS} = 
5.9\times10^{36} \upsilon_8 \rho_{-12} R_{\rm a10}^2 ~{\rm erg ~s}^{-1}. 
\end{equation}
This is the highest X-ray luminosity achievable by a wind-fed neutron star in a SgXB.  
\end{itemize} 

Some of the accretion regimes discussed above have been also investigated through numerical 
simulations, especially the direct accretion regime \citep{Toropin1999,Toropina2012} and the 
supersonic propeller \citep{Romanova2003}.  Strongly magnetized neutron stars were also considered  
\citep{toropina2006}. The general results obtained within these simulations 
are qualitatively in agreement with the theoretical predictions discussed above, 
but a more quantitative comparison would require the usage of full 3D numerical 
simulations that so far have been developed only for the case of disk accretion and for weakly magnetized 
neutron stars \citep[see, e.g.,][and references therein]{Kulkarni2008,Romanova2012}. The issue of extending 
these simulations to the case of strongly magnetized neutron stars is that for the latter the magnetospheric radius 
is located at hundreds of stellar radii, thus making the numerical approach used by the simulations challenging.

\subsubsection{X-rays and stellar wind interaction}
\label{sec:interaction}

In all accretion regimes previously discussed, a conspicuous amount of X-ray radiation is released 
by the compact object and expected to interact with the surrounding environment. One of the most important effects 
resulting from this interaction in SgXBs is the reduction of the radiative force that accelerates the supergiant star wind due to the 
photoionization of heavy ions. The reduction is strongly related to the value of $L_{\rm X}$: for sufficiently high X-ray luminosities 
it is possible that most of the heavy ions in the stellar wind become so largely photoionized that the stellar wind 
is halted already at the supergiant photosphere, practically inhibiting the formation of a proper stellar wind. 
The extension of the region in which the wind velocity is affected by the X-ray radiation (measured from the position 
of the compact object) is called Str\"omgren sphere. Note that the inhibition of the stellar wind acceleration has a potentially 
dramatic effect on the accretion process and the emission of the X-ray radiation, as the latter can only be powered when the 
stellar wind material is channelled on the compact object surface. An elegant analytical approximation describing this feedback 
mechanism between the wind acceleration and the release of the accretion luminosity was proposed in the seminal paper of 
\citet{Ho1987}, and thus we summarise here briefly their approach. 

It is assumed that the wind of the supergiant star is accelerated according to Eq.~\ref{scalerel} up to a distance $r_{\xi}$ from the 
star, where $r_{\xi}$ is the radius of the Str$\ddot{\rm o}$mgren sphere. From this distance onward, the stellar wind moves toward 
the neutron star with a constant velocity (due to the lack of any efficient acceleration) according to the equation: 
\begin{displaymath}
\upsilon_{\rm w}= \left\{ \begin{array}{ll}
\upsilon_\infty (1-R_*/r)^{\beta} & \textrm{if $r$ $<$ $r_{\xi}$}\\
\upsilon_\infty (1-R_*/r_{\xi})^{\beta} & \textrm{if $r$ $\geq$ $r_{\xi}$}. 
\end{array} \right.
\end{displaymath}
Here $r$ is the distance from the supergiant star, and $r_{\xi}$ is defined by the equation
\begin{equation}
\frac{L_{\rm X}}{n(r_{\xi})(r-r_{\xi})^2}=\xi_{\rm cr}. 
\label{eq:xicr} 
\end{equation}
If we assume that $n(r)$ $\simeq$ $\rho_{\rm w}(r)$/$m_{\rm H}$, $\rho_{\rm w}(r)$ is given by Eq.~\ref{eq:rho}, 
$L_{\rm X}$ is given by Eq.~\ref{eq:lx}, $m_{\rm H}$ is the proton mass, and $\xi_{\rm cr}$ is the critical value of the 
so-called ionization parameter $\xi$ at which the wind acceleration is completely cut-off by the X-ray ionization (i.e. 
the stellar wind becomes virtually transparent to UV photons). 
The value of $\xi_{\rm cr}$ is strongly dependent on the chemical composition of the stellar wind and on the details 
of the micro-physics intervening in the ionization process. \citet{Ho1987} assumed $\xi_{\rm cr}\sim10^4$, but such value 
was then revised to $\xi_{\rm cr}\sim3\times10^2$ by \citet{stevens91}. By putting together 
Eq.~\ref{eq:xicr}, \ref{eq:lx}, and \ref{eq:rho}, we obtain 
\begin{equation}
\xi_{cr}=4\pi m_{\rm H} \frac{G^3 M_{\rm NS}^3}{R_{\rm NS}} \frac{1}{a^2\left(\frac{a}{r_{\xi}}-1\right)^2}\frac{1}{\upsilon_{\rm w}^3(r_{\xi})}.  
\label{eq:feedback} 
\end{equation}
This is the so-called feedback equation of \citet{Ho1987} and can be solved to derive the distance $r_{\xi}$ from the NS at 
which its X-ray radiation is able to stop the acceleration of the companion star wind, in turn regulating the mass accretion 
rate onto the compact object and the production of the X-ray emission itself. \citet{Ho1987} showed that Eq.~\ref{eq:feedback} 
has typically two solutions: the first one corresponds to the case in which $r_{\xi}$ $\simeq R_{*}$ and the wind velocity 
is very much reduced compared to that predicted by the CAK model. This is called the high luminosity solution because for such 
low velocity Eq.~\ref{eq:raccr} and \ref{eq:lx} predict a large X-ray luminosity. In the low luminosity solution, $r_{\xi}$ $\simeq a$ and 
the expected X-ray luminosity is much lower due to the higher wind velocity (the wind can be accelerated  up to the neutron star location). 
Originally, the presence of these two branches of solutions was used to explain the differences between brighter (Cyg\,X-1, SMC\,X-1) and 
fainter HMXBs (Vela\,X-1, 4U\,1700-37, 4U\,1538-52). More recently, their applicability 
to the case of wind accreting HMXBs was revised and discussed by \citet{karino14}, who also pointed out the limitations of  
one dimensional calculations in the case of binary systems. \citet{Ducci2010} also presented an extension of the previous 
\citet{Ho1987} calculations, by including in the feedback equation a term describing the reduction in the mass accretion rate onto the 
neutron star surface caused by the wind material that is not sufficiently accelerated after the ionization to escape the supergiant 
companion and to reach the compact object. 

The convincing observational evidences of the disruption of stellar winds in SgXBs \citep[see, e.g.][]{Watanabe2006} provided 
the required momentum to stimulate refined models of the interaction between the X-rays and the surrounding medium in these systems. 
Starting from the pioneering paper of \citet{Blondin1995}, in which the first two dimensional calculations of the stellar wind 
photoionization in HMXBs were presented, a number of recent studies were presented which include a more realistic description 
of the stellar wind composition and the photoionization process. We discuss the results of these papers in more details in 
Sect.~\ref{xraypars}. 

We note that a different way to probe the interaction between the X-rays from the neutron star 
and the surrounding stellar wind is offered by the so-called HM-effect \citep{hatchett77}. The presence 
of the Str{\"o}mgren sphere around the compact object was demonstrated to induce noticeable variations in 
the UV resonance lines produced within the massive star wind and providing measurements of the local 
wind velocity. When applied to the case of Vela\,X-1, the results of detailed spectroscopic observations 
obtained in the UV domain allow us to compare the properties of the photoionized region around the neutron 
star with that determined from complementary investigations in the X-ray domain \citep{vanloon2001,Watanabe2006}. 
The major limitation impacting the exploitation of the HM-effect to study photoionization in wind-fed binaries 
is the availability of high resolution UV spectra of these systems with a sufficiently high signal-to-noise 
ratio. So far the HM-effect could be verified only in the cases of Vela\,X-1 and 4U\,1700-37, as these are 
the closest-by and brightest wind-fed SgXBs. For all other known sources of the same class, no data of the 
required quality could be obtained so far in the UV domain and thus the applicability of the HM-effect 
as remained largely unexplored \citep[see][and references therein]{vanloon2001}.

\section{Stellar and wind parameters of massive stars}
\label{stellarwindpars}

\subsection{Quantitative spectroscopy}
\label{stellarwindpars:spec}

Stellar and wind-parameters of hot massive stars are
conventionally determined by means of {\it quantitative spectroscopy},
i.e., by fitting synthetic spectral energy distributions and
normalized spectra\footnote{Colours are less meaningful (or not usable at all),
because of the high temperatures.} to observations, mostly in the
optical and/or the UV range. Meaningful mass-loss diagnostics are also
provided by observations in the X-ray (see Sect.~\ref{sec:Xraywithmacro}) and 
far-IR/mm/radio domain. 

For calculating synthetic energy distributions, a consistent treatment
of photosphere {\it and} wind requires so-called {\it unified model
atmospheres}, at least if the \newline lines/continua are formed outside the
(quasi-) hydrostatic region. Such unified model atmospheres have been 
introduced by \citet{Gabler89}. By means of a typical velocity law
(with $\beta = 1$, see Eq.~\ref{scalerel}), one can derive a maximum
\mdot\ for which a hydrostatic treatment is still possible, at least
in the optical and most of the UV. Estimating a limiting Rosseland
optical depth $\taur < 10^{-2}$ at the transition between the
photosphere and wind (roughly located at 10\% of the sound-speed), one
finds $\mdot < 6\cdot10^{-8} \msunyr (\Rstar/10\,\rsun)
(\vinf/1000~\kms)$. Comparing with "observed" (i.e., empirically
derived) mass-loss rates, this limit implies that hydrostatic models
are still largely sufficient for late O-dwarfs, B-stars up to
luminosity class II (for early sub-types) or Ib (for mid/late
sub-types), and A-stars up to luminosity class Ib.  Above this limit,
unified model atmospheres have to be used, since the spectral
appearance is considerably affected by wind effects, even if there are
no apparent wind features in a particular observation.  Note that
unified model atmospheres are computationally expensive, because of
the coupling between velocities and frequencies via corresponding
Doppler-shifts. Moreover, all models, even the purely hydrostatic
ones, need to account for NLTE conditions, due to the high
temperatures and comparatively low densities.

Table~\ref{codes} compares present state-of-the-art atmospheric codes
which can be used for the spectroscopic analysis of hot stars. Since
the codes {\sc d}etail/{\sc s}urface and {\sc tlusty} 
calculate occupation numbers/spectra on top of hydrostatic, plane-parallel
atmospheres, they are "only" suited for the analysis of stars with
negligible winds (see above). The different computation times are
mainly caused by the different approaches to deal with
line-blocking/ blanketing. The overall agreement between the various
codes (within their domain of application) is quite satisfactory,
though certain discrepancies are found in specific parameter ranges,
particularly regarding EUV ionizing fluxes \citep{Puls05, SimonDiaz08}
and surface gravities \citep{Massey13}.

\begin{table}
\caption{Comparison of state-of-the-art, NLTE, line-blanketed model
atmosphere codes suited for the analysis of hot massive stars
\label{codes} }
\footnotesize
\tabcolsep1.0mm
\begin{center}
\begin{tabular}{@{}l|ll|lllll}
     \hline
code  &  {\sc d}etail/ &  {\sc tlusty}$^2$ &  {\sc p}o{\sc wr}$^3$ &  {\sc
phoenix}$^4$ & {\sc cmfgen}$^5$ & {\sc wm}-basic$^6$ &  {\sc fastwind}$^7$  \\
      & {\sc s}urface$^1$ &  &  &  &  &  &  \\
\hline
 geometry  &  plane- &  plane- &  spherical &  spherical/ & spherical &  spherical &  spherical \\
       &  parallel  &  parallel &           &  pl.-parallel  &        &     &   \\
     \hline
  blanketing &  LTE &  yes &  yes &  yes & yes &  yes &  approx. \\
     \hline
  diagnostic &  no &  no    &  no  &  no &  no  &  UV      & optical/IR\\
  range    &  limitations &  limitations &  limitations &  limitations &  limitations &    &    \\
     \hline
  major &  BA stars &  hot stars &  WRs, &  cool stars, &  OB(A)- &  hot stars w. & OB-stars,  \\
 application &  with negl.&  with negl.& OB-stars & SNe         &  stars, &  dense winds, & early A-sgs  \\
             &  winds     &  winds     &      &              &  WRs, SNe& SNe &     \\
     \hline
 comments &  no wind  &  no wind & \qquad-     &  no clumping      & \qquad-  & no clumping      &  no X-rays \\
          &           &          &          &  no X-rays        &
	  &     & (in progress) \\
	     \hline
 execution &  few    &   hours   &  hours   &  hours    &  hours &  1 to 2 h & few min. \\
 time     &  minutes  &             &            &     &      &  & to 0.5 h  \\
     \hline          
\end{tabular}
\end{center}
(1)~\citet{Giddings81, ButlerGiddings85}, (2)~\citet{Hubeny98}, 
(3)~\citet{Graefener02}, (4)~\citet{Hauschildt92}, (5)~\citet{HillierMiller98}, 
(6)~\citet{Pauldrach01}, (7)~\citet{Puls05}
\end{table}

The best fitting model-spectra and thus the corresponding stellar and
wind parameters are found by either a simple `by-eye'-inspection
(e.g., \citealt{Repolust04}), or by minimization methods (using a
genetic algorithm, e.g., \citealt{Mokiem05}, or pre-calculated
model-grids, e.g., \citealt{LH2003, Hamann04, Lefever10, SimonDiaz11, SHT2012, Hainich2014}).
It should be noted that a proper comparison between observed
and model spectra can only be automatized to a certain degree as it
requires detailed knowledge about which features are affected by which
parameters. 
Furthermore, such comparisons should be made with as
many lines as possible, as a particular single line can be affected
by a variety of parameters which can lead, in different combinations, 
to the same line shape. 

\subsection{Stellar parameters} Regarding the individual stellar
parameters, the following prime diagnostics are available: effective
temperatures, \Teff, are derived from the ionization
equilibrium\footnote{By fitting the corresponding {\it photospheric}
lines/line-strength ratios from two (or more) neighbouring ionization
stages} of nitrogen (O2, O3, and WN), helium (WR and O-stars up to
O4), silicon (B-stars), and magnesium (A-stars). Typical accuracies are 5\%.
Surface gravities are derived from the wings of the Stark-broadened
Balmer lines ($\pm$ 0.1 dex). This method works only for purely
photospheric absorption lines, i.e., \Hg\ and/or \Hd. Thus, if the
available spectral range consists of emission lines only, there is no
direct spectral diagnostics to infer the effective surface gravity.
Radii and thus luminosities are obtained from comparing the
theoretical and absolute observed fluxes\footnote{If this is done over
a wider range, the term \emph{spectral energy distribution} (SED) is
typically used.}, or from using \Teff, \MV\ and the bolometric
correction. The distance is the major source of error here.  The
abundances of individual elements are finally determined from fits to
their corresponding spectral lines (in the optical and the UV), where,
if possible, photospheric lines should be used, to avoid contamination
from wind-inhomogeneities and wind-embedded shocks. Corresponding
errors depend on the complexity of the line-formation and on the
number of available diagnostic lines. E.g., for BA-stars very accurate
measurements of the abundances are possible for many
elements (0.05 to 0.1 dex, see \citealt{Przybilla08}), whilst for
nitrogen and carbon in O-stars the typical errors are considerably
larger (0.15 to 0.2 dex or higher), due to the complex line-formation
\citep{Rivero12, Martins12}, and the scarcity of lines. Finally, we
note that a complete analysis involves many more parameters, such as
projected rotational velocity, \vsini, micro- and macro-turbulence. A
discussion of these parameters is beyond the scope of this review, just
to mention that the measurement of \vsini\ is quite precise if the
rotational speed exceeds 100 \kms, and it becomes problematic for speeds
below 40 \kms (e.g., \citealt{Sundqvist13a}).

\subsection{Velocity field} Since the measurement of \mdot, one of the
most important parameters in this context, will be detailed in the
next section, we will concentrate here on the velocity field. Terminal
velocities \vinf\ can be quite easily measured from the blue
edges\footnote{Or from the black troughs, if present} of UV P Cygni
profiles (typical accuracy: 10\%), as long as these profiles display
strong absorption (and the UV has been observed!). Weaker resonance
lines (e.g., from late O and B-dwarfs) pose a severe problem in this
regard, since it is quite likely that the apparent edge velocity is
lower than the actual \vinf. When possible, the asymptotic behaviour
of DACs (Sect.~\ref{sec:largescale}) can be utilized here.  
In many cases,
additional spectral modelling is necessary to confirm velocity
measurements from UV resonance lines, since those lines are located in
the iron forest which often dominates the spectral appearance in the
UV.  If, however, the UV has not been observed, scaling
relations derived from a multitude of measurements can be used (e.g.,
\citealt{KP00}), or, in the case of mid/late B- and A-supergiants,
\vinf\ might be derived from \Ha \citep{Kud99} which displays a P
Cygni profile for these spectral types. Finally, the slope of the
velocity law (usually corresponding to the value of
$\beta$\footnote{Sometimes, two velocity laws with different $\beta$
are combined, e.g. \citealt{Todt2010}}) can be determined from the
shape of the UV P Cygni emission and/or the shape of \Ha/\HeII\,4686,
if in emission. We note, however, that particularly the latter optical
diagnostics also depend on the degree and stratification of the wind
clumping, thus introducing a significant dichotomy.  
Since a particular line can easily be affected by
a variety of phenomena, we stress again that all diagnostics should not 
rely on a single line alone, but needs to be performed in combination
with a reasonable fit to the overall available spectrum. Most unified
model atmospheres are applicable for a wide spectral range
(cf.\,Table~\ref{codes}), and the better the overall reproduction of
the observed spectrum is, the less is the chance to mis- or
over-interpret a particular feature.


\subsection{Mass-loss diagnostics and the effects of wind clumping}  
\label{sec:mdotclump}
\smallskip\noindent

The most widely used diagnostics to determine mass-loss rates
from O and WR stars are: {\em i)} thermal radio continuum emission; {\em
ii)} ultraviolet resonance lines, and {\em iii)} UV and optical
emission lines (see \citet{Lamers1999} for details).

Besides these well established methods, other empiric diagnostics of
mass-loss rates have been suggested. E.g., \citet{cohen2010} proposed
to measure mass-loss rates by analysing X-ray emission line profiles.
\citet{pm2013} suggested that the ratio of narrow features observed at
terminal velocities in the UV resonance lines of B-supergiants
provides a clumping independent tool to estimate mass-loss rates in
these objects.  This may be especially important because in B-type
supergiants, the H$\alpha$ line might be optically thick, and thus
lose its diagnostic value \citep{petrov2014}. \citet{Kobulnicky2010}
suggested that IR observations of the circumstellar medium around
massive stars provide a new laboratory for estimating stellar mass
loss rates. This idea was applied in \citet{Hub2011} and further
developed in \citet{Gvar2012}. \citet{Huen2012} suggested that the
X-ray flux may provide a reliable mass-loss diagnostics in late type
O-dwarfs. All this is by far not an exhaustive list of possible
mass-loss diagnostics, but below we will concentrate on already well
established methods.  

The thermal radio emission is easy to interpret
\citep{WrightBarlow1975}, but it can only be observed in a small
number of nearby stars.  The other two diagnostics are based on
spectroscopic models, and their interpretation is not straightforward.
Many investigations are restricted to a semi-empirical modelling of the
line profiles, e.g.\ using the ``Sobolev with exact integration
method'' \citep[SEI,][]{Hamann1981,Lamers1987} and a subsequent 
interpretation of the derived optical-depth parameters. Not being
based on consistent models, the obtained mass-loss rates are highly
uncertain (e.g.\ because of the unknown degrees of ionization).
Moreover, most resonance lines in the UV range (e.g.\ N\,{\sc v},
C\,{\sc iv}) are often saturated and therefore not suitable for
precise $\dot{M}$ determinations. The far-UV (FUV) includes the
P\,{\sc v} resonance line which is very useful because it is
unsaturated in typical O stars \citep{bouret2005,Fullerton2006}, due
to the low phosphorus abundance. However, since only a few instruments
-- such as the FUSE satellite -- have been capable to observe in the FUV
range, such data are rare.

A quantitative analysis of the emission\footnote{for low
wind-densities, the wind emission `only' fills in the photospheric
absorption} lines (e.g.\ H$\alpha$), as well as a reliable
interpretation of the UV resonance lines, requires a detailed modelling
of the stellar atmosphere and wind.  The conditions in the supersonic
winds from hot stars deviate extremely from local thermodynamical
equilibrium, therefore the spectral analysis has to be conducted by
means of appropriate stellar atmosphere models (see 
Table\,\ref{codes}). The advances in the radiative transfer offered by
these NLTE codes are huge, however, the geometries that are adopted
are rather simple. Typically, only spherically symmetric stellar winds
are described by the NLTE stellar atmospheres. Consequently, wind
clumping can be included in these sophisticated codes only using some
approximations.

\subsection{Microclumping}

The most stringent possible approximation is that clumping does not
directly\footnote{but note that, e.g., optical depths can change, see
below} affect the radiative transfer in a stellar wind. This would be
the case when wind clumps are assumed to be optically thin at {\em
all} frequencies. This approximation is called ``microclumping'' (or
optically thin clumping) and was implemented in such codes as PoWR and
CMFGEN already back in the 1990s.

It is assumed that inside clumps the density is uniform, and
enhanced by a factor $D$ compared to a smooth model with the same
mass-loss rate \mdot. The volume filling factor of the clumps is
$f_{\rm V} = D^{-1}$, because the interclump medium is assumed to be
void.  Thanks to the latter assumption, the rate equations have to be
solved only for the clump medium, 
i.e.\ for the enhanced density $\rho_{\rm C} =
D\rho$. From the obtained population numbers, the non-LTE opacity and
emissivity of the clump matter, $\kappa_{\rm C}(D\rho)$ and $\eta_{\rm
C}(D\rho)$, can be calculated.

In the radiative transfer equation, the smooth-wind opacity and
emissivity $\kappa(\rho)$ and $\eta(\rho)$ must be replaced for a
clumped wind by 
\begin{equation}
\kappa_{\rm f} = f_{\rm V}\ \kappa_{\rm C}(D \rho) ~~~\mathrm{and}~~~ 
\eta_{\rm f} = f_{\rm V}\ \eta_{\rm C}(D \rho)\   .
\label{eq:kappa_D}
\end{equation}

The atomic transition rates contributing to the opacity and emissivity
scale with different powers of the density depending on their respective
nature. For processes linear in density, $f_{\rm V}$ and $D$ cancel,
but contributions scaling with the square of the density (bound-free
emission, or free-free absorption and emission) are effectively enhanced
by a factor of $D$, which may depend on the  location in the stellar
wind, i.e.\ on the radial coordinate $r$.

Empirical mass-loss diagnostics are widely based on processes that scale
with the square of the density. The wind emission lines in Wolf-Rayet
and O stars, including H$\alpha $, form in de-excitation cascades that
are fed by radiative recombination. The thermal radio emission from
stellar winds is due to the free-free process. Consequently, when a
given (radio or line) emission is analysed with a model that accounts
for microclumping, the derived mass-loss rate will be lower by a factor
of $\sqrt{D}$ relative to a smooth-wind model.

Fitting the observed electron scattering wings of strong emission
lines in Wolf-Rayet spectra can be used to determine the clumping
factor $D$ and its radial dependence \citep{hil1991,ham1998}. For WR
stars this method yields typical clumping factors $D$ between 4 and 10
\citep[e.g.,][]{ham1998}. Unfortunately, the method is not applicable
to O stars, since their spectra do not show suitable emission lines.
In this case, only the resonance lines remain as a density-linear
diagnostic among the mass loss indicators, especially the unsaturated
P\,{\sc v} doublet. 

For these stars too, the mass-loss rate derived from smooth
models and density-squared diagnostics (H$\alpha$, infra-red and radio
emission) needs to be scaled down by the square root of the clumping
factor.

Strong clumping and correspondingly low mass-loss rates were obtained
by \citet{bouret2005} for two O stars analysed with non-LTE model
atmospheres. \citet{Fullerton2006} exploited the fact that the far-UV 
spectral range observed with FUSE contains the P\,{\sc v} resonance
line, which is typically unsaturated in O-star spectra, and analysed
this line in spectra of 40 O-type stars with the SEI method. They
found a large discrepancy between mass-loss rates obtained from
$\rho^2$ diagnostics (such as the analysis of H$\alpha$ line or radio
emission) based on unclumped models and mass-loss rates from P{\sc v} 
lines. To overcome this discordance it was suggested that the clumping
is very strong ($f_{\rm V}\lsim 0.01$) and therefore mass-loss rates
have to be reduced by orders of magnitude.

The drastic revisions of stellar mass-loss rates because of clumping
suggested from FUV diagnostics were in apparent agreement with
insights gained from the analysis of high-resolution X-ray spectra of
O-stars. According to the widely accepted scenario 
(e.g., \citealt{Feldmeier97b}, see Sect.~\ref{sec:smallscale}), 
X-rays from hot-star winds are produced in
shock-heated gas, while the bulk of the stellar wind is cool and
absorbs part of the X-rays before they can emerge.  The X-rays are
mainly emitted in spectral lines, while the absorption is continuous
(K-shell photoionization). This decoupling of emission and absorption
greatly simplifies modelling of the transfer of X-rays in stellar wind.
The shape of emission line profiles for such a situation was predicted
by \citet{macf1991}.

\begin{figure}[t]
\centering \includegraphics[width=8cm]{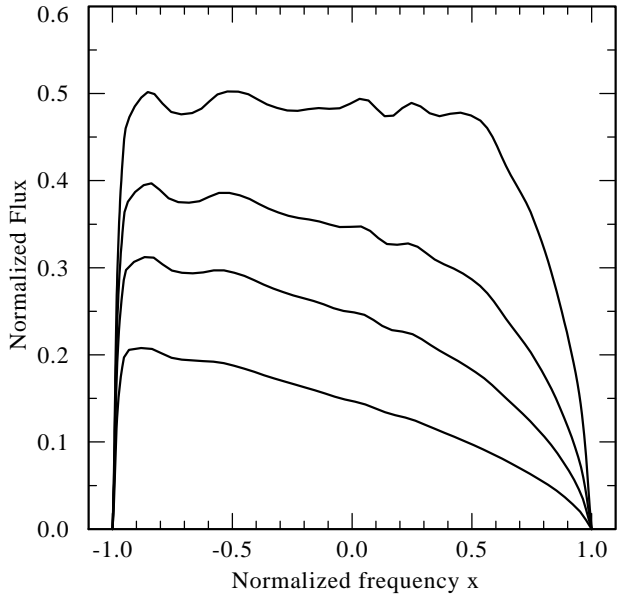}
\caption{The X-ray emission line profiles based on \citet{macf1991}
assuming a smooth wind with constant wind velocity.  For convenience
we display the computed line profiles as a function of the dimensionless
frequency $x$, measured relative to the line centre and in Doppler
units referring to the terminal wind velocity. The negative $x$ refers
to blue frequencies. From the top to the bottom line profile, 
$\kappa_\lambda=1,\,10,\,20,\,40$.  Occultation by the stellar core is
included in the model. The lines are not smooth because the X-ray
emission is assumed to originate from a limited number of small parcels
of hot gas randomly distributed in the wind (see details in \citet{osk2004}).}
\label{fig:smoothlines}
\end{figure}

\citet{macf1991} considered the optically thin emission from a
radially expanding shell of hot gas which suffers absorption in the
smooth cool stellar wind.  When this absorption is small, the line is
broad and has a box-like shape. For stronger wind absorption, the line
becomes more skewed (see Fig.\,7 in \citet{macf1991}). The
optical depth along the radial line of sight is given by
$\tau_\lambda=\int \chi_\lambda(r){\rm d}r$, where the atomic opacity 
($\chi_\lambda = \rho_{\rm w} \kappa_\lambda$) is the product of the
mass absorption coefficient  ($\kappa_{\lambda}$ [cm$^2$\,g$^{-1}$]), and
the density of the cool wind ($\rho_{\rm w}$). The latter is obtained 
from the continuity equation $\dot{M}=4\pi\rho_{\rm w}(r)
\upsilon(r)r^2R_\ast^2$, where $r$ is the radial distance in units of
$R_\ast$, and $\upsilon(r)$ is the velocity law that is prescribed by the
expression
(see also Eq.~\ref{scalerel})
$\upsilon(r)=\upsilon_\infty(1-1/r)^\beta
\equiv \upsilon_\infty w(r)$. 
Neglecting the radial dependence of the mass 
absorption coefficient, the optical depth can be parameterized as 
$\tau_\lambda=\tau_0\int w(r)^{-1}r^{-2}{\rm d}r$, where  
\begin{equation} 
\tau_0=\frac{\kappa_\lambda \dot{M}}{4\pi \upsilon_\infty R_\ast}.  
\label{eq:t0}  
\end{equation}  
Assuming that the visible emission most likely originates far from the
stellar core in the wind regions with constant velocity,
\citet{macf1991} adopted a constant velocity ($\upsilon_\infty$). The line
shape is largely determined by the parameter $\tau_0$.
\citet{macf1991} noticed that when $\tau_0$ increases, the red-shifted
part of the line ($\Delta\lambda>0$) becomes significantly more
attenuated than the blue-shifted part (see
Fig.\,\ref{fig:smoothlines}). They suggested that evaluating the line
shape can be used to determine $\tau_0$ and the wind density. 

The K-shell opacities vary with wavelength with a power between 2 and
3 (e.g., \citealt{verner1996b}). Therefore $\tau_0$ should change by
orders of magnitude in the X-ray band. Consequently, the X-ray
emission line shape at shorter and longer wavelengths should be
different.

The follow-up works by \citet{ign2001} and \citet{oc2001}
applied the \citet{macf1991} results also for X-ray emission lines,
where the latter improved upon \citet{macf1991}'s model by, e.g.,
assuming a spatially distributed emitting plasma with velocity
kinematics (used also to calculate the cool wind density) according to
the "beta-law" given above (which can lead to quite different
line-profiles than when assuming emission only at a specific
location). 

%

The first high-resolution X-ray spectrum of an O star was obtained by
\citet{wc2001}. Their analysis of the MEG spectrum of the O9.7Ib star
$\zeta$\,Ori\ revealed that the hot plasma is located relatively close
to the stellar photosphere and that the line profiles are nearly
symmetric and not skewed. \citet{wc2001} expanded the single-shock
model of \citet{macf1991} and considered emission from spherically
symmetric shocks equally distributed between 0.4\vinf\ and 0.97\vinf\,
with temperatures ranging from 2 to 10\,MK. They found that to explain
the nearly symmetric X-ray emission line profiles, the mass-loss rate
from the O-type supergiant $\zeta$\,Ori\ must be an order of magnitude
lower than derived from radio and UV-diagnostics \citep{Lamers1976,
Abbott1980}.

Similarly, much less skewed than expected X-ray line profiles were
measured from other O-type stars. Fitting observed lines with the
\citet{oc2001} model generally yields quite low values of $\tau_0$, 
with weaker dependence on wavelength than expected from a smooth wind
with corresponding mass-loss rates derived from, e.g., H$\alpha$ using
smooth wind models \citep{cas2001, kahn2001, kramer2003}. One possible
explanation of these results is a thinner wind than predicted by 
``standard'' theoretical wind models like \citet{Vink00}. 

\subsection{New evolutionary scenario for massive stars?}

Thus, about a decade ago the massive star community encountered a
severe problem -- based on UV and X-ray diagnostics, the empirically
estimated mass-loss rates from O-type stars were significantly lower
than predicted by the theory. These low mass-loss rates were the basis
for \citet{so2006} to conclude that the ``reduced mass-loss rates mean
that steady winds are simply inadequate for the envelope shedding
needed to form a WR star''. Hence, they proposed a new scenario
according to which the stars with initial mass above $40 {\ldots} 50\,M_\odot$
loose the bulk of their mass not via radiation line-driven
stellar winds, but via optically thick, continuum-driven outbursts (or
by hydrodynamic explosions). It was also suggested that Wolf-Rayet
type stars may not be the descendants of the most massive stars, with
the latter exploding as SNe at the LBV evolutionary stage.

This far reaching scenario challenged stellar evolution calculations 
\citep[see also][]{groh2014} but was largely grounded on the empirical
mass-loss rates derived using the assumption of microclumping.

\subsection{Macroclumping}

While the microclumping approximation is extremely convenient it was
always clear that this approach is too stringent to describe realistic
winds. 

\citet{Massa2003} and \citet{Fullerton2006} realized that for strong
lines  in dense winds microclumping is not justified. They  discussed 
that correctly accounting for clumping in UV resonance lines can affect 
the formation of P Cygni lines and, in extreme cases,  produce an
apparently  unsaturated profile for a line that would be  extremely
saturated if the  wind material were distributed smoothly. 
\citet{pm2010} studied B-supergiant winds and found wide-spread
signatures of optically thick clumping in spectra of these stars.  

\citet{osk2007} demonstrated that the microclumping approximation is
not suitable for empirical mass-loss diagnostics based on UV resonance
lines formed in massive star winds.  They developed a radiative
transfer technique that describes realistic stellar winds without
relying on the microclumping approximation. This approach, which is
called ``macroclumping'' (using the physical analogy with micro- and
macroturbulence) was successfully used to demonstrate the concordance
of empirical mass-loss rates derived from $\rho$ and $\rho^2$ 
diagnostics. 

It was shown that while microclumping is not suitable 
for UV-resonance lines, it still can be used for the H$\alpha$-line in
O-supergiants. As a result, the mass-loss rate should be reduced by a
factor $\sqrt{D}$ (see Eq.\,\ref{eq:kappa_D}) compared to the 
unclumped models. In case of $\zeta$ Puppis, a satisfactory fit to 
H$\alpha$ was produced with $D=10$. Consequently, the mass-loss rate
was shown to be lower by a factor 2-3 than in unclumped models. 
Independent work by other authors (see below) largely
confirmed this approach also in other O-supergiants. 

Waiving the microclumping approximation in modelling UV resonance lines
requires to account for clumps of arbitrary optical thickness in line
radiative transfer. For radiative transfer in lines, a clump opacity
depends on its geometry, abundances, and velocity field. The latter 
is because with lower velocity dispersion within a clump, the Doppler
broadening is smaller, and the line absorption profile is narrower but
peaks higher. Thus, the clump optical depth in the line core becomes
larger, while it is smaller in the line wings. Higher optical depth in
the line core leads to a reduction of the effective opacity and a
weakening of the line \citep[see Fig.\,5 in][]{osk2007}. In a
supersonically expanding medium, rays of a given (observer's frame)
frequency can only interact with clumps near the surface of constant
radial velocity (CRVS). The line opacity of the other clumps is
Doppler-shifted out of the resonance. The porosity effect for lines is
very pronounced \citep{ham2008}, and this porosity stems from the
fact that the clumps are not covering the complete \textit{velocity
field}, rather than from clumps that are spread out \textit{spatially}
in the wind. As such, other authors \citep{Owocki2008, Sundqv2010,
sund2014} explicitly differentiate between porosity in physical and
velocity space. 

To implement the macroclumping approach in PoWR NLTE model
atmospheres, a one-component wind model was considered, where the wind
consists only of clumps (assumed to follow the "$\beta$-law" velocity
law, see above) and is void between them. Moreover, the clumps were
treated statistically, with two key parameters describing clumped wind
properties, $L_0$ and $\upsilon_{\rm D}$, where the former is a parameter
setting the average separation between clumps in the wind, and the
latter describes the velocity distribution within clumps.  The
macroclumping formalism was implemented in the formal solution of the
radiative transfer equation. The synthetic spectra were compared to
the observed, and it was shown that mass-loss rates inferred from
optically thin emission, such as the H$\alpha$ line in O stars, are
not influenced by macroclumping. On the other hand, the strength of
optically thick resonance lines, such as P\,{\sc v}, was strongly
reduced because of macroclumping effects. Thus it was demonstrated
that the microclumping approximation is not automatically valid for
resonance lines. Consequently, relying on this approximation can lead
to underestimating empirical mass-loss rates. 

%
\begin{figure}[t]
\centering
\includegraphics[width=12cm]{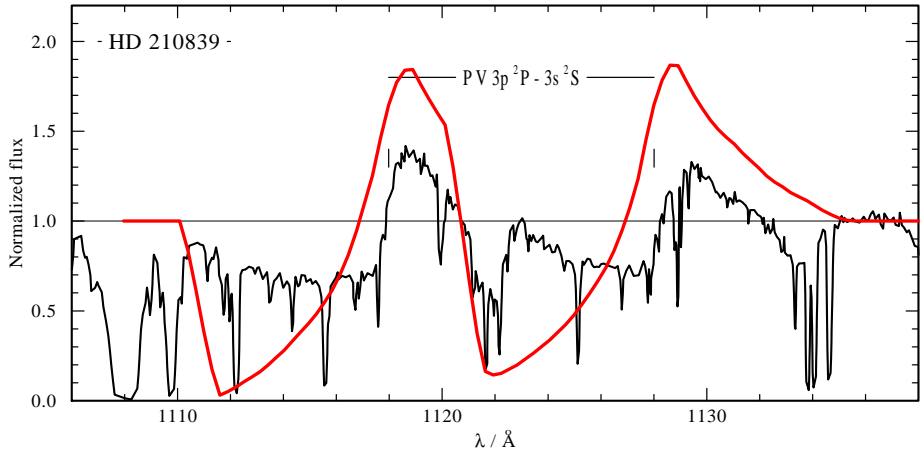}
\caption{Comparison of the observed P\,{\sc v} doublet in the UV spectrum 
of $\lambda$\,Cep (thin black line) and the model line calculated for
$\dot{M}=1.6\times 10^{-6}\,M_\odot$\,yr$^{-1}$ using the microclumping 
approximation (figure courtesy B.\,Kubatova)}.
\label{fig:pvsm}
\end{figure}

%
\begin{figure}[t]
\centering
\includegraphics[width=12cm]{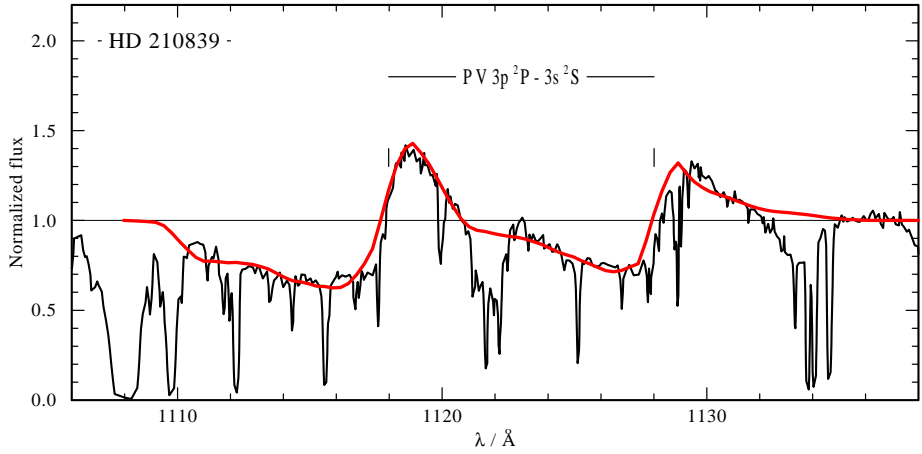}
\caption{The same as in Fig.\,\ref{fig:pvsm}, except that the model that is computed
is based on a 3-D macroclumping wind model with the same mass-loss rate 
$\dot{M}=1.6\times 10^{-6}\,M_\odot$\,yr$^{-1}$ (figure courtesy B.\,Kubatova)}.
\label{fig:pvcl}
\end{figure}

The shortcomings of the statistical approach used in \citet{osk2007}
were overcome by Monte Carlo radiative transfer models for
inhomogeneous expanding stellar winds \citep{sur2012}.  While still
relying on a phenomenological description of the wind (i.e. the
hydro-equations are not being solved), this code provided a full 3-D
description of the dense as well as the tenuous wind components
enabling the modelling of resonance lines in a clumped stellar wind.
The code also accounts for non-monotonic velocity fields.  

\citet{sur2013} analysed the observed optical and UV spectra of O-type
stars. As a first step, the observed spectra were modeled with PoWR
model  atmospheres, and the empirical mass-loss rates were estimated
from fitting  the observed H$\alpha$ emission lines. As a next step, 
the unsaturated UV resonance lines (i.e., P\,{\sc v}) were modeled
using a 3D Monte-Carlo code. The ionization stratification and underlying
photospheric spectra were adopted from the PoWR models. This allowed to
constrain the properties of the wind clumps by fitting the
observed resonance line profiles. It was shown that the UV resonance
lines, such as the unsaturated doublet of P\,{\sc v}, can  be easily
reproduced when macroclumping is taken into account. 

Including a velocity dispersion within the clumps helps to improve
the fitting of the line wings and, thus, allows a better agreement with
the observations. 

To illustrate the effects of macroclumping, in Fig.\,\ref{fig:pvsm} we
compare the P\,{\sc v} line observed in the O-type star $\lambda$\,Cep
and the model spectrum with $\dot{M}=1.6\times 
10^{-6}\,M_\odot$\,yr$^{-1}$ of the same line computed with a standard
model and a microclumping approximation. Clearly, the model severely
over-predicts the line strength. In order to match the observed and
the model lines using the microclumping approximation one would need
to significantly reduce the mass-loss rate. Alternatively, the
abundance of phosphorus could be severely reduced as well as some
fitting parameters for the radial distribution of the volume filling
factor could be introduced \citep[e.g.,][]{bouret2012}. However,
Figure\,\ref{fig:pvcl} illustrates that these problems are easily 
overcome when the microclumping approximation is waived, and realistic
3-D macroclumping models \citep{sur2012} are used.

Independent from the work by \citet{sur2012,sur2013}, Sundqvist and
collaborators, in a series of publications
\citep{Sundqv2010,Sundqv2011,sund2014}, solved the problem in a
semi-analytic way, providing expressions that can be implemented in
present-day NLTE codes, thus accounting for optically thick clumping
not only in the formal solution, but also in the NLTE rate equations. 
Thus, they provide a tool that accounts for the back-reaction of such
clumping onto the ionization balance and excitation, and which has
been benchmarked using multi-D models \citep[e.g.,][]{Sundqv2011}
similar to those described above by \citet{sur2012, sur2013}. 

Another independent study was performed by \citet{Mui2011} and
collaborators who combined non-LTE model atmospheres and a Monte Carlo
method to compute the transfer of momentum from the photons to the gas
in an atmosphere with clumping and porosity. Effects of clumping at
the photosphere were also considered. It was shown that the line force
increases as a result of increased recombination in a clumped wind.
However, the line force decreases because of porosity (or
macroclumping), simply because photons may travel in between the
clumps, avoiding interactions with the gas.  The effects of clump size
on the mass-loss rate were demonstrated, e.g.\ it was shown that
different predicted mass-loss rates are expected for different clump
sizes. E.g.\ very high overdensities of gas in large clumps may result
in the predicted $\dot{M}$ to decrease by factors 10 to 100 compared
to winds with less dense and smaller clumps. 

Recently, \citet{Noe2015} have introduced a new approach to solve
line-driven stellar winds self-consistently by using a Monte
Carlo-based radiation hydrodynamics.  While a number of
simplifications were adopted (e.g. Sobolev approximation), this Monte
Carlo-based technique can be generalized to multidimensional 
geometries, and is planned to be applied for future studies of 
inhomogeneous outflows. 

Thus, continuing the studies by \citet{osk2007} on the physics of radiative
transfer in resonance lines formed in clumped stellar winds and its
applications for the empirical mass-loss rate diagnostics,
significant work by different independent groups was done. Those
studies largely confirmed and further developed the macroclumping
(sometimes called porosity and vorosity) approach.  Allowing for
optically thick clumping in modelling resonance line in the UV spectra
of massive stars resolves the previously reported discrepancy between 
mass-loss estimates based on $\rho$ and $\rho^2$ based diagnostics.  

As a common result, \citet{osk2007, ham2008}, \citet{sur2012, sur2013}
and \citet{sund2012, sund2014} agree that the mass-loss rates derived
from optical and UV are a factor of a few (2 {\ldots} 3) lower than
predicted by the standard mass-loss recipe from \citet{Vink00}.

Interestingly, in a recent work by \citet{shenar2015}, where the
X-ray, UV, and optical spectra of the O-type star $\delta$~Orionis
were analysed consistently, it was shown that the empirically derived
mass-loss rate $\dot{M}\approx 4\times 10^{-7}\,M_\odot$\,yr$^{-1}$ is
only slightly below the predicted value ($\dot{M}_{\rm
Vink}\approx 5 \times 10^{-7}\,M_\odot$\,yr$^{-1}$). This demonstrates that
the analysis of stellar spectra is a complex task, and that
it will become difficult (or even impossible) to provide a universal,
exact and final number for the mass-loss reduction factor.  
 
\bigskip 
\noindent 
To conclude, macroclumping elevates a problem with {\it severely}
reduced mass-loss rates from O-type stars. Abolishing the
micro-clumping approximation from the modelling of UV and X-ray spectra
might allow to restore the standard picture, where the steady
radiatively driven mass-loss from massive O-type stars could be
sufficient to drive stellar evolution.  However, it may be noted that
already a factor of 2 to 3 lower mass-loss rates (compared to the
mass-loss recipe by \citet{Vink01} that is used in evolutionary models)
as derived in many of the above investigations (see also the next
sections) might be sufficient to modify present evolutionary
predictions already at early phases, particularly regarding the
evolution of the rotational velocity \citep{Keszthelyi15}.


\subsection{X-ray spectroscopy with macroclumping}  
\label{sec:Xraywithmacro}

Macroclumping may also elegantly explain the shapes of emission lines 
observed in X-ray spectra of O-type stars.

\begin{figure}[t]
\centering
\includegraphics[width=10cm]{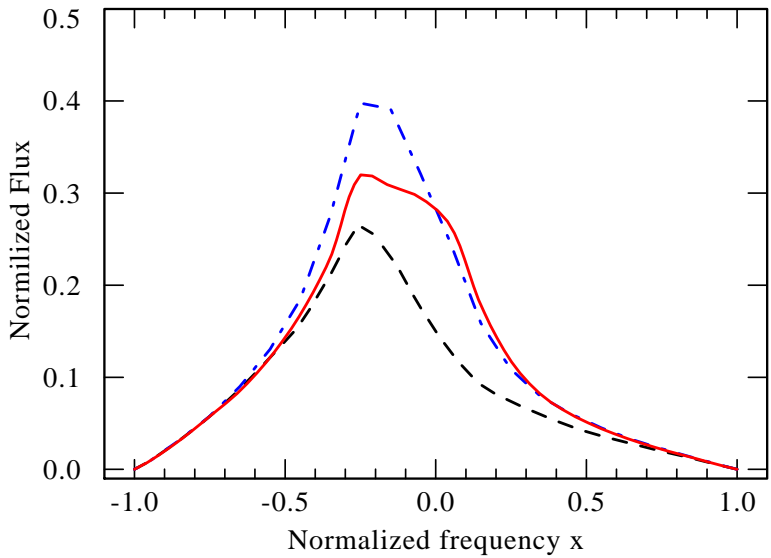}
\caption{Comparison of model X-ray line profiles. Three models are shown, 
the black dashed line is a smooth wind model, the blue dash-dotted line is a 
clumped wind model with spherical clumps, and the red solid line model with 
flat `pancake' clumps. 
The computed line profiles are displayed as a function of dimensionless 
frequency $x$, measured relative 
to the line center and in Doppler units referring to the terminal wind 
velocity.  The negative $x$ refers to blue frequencies. 
The stellar, wind, and model parameters are for the O-type star 
$\delta$\,Ori \citep[see][for details]{shenar2015}. The lines are 
normalized such that in the absence of any wind absorption the integrated line 
flux would be unity. }
\label{fig:xlinemod}
\end{figure}

There is strong and direct observational evidence for wind
clumping in O-type stars \citep[e.g.,][]{ev1998}. It was suspected
since long that stellar wind inhomogeneities affect the propagation of
X-rays in stellar winds.  The {\it Einstein} observatory discovered
that OB stars are quite soft X-ray sources. Given that massive stars
have dense stellar winds, it was apparently clear that the observed
X-rays cannot originate from the stellar surface, since they would be
completely absorbed in the overlaying dense stellar wind if it were
smooth, and never reach the observer. Thus, the X-rays have to be
generated in those wind regions, from where they can escape 
\citep{hil1993}.
%
\begin{figure}[t]
\centering
\includegraphics[width=10cm]{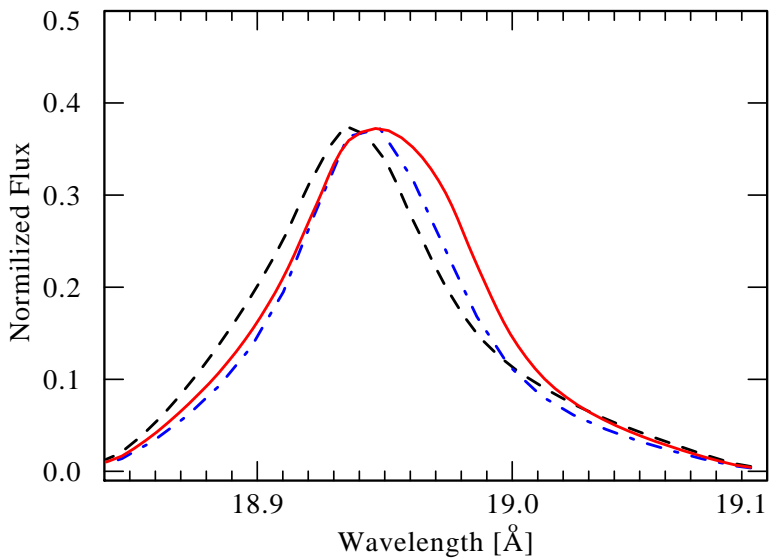}
\caption{The same as in Fig.\,\ref{fig:xlinemod}, but now the computed 
lines are displayed as a function of wavelength close to the O\,{\sc viii} 
L$\alpha$ line. The lines are re-normalized to the 
maximum as typically done in the analysis of the observed X-ray spectra, and 
convolved with the instrumental response of the {\em Chandra} HETGS MEG 
detector.}
\label{fig:xlinewave}
\end{figure}

\citet{StewartFabian1981} used {\em Einstein} spectra of O-stars to
study  the transfer of X-rays through a uniform stellar wind as a means
to determine stellar mass loss rates. They used a photoionization code
to find the wind  opacity. Using \mdot\ as model parameter, they found
from matching the model and the observed X-ray spectrum of
$\zeta$\,Pup\ that the X-ray based mass-loss rate  from this star 
is a factor of a few lower than the mass-loss rates obtained  from 
fitting the H$\alpha$ line and the radio and IR excess. As a most
plausible explanation for this discrepancy they suggest the neglected 
clumping in the stellar wind. 

The first high-resolution spectra of O-type stars \citep{wc2001}
showed nearly symmetric lines that can be well fitted by Gaussian
functions, and do not show shapes predicted by \citet{macf1991} (see
Fig.\,\ref{fig:smoothlines}). As a potential solution to this
discrepancy \citet{wc2001} suggested inhomogeneous or non-symmetric
winds. The idea that clumping may reduce the wind opacity and lead to
more symmetric line profiles was also discussed by
\citet{oc2001,kramer2003}. However, the first quantitative models for
the shapes of emission lines emerging from a clumped stellar wind
were computed by \citet{feld2003}.

In \citet{feld2003} and \citet{osk2004,osk2006} a general formalism
for X-ray radiative transfer in a clumpy stellar wind was developed; 
while this formalism uses a phenomenological description of the 
clumpy wind, it attempts to take into account the basic picture
emerging from 1-D LDI simulations as those discussed in 
Sect.~\ref{sec:smallscale}.

The basic picture emerging from radiation hydrodynamic
simulations of the non-linear evolution of instabilities in stellar
winds is that of strong reverse shocks, which arise when a high-speed,
rarefied flow impacts on slower material that has been compressed into
dense shells (e.g., \citealt{Owocki88, Feldmeier97b},
see also Fig.\,\ref{Fig:ldi}). 

Building on this, the phenomenological 'broken-shell' model by
\citet{feld2003} showed that the asymmetry of such broken shells lead
to a characteristic shape of X-ray emission lines -- the blueshifted
part of X-ray line profiles remains flat-topped even after severe wind
attenuation, whereas the red part shows a steep decline \citep[see
Fig.\,5,6 in][]{feld2003}. 

The emission line models of \citet{feld2003} do not constrain clump
optical depths -- three cases were considered, optically thin clumps,
optically thick clumps, and clumps with arbitrary optical depth
(bridging between optically thin and thick cases).  \citet{feld2003}
considered a general structured stellar wind model - they did not use
the outcome of a specific hydro-simulation, but rather concentrated on
novel radiative transfer effects. On the other hand, \citet{sund2012}
applied a specific 1D LDI simulation \citep{Feldmeier97b} that was phased 
randomly among 3D patches, and computed parameterized emergent line
profiles for a pre-specified value of $\tau_0$ by calculating the absorption 
directly from the actual density structure of the radiation-hydrodynamic 
model. They did not find significant porosity effects in their simulations.



\citet{osk2004} used a 2.5-D stochastic wind model to study the
propagation of X-rays in structured winds. They considered a
stochastic ensemble of clumps, allowing for a randomized mixture of
clump optical depths at each radius and direction in the wind.  It was
shown that the clumped wind is much more transparent for the X-ray
emission, even when the covering factor of the absorbing material is
unity (i.e. when there are no gaps in the wind). They considered the
limiting cases of optically thin and optically thick clumps, as well
as the bridging case of arbitrary clump optical depths. A concept of 
effective opacity was introduced, and a thorough comparison with
analytical models was done. It was shown that in the limiting case of
optically thin clumps, the macroclumping formalism is identical to the
microclumping or smooth wind formalism.  \citet{osk2006} compared the
model X-ray line profiles with those observed in high-resolution X-ray
spectra of O-type stars, and concluded that wind clumping explains
well the observed line shapes. They showed that macroclumping reduces
the wavelength dependence of opacity, and in case of optically thick
clumps, the opacity becomes grey.  Importantly, \citet{osk2006}
demonstrated that macroclumping is a wavelength dependent effect --
while for lines at higher energies the macroclumping is negligible,
for lines at longer wavelength the macroclumping effects are
important \citep[see Fig.\,8 and Fig.\,9 in][]{osk2006}.  
%
\begin{figure}[t]
\centering
\includegraphics[width=10cm]{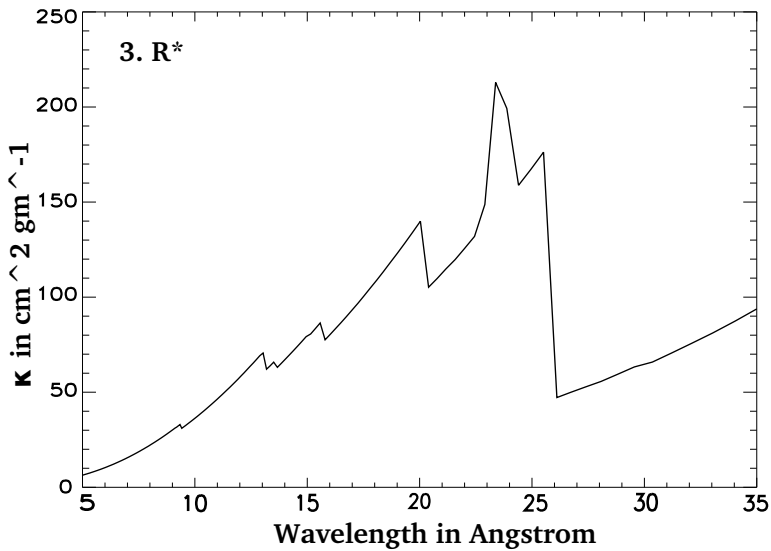}
\caption{Wavelength dependence of the mass absorption coefficient in
the wind of the O-supergiant $\zeta$ Puppis 
at the distance in the wind $r \approx 3\,R_\ast$.
Adopted from \citet{herve2013}.}
\label{fig:kaplam}
\end{figure}

\citet{shaviv1998} considered grey opacity in a stellar atmosphere
where the photon mean free path does not exceed the scale of wind
inhomogeneities. \citet{shaviv2000} coined the term "porous
atmosphere'' for massive stars to describe a multi-phase medium that
allows more radiation to escape while exerting a weaker average force.
Thus, in such atmosphere a considerably lower radiative acceleration
results from the same total luminosity and, hence, can explain the
apparent existence of super-Eddington stars \citep{shaviv2000}. 
  
\citet{OGS04} introduced a ``porosity-length'' formalism and
applied it to a continuum dominated by electron scattering.
In this approach, the porosity length turns out to be a photon mean
free path for a medium consisting of optically thick clumps. This
formalism was then used to obtain an associated scaling for the
continuum-driven mass-loss rate from stars that formally exceed the
Eddington limit, such as $\eta$\,Car during its giant eruption in the
19th century. 

\citet{oc2006} applied the porosity formalism to the case of
wavelengths dependent K-shell opacity \citep[i.e.\ to the same problem
as was considered in][]{feld2003}.  They considered spherical clumps,
and developed an effective opacity law bridging the limits of
optically thin and thick clumps; while the bridging law in
\citet{oc2006} was chosen to allow for simple computations, it later
turned out \citep{sund2012} that this form in fact represents that of
an ensemble of clumps with local scales set by Markovian statistics. 

As illustrated in Figures~\ref{fig:xlinemod} and \ref{fig:xlinewave},
the clump shape, i.e.\ the angular dependence of the clump
cross-section, affects the shape of line profiles. \citet{osk2006}
showed that in case of spherically symmetric clumps, the {\em line
profiles are identical} to those emerging from the smooth wind, albeit
the same line shape would be achieved for larger $\dot{M}$ in case of
a clumped wind as compared to the smooth wind. Thus, the fitting of
line shapes alone cannot discriminate between the clumped and
the smooth wind models. Note that this conclusion is in conflict with
that reached by \citet{leu2013}, who by fitting X-ray lines of $\zeta$~
Pup showed that the profiles of smooth and isotropic porosity-models
are only degenerate in the case of small porosity lengths, in which
case the mass-loss rate correction from porosity is in any case
marginal.  

\begin{figure}[t]
\centering
\includegraphics[width=6.5cm, angle=-90]{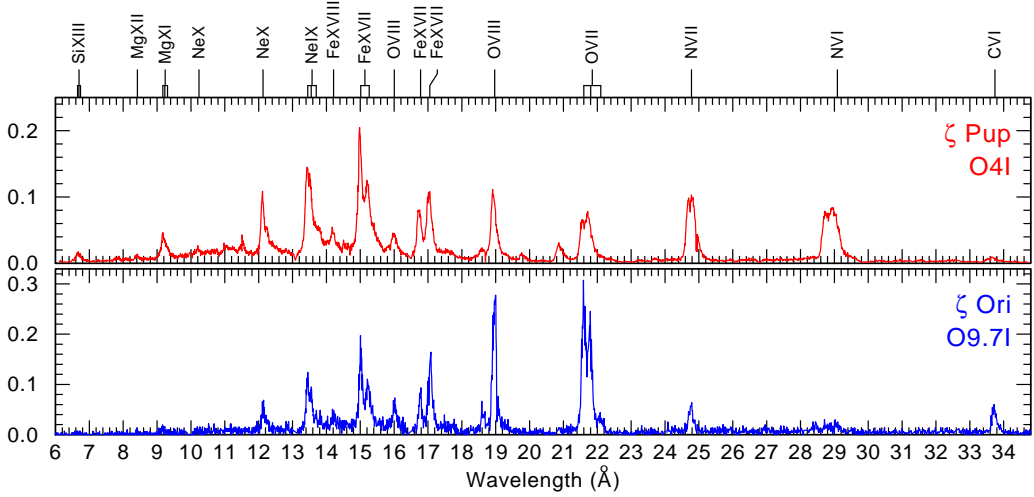}
\caption{An example of {\em XMM-Newton} RGS spectra of two prominent
O-supergiant stars.}
\label{fig:rgso}
\end{figure}

Macroclumping theory predicts that the wavelength dependence of
stellar wind opacity for X-rays is reduced compared to
smooth/microclumping models, i.e., winds are more grey for X-ray
radiation \citep{osk2006}. 
Actually, this is not only true for X-rays, but in all those cases
where the clumps become optically thick to the specific radiation
\citep{OGS04}.

Such a reduced wavelength dependency has been found in
O-supergiant X-ray spectra by e.g.\ \citet{kramer2003}, who noticed that the
wind opacity is nearly grey for X-rays in the wind of $\zeta$\,Pup, while 
\citet{rauw2015} found 
a rather modest wavelength dependence (at least longwards of 16~\AA, 
see Sect.~\ref{no-macro}) of $\tau_0$ at $\approx 2R_\ast$ in their
analysis of $\lambda$\,Cep \footnote{\citet{coh2014} find wavelength
dependences of $\tau_0$ in their sample of O stars that are consistent
with models without porosity; this is not surprising, since in the
high frequency regime considered by \citealt{coh2014} (below
$\sim$15-20 \AA) porosity is much less effective due to the lower
opacity.}.  

\citet{wc2007} presented an analysis of a broad collection of OB
stellar line profile data to search for morphological trends. It was
found in the majority of the OB-stars that the emission lines are
symmetric. But there is evidence for small, finite, blueward 
line shifts that also increase with stellar
luminosity. This may be evidence for macroclumping
playing an increasing role in OB-supergiant winds, compared to the
stars of lower luminosity classes. 
However, it may also simply be an effect of the increasing importance
of absorption effects due to the higher supergiant mass-loss rates.  

\subsection{The size of clumps that attenuate X-rays}\label{sec:size}

In the context of this review it is useful to provide estimates for
the size of clumps that would attenuate X-rays, especially to make
sure that such clumps indeed can be present in stellar winds in
accordance with our knowledge of OB-supergiants. To obtain these
estimates of clump sizes we follow the treatment discussed in
\citet{osk2011} and \citet{osk2012}.  

Using Eq.\,\ref{eq:t0} the optical depth of an ideal spherical clump is
     
\begin{equation}
\tau_{\rm clump}(\lambda)
= \rho_{\rm w}\kappa_{\lambda}D d_{\rm clump}\Rstar\,=\, 
\tau_0(\lambda)
\frac{d_{\rm clump}}{f_{\rm v}}
  \cdot\frac{1}{(1-\frac{1}{r})^{\beta}r^2},
\label{eq:tc}
\end{equation}
where $r=R/\Rstar$ and $d_{\rm clump}$\, is the characteristic
geometrical size of the clump expressed in \Rstar.  Note further that
here the porosity length (mean free path between clumps), $h = d_{\rm
clump}/f_{\rm V}$.  The strong wavelength dependence of
$\kappa_\lambda$ (see an example in Fig.\,\ref{fig:kaplam}) implies
that the clump optical depth is wavelength dependent.

It is convenient to express $\tau_0(\lambda)$  (see Eq.\,\ref{eq:t0}) as
\begin{equation}
\tau_0(\lambda)\,\approx\,722\frac{\mdot_{-6}}{\upsilon_\infty {\cal 
R}_\ast}\cdot\kappa_\lambda,
\label{eq:tatau}
\end{equation}
where $\mdot_{-6}$ is the mass-loss rate in units $10^{-6}$\,\myr,
$\upsilon_\infty$ is the terminal velocity in [km\,s$^{-1}$], and ${\cal
R_\ast}=\Rstar/R_\odot$. If at some radius $r$, the clump size is
larger than $d_{\rm clump}^{\tau=1}$, such a clump is no longer
optically thin for the X-ray radiation at wavelengt $\lambda$.  Assuming again a
beta-velocity law, the size of a clump with optical depth $\tau=1$
at wavelength $\lambda$ is
\begin{equation}
d_{\rm clump}^{\tau=1}=\frac{f_{\rm V}}{\tau_0(\lambda)}
\cdot\left(1-\frac{1}{r}\right)^{\beta}r^2. 
\label{eq:dc}
\end{equation}
The microclumping approximation is valid only for significantly smaller 
clumps. 

Let us estimate the geometrical size of a clump which has optical depth
unity. We consider the O-type star $\zeta$ Pup\ and use its parameters
from \citet{bouret2012}: 
$\mdot_{-6}=1.9$, $\upsilon_\infty=2300$, $R_\ast = 18.8 R_\odot$, 
$\beta=0.9$, and a filling factor $f_{\rm V} = 0.05$.  Then 
$\tau_0(\lambda)\,\approx\,0.03\kappa_\lambda$. 

Assuming a constant volume filling factor $f_{\rm V} = 0.05$, and
using values of $\kappa_\lambda$ at $3\,R_\ast$ as shown in 
Fig.\,\ref{fig:kaplam}, the geometrical size of a clump optically
thick for X-ray radiation at $\lambda = 10$\AA\ and $r=3\,R_\ast$ is 
$d_{\rm clump}\approx 0.35\,R_\ast$, while at $\lambda = 24$\AA\
$d_{\rm clump} \approx 0.06\,R_\ast$. 

Though these estimates seem to be reasonable, 
the clump size and volume filling factor are actually not
mutually exclusive, but related physically via the mean-free-path
between clumps (i.e. via the porosity length $h = d_{\rm clump}/f_{\rm
vol}$). Inserting the above numbers, one finds $h \approx 6.9 R_\ast$ at
10 \AA \ and $h \approx 1.2 R_\ast$ at 24 \AA. This thus means that
for clumps to become optically thick at 10 \AA, they have to be
individually separated by a (physical) mean free path $\sim$7 $R_\ast$
(which is even much larger than the 3 $R_\ast$ at which the
computation were made).  Thus, the clumps considered above {\it
cannot} become optically thick at 10~\AA, and their actual sizes will
be smaller than calculated from the condition of unity optical depth.

SgXBs are quite hard X-ray sources, e.g.\ the maximum of the
spectral energy distribution in Vela~X-1 is at $\lambda \approx
4$\,\AA, \citep[e.g.,][]{Doroshenko:2011}. As shown in \citet{osk2012}, at a
distance of $2\,R_\ast$ from the stellar core, only geometrically large
clumps with diameters larger than $\sim 0.5\,R_\ast$ are optically
thick for the ionizing radiation below 10\,\AA. 

 
The analysis of the X-ray emission from the hard state of Cyg X-1 (a
binary system hosting a supergiant star with a blackhole companion)
presented by \citet{grin2015} showed that an optical depth of order
unity at 3 keV can only be achieved if $h \approx 60 R_\ast$ (adopting
the same parameters as above and the approximate scaling
$\kappa_\lambda \approx 30 E_{\rm keV}^{-2}$, with $E_{\rm keV}$ the
energy in keV, e.g., \citealt{Ow2013}).This illustrates that clumps can
hardly be optically thick at those energies, similarly to what was
concluded by \citet{osk2012}.
 
Moreover, and as also demonstrated above, due to the strong
wavelength-dependence of the opacity, effects of macroclumping might
indeed start to become relevant at longer wavelengths, and in
particular when approaching the EUV part of the spectrum.

\subsection{X-ray spectroscopy neglecting macroclumping}
\label{no-macro}

There are other approaches that allow to fit observed X-ray line
profiles. In this section we briefly review these works. 

As was pointed out in \citet{osk2006}, X-ray lines in O-star spectra 
are differently affected by clumping. Importantly, the best X-ray
lines to probe the stellar wind {\it structure}
are at the wavelengths where the wind opacity achieves
the largest value. This is because the same wind clump would be optically thin
at shorter wavelength and optically thick at longer ones, see
Eq.\,\ref{eq:tatau}. 

Due to its large effective area and softer response, the {\em XMM-Newton}
RGS spectrograph is excellently suited to probe such lines.
Figure\,\ref{fig:rgso} shows an example of X-ray spectra of two O-type
stars. The wavelength dependence of the mass-absorption coefficient in the
wind of $\zeta$\,Pup is illustrated in Figure\,\ref{fig:kaplam}. As
can be easily seen from a comparison of these two figures, the strongest
lines at the wavelength where the cool wind opacity is largest and
hence the lines that are best suited to probe wind clumping are
O\,{\sc viii} $\lambda 18.97$\,\AA, O\,{\sc vii} $\lambda 21.6$\,\AA,
N\,{\sc vii} $\lambda 24.78$\,\AA\ and N\,{\sc vi} $\lambda
28.78$\,\AA.

\citet{leu2013} fitted X-ray emission line profiles from high
resolution {\em XMM-Newton} and {\em Chandra} grating spectra of the
early O supergiant $\zeta$\,Pup, to evaluate the relative importance
of porosity and mass-loss rate in affecting the line shapes.  Their
method relies on fitting the \citet{oc2001} model mentioned above,
with extensions to account for porosity and macroclumping according to
\citet{sund2012} and resonance line scattering according to
\citet{leu2007}.  The model is specified by a rather small number of
free parameters and implemented as local models in the global fitting
software {\em XSPEC}, and the match to observed lines is then
evaluated by a standard goodness of fit. 

The conclusions are based on the formal 68.3\% confidence limits
obtained during the fitting procedure. \citet{leu2013} excluded from
their analysis the diagnostic lines most sensitive to wind clumping, 
since they were aiming to primarily investigate the reliability of
deriving mass-loss rates from (relatively high frequency) lines. Only
lines blueward of $\lambda 20.91$\,\AA, i.e.\ probing relatively low
wind opacities, were thus considered. As already discussed in the 
previous paragraph, this might be a reason why only modest or no
porosity effects were noticed in their analysis. 

\citet{leu2013} demonstrated that neglecting macroclumping may lead 
to about 40\%\ underestimate of mass-loss rates (see their section
4.3) for a porosity length $h\sim R_{\ast}$ when porosity effects are
important, which, as outlined above, is a question of wavelength,
mass-loss rate, velocity law, etc.  Interestingly, such porosity
lengths were recently derived from the analysis of X-ray variability
in Cyg~X-1 \citep{grin2015}.  

Indeed, the wavelength dependence of wind opacity implies that lines
at shorter wavelengths ($<20$\,\AA) are less affected by wind
absorption/clumping than the lines at longer wavelength. Thus, these
lines are well suited to study the dynamics, emission measure, and
spatial distribution of the hot plasma.  

Consistent analyses of spectral lines, including important wind
diagnostic lines of O and N, in the X-ray spectra of $\zeta$ Pup 
(O4\,Ief) and its spectroscopic twin $\lambda$ Cep (O6\,Ief) were 
performed by \citet{herve2013} and \citet{rauw2015}. Both these works
neglected the influence of porosity on the analysed X-ray emission
lines (based on previous results -- see above -- for lines at not too
long wavelengths), and yet were able to achieve excellent fits to the
observed X-ray spectra for realistic stellar wind parameters. This was
achieved by introducing a new free parameter - the radial distribution
of the hot plasma filling factor.  

It is important to note that \citet[e.g.,][]{coh2010} tested models with
varying filling factors by fitting them to observed X-ray emission line
profiles. They concluded that including a radial dependence of the
filling factor does not improve the line fits. Hence, they suggested that
a radial dependence of the filling factor can be neglected. 


In their analysis \citet{rauw2015} employ the cool wind opacity as
computed using the non-LTE stellar atmosphere model CMFGEN. This numeric
stellar atmosphere provides theoretical predictions for the parameter 
$\tau_0(\lambda,r)$ (see Eq.\,\ref{eq:t0}). The same parameter can be
empirically estimated by fitting the \citet{oc2001} model to the
observed line profiles. Then, the predictions from the NLTE wind model 
and  the empirically estimated values can be compared. 


Studying the X-ray emission line profiles, \citet{rauw2015} 
demonstrated that the \citet{oc2001} models cannot easily explain the
morphology of the N{\sc vii} Ly$_\alpha$ line, though some other used
lines (Ne{\sc X}, Fe{\sc XVII} and O{\sc VIII}, with $\lambda <$
20~\AA) were matched almost perfectly. Notably, the wind opacity
derived using this formalism is smallest at the wavelength of the
N\,{\sc vii} $\lambda 24.78$\,\AA\ line (see Table 4 in
\citet{rauw2015}), in contradiction to the expectations from the NLTE
stellar wind models.

To achieve satisfactory fits for {\it all} lines of the X-ray spectrum
of $\lambda$ Cep, \citet{rauw2015} introduced a model, where the
stellar wind is divided into a grid of concentric shells, each with
own temperature and emission measure of X-ray emitting gas. The
temperature and the emission measure filling factor is then varied
until the model is a good agreement with observations. As previously,
the cool wind opacity is provided by the NLTE atmosphere with
specified mass-loss rate, velocity field and other stellar parameters.
The good correspondence between the model and the observations is
achieved for mass-loss rates in agreement with the one obtained from
fitting the UV lines with the models that account for macroclumping
(see Figs.\,\ref{fig:pvsm},\ref{fig:pvcl}) as derived by
\cite{sur2013, sund2014}. Remember, that macroclumping effects needed
to be included in the UV-analysis (optically thick clumps with respect
to UV-lines), but were excluded from the X-ray analysis. The best fit
model constrains the location of the X-ray plasma in $\lambda$ Cep
between $1.1\,R_\ast$ and $2.5\,R_\ast$, with a non-monotonic radial
distribution of the filling factor. 

Despite the fact that $\lambda$ Cep is a spectroscopic twin of $\zeta$
Pup in the optical and in X-rays, there are very different conclusions
about the filling factor distribution and the location of hot gas in
these stars.  While the results from \citet{herve2013} and
\citet{leu2013} allow the presence of hot plasma at very large
distances from the star, the results of \citet{rauw2015} require a
finite radius for the X-ray emitting gas. In this respect, this is
similar to the approach of \citet{osk2006}.

It seems that numeric hydrodynamic simulations could accommodate
non-monotonic filling factors, but it is difficult to reconcile the
hydrodynamic models with such a sharp cut-off of X-ray emitting gas at
$2.5\,R_\ast$ in $\lambda$ Cep (see \citealt{krt2009}). 

\citet{wc2007} studied the spatial distribution of hot gas in OB stars
from an analysis of their X-ray spectra. They found that the highest
temperatures occur near the star and steadily decrease outward. This
trend seemed to be most pronounced in OB supergiants. For the lower
density wind stars, both high and low X-ray source temperatures may
exist near the star.  They called this intriguing temperature distribution 
the ``near-star
high-ion problem'' for OB stars. By invoking the traditional OB
stellar mass-loss rates, \citet{wc2007} found a good correlation
between the spatial onset of X-ray radiation and X-ray continuum
optical depth unity radii. On the other hand \citet{leu2013} used
the observed flux in the far wings of various X-ray lines, and from
this inferred hot gas at high temperatures far away from the star
directly from observations of $\zeta$ Pup. 

Observational constraints on the radial stratification of the clumping
filling factor of the cool wind material were provided in
\citet{puls2006}. From the study of own and archival data for
H$\alpha$, IR, mm and radio fluxes, and using approximate methods,
calibrated to more sophisticated models, it was shown that the minimum
clumping factor (except for the lowermost unclumped wind with $r < 1.1
R_{\ast}$) is found in the outermost, radio-emitting region
\footnote{The clumping factor used in this (and other) work
corresponds to the over-density $D = f_V^{-1}$ introduced in Sect.~\ref{basicsldw}.}.
Thus, the radio mass-loss rates would be the lowest ones,
compared to those derived from H$\alpha$ and the IR, when analysed by
means of unclumped models. These radio rates (assuming $D=1$ in the
outermost wind) agree well with those predicted by \citet{Vink00}, but
are only upper limits, since the absolute value of $D$ could not be
constrained from this pure micro-clumping investigation, and might be
larger than unity. For denser winds, it turned out that the inner wind
region (from $r > 1.1 R_{\ast}$) is more strongly clumped than the
outermost one (with a normalized clumping factor of $\approx 4$),
whereas thinner winds have similar clumping properties in the inner
and outer regions.

\begin{figure}[t]
\centering
\includegraphics[width=7cm, angle=-90]{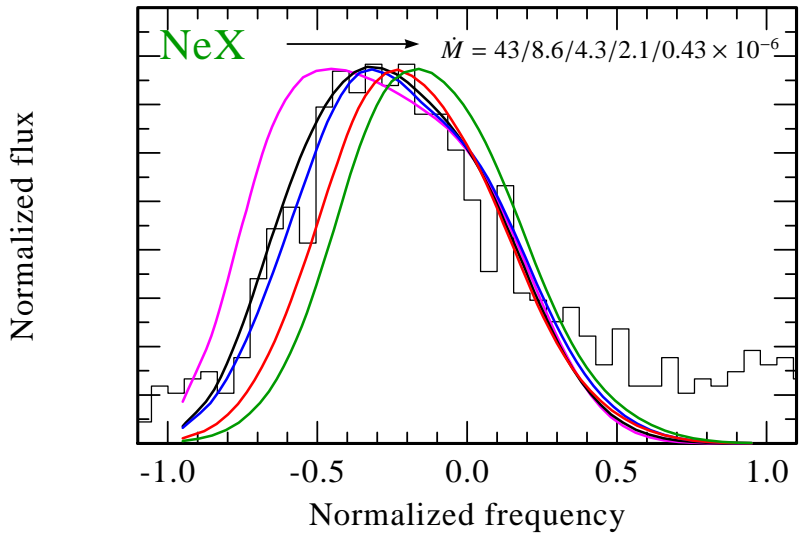}
\caption{Observed (histogram, {\em Chandra} X-ray telescope) and 
modelled Ne\,{\sc x} line in $\zeta$\,Pup. From left to right,
model line profiles are calculated assuming $\dot{M}
= 43, 8.6, 4.3, 2.1, 0.43\times 10^{-6} M_\odot$yr$^{-1}$. 
All synthetic lines have been normalized to the observed maximum flux.
Adopted from \citet{osk2006}}.
\label{fig:nemdot}
\end{figure}

An interesting application of X-ray spectroscopy of stellar winds is
its usefulness to constrain the absorbing column density of cool wind
material. Indeed, the stellar mass-loss rate has a direct
influence on the shape of the X-ray emission lines (see
Fig.\,\ref{fig:nemdot} adopted from \citet{osk2006}). In this respect,
\citet{osk2006,herve2013,shenar2015,rauw2015} demonstrated that X-ray
spectroscopy is a very useful tool to constrain stellar mass-loss
rates, especially when used in combination with NLTE modelling of
UV/optical spectra. 

A different approach was used by \citet{coh2014}. Based on their
earlier work, they suggested to directly measure mass-loss rates from
the observed X-ray line profiles, which should result in an unbiased
value from all lines with negligible porosity (see above). By this
method, 
they found mass-loss
rates for O-stars which are, interestingly, also a factor of 2 to 3
(or even more) lower than the corresponding rates from \citet{Vink01}, in
accordance with most results from macro-clumped UV-diagnostics (see above), 
and also consistent with optical H$\alpha$ when clumping is accounted for 
(see \citealt{coh2014}). Future work shall demonstrate how useful and 
robust this method is. 

\bigskip
\noindent
In summary, this brief review does not cover all available models and
approaches to model X-ray emission lines and spectra. Mainly, we
concentrated here on the influence of clumping/porosity on UV and
X-ray line profiles. Some of the models we considered do not favour
the presence of optically thick clumping for X-rays in stellar winds.
Others provide a satisfactory description of observed spectra allowing
variable filling factors of hot gas. In another approach the optical
depth of clumps in stellar wind is not constrained, and clumps can be
optically thick for X-rays, depending on wavelength and mass-loss
rate. Future observations and modelling work will either allow to
discriminate between the importance of non-monotonic X-ray filling factors
and clumping in O-star winds, or provide evidence for the co-existence
of both effects. 

\begin{table*}
\setlength{\tabcolsep}{2pt}
 \centering
  \caption{Wind-accreting Supergiant X-ray Binary systems 
  with neutron stars:  (A) SgXBs (B) SFXTs.}\label{tab:windfed}
\begin{footnotesize}
  \begin{tabular}{@{}lllllll@{}}
\hline
   Name & Companion & Distance & Orbital              &   Spin            &  Super-Orbital      & Current \\
        &           &  kpc     &  Period (d)          &  Period (s)       &  Period (d)      & Class.\\
\hline
1A\,0114+650      & B0.5I$^{\,1}$ & 7.2$^{\,2}$ & 11.59$^{\,1}$  &  9475$^{\,3}$ &  30.7$^{\,4}$ & A \\
Vela~X-1         & B0.5Ib & 1.7-2.1$^{\,5}$ & 8.964$^{\,6}$ & 283$^{\,6}$ & --- & A \\
1E\,1145.1-6141   & B2Iae$^{\,7}$ &8.2$^{\,7}$ & 14.365$^{\,8}$ & 297$^{\,9}$ & --- & A \\
GX\,301-2         & B1 Ia+$^{\,10}$ & 3-4$^{\,10}$& 41.492$^{\,11}$ & 675700$^{\,10}$ & --- & A \\
4U\,1538-522      & B0I$^{\,12}$ & 5.5$^{\,12}$ & 3.728$^{\,13}$ & 528-530$^{\,14}$ & --- & A \\
IGR\,J16318-4848  & sgB[e]$^{\,15}$ & 0.9-6.2$^{\,15}$ & --- & --- & --- &  A \\
IGR\,J16320-4751  & O8I$^{\,16}$ & 3.5$^{\,16}$ & 8.986$^{\,17}$ & 1309$^{\,18}$ & --- & A  \\
IGR\,J16393-4641  &OB?$^{\,19}$ & $>$10?$^{\,19}$ & 4.24$^{\,20}$ & 912.0$^{\,21}$ & --- &  A \\
IGR~J16465-4507  & O9.5Ia$^{\,22}$ & 9.5$^{\,23}$  & 30.243$^{\,24,25}$ & 228$^{\,26}$ & --- & A \\
IGR\,J16493-4348  &B0.5 Ib$^{\,27}$ & $>$6$^{\,27}$ & 6.782$^{\,28}$ & 1093$^{\,28}$ & 6.8$^{\,29}$ & A \\
OAO\,1657-415     & Ofpe/WN9$^{\,30}$ &4-8$^{\,30}$ & 10.448$^{\,30}$ & 38.2$^{\,31}$ & --- &  A \\
4U\,1700-37       &O6.5 Iaf+ $^{\,32}$ &1.9$^{\,33}$ & 3.412$^{\,34}$ & --- & --- & A \\
EXO\,1722-363     &B0-B1Ia$^{\,35}$ & 6-10.5$^{\,36}$ & 9.742$^{\,37}$ & 413.89$^{\,38}$ & --- & A \\
IGR\,J18027-2016  &B1 Ib$^{\,39}$ & 12.4$^{\,39}$ & 4.469$^{\,39}$ & 139.612$^{\,40}$ & --- & A \\
XTE\,J1855-026    & B0 Iaep$^{\,41}$ & --- & 6.0724$^{\,42}$ & 360.7$^{\,43}$ & --- & A \\
4U\,1907+097      & O8Ia,O9Ia$^{\,44}$ & 5$^{\,44}$ & 8.3753$^{\,44}$ & 437-441$^{\,45}$ & --- & A \\
4U\,1909+07       & B1-B3 (I)$^{\,46}$ & 4.85 $\pm$ 0.50$^{\,46}$ & 4.4$^{\,47}$ & 605$^{\,46}$ & 15.180$^{\,39}$ & A  \\
IGR\,J19140+0951  &  B0.5Ia/d$^{\,47}$ &2-5$^{\,47}$ & 13.55$^{\,48}$ & --- & --- & A \\
IGR J16207-5129   &  B0I$^{\,42}$  &  6 $^{\,49}$              &     9.726$^{\,50}$   &   ---         & ---  & A \\
\hline
\hline
IGR~J16418-4532 &  OB Sg  & 13$^{\,51}$ &  3.753$^{\,52}$ & 1212$^{\,53}$ & 14.6842$^{\,54,55}$ &  A or B$^{\,56}$\\
\hline
\hline
XTE~J1739-302   & O8Iab(f)$^{\,57,58}$   & 2.7$^{\,58}$ & 51.47$^{\,59}$    & --- &  --- & B\\
IGR~J17544-2619 & O9Ib$^{\,60}$         &  3.0 $\pm$ 0.2 $^{\,60}$  & 4.926$^{\,61}$   &  71.49/11.6$^{\,62}$ & --- &  B\\
SAX~J1818.6-1703 & $\sim$B0I$^{\,63,64}$ & 2, 2.1$^{\,65,66}$ & 30.0$^{\,67,68}$  & --- &  ---  & B\\%
IGR~J16479-4514 & O8.5I, O9.5Iab$^{\,58,42}$& 4.9, 2.8$^{\,58,42}$ & 3.3194$^{\,68,69}$  &     ---                        &   11.880$^{\,54,55}$ &   B\\
IGR~J18483-0311 &  B0.5Ia $^{\,70}$        & 3$^{\,70}$          &  18.55$^{\,71}$     &  21.0526$^{\,72}$  &   --- &  B\\
IGR~J18450-0435 & O9.5I$^{\,73}$           & 3.6$^{\,73}$        &  5.7195$^{\,74}$   &       ---                      &  ---  & B\\
IGR~J18410-0535 &  B1Ib$^{\,14}$         &   3.2$^{\,75}$   &    ---                          &    ---               & ---  &  B\\
IGR~J08408-4503  & O8.5Ib$^{\,76}$       &   2.7$^{\,77}$                     &    9.5436$^{\,78}$   &     ---      & 285 $\pm$ 10$^{\,78}$  &  B \\
IGR~J11215-5952  &  B0.5Ia$^{\,79,80}$    &  6.2, 8$^{\,81,79}$   &  164.6$^{\,82,83}$   &  186.78$^{\,84}$  & ---  & B\\
IGR J16328-4726  &   O8Iafpe$^{\,85,86,87}$  &  3-10 $^{\,87}$              &     10.068$^{\,88,89}$  &   ---         & ---  & B\\
IGR J18462-0223   &  & & & &  & B$^{\,90}$\\ 
\hline
\multicolumn{7}{@{}p{0.95\textwidth}}{\vspace*{0.2\baselineskip}%
(1)~\citealt{Crampton1985};
(2)~\citealt{Reig1996};
(3)~\citealt{Wang2011};
(4)~\citealt{Farrell2006}; 
(5)~\citealt{Nagase1986};
(6)~\citealt{Quaintrell2003};
(7)~\citealt{Densham1982};
(8)~\citealt{Ray2002};
(9)~\citealt{White1980};
(10)~\citealt{Kaper2006}; 
(11)~\citealt{Koh1997}; 
(12)~\citealt{Reynolds1992}; 
(13)~\citealt{Clark1994}; 
(14)~\citealt{Clark2000}; 
(15)~\citealt{Filliatre2004};
(16)~\citealt{Rahoui2008}; 
(17)~\citealt{Corbet2005}; 
(18)~\citealt{Lutovinov2005}; 
(19)~\citealt{Bodaghee2012}; 
(20)~\citealt{Perlman2011}; 
(21)~\citealt{Bodaghee2006}; 
(22)~\citealt{Coe1996};  
(23)~\citealt{Clark2010};  
(24)~\citealt{LaParola2010};  
(25)~\citealt{Walter2006};
(26)~\citealt{Nespoli2010}; 
(27)~\citealt{Pearlman2013};
(28)~\citealt{Corbet2010b}; 
(29)~\citealt{Mason2012}; 
(30)~\citealt{White1979}; 
(31)~\citealt{Jones1973}; 
(32)~\citealt{Ankay2001}; 
(33)~\citealt{Corbet2010a}; 
(34)~\citealt{Mason2009};
(35)~\citealt{Manousakis2011}; 
(36)~\citealt{Masetti2008}; 
(37)~\citealt{Torrejon2010a};
(38)~\citealt{Mason2011}; 
(39)~\citealt{Negueruela2008a}; 
(40)~\citealt{Corbet1999}; 
(41)~\citealt{Corbet2002}; 
(42)~\citealt{Nespoli2008}; 
(43)~\citealt{Zand1998}; 
(44)~\citealt{Cox2005};
(45)~\citealt{Fritz2006};
(46)~\citealt{martinez-nunez2015}; 
(47)~\citealt{Hannikainen2007}; 
(48)~\citealt{Corbet2004};
(49)~\citealt{Negueruela2007}.
(50)~This period has been reported by \citealt{Jain2011} but never confirmed; 
(51)~\citealt{Chaty2008};  
(52)~\citealt{Corbet2006};  
(53)~\citealt{Sidoli2012b};  
(54)~\citealt{Corbet2013};  
(55)~\citealt{Drave2013Atel}; 
(56)~This source has been considered as SgXB by \citealt{Romano2014};\citealt{Drave2013} and as SFXT
by \citealt{PaizisSidoli2014} since it displays hard X-ray properties, in term of cumulative luminosity 
ditribution of the flares, compatible with other SFXTs;
(57)~\citealt{Negueruela2006};    
(58)~\citealt{Rahoui2008};    
(59)~\citealt{Drave2010}; 
(60)~\citealt{Gimenez2016};
(61)~\citealt{Clark2009};  
(62)~A possible spin period of 71.49$\pm{0.02}$ was reported by \citealt{Drave2012} but this result 
was not confirmed and questioned by \citep{Drave2014}; \citealt{Romano2015} reported instead a marginal 
evidence for a periodicity at $\sim$11.6~s which could be associated with the neutron star spin period;
(63)~\citealt{Negueruela2006:aTel831} 
(64)~\citealt{Torrejon2010b}; 
(65)~\citealt{Negueruela2008int};  
(66)~\citealt{Zurita2009}; 
(67)~\citealt{Bird2009};  
(68)~\citealt{Jain2009}; 
(69)~\citealt{Romano2009}; 
(70)~\citealt{RahouiChaty2008}; 
(71)~\citealt{Levine2006}; 
(72)~\citealt{Sguera2007} but note that their results were criticized by \citep{Ducci2013}; 
(73)~\citealt{Coe1996};  
(74)~\citealt{Goossens2013}; 
(75)~\citealt{Walter2006};
(76)~\citealt{Barba2006};
(77)~\citealt{Leyder2007}; 
(78)~\citealt{Gamen2015};
(79)~\citealt{Negueruela2005hd}; 
(80)~\citealt{Lorenzo2010};  
(81)~\citealt{Masetti2006};
(82)~\citealt{Sidoli2006};  
(83)~\citealt{Sidoli2007};  
(84)~\citealt{Swank2007}; 
(85)~\citealt{Hanson1996}; 
(86)~\citealt{Coleiro2013};
(87)~\citealt{Fiocchi2010};
(88)~\citealt{Corbet2010};
(89)~\citealt{Fiocchi2013};
(90)~\citealt{Sguera2015}
}
\end{tabular}
\end{footnotesize}
\end{table*}


\section{Stellar wind parameters from X-ray observations of classical SgXBs and SFXTs}
\label{xraypars}

The class of high mass X-ray binaries (HMXBs) with supergiant companions,
is composed by two sub-classes: the classical systems or SgXBs, like Vela~X-1 already mentioned earlier in this review, 
and the Supergiant Fast X-ray Transients (SFXTs). While these systems share a number of common properties, their variability behavior 
in the X-ray domain is significantly different. As we discuss in the following sections, observations 
of these systems can provide us with information on the physical properties of massive star winds. For instance, recently 
a comparative analysis of the optical companion winds between the SgXBs Vela X-1 and the SFXT IGR J17544-2619, pointed 
to a substantial difference in their terminal velocities, being $\sim 1500$ km s$^{-1}$ in the
case of IGR J17544-2619 and $\sim 700$ km s$^{-1}$ in Vela X-1 \citep{Gimenez2016}.

We summarise shortly below the present knowledge about classical SgXBs and SFXTs. We refer the 
reader to \citet{Walter15} for a recent and more extended review of these systems. 
\begin{figure}
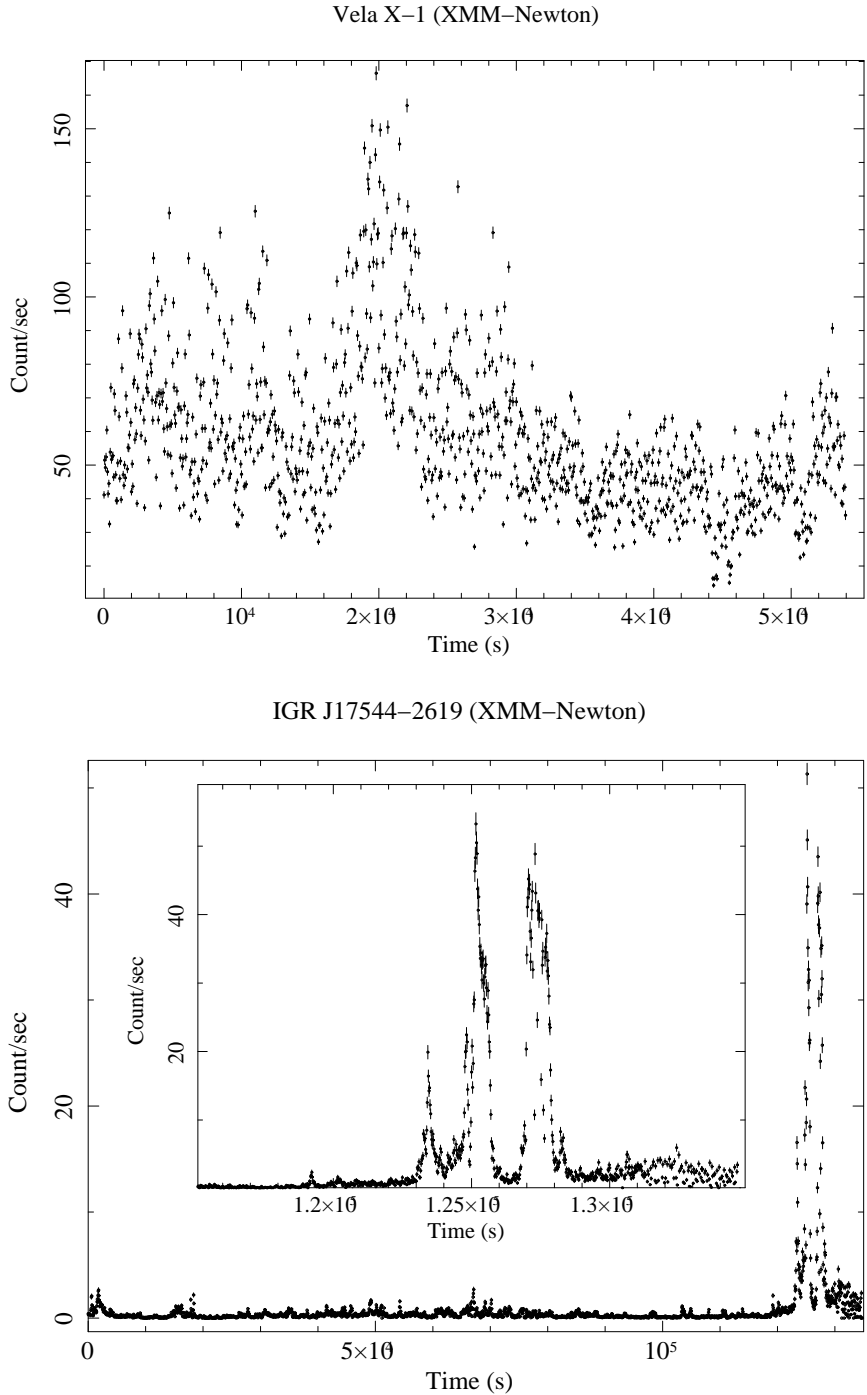

   \begin{tikzpicture}
     \useasboundingbox (0,0) -- (12,0) -- (12,18.5) -- (0,18.5);
     \node at (5.5,14) {\includegraphics[scale=0.48,angle=-90]{lcurve_vela.ps}};
     \node at (5.45,4.5) {\includegraphics[scale=0.51,angle=-90]{J17544_xmm_lc.ps}};
     \node at (5.5,4.85) {\includegraphics[scale=0.45,angle=-90]{J17544_xmm_lc_zoom.ps}};
  \end{tikzpicture}
  \caption{\textit{Top}: An example of a classical SgXB lightcurve as observed by \emph{XMM-Newton} (0.5-10 keV). 
  In this case the source is Vela\,X-1 (see Table~\ref{tab:windfed}). The observation (ID.~0111030101) was carried out on November 2, 2000 for about 60~ks. The flaring behaviour is evident, 
  with X-ray flux variations by a factor of $\sim$3-10 during the entire observation \citep[see][for more details on the \emph{XMM-Newton} data 
  analysis and results for this source]{martinez-nunez2014a}. This variability 
  can be reasonably well explained by assuming that the accretion on the neutron star hosted in this system is taking place from a 
  highly structured and clumpy stellar wind. 
  \textit{Bottom}: an example of an SFXT lightcurve as observed by \emph{XMM-Newton} (0.5-10 keV). 
  In this case the source is the SFXT prototype IGR\,J17544-2619 and the lightcurve is obtained from the longest continuous observational 
  campaign performed in X-rays on one of these sources (ID.~0744600101; \citep{Bozzo2016}).  
  The more extreme variability compared to the classical SgXB can be immediately seen by comparing this lightcurve with that on the top panel. 
  A simple clumpy wind accretion model is not able to fully explain the SFXT behaviour in X-rays, and additional complications 
  have to be taken into account (see Sect.~\ref{SgXB:SFXT}).}    
  \label{fig:velax1lc}
  \vspace{-0.5cm}
\end{figure}

Supergiant X-ray Binaries (SgXBs) are among the first detected galactic X-ray sources
e.g., Cyg~X-1 or Vela~X-1. They consist of OB supergiant mass donors and a compact
object accreting from the strong stellar wind. Table~\ref{tab:windfed} lists 
all known SgXBs harbouring as a compact object a neutron star. 

\begin{figure}
  \begin{tikzpicture}
     \useasboundingbox (0,0) -- (12,0) -- (12,18) -- (0,18);
     \node at (5.4,13.2) {\includegraphics[scale=0.45,angle=-90]{vela_spe.ps}};
     \node at (6,4.3) {\includegraphics[scale=0.75]{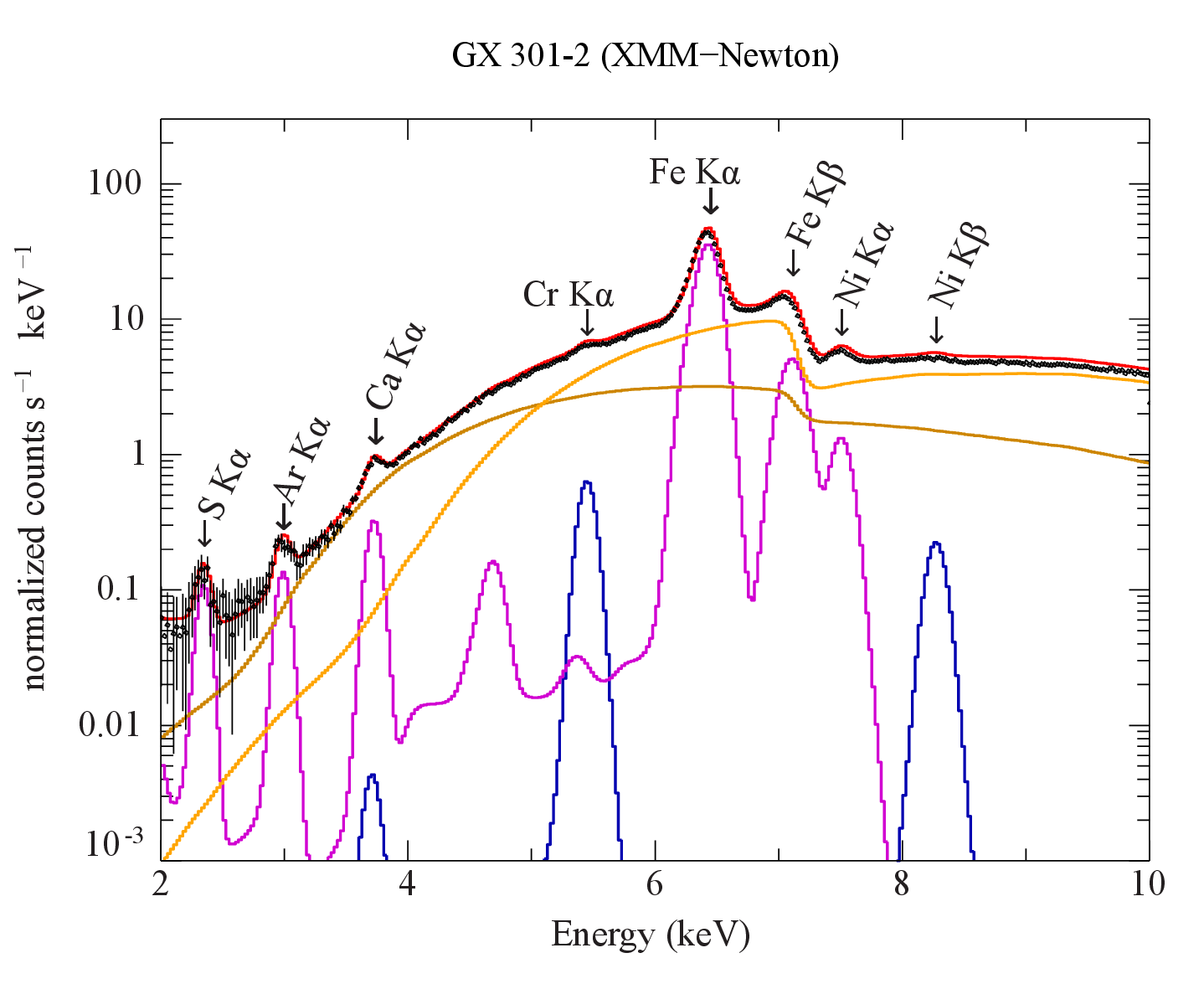}};
  \end{tikzpicture}
  \caption{\textit{Top}: An example of an X-ray spectrum of Vela\,X-1. This corresponds to the average 
  source spectrum extracted during the \emph{XMM-Newton} observation ID.~0111030101 (see Fig.~\ref{fig:velax1lc}). 
  The continuum of the spectrum is preliminarily described by the 
  sum of two power-law components (one including a high energy cut-off) and a prominent iron K$\alpha$ line line at 6.4~keV 
  (see Sect.~\ref{xraypars:feline}). The latter can be clearly seen in the residuals from the fit when 
  it is not included in the spectral model (bottom panel). The average absorption column density measured from Vela\,X-1 
  during the \emph{XMM-Newton} observation is $N_{\rm H}$$\sim$2$\times$10$^{23}$~cm$^{-2}$. \newline
  \textit{Bottom}: Another example 
  of an X-ray spectrum from a classical SgXB. In this case the source is GX\,301-2, the SgXB characterized by one of the  
  highest intrinsic absorption column density ($N_{\rm H}$$\sim$10$^{24}$~cm$^{-2}$). The continuum emission 
  is described by a partial covering plus an absorbed power-law model. A number of prominent emission lines are visible above the continuum. 
  These are the Fe K$\alpha$, Fe K$\beta$, Ni K$\alpha$, Ca K$\alpha$, Ar K$\alpha$, and S$\alpha$ fluorescence lines 
  \citep[see][for details about the data analysis and results]{Fuerst2011}. Reproduced with permission from \citet{Fuerst2011}.}     
  \label{fig:velax1spe}
  \vspace{-0.5cm}
\end{figure}

The classical SgXBs are persistent systems in X-rays, displaying a moderate X-ray 
luminosity achieving 10$^{36}$-10$^{37}$~erg~s$^{-1}$. 
The X-ray emission from SgXBs is usually characterised by a typical short-term variability, comprising 
flares and ``off-states'' that can reach a dynamical range of $\sim$10-100 over time-scales of few to hundred of seconds. 
As briefly anticipated in Sect.~\ref{sec:intro}, this variability is ascribed to the presence of inhomogeneities in the 
wind material that is accreted onto the neutron star or to switches between different accretion regimes. 
We discuss these aspects in more details in Sect.~\ref{xraypars:clumps}.  
As an example of the SgXBs X-ray variability, we show in the left panel of Fig.~\ref{fig:velax1lc} 
a typical lightcurve from Vela~X-1 as observed in the soft X-ray domain. The corresponding X-ray spectrum is 
reported in the left panel of Fig.~\ref{fig:velax1spe} (together with a second example 
of an X-ray spectrum for the classical SgXB GX\,301-2). 

The spectrum of Vela~X-1, as for other SgXBs, displays a large absorption column density 
($N_{\rm H}$ $\gg$ 10$^{22}$~cm$^{-2}$) that is usually ascribed to the dense 
wind material from the supergiant companion filling the neutron star surroundings. 
Monitoring observations of these sources at different orbital phases can thus be used 
to measure changes in the absorption column density and probe the massive star wind on large scales 
(see Sect.~\ref{xraypars:nh}). These observations are also particularly useful  
to reveal the presence of long-lived dense structures around the compact objects that 
are usually associated to the disruption of the stellar wind by the gravitational field and 
X-ray radiation from the compact object (see Sect.~\ref{xraypars:wakes}). 

The relatively complex continuum emission can be described by using one or more power-law components, 
typically interpreted in terms of Comptonization of the thermal photons from the neutron 
star surface in its accretion column. Several emission lines with energies comprised 
in the range 2-8~keV are also usually detected in the spectra of Vela~X-1 and other classical 
SgXBs. These are produced as a consequence of the fluorescent emission of X-rays from the neutron star 
on the surrounding wind material. The most prominent line is the Fe K$\alpha$ from neutral iron 
and/or low-ionization stages, which has a centroid energy of 6.4~keV. Fluorescence from Fe ions at higher ionization states 
can give rise to emission lines in the range 6.7-6.9~keV. The weaker line detected usually around 
7.1~keV corresponds to the Fe K$\beta$ line. In sources with particularly high absorption column densities 
($N_{\rm H}$ $\gg$ 10$^{23}$~cm$^{-2}$) other fluorescent lines can also be detected (e.g., the Ni K$\alpha$, 
Ca K$\alpha$, Ar K$\alpha$, and S K$\alpha$; see right panel of Fig.~\ref{fig:velax1spe}). 
We exploit the usability of these fluorescence features as probes of the 
massive star wind in Sect.~\ref{xraypars:feline}. 

High resolution X-ray spectra of classical SgXBs carried out with 
the gratings on-board {\em XMM-Newton} and {\em Chandra} revealed also the presence of other much less prominent emission lines 
at energies of $\lesssim$3~keV that are typically produced by radiative decays of 
highly ionized ions \citep[e.g., S, Si, Mg, Ne; see][and references therein]{Watanabe2006}. 
These features are particularly important to investigate the properties of the stellar winds, as the measured Doppler 
shifts in their centroid energies allow 
us to derive accurate estimates of the relative velocity between the neutron star and the surrounding material. This is discussed 
in Sect.~\ref{xraypars:wakes}.  

A peculiar sub-class of SgXBs is that of the SFXTs \citep{Sguera2005, Negueruela2005a, Sguera2006}. 
At present, about 10 SgXBs belong to this sub-class, with the prototype being IGR~J17544$-$2619 
\citep{Sunyaev2003}. An overview of the confirmed SFXT sources is reported in Table~\ref{tab:windfed}.  
Although the SFXTs host neutron stars accreting from similar supergiant stars as those in classical 
SgXBs, they display a much more pronounced variability in X-rays, with a typical dynamical range that 
can reach 10$^{5}$-10$^{6}$, compared with a factor 10 in Vela X-1 during a giant flare. The SFXTs 
achieve an X-ray luminosity as high as the typical average value 
of classical SgXBs only sporadically during their peculiar short outbursts. 
These events last a few thousand of seconds at the most and reach 10$^{37}$-10$^{38}$~erg~s$^{-1}$ 
\citep[see, e.g.][]{Rampy2009,Romano2015,Bozzo2011}. We show an example of an SFXT lightcurve in the soft X-ray domain (0.5-10~keV) 
in the lower panel of Fig.~\ref{fig:velax1lc}, where a bright flare is suddenly detected after an extended period of 
low X-ray activity. 

The amount of time that the SFXTs spend in these bright flares is only a few \% \citep{PaizisSidoli2014,Romano2014b}. 
For most of the time they are detected at an intermediate emission state with a typical luminosity of 
10$^{33}$-10$^{34}$~erg~s$^{-1}$, during which less prominent flares are observed. The latter are reminiscent of 
the brightest outburst (i.e. they occur on the same time-scales), but their peak luminosity is  
$\lesssim$10$^{35}$~erg~s$^{-1}$ \citep{Sidoli2008,Bodaghee2011,Sidoli2010,Bozzo2010}. 
A number of SFXTs were also observed in a quiescent state, with luminosities down to 
10$^{32}$~erg~s$^{-1}$ \citep{Gonzalez-Riestra2004,Zand2005,Leyder2007,Bozzo2010,Bozzo2012,
Drave2014}. Overall, these sources appear to be substantially sub-luminous 
compared to the classical SgXBs \citep{Bozzo2015}. 

Given several similarities in the spectral shape between SFXTs (especially when in outburst) and classical SgXBs, 
it has always been assumed that the bulk of the X-ray emission from the two classes of systems has a common origin and 
should be produced by the accretion of the structured supergiant star wind onto the neutron star. Clumps are thus expected 
to drive at least part of the X-ray variability of the SFXTs. However, the physical mechanisms responsible for the more extreme 
behaviour of these sources is still not fully understood and remains highly 
debated \citep{Sidoli2012,PaizisSidoli2014,Walter15}. We discuss this issue more extensively    
in Sect.~\ref{SgXB:SFXT}.

\subsection{X-ray variability and clumpy stellar winds: flares and off-states}
\label{xraypars:clumps}

It was shown (see Sect.~\ref{sec:bondi}) that X-ray luminosity \Lx\ recorded from SgXBs and SFXTs traces reasonably 
well the mass per unit time that is accreted by the neutron star. It was also shown in that section how 
changes in the mass accretion rate of a wind-fed system can lead to almost immediate variations of its 
X-ray luminosity, as there is no accretion disk mediating the transfer of material from the supergiant companion 
to the compact object. In the simplest accretion scenario depicted in Sect.~\ref{sec:bondi}, where all complications 
related to the presence of the neutron star magnetic field and spin rotation are neglected, we can obtain
an expression connecting directly the X-ray luminosity with the properties of the stellar wind merging
Eq.~\ref{eq:acrrate} and \ref{eq:lx_base}:

\begin{equation}
L_{\rm X} = 4 \pi \zeta \rho(a,t) \frac{(G M_{\rm NS})^3}{R_{\rm NS}\upsilon^3_{\rm w}},  
\label{eq:lx_nswind}
\end{equation} 
where we also applied the simplification $\upsilon_{\rm rel}$ $\sim$ $\upsilon_{\rm w}$. 
From this equation, it can be seen that a change in the density of the wind 
can lead to a comparable variation of the X-ray luminosity and the latter can be boosted even more 
in case of variations in the wind velocity. As we have seen in Sect.~\ref{sec:smallscale}, 
changes in density and velocity of the wind at the neutron star location 
can be induced by the presence of moderately dense clumps or inhomogeneities. It thus turns out 
from Eq.~\ref{eq:lx_nswind} that a dynamic range up to 10-100 in the X-ray luminosity of wind-fed binaries 
can be easily achieved by assuming that the neutron star is accreting from a clumpy and structured wind. 
It is worth remarking here that Eq.~\ref{eq:lx_nswind} is derived in the simplest case of the direct accretion regime, 
thus neglecting any factor that might arise due to the presence of a centrifugal/magnetic gate and the onset 
of a possible settling accretion regime. 

Under the assumptions above, one could thus use the observed X-ray variability and continuous flaring behaviour 
in classical SgXBs to estimate the properties of clumps needed to regulate it. In the simplified approach proposed by 
\citet{walter2007a}, it is assumed that the neutron star accretes only a certain fraction of the clump 
mass (\Mcl$\gtrsim$\Macc) and that its radius is larger than the NS corotation radius (\Rcl$\gtrsim$\Racc; see 
Eq.~\ref{eq:raccr}). In this case we can define: 
\begin{equation}
\Mcl = (\Rcl/\Racc)^2\,\Macc.  
\end{equation}

If we further consider that the duration of a flare \tfl is proportional to the time that the neutron star 
needs to go through the clump and accrete (at least) a part of it, we can estimate

\begin{equation}
\Rcl \sim \vcl\times\tfl .   
\end{equation}
Here the velocity of the clump \vcl=$f$$\upsilon_{\rm w}$ is assumed to be a fraction $f$ of the stellar 
wind velocity $\upsilon_{\rm w}$. The mass of the clump is thus given by: 
\begin{equation}
\Mcl = \frac{L_{\rm X} R_{\rm NS}}{(GM_{\rm NS})^3}\tfl^3 f^2 \upsilon_{\rm w}^6.
\end{equation}
For typical parameters of flares in classical SgXBs (see, e.g., the right panel of Fig.~\ref{fig:velax1lc})  
and considering a standard neutron star ($R_{\rm NS}$=10~km, $M_{\rm NS}$=1.4~$M_{\odot}$), 
one obtains:   
\begin{equation}
\Mcl \simeq 5\times10^{18}\,\text{g} \left(\Lx \over {10^{36} \text{erg/s}}\right)
       \left(\tfl \over {0.5\,\text{ks}}\right)^3
       \left({f} \over {0.5}\right)^2 \left({\upsilon_{\rm w}} \over {1000\,{\rm km~s^{-1}}}\right)^6
       \label{eq:clmass}
\end{equation}
that yields a reasonable value compared to the clump mass of $\sim$10$^{18}$~g (at the most) inferred from LDI numerical simulations of 
hot star winds (see Sect.~\ref{sec:smallscale}). \citet{walter2007a} also noted that through Eq.~\ref{eq:clmass} particularly 
long (\tfl$\gg$1~ks) and structured flares could still be interpreted in terms of accretion from an inhomogeneous wind, 
if each individual peak of intensity during the flare is associated with the accretion of a small clump being part of 
a clustered structure. 

Refined calculations to link X-ray flares to the physical properties of the clumps were 
presented by \citet{Ducci2009}, including the orbital characteristics of SgXBs and a distribution in radius and mass 
for the clumps. The effect of different impact parameters for the clump accretion onto the 
neutron star was also included into the calculation, thus permitting more reliable estimates for the clump 
physical properties from the observation of X-ray flares that those inferred from the simplified calculations presented above.  
Assuming the standard direct accretion scenario (Sect.~\ref{sec:regimes}), \citet{Ducci2009} derived synthetic X-ray lightcurves of several 
classical SgXBs, and comparing these results with those obtained from X-ray observations, inferred ranges for the clump masses of 
$5\times10^{18}\,\text{g}$ to $5\times10^{21}\,\text{g}$ in Vela\,X-1, $5\times10^{16}\,\text{g}$ to 
$2\times10^{19}\,\text{g}$ in 4U\,1700-37, and $5\times10^{16}\,\text{g}$ to 
$2\times10^{21}\,\text{g}$ in IGR\,J16418-4532 \citep[see also][]{Romano2012}. 
In some cases, these calculations thus suggest the presence of clumps with 
a significantly larger mass than that expected according to LDI simulations of massive star winds (see above). 

Similar results were also independently obtained by other authors with different methods.  
\citet{Furst2010} studied the  lightcurve of Vela~X-1 as observed by the \emph{INTEGRAL} satellite and showed that the 
log-normal distribution of the brightness of individual flares could be well reproduced in a simulation in which 
the direct accretion process onto the neutron star is taking place from a clumpy rather than a smooth 
stellar wind. They found that clumps in Vela~X-1 should have masses in the range 
$5\times10^{19}$-$10^{21}\,\text{g}$ in order to explain the X-ray data \citep[see also][]{Fuerst2014}. 
Similarly, \citet{martinez-nunez2014a} and \citet{Fuerst2014} could only explain the properties of some X-ray flares 
detected during dedicated observations of Vela\,X-1 by assuming clump masses of $\gtrsim$10$^{21}\,\text{g}$. 

On one hand, it thus seems that the scenario of the accretion from a clumpy wind can explain reasonably well 
the X-ray variability of classical SgXBs. In particular, it can reproduce the main features of their high energy emission 
and the typical shape of the observed lightcurves. On the other hand, a number of published results  
suggest the presence of a population of clumps in these sources 
that are a factor of $\sim$100-1000 more massive than those expected from the LDI simulations 
(see Sect.~\ref{sec:smallscale}). One possibility is that such a discrepancy arises as a consequence of the 
currently known limitations of LDI simulations, which cannot yet account for complex interactions 
between these structures when multi-dimensional approaches are used (and might thus lead to the formation of 
larger structures than currently foreseen; see Sect.~\ref{sec:smallscale}). 
However, it is also likely that the presence of centrifugal/magnetic gates or the onset of a settling 
accretion regime could play an important role in the determining the release of the X-ray luminosity, thus affecting 
the estimates of the clump masses derived with the techniques described above.  
As shown in Sect.~\ref{sec:regimes}, the switch between one regime of accretion 
to another can boost the variation of the system X-ray luminosity dramatically, even in case of modest changes in  
the wind density and/or velocity. In these cases, the X-ray luminosity is no longer an obvious tracer 
of the mass accretion rate, and small clumps can easily become the cause of large swings in the X-ray 
emission properties of the system. Neglecting these effects can thus potentially led to 
an overestimate the sizes and masses of clumps. A more quantitative analysis is currently hampered 
by the uncertainties on the physical processes occurring in the different accretion regimes and the poorly known 
properties of the neutron stars hosted in many of the SgXBs. The development of a clumpy wind accretion 
model accounting for different accretion regimes is still under way. 

Clumpy wind accretion models encounter more difficulties in explaining the extreme variability of the SFXT sources. 
By looking at Eq.~\ref{eq:lx_nswind}, one would immediately conclude that unreasonably large and dense clumps 
would be needed to achieve a dynamic range in the X-ray luminosity comparable to that of the SFXTs 
(up to 10$^5$-10$^6$). Furthermore, the typical lightcurve of an SFXT is significantly different from that of a classical 
system, as it is characterized by sporadic isolated bright flares between which the source spend a long time in 
a much lower and less variable luminosity regime (see the right panel of Fig.\ref{fig:velax1lc}). 
As we discuss in Sect.~\ref{SgXB:SFXT}, the models proposed so far to interpret the peculiar 
behaviour of the SFXTs are considering a number of 
different mechanisms to regulate their variability, but in all cases clumps still play an important role 
in achieving the overall X-ray dynamic range. 

Convincing evidences for the presence of clumps in the stellar winds of SFXT supergiants 
have been found especially by searching for absorption events in the soft X-ray 
lightcurves of these objects. The role of dense clumps is indeed two-fold. Clumps 
that lead to an increased mass accretion rate produce X-ray flares as discussed above, but during the 
ingresses (egresses) from these flares the clump approaching (moving away from) the neutron star will also   
cause a dimming or even obscuration of the X-ray source. These events lead to remarkable increases of 
the absorption column density in the direction to the sources and are expected to be even more frequent than 
the X-ray flares (all clumps simply passing through the line of sight to the observer even without being 
accreted contribute to increase the $N_{\rm H}$). 
If we use again the simple equations presented earlier in this section, we can estimate the 
absorption column density associated to the passage of a clump along the line of sight to the observer as: 
\begin{equation}
\begin{split}
\nh \simeq \frac{\Mcl}{\Rcl^2 m_{\rm p}} = \frac{L_{\rm X} R_{\rm NS}}{(GM_{\rm NS})^3}\tfl \upsilon_{\rm w}^4 = \\
9 \times 10^{22}\,\text{cm}^{-2} \left( \frac{L_\text{fl}}{10^{36}\,\text{erg}\,\text{s}^{-1}}
\frac{t_\text{fl}}{0.5\,\text{ks}}\left(\frac{\upsilon_{\rm w}}{1000\,{\rm km~s^{-1}}}\right)^4\right)
\label{eq:nh_cl}
\end{split}
\end{equation}
From Eq.~\ref{eq:nh_cl} we note that clumps usually remain optically thin in the X-rays
(see also Sect.~\ref{sec:size}) and that the 
increase in the absorption column density is in the range fully accessible through soft X-ray 
observations (e.g. carried out with \emph{XMM-Newton}, \emph{Swift/XRT}, \emph{Suzaku}, and 
\emph{Chandra}). \citet{Rampy2009} were the first to report the evidence of clumps in the winds of the 
SFXTs supergiant companions by using this technique. These authors detected an absorption 
events in the \emph{Suzaku} lightcurve of the SFXT IGR\,J17544-2619 that lasted 300~s 
and estimated that such effect could have been produced by a clump as large as 
$\sim$4.2$\times$10$^9$~cm and with a mass of $\sim$1.5$\times$10$^{18}$~g. The increase 
of the absorption column density in the direction of the SFXT IGR\,J18410-0535 during a bright flare 
from the source led \citet{Bozzo2011} to conclude that the event could have been caused by a clump 
with \Rcl$\sim$8$\times$10$^{11}$~cm and \Mcl$\sim$1.4$\times$10$^{22}$~g. 
By applying the refined clumpy wind accretion model developed by \citet{Ducci2009} to the SFXTs  
IGR\,J11215$-$5952 and IGR\,J18483-0311, clumps in the mass range 
$10^{17}\,\text{g}$ to $5\times10^{20}\,\text{g}$ and $10^{18}\,\text{g}$ to $5\times10^{21}\,\text{g}$ 
were obtained \citep[see also][]{Romano2010}.  

As for the classical SgXBs, the estimated properties of the clumps presented above were all obtained by 
assuming the simplest direct accretion regime and neglecting possible systematic uncertainties affecting the  
luminosity and absorption column density determined from the fit to the X-ray spectra of these sources. 
For this reason, caution should be taken when comparing the derived clump properties in the SFXTs with 
those inferred from simulations and observations of isolated supergiant stars. 

Furthermore, it should be noted that in both cases in which clump parameters are derived from flares of 
SgXB sources or absorption events in their lightcurves, there are systematic uncertainties on the 
values of the X-ray luminosity, $L_{\rm X}$, and the amount of absorbing neutral material, 
$N_{\rm H}$, that are usually not included in the calculations. We discuss separately these issues 
in Sect.~\ref{xraypars:nh} and \ref{susec:ns_bolum}.    

X-ray flares are not the only way in which we can probe the presence 
of small scale structures in the winds of supergiant stars in SgXBs. 
At least four classical SgXBs (Vela\,X-1, 4U\,1700-37\footnote{A source that we do not deeply discuss 
in this review as the nature of its accreting compact object is still debated.}, 
4U\,1907+09 and GX\,301$-$2) show occasional ``off-states'' (see Table~\ref{tab:offstates}), which have been often interpreted in 
terms of the clumpy properties of the stellar wind. 
During the off-states the X-ray luminosity of these sources reaches very
low levels, often below the observable limit, and sometimes
also other changes to the X-ray emission properties are detected. 
It is important to note here that different works in the literature use the term off-state 
for different levels of luminosity even in the same source, as the effective limit 
is often driven by the sensitivity and energy range of the observing instrument 
rather than by physical considerations. A variety of mechanisms has been proposed 
to explain these low luminosity events, and we summarise below all relevant findings from 
the literature. 

\begin{figure}
\begin{minipage}{.55\textwidth}
  \includegraphics[scale=0.55]{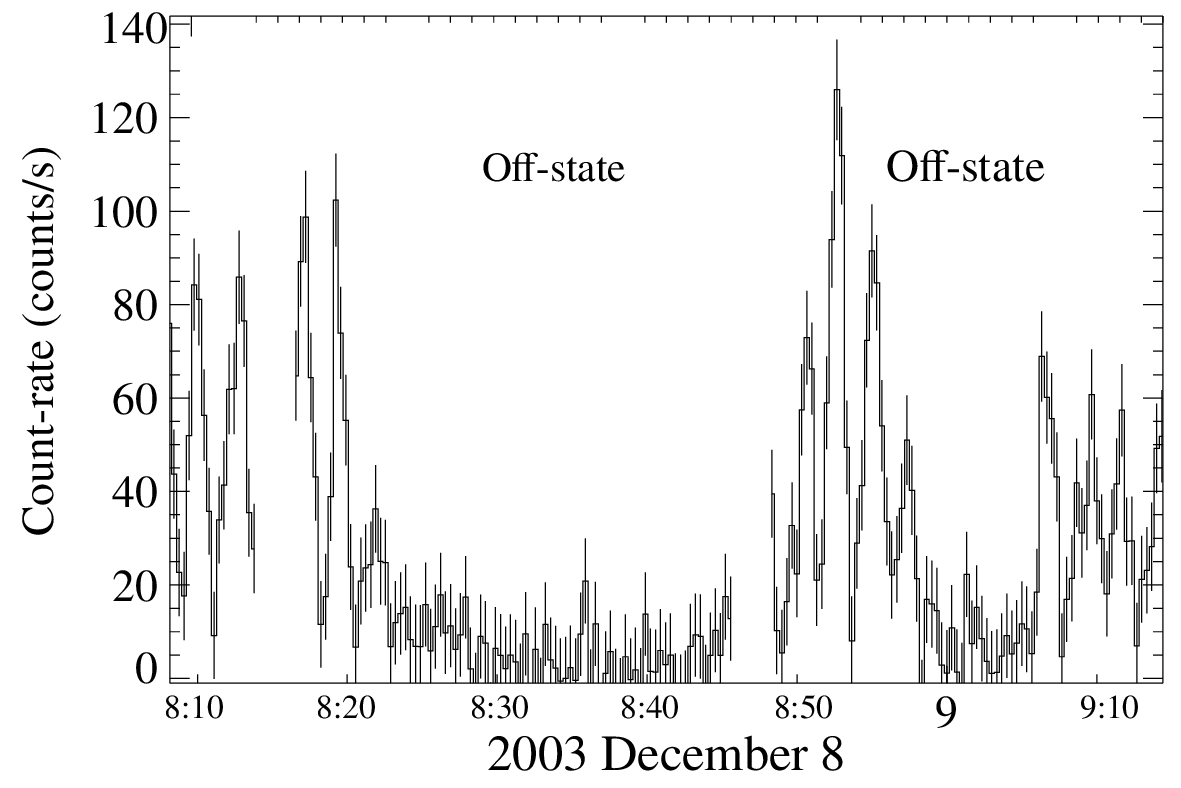}
\end{minipage}\hfill\begin{minipage}{.4\textwidth}
  \caption{An example of an off-state in Vela\,X-1 as observed by IBIS/ISGRI 
  on-board \emph{INTEGRAL} (20-60 keV). Due to the relatively low sensitivity of ISGRI 
  to faint sources, no residual flux from Vela\,X-1 is recorded by \emph{INTEGRAL} during these 
  off-states. However, more sensitive instruments revealed that the source is still emitting in X-rays 
  and pulsating during these events, albeit at a much more reduced level than usual (see text for details).
  Reproduced with permission from \citet{krey2008}.}
  \label{fig:velaOFF}
\end{minipage}
\end{figure}

As in many other areas, the best studied source for low and off-states is Vela~X-1, 
due to its relatively small distance and high X-ray flux (see Table~\ref{tab:windfed}). 
The possibly first observation of such a state was reported by \citet{KallmanWhite:82}, 
and described as a ``brief absorption event'' lasting for $\sim$6~minutes 
with no evident change in the source spectral properties.  
In 1983, a \emph{Tenma} observation showed a drop in intensity lasting for less than 80 minutes
and during which pulsations disappeared as well. \citet{Hayakawa:84} discussed both of the 
preceding observations as being possibly caused by a planet eclipsing the source, while \cite{Inoue:84},
analysing the 1983 observation in detail offered a sudden choking of the accretion as alternative.
\citet{Lapshov:92} reported on an extended low state in Vela\,X-1 of more than 10~hours duration
outside the source X-ray eclipse observed by Granat/Watch. In this case the off-state was 
explained in terms of an abrupt reduction in the accretion rate.
At the start of a \emph{RXTE}/PCA observation in 1996, \citet{Kreykenbohm:99} noted an interval of
at least 550~s duration with low count rates and no observable pulsations from the source. 
In a somewhat different case, \citet{Kretschmar:CGRO99} found a sudden flux
decrease in January 1998 with \emph{RXTE} with disappearing pulsations 
while significant non-pulsed source flux remained for more than an hour. As an explanation, a 
massive ($\tau\approx 1.6$) and very large ($\sim10^{13}$~cm) clump 
obscuring the pulsar and destroying coherent pulsations by scattering was considered. 
Using deep \emph{INTEGRAL} observations of Vela~X-1 from late 2003 \citet{krey2008} found five occurrences of
off-states with durations between 520 and 1990~s (an example of an off-state from Vela\,X-1 
is shown in Fig.~\ref{fig:velaOFF}). 
For four of the five instances the preferred explanation is a combination of strong density fluctuations in the wind
and the onset of a centrifugal gate, while for the last and longest, again
a massive clump was found to best explain the data. 
\citet{Doroshenko:2011} analysed a 100~ks observation of Vela~X-1 by \emph{Suzaku} and
found three off-states. At odds with previous results, mostly obtained with
less sensitive instruments, they found that pulsations continued also during the
off-states, albeit with clear changes in the spectra and pulse profiles. They explained
their results by invoking a gated accretion, in which the magnetospheric boundary becomes 
stable with respect to the RTI but remains unstable with respect to the KHI (that usually 
allows accretion to occur at a somewhat reduced rate; see Sect.~\ref{sec:regimes}).  
\citet{Sidoli:2015} combined data over 10~years of INTEGRAL observations searching
for low or off-states in the hard X-ray lightcurve of Vela~X-1. They found that while some 
off-states can appear at any orbital phase, other are more common close to the eclipses and with 
different distributions for the eclipse egress and ingress. This asymmetric distribution is 
explained by scattering in ionized material, compatible with the photoionisation 
wake (see, e.g., the geometry in the Vela~X-1 system outlined by \cite{Kaper1994}). The cases of 
off-states evenly distributed around the orbit could be caused either by exceptionally low 
matter density or gated accretion or by changes in accretion regime within 
the spherical settling accretion model \citep{Shakura2013}. In order to test possible explanations of the 
observed off-states,  \citet{ManousakisWalter:2015} created a hydrodynamical model of the Vela~X-1 system.
They found that low-density bubbles formed close to the neutron star, due to 
unstable hydrodynamic flows, lead to off-states with typical durations between 
5-120~minutes. As in the simulations presented by these authors   
the stellar wind is assumed to follow the simplest CAK approximation (see Sect.~\ref{basicsldw:CAK}),  
these findings opened the new interesting possibility that (at least) not all off-states could  
be related to clumps and/or gated accretion regimes.  

In the case of 4U~1907+09, a less deeply studied source than Vela~X-1, \citet{intZand:1997}
first noted a peculiar dipping behaviour in several \emph{RXTE} observations, 
finding at least 11 dips longer (up to $\sim$1.5~h) than the pulse period ($\sim$600~s). 
Voids in the inhomogeneous medium surrounding the neutron star were considered the most likely 
possibility to interpret the drops in the accretion, and the possibility was considered 
that voids might also be created due to photoionization.
\citet{Roberts:2001} found 4U~1907+09 in a very low state during two out of the four performed 
\emph{ASCA} observations spread roughly equally around the 8.4~d orbital period. These low states
of relatively long duration ($\sim 10$~ks) showed no signs of increased absorption, and were interpreted 
in terms of magnetically inhibited accretion. 
\citet{Rivers:2010} noted dips in the \emph{Suzaku} observations of 4U~1907+09, which 
\citet{Doroshenko:2012} analysed later in more detail. These observations included
four off-states of about 700-900~s duration. Similar to the 
case of Vela~X-1, \citet{Doroshenko:2011} demonstrated the presence of a weak but significant 
flux and pulsations during all these off-states and interpreted them in a coherent way for the 
two sources.

GX\,301-2 is a rather peculiar system with an eccentric orbit ($e=0.462$) and a marked
orbital flux variation with a strong pre-periastron flare. 
\cite{Gogus:2011} noted a peculiar dip in one of the \emph{RXTE} lightcurve of the source 
during which X-ray pulsations declined, disappeared for one spin cycle (686~s), and
then reappeared again. Spectral changes were similar to those observed
during the dips in Vela~X-1 \citep{Kreykenbohm:99}. A brief cessation of accretion 
to the magnetic poles was considered the most probable explanation.
During an observation of part of the source pre-periastron flare with \emph{XMM-Newton},  
\citet{Fuerst2011} found another interval of low flux in which the pulsations ceased almost completely
for several spin cycles, without any indication of increased absorption or other spectral changes. 
Investigating this dip in detail, they argued that it is most likely that during the dip the accretion ceased 
and the afterglow of fluorescent iron around the neutron star accounted for the main portion of the 
residual X-ray flux. 

Two possible off-states were also identified in the lightcurves of the classical SgXB
IGR\,J16418-4532 and the SFXT IGR\,J16328-4726 by \citet{Drave2014} and \citet{Bozzo2012}, respectively. 
In both cases the statistics of the X-ray data was far too low to perform a detailed spectral 
and timing analysis of the two events, but Table~\ref{tab:offstates} shows that the luminosity 
measured during the off-state was roughly comparable to those obtained in the brighter SgXBs 
mentioned above. 

Although the origin of the off-states is still unclear and highly debated, all possibilities 
described above are particularly interesting to use these events as additional probes 
of stellar winds in SgXBs. If off-states are due to voids between clumps, observations of these events could 
be used in combination with the detection of X-ray flares to build a more complete map 
of the stellar wind and compare this with the outcomes of LDI or more advanced simulations 
(see Sect.~\ref{sec:smallscale}).A complication to this simple scenario comes from the fact that, as 
pointed out by several authors, off-states might also be (at least) partly related to a change in the 
accretion regime, or to the onset of a centrifugal and/or magnetic inhibition of accretion, or to the scattering 
effect produced by photo-ionised large-scale wind structures. It is particularly tricky to disentangle these effects, because any  
drop in the mass accretion rate implies an increase in the size of the neutron star magnetosphere and thus 
a large possibility to switch toward a propeller-like accretion regime (see Sect.~\ref{sec:regimes}). 
Additionally, as for the X-ray flares discussed above, the properties of the stellar 
winds inferred from the study of the off-states would also be affected by the known  
systematic uncertainties on the source luminosity and absorbing material derived from the X-ray 
data (see Sect.~\ref{xraypars:nh} and \ref{susec:ns_bolum}). 
\begin{table}
\caption{Overview of off-state or low flux observations in SgXBs. 
In some cases luminosities have been estimated based on the information in the original 
work and the newer distance estimates became available in the literature 
(see Table~\ref{tab:windfed}).}
\label{tab:offstates}
\begin{footnotesize}
\begin{center}
\scriptsize
\begin{tabular}{@{}lrcrrl@{}}
\hline
Source   & \multicolumn{1}{c}{Date} & Instrum. &Energy range&Off-state luminosity  &References \\ 
\hline
\hline
Vela\,X-1 &  9 May 1979 & Einstein & 2-10~keV & \emph{not quantified} & KW82, Ha84 \\
Vela\,X-1 & 12 Mar 1983 & Tenma & 3-9~keV & \emph{$<$10\% of preceding} & In84, Ha84 \\
Vela\,X-1 &  9 Jan 1991 & Watch & 8-15 keV & \emph{not quantified} & La92 \\
Vela\,X-1 & 23 Feb 1996 & RXTE/PCA & 3-30~keV & \emph{$<$15\% of normal} & Kr99 \\
Vela\,X-1 & 22 Jan 1998 & RXTE/PCA & 2-60~keV & \emph{not quantified} & Kr00 \\
Vela\,X-1 &  8 Dec 2003 & ISGRI &  20-40~keV & \emph{not quantified} & Kr08 \\
Vela\,X-1 & 17-18 Jun 2008 & Suzaku/XIS & 0.4--70~keV & $\sim 2.4\times10^{35}$~erg~s$^{-1}$ & Do11 \\
Vela\,X-1 &       2002--2012 & ISGRI &  22-50~keV & $\lesssim 3\times10^{35}$~erg~s$^{-1}$  & Si15 \\
\hline
4U\,1907+09 & 17-23 Feb 1996 & RXTE/PCA  & 2-15~keV & \emph{up to 98\% decrease} & iZ97\\
4U\,1907+09 & 14-16 Oct 1996 & ASCA  & 2-10~keV & $\sim 1.6\times10^{34}$~erg~s$^{-1}$ & Ro01\\
4U\,1907+09 &  2--3 May 2006 & Suzaku/XIS &    & $\sim 10^{35}$~erg~s$^{-1}$ & Ri10, Do12 \\
\hline
4U\,1700-37 & 17-18 Feb 2001  & XMM-Newton & 0.5-10~keV &  $2\times10^{35}$~erg~s$^{-1}$ & vdM05 \\
\hline
GX\,301-2 & 28 May 2010 & RXTE/PCA & 3-25~keV & \emph{not quantified} & Go11 \\
GX\,301-2 &  & XMM-Newton & 2-10~keV & $\lesssim 4\times10^{35}$~erg~s$^{-1}$ & Fu11 \\
\hline 
IGR\,J16418-4532 &  & XMM-Newton & 0.5-10~keV & $\lesssim 8\times10^{34}$~erg~s$^{-1}$ & Dr13 \\
\hline
IGR\,J16328-4726 &  & XMM-Newton & 0.5-10~keV & $4.1\times10^{35}$~erg~s$^{-1}$ & Bo12 \\
\hline
IGR\,J17544-2619 & 16 Sept 2012 & XMM-Newton & 0.5-10~keV &  $4\times10^{32}$~erg~s$^{-1}$ & Dr14  \\
\hline
\smallskip \\ 
\multicolumn{2}{@{}l}{Bo12: \citet{Bozzo2012}}       & \multicolumn{2}{l}{Ha84: \citet{Hayakawa:84}}     & \multicolumn{2}{l}{La92: \citet{Lapshov:92}}\\ 
\multicolumn{2}{@{}l}{Do11: \citet{Doroshenko:2011}} & \multicolumn{2}{l}{In84: \citet{Inoue:84}}       & \multicolumn{2}{l}{Ri10 \citet{Rivers:2010}}\\
\multicolumn{2}{@{}l}{Do12: \citet{Doroshenko:2012}} & \multicolumn{2}{l}{iZ97: \citet{intZand:1997}}   & \multicolumn{2}{l}{Ro01 \citet{Roberts:2001}}\\
\multicolumn{2}{@{}l}{Dr13: \citet{Drave2013}}       & \multicolumn{2}{l}{KW82: \citet{KallmanWhite:82}}& \multicolumn{2}{l}{Si15: \citet{Sidoli:2015}}\\
\multicolumn{2}{@{}l}{Dr14: \citet{Drave2014}}       & \multicolumn{2}{l}{Kr00: \citet{Kretschmar:CGRO99}}& \multicolumn{2}{l}{vdM05: \citet{vdm2005}}\\
\multicolumn{2}{@{}l}{Fu11: \citet{Fuerst2011}}      & \multicolumn{2}{l}{Kr99: \citet{Kreykenbohm:99}} & \\
\multicolumn{2}{@{}l}{Go11: \citet{Gogus:2011}}      & \multicolumn{2}{l}{Kr08: \citet{krey2008}}       & \\
\end{tabular}
\end{center}
\end{footnotesize}
\end{table}

\subsection{Disrupting the stellar wind: photoionization, accretion wakes, and super-orbital modulations}
\label{xraypars:wakes}

Observations and hydrodynamic simulations of wind-fed systems have convincingly shown 
that these stellar winds are heavily affected by the X-ray 
radiation emitted by the neutron star 
\citep[see, e.g.,][and references therein]{Blondin1991}. If the luminosity of the X-ray source 
is sufficiently intense (typically $\gtrsim$10$^{35}$-10$^{36}$~erg~s$^{-1}$), the 
photoionization of the wind changes the state of a large fraction of the heavy ions within the so-called 
Str\"omgren sphere (they become ionized to a higher degree, with corresponding lines only in the
high energy, low-flux spectral region of the star), and the driving acceleration mechanism of the wind
is cut off (see Sect.~\ref{basicsldw}). The wind is not further accelerated, it is accumulated at the neutron star location, 
giving rise to a ``photoionization wake'' \citep[see, e.g.,][]{Fransson1980}. 
As the neutron star moves along its orbit, the photoionization wake is displaced 
together with the compact object and the additional material halted at the bow-shock 
(see Sect.~\ref{sec:bondi}). During the neutron star revolution, the latter give rise to 
an elongated dense structure that is usually termed ``accretion wake'' \citep[see, e.g.,][]{Blondin1990}.   
A substantial theoretical effort was devoted in the past to calculate 
the details of the interaction between the X-ray radiation and the surrounding environment, 
starting from the pioneering works of \citet{Tarter1969} and \cite{Krolik1984} and arriving 
to the more specialized calculations of \citet{Friend1982} and \citet{Ho1987} to the case of  
wind-fed binaries  (see also, more recently, \citet{Ducci2010}). However, only with the most recent hydrodynamical simulations it has been 
possible to follow in detail the photoionization of the wind \citep[see, e.g.,][]{Krticka2012,Krticka2015} 
and the formation of time-dependent accretion/photoionization wakes \citep{ManousakisWalter:2015,Cechura2015}. 

As we discuss below, the photoionization of the wind and the formation of accretion/photoionization wakes 
are all additional tools to probe the characteristics of the stellar winds in SgXBs. In particular, they can 
be used to indirectly infer the physical properties of the unperturbed winds in these systems by comparing observational 
results with numerical simulations of the interaction between the supergiant outflow and the X-ray radiation 
from the compact object.  

Among the different spectral features produced by the photoionization of the stellar wind and thus arising especially 
in the photoionization wake, the Lyman series lines from the H and He-like ions are the most appropriate features
to probe the dynamics of this medium. These lines are formed by radiative decays following either 
photoexicitation or radiative recombination, and their properties can be accurately calculated from theory. 
Shifts in the centroid energies of these lines can thus be easily evaluated from X-ray observations, providing 
a direct measurement of the wind velocity. Together with these lines, the width in energy of the radiative recombination 
continuum, which is formed by the photons generated in recombination and distributed into a continuum 
(typically line-like in the case of the SgXBs), provides an estimate of the wind electron temperature. 
These diagnostic techniques have been successfully exploited, as an example, in the case of Vela\,X-1 
by \citet{Watanabe2006}. The red and blue velocity shifts measured at different orbital phases from the emission lines 
of several elements (Si, S, Ca, Al, Mg, Ne) at energies $\sim$1.9-2.7~keV revealed that the supergiant wind 
close to the neutron star location has a velocity that is a factor of a few lower than $\sim$2-3 expected.   
By using Monte Carlo simulations, \citet{Watanabe2006} 
concluded that the reduced velocities were due to the effect of the wind photoionization by the neutron star 
X-ray emission that led to a substantial drop in the radiatively driven acceleration of the stellar wind. 
The parameters of the unperturbed wind reconstructed from their Monte Carlo simulations provided a mass-loss rate
of (1.5-2.0)$\times$10$^{-6}$~$M_{\odot}$~yr$^{-1}$ and a terminal wind velocity of 1100~km~s$^{-1}$, compatible 
with independent estimates obtained from the P Cygni profile of UV resonance lines and expected for a 
B0.5\,Ib supergiant (as the one hosted in Vela\,X-1; see Table~\ref{tab:windfed}). 
Similar results have been presented also for the bright 4U\,1700-37 system, a source that we do not 
deeply discuss in this review as the nature of the accreting compact object in this system is still 
debated \citep{Boroson2003}.

In general, despite the interest and the important achievements of such studies, 
they have been carried out so far only for a few close-by SgXBs. This is mainly due to the fact that 
SgXBs are heavily absorbed in the soft X-ray domain and the sensitivity of the currently available X-ray 
gratings, the sole instruments that permit to analyse spectral lines at $\lesssim$3~keV with the 
required accuracy, limits the possibility to carry out these studies only to tentatively
bright objects.       

\begin{figure}
\begin{minipage}{.55\textwidth}
  \includegraphics[scale=0.30,angle=-90]{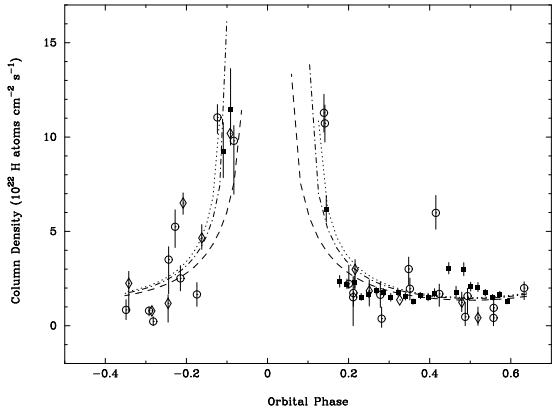}
\end{minipage}\hfill\begin{minipage}{.33\textwidth}
  \caption{Orbital modulation of the $\nh$ as a function of the orbital phase in the eclipsing SgXB 4U~1538$-$522, 
  \citep[see][reproduced with permission]{Mukherjee2006}. The authors assume a circular orbit with $a_x\,\sin(i)=53.1$~lt-s and three different 
  values for the inclination: $i=65^{\circ},75^{\circ},85^{\circ}$ (dashed, dashed-dotted, and dotted lines respectively). 
  The following wind parameters are also assumed: $\beta=0.5$ (Please note that the assuming $\beta$ is 
  too small, since considering the so-called finite-cone-angle effect implies a minimum $\beta$ value of $\sim$ $0.7-0.8$), 
  $\dot{M}=10^{-6}\,M_{\odot}/yr$, $\upsilon_\infty=1000$~km/s. They use observations from \textsl{RXTE} in 1997 (diamonds), 
  \textsl{BeppoSAX} in 1998 (squares), and \textsl{RXTE} in 2003 (circles).}
  \label{NH_modu}
\end{minipage}
\end{figure}

The presence of an accretion wake can be observationally inferred in many sources 
by studying the different shapes of the energy resolved lightcurve folded on the system orbital period 
\citep[see, e.g.,][for examples]{Falanga:2015X}, or by measuring the variation of the $N_{\rm H}$  
at different orbital phases. The absorption column density can easily reach values as high as 
$N_{\rm H}$$\gtrsim$10$^{23}$-10$^{24}$~cm$^{-2}$ when the accretion wake is located 
along the line of sight to the observer. In a system where the wind can be described by 
a $\beta$-law, it follows the CAK approximation 
(Sect.~\ref{basicsldw}), the hydrogen number density $n_\text{H}(r)$ as a function of 
the distance $r$ from the supergiant is given by: 
\begin{align}
n_\text{H}(r) = \frac{X_\text{H}\,\dot{M}_{\rm w}}{m_\text{H}\,\upsilon_{\infty}\,(1- R_{\star}/r)^{\beta}\,4\,\pi \,r^2}
\label{eq:nhr}
\end{align}
Here $X_\text{H}$ is the hydrogen mass fraction, $\dot{M}_{\rm w}$ is the supergiant mass loss rate, $\upsilon_{\infty}$ the terminal 
velocity of its wind and $R_{\star}$ its radius. The corresponding value of the \nh can be calculated 
by integrating $n_\text{H}(r)$ along the observer line of sight and depends on the system 
inclination $i$, its semi-major axis $a$, its eccentricity $e$, and the longitude of its periastron $\omega$. 
Any deviation in the measured profile of the \nh compared to the one in Eq.~\ref{eq:nhr}, can be ascribed to the presence 
of massive and long-lived structure in the system, and thus most likely to an accretion wake. 
We show an application of this method in Fig.~\ref{NH_modu}. Note that the observations used to study the orbital 
profile of the absorption column density in a system are usually collected over time scales much longer than the 
orbital period and thus the \nh enhancement cannot be caused by clumps or other short lived structures. 
It is indeed expected that the contribution of these fast moving structures is averaged away by the long 
integration times.  

As the characteristics of the accretion wake depend on the compact object and wind properties, 
some authors have been using measurements of the averaged absorption column density profile to indirectly infer 
the supergiant mass low rate and the wind velocity from the comparison between observational results and 
numerical simulations (i.e., as done in the case of the emission lines from photoionized plasmas described above). 
This technique has been applied, for example, to estimate the wind velocity of the classical SgXB 
IGR\,J17252-3616 \citep{Manousakis2011}. 

A relatively recent and puzzling discovery is the detection in several SgXBs 
of a super-orbital variability, i.e., a modulation of their X-ray flux on a time-scale that 
is significantly longer that the orbital period. The super-orbital variability is a well-known phenomenon  
of disk-accreting systems and it is usually ascribed to the presence of a precessing warp in the disk that periodically 
obscures the X-ray source \citep{Ogilvie2001}. As in all SgXBs of interest for this review accretion takes place from the stellar 
wind, it could be possible that the periodical obscuration of the X-ray source is related to other dense structures, 
e.g., the accretion wake. However, \citet{Farrell2008} showed that at least in the case of 
the classical SgXB 1A\,0114+650, no significant changes of the absorption column density in the direction of the source could 
be measured at different super-orbital phases. These authors suggested that a modulation of the 
accretion rate on the super-orbital time-scale could better explain the data, but so far no mechanism has been proposed to 
drive such changes. \citet{Corbet2013} reported on several additional discoveries of super-orbital modulations 
both among the classical SgXBs as well as in SFXTs. In all these systems, there is a clear correlation 
between the orbital and the super-orbital periods, and thus it was suggested that the mechanism 
modulating the mass accretion rate should be related to the separation between the compact object and the supergiant star. 
As the mass accretion rate in all relevant systems is regulated by the wind of the massive star and only a few systems among 
both the classical sgXBs and SFXTs show superorbital modulations, it is likely that the involved mechanisms are related 
to some specific properties of the mass donor star rather than of the compact object. An example is the presence of an 
off-set between the orbital plane and the rotation axis of the primary, which could modulate the rate at which the stellar 
wind arrives close to the compact object on a time scale related (but longer) that the orbital period of the system. The 
presence of an (even small) eccentricity could affect and complicate the period of the resulting super-orbital modulation 
by enhancing the mass accretion rate toward periastron and decreasing it at apastron 
\citep[see,][and references therein]{Corbet2013}.

\subsection{Fluorescence emission lines from the stellar wind and the neutron star surroundings}
\label{xraypars:feline}

An important emission line formation process in wind-fed systems is the fluorescence from ions in a low 
charge state, produced due to the illumination by the X-ray source of the circumstellar 
material. As the intensity of a fluorescence emission is proportional to the fluorescence yield and the latter 
increases with the atomic number of the ion, the \feka fluorescent line is usually the most prominent 
among these features in the spectra of SgXBs \citep[see, e.g.][for a recent work]{Gimenez2015}. 
The energy of fluorescence emission is affected in a non-trivial way by the charge state of the ion, 
and thus it is more difficult to probe the physics of the emitting medium with the corresponding lines  
compared to the emission lines from the H and He-like ions discussed in Sect.~\ref{xraypars:wakes}. 
However, as we discuss below, the fluorescence lines can be used to investigate the distribution 
of cold matter around the neutron star. 

\begin{figure}
 \centering
  \begin{tikzpicture}[]
     \node at (-10,0.0) {\includegraphics[scale=0.25,angle=-90]{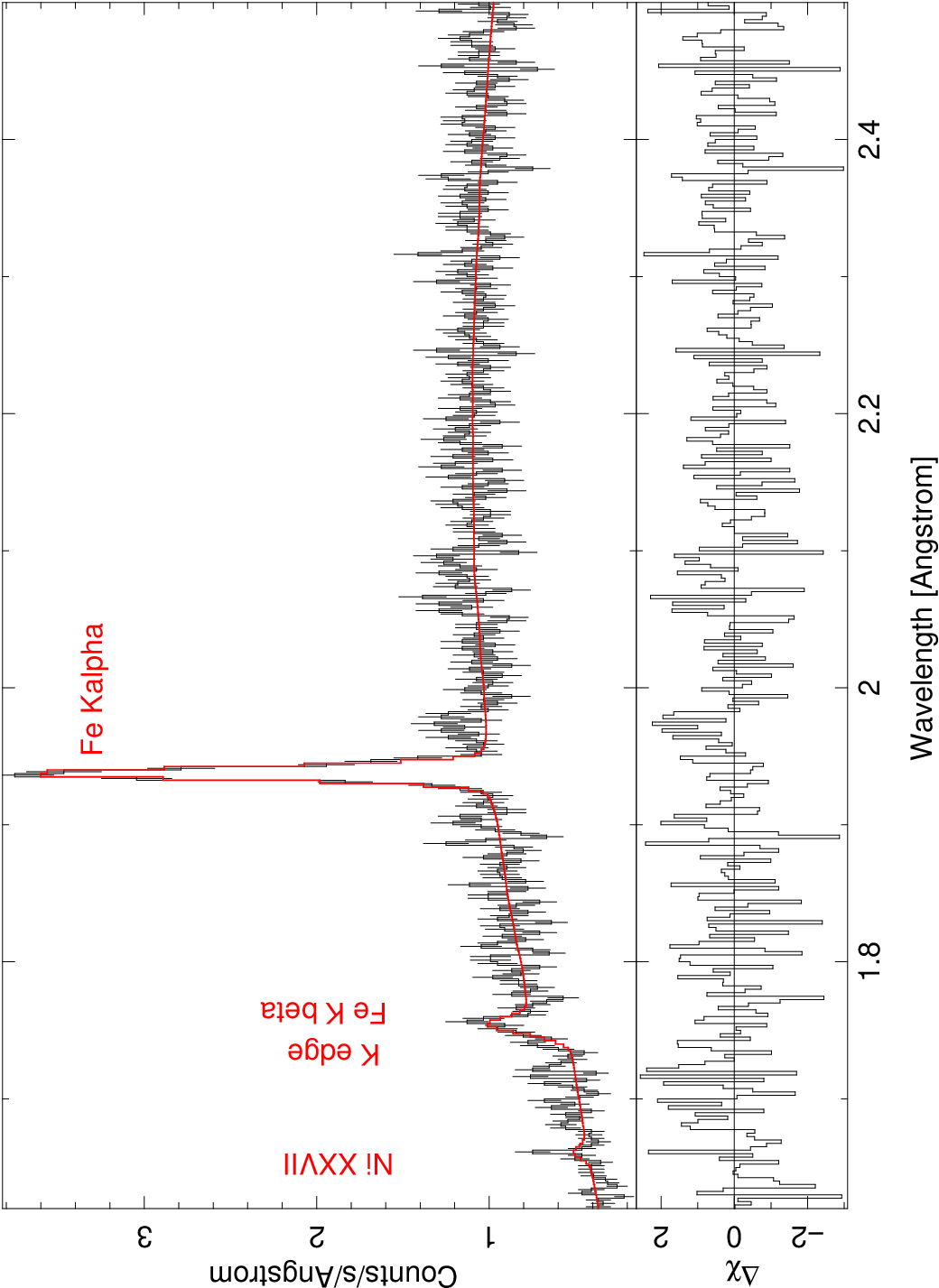}};
     \node at (-4,0) {\includegraphics[scale=0.25]{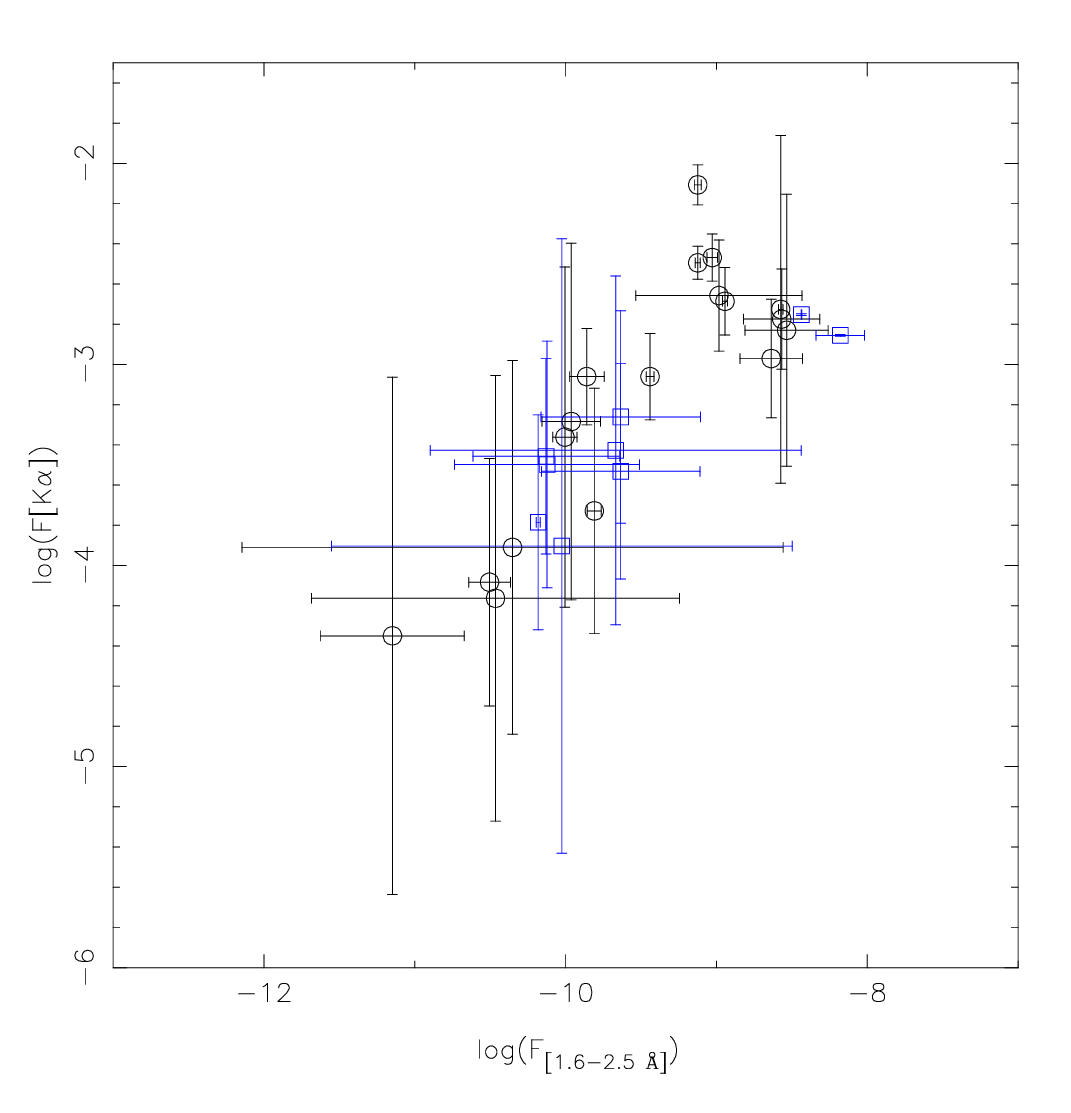}};
  \end{tikzpicture}
\caption{{\it Left}: plot reproduced with permission from \cite{Torrejon2010b} . \emph{Chandra} HETG spectrum of the HMXB Vela~X-1. We can 
see a prominent \feka line together with FeK$\beta$. {\it Right}: plot reproduced with permission from \cite{Torrejon2010b}. 
The logarithm of the flux of \feka (ph/cm$^2$/s) is well 
correlated with the logarithm of the continuum flux (erg/cm$^2$/s).  
The black circles correspond to HMXBs, while blue squares indicate low mass X-ray binaries (LMXBs).}
\label{Intensities}
\end{figure}

The profile of the \feka$ $ fluorescent line in SgXBs can usually be well characterized by using a simple 
Gaussian profile. Although this is known to be a simplistic approach, fits to the X-ray spectra with such a model 
generally permit us to estimate the intensity, width, centroid energy, and equivalent width (EW) of the line with 
a reasonable accuracy, as well as to constrain the time variability of these parameters. 
High resolution spectra, obtained with the gratings instruments on-board \emph{Chandra},  
have shown that in wind-fed SgXBs the \feka line is a narrow feature (FWHM~$<$~5\,m\AA) and it is 
centred on average at $\lambda = 1.9387 \pm 0.0016$\,\AA\ (see the left panel 
of Fig.~\ref{Intensities} and \citet{Torrejon2010b}). 
The general finding that the intensity of the \feka line directly correlates with the intensity of the source 
confirms that this feature is produced as a consequence of the illumination of the stellar wind material 
by the X-ray emission from the compact object (see right panel of Fig.~\ref{Intensities}). This has recently been also  
reconfirmed through an in-depth analysis of all available \emph{XMM-Newton} data on SgXBs \citep{Gimenez2015}. 

\begin{figure}
 \centering
  \begin{tikzpicture}[]
     \node at (-10,0.0) {\includegraphics[scale=0.30]{CurveGrowth.ps}};
     \node at (-4,-0.5) {\includegraphics[scale=0.28]{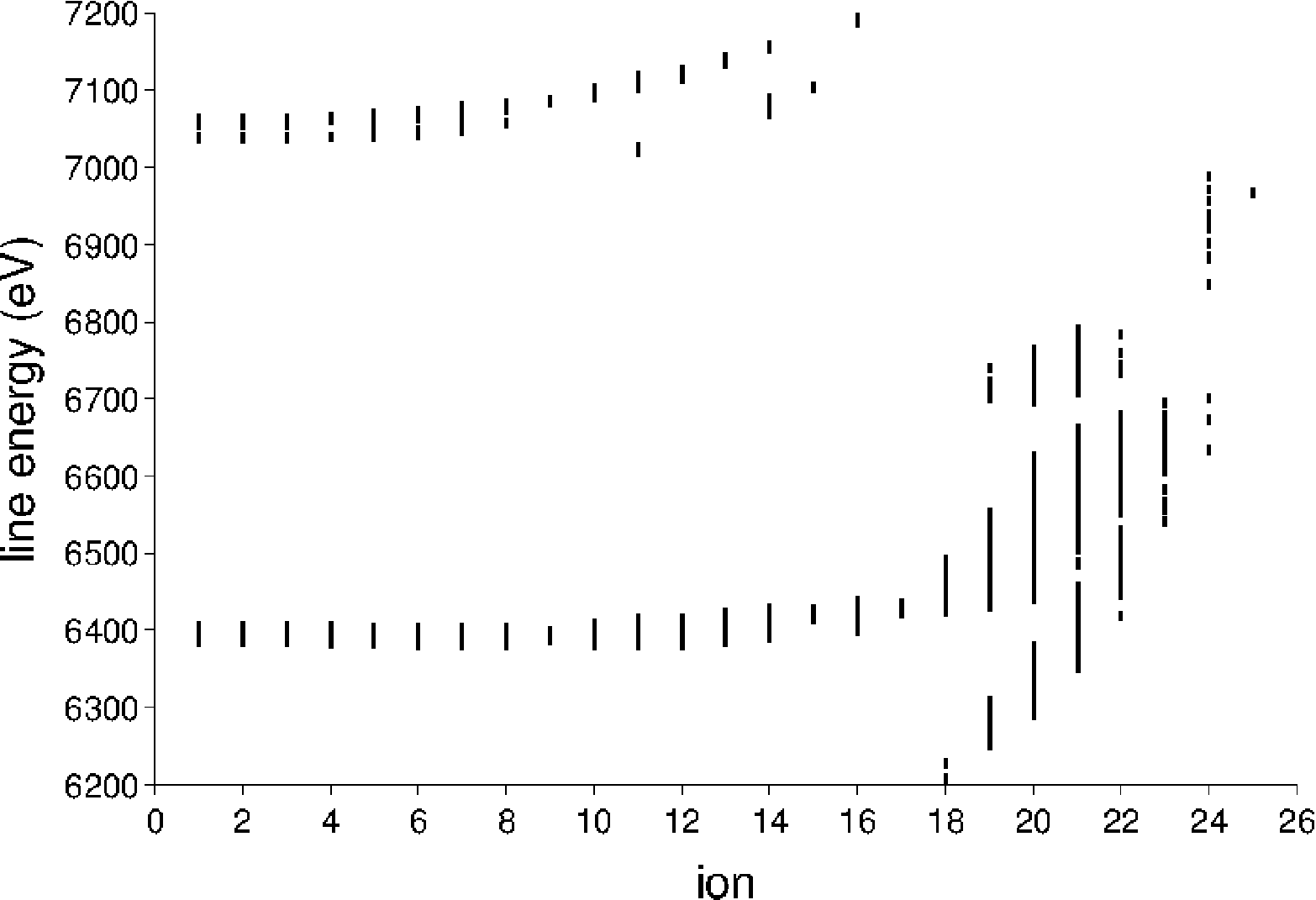}};
  \end{tikzpicture}
  \caption{{\it Left}: EW of \feka against \nh \citep[plot from][reproduced with permission]{Torrejon2010b}. 
The EW of the line increases with the column density. Black circles are HMXBs, blue squares are 
LMXBs, and black triangles are eclipse data. In eclipse the EW of \feka is greater because a 
significant fraction of the continuum is blocked. The black dashed line represents the prediction 
of the theoretical model in \citet{Kallman2004}. {\it Right}: in the lower part of the plot the energy of 
\feka is shown as a function of the ionization state of Fe ions \citep[reproduced with permission]{Kallman2004}.}
\label{Ionstate}
\end{figure}

The size of the fluorescence emitting region in SgXBs has been estimated for a number of sources 
using different methods. Observations of eclipsing wind-fed SgXBs show that the EW 
of the \feka is generally higher when the systems are in eclipse (see the left panel of Fig.~\ref{Ionstate}). 
This indicates that at least in some cases the reprocessing region is extending beyond the 
radius of the massive companion. In other cases it was shown that the bulk the \feka emission comes 
from the the immediate surroundings of the compact object, and thus that the 
accretion wake is likely playing a key role in the \feka line production (see Sect.~\ref{xraypars:wakes}). 
This is the case, for example, of Vela\,X-1 \citep{Watanabe2006}. 
In IGR~J17252$-$3616, \citet{Manousakis2011} inferred the existence of a dense cocoon of cold material with a radius of 
$\sim 10$~R$_\odot$ around the neutron star by analysing the variation of \feka during the eclipse.  
The size of this structure is thus $\sim$0.5 times smaller than the stellar radius estimated for the supergiant  
hosted in this system \citep{Mason2009}. \citet{Audley2006} 
used the time delay between the pulsar eclipse and the \feka eclipse in OAO~1657$-$415 to prove that  
80\% of \feka is emitted in this source within a region of $R<8$~R$_\odot$. \cite{Fuerst2011} 
analysed the delay in the arrival times of the pulses in the \feka energy range to conclude 
that the fluorescence region in GX~301$-$2 has a radial extension $R<1$~R$_\odot$. Different results 
were reported for this source by \citet{Endo2002}. The latter authors assumed that the width of the 
line was caused by the Doppler broadening of the free-fall motion of the accreting matter onto the neutron star and  
inferred a size of the fluorescence emitting region of $R<0.1$~R$_\odot$. This is significantly smaller than previous 
estimates. A plausible reason is that they neglect other broadening mechanisms, such as the blend of multiple fluorescence lines from 
different ionization states of iron \citep{Fuerst2011}. Finally, the alternative method 
employed by \cite{Rodriguez2006}, which takes into account multiple ionization states of iron 
and the impact of a ionization parameter that changes at different distances from the neutron star, 
gave an extension of the reprocessing region in the SgXB IGR~J16320-4751 of $R>1.5$~R$_\odot$.

So far, the observations of wind-fed SgXBs have been supporting reasonably well 
the theoretical prediction that the measured equivalent width (EW) of the \feka 
line due to X-ray transmission through a optically thin medium is related to the hydrogen 
column density \nh measured in the direction of a source. The latter is theoretically estimated through 
Eq.~5 in \cite{Kallman2004}:
\begin{equation}
EW \simeq 300\,\textrm{\footnotesize eV} \frac{N_\textrm{\footnotesize H}}{10^{24}\,\textrm{\footnotesize cm}^{-2}}. 
\label{eq:FeEW2NH}
\end{equation} 
We show in Fig.~\ref{Ionstate} (left panel) the currently observationally measured correlation 
between the EW of the \feka and the \nh in different SgXBs \citep[see also][]{Gimenez2015}. 
In these sources the region reprocessing most of the neutron star X-ray emission is thus also providing a dominant 
contribution to the material absorbing this radiation along the observer line of sight.  

The centroid energy of the \feka is one of the measurements of this spectral feature that provides a 
direct indication of the physical conditions in the wind, as it depends on the ionization state of the Fe ions.  
In Figure~\ref{Ionstate} (right panel) we can see that the centroid energy of the line significantly 
increases with luminosity, especially when the iron is highly ionized \citep{Kallman2004}.
\textsl{Chandra} measurements revealed the presence of a number of Fe ions, ranging from 
\FeI\ to \FeX\ \citep{Torrejon2010b}. This is compatible with the properties of an O-B supergiant companion, 
as we expect to find Fe\,{\sc ii-iii} in the wind of a BI star and
Fe\,{\sc iv-vi} in the wind of an O-star \citep[see, e.g.,][]{KP00}.
As can be seen in Fig.~\ref{Ionstate} (right panel) the centroid energy of the line 
depends only weakly on the ionization state of the iron ions below $\sim$Fe\,{\sc xviii}. 
For the low Fe ionization states, a more sensitive tracer of the conditions in the donor star wind 
is the ratio between the fluxes of the \feka and the \fekb line \citep[see, e.g.][]{palmeri03a}. For neutral iron we expect a 
value of \fekb/\feka = 0.125 \citep{kaastra93a}. This ratio increases to $\sim$0.2 for \FeVIII . 
However, as generally the flux of the \fekb is only 10-20\% of the \feka line, 
this technique requires data with particularly high statistics in order to perform sufficiently 
accurate measurements \citep[see, e.g., the case of GX~301-2;][]{Fuerst2011}. 
When calculating the ratio between the fluxes of the \feka and the \fekb line in sources characterized by 
an absorption column density $>10^{23}$\,cm$^{-2}$, self-absorption is also  
expected to play an important role. Theoretically, the effect of self-absorption is different for the two 
lines and might lead to artificially high evaluated flux ratios. However, it shall be noted that 
the effect of self-absorption could not be robustly tested yet against high quality observations and thus 
its anticipated impact on the \feka and the \fekb flux ratio remains to be confirmed. 

Another characteristic of the \feka line that can be used to probe stellar winds is its width in energy. 
This parameter provides an upper limit on the velocity of the X-ray reprocessing material around the 
neutron star, as this is generally considered to be the main broadening mechanism for the line 
(but other mechanisms, such as line blending, cannot be excluded).    
\emph{Chandra} observations proved so far that the \feka line in SgXBs is relatively narrow 
(FWHM$<5$~m\AA), leading to upper limits on the velocity of the X-ray reprocessing material of $\lesssim$1000~km/s 
\citep{Hanke2009,Torrejon2010b,Torrejon2015}. This is in agreement with the typical velocities of O-B star winds 
in SgXBs but would also be compatible with the free fall velocity of material located at a distance from the neutron star 
where the \feka line is produced \citep{Endo2002}. The two processes are generally difficult to disentangle, unless  
accurate measurements of the width of the line are available during different phases of the neutron star orbit 
(the accretion flow is not expected to be as isotropic as the stellar wind).

In a few SgXBs, the \feka line has been also observed to show a Compton-shoulder extending down to $\sim$6.2~keV. 
A Compton shoulder forms when X-ray photons propagating through a low temperature medium ($\lesssim$10$^5$~K) transfer 
part of their energy to the free electrons via Compton scattering. If the Compton optical depth of the medium is 
$\gtrsim$0.1, the probability of those scatterings to occur is not negligible. The resulting effect is more pronounced  
on emission lines at higher energies, and thus particularly on the \feka line, due to the energy dependence of 
the Compton scattering opacity with respect to that of the photoionization. 
As the shape of the Compton shoulder is sensitive to the electron column density and temperature of the 
scattering medium, it provides an additional tool to study the physical properties of the cold material around 
the neutron star. So far, this feature could be clearly detected in the two classical SgXBs characterized by the 
highest absorption column densities, i.e., GX\,301-2 \citep{Watanabe2003} and IGR\,J16318-4848 \citep{Ibarra2007}.  
  
The \feka line is not the sole fluorescent line that can be detected in the X-ray spectra of SgXB and SFXT sources. 
As an example, \cite{Fuerst2011} reported the detection of the NiK$\alpha$, NiK$\beta$, CrK$\alpha$, CaK$\alpha$, ArK$\alpha$, 
and SK$\alpha$ fluorescent lines in GX\,301-2. Due to the lower fluorescence yield of all these elements compared with iron, 
these additional spectral features are usually much fainter than the \feka line and thus it is more difficult to use them 
as probes of the material around the neutron star.

\subsection{X-ray observations of SgXBs and SFXTs: caveats}
\label{xraypars:caveats} 

As anticipated in Sect.~\ref{xraypars:clumps}, we discuss below 
the effect of systematic uncertainties affecting two of the key  
observables obtained from the X-ray data of SgXBs and used to infer stellar wind 
parameters: the absorption column density in the direction of a source and 
its bolometric X-ray luminosity. 

\subsubsection{The absorption column density}
\label{xraypars:nh}

The absorption of X-rays in the interstellar medium or close to the
X-ray source itself plays a key role in the analysis of X-ray
spectra \citep{wilms2000a, hanke2011a}. 
As an observational fact, the flux of any kind of
X-ray source located within the Milky Way or beyond is significantly suppressed  
below $\sim$5\,keV. This is due to photo-ionization of matter by X-ray
photons along the line of sight to the source
\citep[see, e.g.,][for reviews about the interstellar extinction]{dickey1990a,mathis1990a}. 
The effect of absorption on the source X-ray flux $I_\text{src}(E)$ at a certain energy $E$ 
can be described by: 
\begin{align}
  I_\text{obs}(E) = I_\text{src}(E)\, e^{-\sigma(E) \nh}
\end{align}
where $ I_\text{obs}(E)$ is the observed source flux and $\sigma(E)$ is the energy dependent 
cross-section of the absorbing medium. In SgXBs, the latter comprises the contribution 
from the inter-stellar medium (ISM) and from the medium local to the source. The latter 
is by far the most dominant, and generally the ISM contributes only for a few 
to $\sim$10-30\% (at the very most). However, in those cases in which an accurate estimate 
of the absorption column density is needed, both contributions have to be taken into account. 

In most cases, the estimates of the ISM absorption column density in the direction of a source 
are assumed to coincide with the measurements of the neutral hydrogen column densities 
obtained from the Galactic surveys carried out at 21\,cm \citep{kalberla2005a,dickey1990a}. 
As pointed out by \cite{wilms2000a}, an uncertainty behind this assumption is that the contribution of 
the Galactic H$_2$ molecular clouds to the total \nh is not considered. These clouds are not uniformly distributed 
in the Milky Way \citep{shull1982a} and thus the assumed abundance of H$_2$ relative to neutral 
hydrogen should be, in principle, carefully evaluated for each source.  

Concerning the material absorbing the X-ray emission local to the source, the problem is even more 
complicated as the abundances of high $Z$-elements (e.g., oxygen and iron which are responsible for most of 
the absorption in X-rays) have to be known accurately. These abundances might change among different 
X-ray sources, as they are expected to be tightly related to the specific binary system environment. 
The latter is known to be heavily affected by the outflowing 
material from the massive companion, as well as from the supernova explosion that 
led to the formation of the neutron star. As these processes of chemical enrichment are not  
known yet with a reasonably good accuracy, it is typically assumed that the surroundings of 
a SgXB are characterized by a solar or ISM-like composition. For the latter, updated models 
are periodically made available, following our progresses in the understanding of the Milky Way chemical composition. 

Measurements of the absorption column density can thus be derived from the assumed  
abundances by using the cross-sections and energies for the X-ray absorption on the 
shell electrons of each of the involved elements. In most of the X-ray sources, the effect of Thomson 
scattering has also to be taken into account when computing the total X-ray absorption, as part of the material 
close to the compact object could be significantly ionized and contribute differently to enhance the \nh compared 
to the neutral material (see Sect.~\ref{xraypars:wakes}). At present, not all cross-sections and energies for 
the Thomson scattering can be derived precisely from the atomic physics and thus most of the absorption models 
have to rely on specific cross-section tables. Furthermore, when the density of the ionized absorbing material 
reaches values of $\gtrsim$10$^{24}\,\text{cm}^{-2}$, Compton-scattering effects also come into play and have to 
be included in the estimate of the total \nh \citep{eikmann2012a}.
An overview of all the currently available absorption models, abundances, and cross sections for the different processes 
is presented in Table~\ref{tab:absorbmodels}, together with their  identification code in the two most used 
software packages for the fitting of X-ray data (i.e., \texttt{ISIS}\footnote{http://space.mit.edu/asc/isis/.} 
and \texttt{XSPEC}\footnote{https://heasarc.gsfc.nasa.gov/xanadu/xspec/.}).  
\begin{table}
  \caption{List of photo-ionization cross-sections$^{\rm a}$ and abundances$^{\rm b}$ as available
in \texttt{ISIS} or \texttt{XSPEC}, as well as absorption models. 
\texttt{ISIS} and \texttt{XSPEC} are two of the most used fitting tools 
for accurate X-ray spectral analysis available to date.  
In the table below, the asterisk identify the default \texttt{XSPEC} configuration (version 12.8.2).}
  \label{tab:absorbmodels}
  \scriptsize
  \begin{tabular}{lll}
    \hline
    \multicolumn{2}{l}{Cross-sections} & Comment \\
    \hline
    obcm  & \cite{balucinska-church1992a} & \\
    bcmc* & \cite{balucinska-church1992a} & H$_2$-cross-section from \cite{Yan1998a} \\
    vern  & \cite{verner1996b} & \\
    \hline
    \multicolumn{2}{l}{Abundance vector} & Comment \\
    \hline
    aneb  & \cite{anders1982a} & solar \\
    angr* & \cite{anders1989a} & solar \\
    feld  & \cite{feldman1992a} & solar \\
    grsa  & \cite{grevesse1998a} & solar \\
    wilm  & \cite{wilms2000a} & ISM \\
    lodd  & \cite{lodders2003a} & solar \\
    aspl  & \cite{asplund2009a} & solar \\
    \hline
    \multicolumn{2}{l}{Absorption models} & Contributions \\
    \hline
    wabs & & neutral, atomic, thin gas \\
         & & (abundances fixed to aneb) \\
    phabs & & neutral, atomic, thin gas \\
    tbabs & \cite{wilms2000a} & neutral, atomic, thin gas \\
          & & + neutral H$_2$-molecules \\
          & & + spherical, chemical homogeneous dust grains \\
          & & (cross-sections fixed to vern) \\
    tbnew$^\text{c}$ & Wilms priv.comm. & improved version of tbabs including high \\
          & & resolution cross-sections at important edges \\
    cabs & & optically-thin Compton scattering \\
    ctbabs & \cite{eikmann2012a,eikmann2014a} & full Compton scattering \\
         & & + fluoresence line emission \\
    warmabs & & uniform, collisional ionization (warm stellar winds) \\
         & & based on a fixed, ionization-balanced, thin gas \\
    \hline
  \end{tabular} \\
  \footnotesize{$^\text{a}$ http://heasarc.nasa.gov/xanadu/xspec/xspec11/manual/node36.html \\
$^\text{b}$ https://heasarc.gsfc.nasa.gov/xanadu/xspec/manual/XSabund.html\\
$^\text{c}$ http://pulsar.sternwarte.uni-erlangen.de/wilms/research/tbabs}
\end{table} 

\begin{figure}
 \centering
  \begin{tikzpicture}[]
     \node at (-10,0.0) {\includegraphics[scale=0.40]{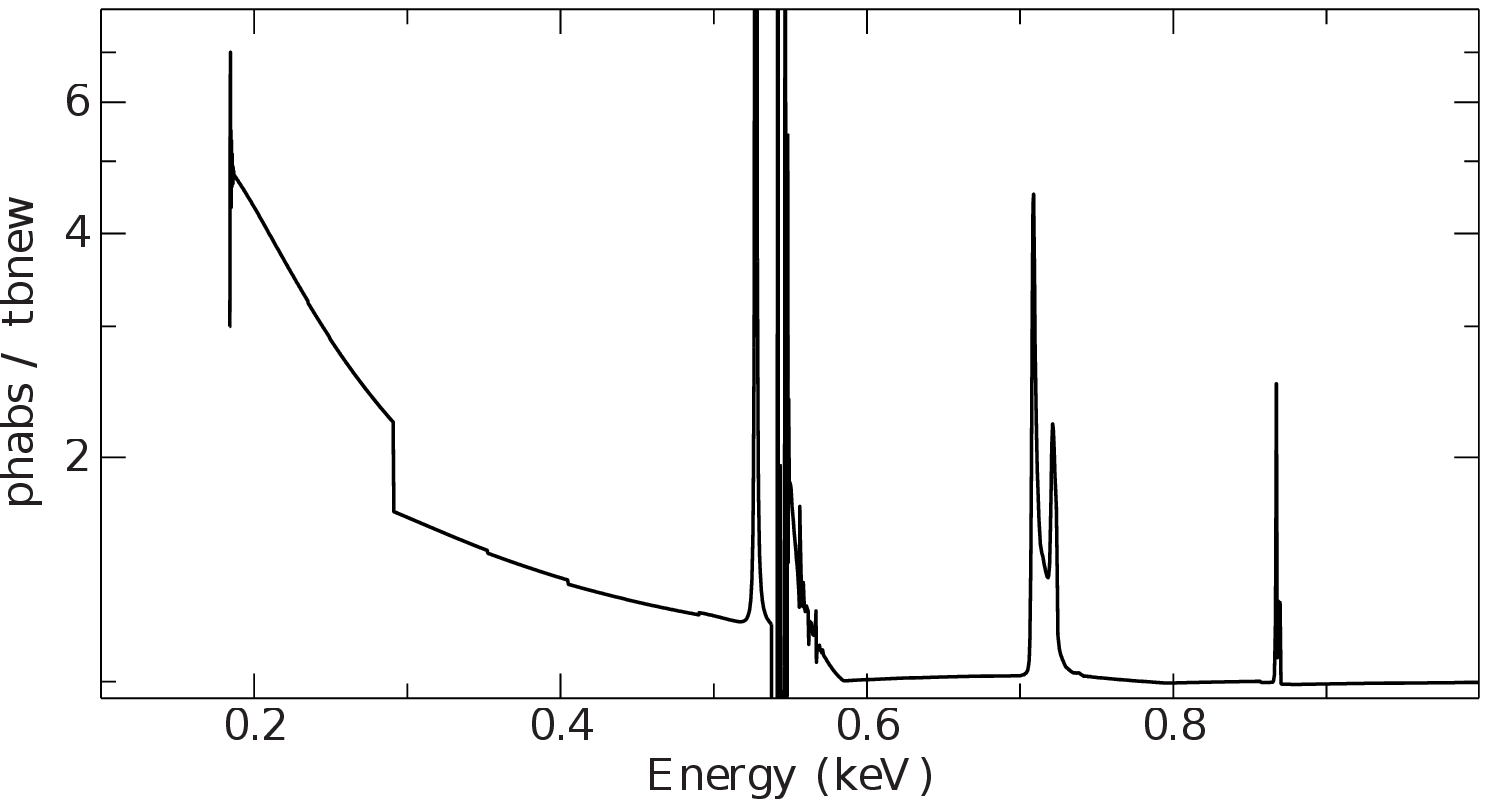}};
     \node at (-4,0) {\includegraphics[scale=0.35]{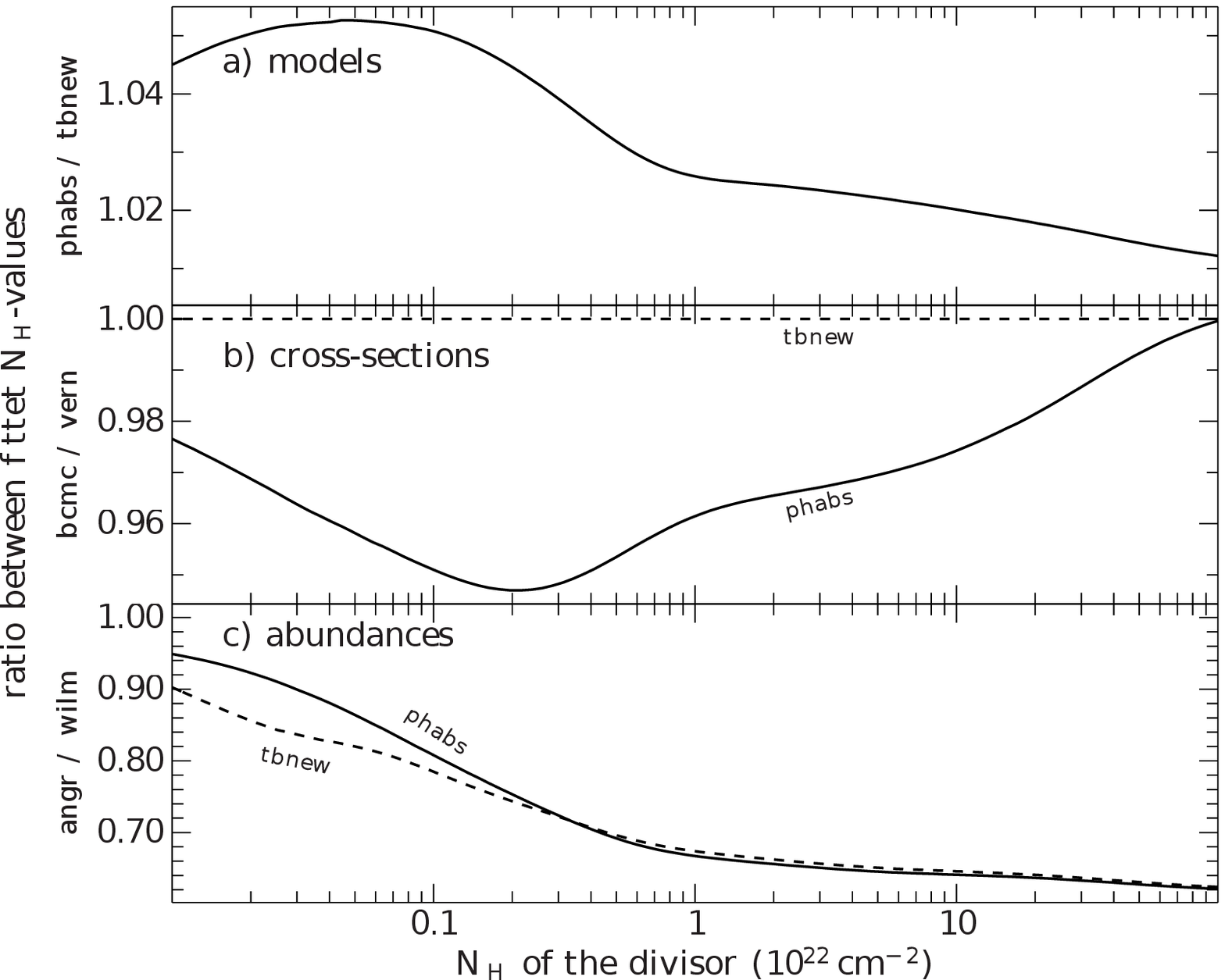}};
  \end{tikzpicture}
\caption{{\it Left}: ratio between the \texttt{phabs}- and the
\texttt{tbnew}-absorption-model. An $\nh$ of
$1 \times 10^{22}\,\text{cm}^{-2}$ was used to calculate the
absorbed spectra featuring a constant photon flux. The
cross-sections were set to \texttt{vern} and the abundances to \texttt{wilm}. 
{\it Right}: comparison of derived $\nh$-values from
X-ray spectra based on a) different models, b) cross-sections,
and c) abundances. The assumed $\nh$ using a
specific configuration is plotted against the $\nh$-ratio
when fitted with a different configuration. The cross-sections and
abundances for a) were set to vern and wilm, respectively. In b)
and c), the results using either the \texttt{phabs}- (solid line) or
the \texttt{tbnew}-absorption-model (dashed line) are also shown.
References to the photo-ionization cross-sections, abundances
and absorption models given in Table~\ref{tab:absorbmodels}.}
\label{fig:Nh_comparisons}
\end{figure}

As we can see from Table~\ref{tab:absorbmodels}, a number of different absorption models have been published 
in the literature, each using different assumptions for the ISM cross-sections 
and abundances of the different elements. To illustrate how the value of the absorption column 
density in the direction of a source changes when different models are used, 
we show in Fig.~\ref{fig:Nh_comparisons} (left panel)  
the ratio of the two absorption models \textsl{phabs} and \textsl{tbnew} 
as a function of the energy for $\nh$=$1 \times 10^{22}\,
\text{cm}^{-2}$. The main differences between the two models can be noted 
around the energies corresponding to the K-edges of oxygen (0.53-0.58\,keV), neon
(0.87\,keV), and the L-edge of Fe (0.71\,keV-0.73\,keV).
Furthermore, \textsl{phabs} absorbs the X-ray spectrum less than
\textsl{tbnew} for energies below 0.6\,keV (up to a factor of $\sim$5 at 0.2\,keV), 
and thus the \nh values derived with \textsl{phabs} are 
systematically higher than those derived with \textsl{tbnew}. 
This is illustrated in the right panel of Fig.~\ref{fig:Nh_comparisons}, where 
several spectra were simulated over a wide range of \nh values by using 
\textsl{tbnew} and fit with \textsl{phabs}. The
largest discrepancy between the two models is about 5\% at $5 \times 10^{20}\,
\text{cm}^{-2}$. Similarly, the derived value of the \nh depends on
the cross-sections and abundances used. Panel b) and c) on 
the right side of Fig.~\ref{fig:Nh_comparisons} show these effects as a function
of \nh. Outdated cross-sections, as \textsl{bcmc}, underestimate the
column density up to 5\% (note that the cross-sections in
\textsl{tbnew} are fixed to \textsl{vern}). The most dramatic differences in
values of the inferred \nh are obtained when models adopting solar and ISM-abundances 
are compared: in the first case, values of the absorption column density exceeding 
$10^{21}\,\text{cm}^{-2}$ are all underestimated by more than
20\%. This effect is more pronounced for higher values of \nh. 

From the above results, it is clear that any measurement of the absorption column density 
in a X-ray source, and particularly in SgXBs, is significantly dependent on the chosen 
absorption model, cross-section, and abundances prescription. In case \nh values derived from the 
X-ray observations are used to infer the physical properties (e.g., density, mass, and size) 
of structures in the stellar wind around a neutron star, it would be a good practice to clearly state the models and 
prescriptions used. In this way, possible discrepancies between different findings can be investigated 
in terms of systematic uncertainties among the different absorption models or ascribed to intrinsic 
changes in the properties of the absorbing medium local to the source.

\subsubsection{The neutron star bolometric X-ray luminosity}
\label{susec:ns_bolum}

The intrinsic bolometric X-ray luminosity $L_\mathrm{X}$ of an SgXB, that as
we saw in the previous sections is a key parameter to infer the stellar wind
parameters and the details of the accretion processes, is commonly derived
from the measured X-ray flux $F_\mathrm{X}$ of the source by using the
relation:
\begin{equation}
  L_\mathrm{X} = 4 \pi d^2 F_\mathrm{X}.
\end{equation}
Here, it is assumed with the $4\pi$ factor that the neutron star's X-ray
radiation is emitted isotropically. In most of the cases of interest for this
review, the larger source of uncertainty on $L_\mathrm{X}$ is due to the
poorly known distance to the system, $d$. As it can be seen in Table 2, this
distance is often known only within an accuracy of a factor ${\sim}$2--3 and
for some of the SgXBs contradicting measurements have been reported.

The rough assumption about the isotropic nature of the X-ray emission
introduces in all cases an additional source of uncertainty. This assumption
does not generally apply to accreting neutron stars: the bulk of the X-rays
are produced in the accretion columns near the two magnetic poles, where
accreting matter hits the neutron star surface \citep[see][and references
therein]{becker2007a}. Furthermore, the emission profiles of the accretion
column (or the hot spots on the surface) are not known and might be a complex
combination of a fan- or pencil-beam \citep[see, e.g.,][]{Harding1994}.

\begin{figure}
  \includegraphics[width=\textwidth]{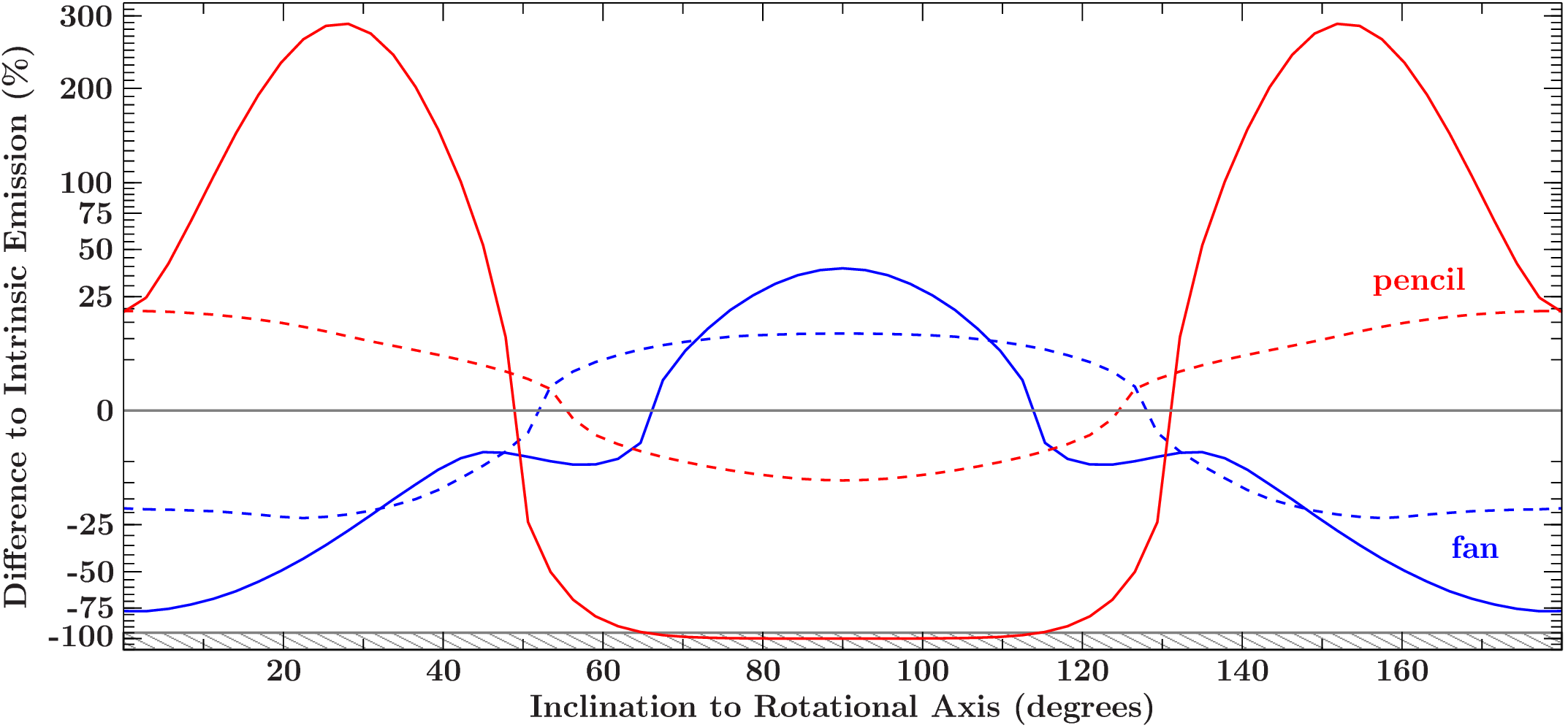}
  \caption{Relative difference of the derived luminosity to the real intrinsic
  luminosity depending on the inclination to the neutrons star's rotational
  axis. The derived luminosity is calculated from the observed flux assuming
  isotropic emission ($4 \pi$-factor). The intrinsic luminosity assuming a
  Gaussian (solid lines) or no beam pattern (dashed lines) is calculated for a
  \textit{pencil beam} (blue) and a \textit{fan beam} geometry (red). The
  dashed area corresponds to observed fluxes $\le5\%$ off the intrinsic flux,
  which would result in an apparently dim pulsar.}
  \label{fig:4Pi}
\end{figure}

The emission patterns are also distorted by general relativistic effects in
the strong gravitational field of the neutron star. In most cases, both
magnetic poles are visible over one full rotation of the neutron star
\citep{beloborodov2002a}. Depending on the geometry and the position of the
observer, the total flux might be beamed and boosted towards the observer, or
pointed in a different direction and lowered in intensity. To provide an
estimate of this effect we have used an extended light-bending code, which is 
based on work by \citet{Kraus2001}. The new code, allowing for more complicated
accretion column and emission geometries, has been used already to explain phase-shifts
at the cyclotron line energies seen in the pulse-profiles of some accreting
neutron stars \citep{schoenherr2014a}. We have assumed standard neutron star
parameters for the mass and radius, and a magnetic axis going
through the star's center  inclined by 30$^\circ$
with respect to the rotational axis. In Fig.~\ref{fig:4Pi} we show, as a function
of the inclination angle to the observer, the difference between the computed
emission and the real intrinsic emission in the case of: (i) a \textit{pencil
beam} with two hotspots of 1\,km radius located at the neutron star magnetic poles,
(ii) a \textit{fan beam} with two accretion columns of 1\,km in radius and
2\,km in height located at the neutron star magnetic poles. In both cases we
have simulated a non-beamed emission from the emission regions and a Gaussian
beam pattern with a half opening angle of 10$^\circ$ \citep[see,
e.g.,][]{BaskoSunyaev:75}. This beam pattern accounts for the angular
distribution of the emerging photons, which boosts the photons leaving (i) the
hot spots perpendicular to the surface in case of a \textit{pencil}-beam and
(ii) the accretion columns sideways in the \textit{fan}-beam case. The calculated
difference in Figure~\ref{fig:4Pi} can be considered as the error that would be
introduced on the estimate of the source luminosity by assuming a simplified
isotropic emission pattern instead of the correct one. The introduced error is
larger in the pencil beam geometry, where the beamed-case easily leads to
an over-estimation as large as ${\sim}300\%$. There is a distinct range of
inclinations between 50$^\circ$ and 130$^\circ$ where the derived luminosity is
drastically under-estimated. The pulsar is even not detectable between
70$^\circ$ and 110$^\circ$ since the pencil beam never ``strikes'' the observer
(dashed region in Fig.~\ref{fig:4Pi}). In the fan beam geometry the effect is more
modest, with the largest recorded over-estimation reaching ${\sim}40\%$. At
inclinations between -20$^\circ$ and +20$^\circ$ the luminosity is
underestimated, however, by -40\% down to -75\%. For both considered emission
cases, the error on the estimated luminosity is limited to a maximum of
$\pm25\%$ once no beam pattern is assumed (dashed lines in Fig.~\ref{fig:4Pi}),
which can be interpreted as a lower limit. We stress that, however, all these 
estimated differences strongly depend on the assumed emission geometry, like 
the location of the magnetic poles. The differences can become even larger if: 
(i) a less favourable geometry is considered
for the inclination of the neutron star's magnetic axis with respect to its
rotational axis and for the emission pattern; (ii) more extreme neutron star
spin periods and magnetic fields are explored.

Finally, a last source of uncertainty comes from the source X-ray flux
obtained form the observational data. Beside uncertainties due to the calibrations
of the X-ray instruments used to carry out the observations \citep[see,
e.g.,]{Guever15}, the bolometric flux of a source is usually measured on a
relatively small energy band (e.g. 0.5-10\,keV) and then extrapolated to a much wider
range to obtain the bolometric luminosity (typically 0.1-100\,keV). This
procedure can lead to a significant over- or under-estimation of the source bolometric 
flux, especially for those cases in which the spectral energy distribution is not
accurately known and fits to the observed X-ray spectra are performed with
phenomenological rather than physically justified models \citep[see the discussions
in, e.g.,][]{becker2012a,Farinelli2012b,Farinelli2012}.

It thus seems that, whenever accurate measurements are required to derive the
physical parameters of stellar winds or details on the accretion processes in
SgXBs from an X-ray observation, caution should be taken on the estimated
values of the source luminosity.
\subsection{The peculiar case of the SFXTs}
\label{SgXB:SFXT}

SFXTs and classical SgXBs are known to have optical counterparts with similar properties 
\citep{Masetti2006, Chaty2008, Rahoui2008}, even though dedicated optical/IR/UV spectroscopic 
monitoring campaigns of all these sources have been performed only rarely 
\citep[especially for the SFXTs; see, e.g.][]{Lorenzo2010, Lorenzo2014}. In all cases, 
these campaigns could only provide limited information on stellar winds, as all SgXBs 
are located several kpc away and the high extinction toward their directions makes it 
impossible to carry out detailed spectroscopic investigations as it is done for close-by 
supergiants (see Sect.~\ref{stellarwindpars}). 

As we discussed in Sect.~\ref{xraypars:clumps}, if one assumes that the flares in SFXTs are produced by direct 
accretion (see Sect.~\ref{sec:regimes}) of wind clumps onto
the neutron star \citep[as originally suggested by][]{Zand2005}, the derived clump masses can be as large as  
$10^{23}$~g and the required density ratio between the clumps and the intra-clumps material 
should be of the order of 10$^4$-10$^5$ \citep[see Sect.~\ref{xraypars:clumps} and][]
{walter2007a, Leyder2007, Ducci2009, Ducci2010, Romano2010, Romano2012, Bozzo2011}. These values are far too extreme 
compared to the more moderate clumpy winds that are expected according to the results of the 
latest observational campaigns and numerical simulations of hot stellar winds 
(see Sect.~\ref{xraypars:clumps}). Therefore, it seems that a simple clumpy wind scenario is not 
able to account for the variability of the SFXTs, at odds with what has been found for the classical 
SgXBs in Sect.~\ref{xraypars:clumps}. Other mechanisms have thus to be invoked.  

The similarity in the orbital periods of most SFXTs and classical SgXBs complicates the picture even more, 
so that models comprising only the accretion from the clumpy supergiant wind and peculiar 
orbital geometries \citep{Negueruela2005a} are not able to fully explain the difference between the two classes of 
systems. However, the high proper motion recently measured in the prototypical SFXT IGR~J17544-2619 supports 
the suggestion that at least some SFXTs might have highly eccentric orbits \citep{Maccarone2014}. 
This would lead to a slight decrease of the averaged X-ray luminosity and an increase in the achievable 
dynamic range of the system, as a neutron star on an eccentric orbit spends a long time away from its companion 
in regions where the stellar wind (and thus the accretion) is lower \citep{Negueruela2005a}. 
With the increasing observing time targeted on SFXTs in the last 10 years, it became 
evident\footnote{We are not considering here in detail the case of the peculiar SFXT IGR~J11215-5952, where 
the outbursts are strictly periodic \citep{Sidoli2006,Sidoli2007} and 
they occur at each periastron passage \citep{Romano2009b}.} that 
the case of IGR~J17544-2619 might not be unique and it was found that bright flares in 
a number of SFXTs appear clustered at certain orbital phases, assumed to be the periastron passage of a somewhat elongated orbit 
\citep{Drave2010, Romano2014}. However, even in these cases, flares do not necessarily occur at every periastron 
passage \citep{Sguera2007, Drave2014} and intense periods of X-ray activity have been recorded by the same sources 
at larger distances from the supergiant \citep[see, e.g.,][]{Goossens2013} or spread along the entire orbit \citep{Smith2012}. 
Controversial results about the distribution of flares along the orbit of XTE~J1739-302 have 
been obtained by different teams, although probably due only to different instrumental sensitivity 
and observing strategy \citep{Smith2012, Drave2010}.

The long term monitoring observations of many SFXTs, that were carried out mainly to search for 
the outbursts of these sources and the most likely orbital phases at which these events are occurring, clearly 
evidenced that these systems are significantly sub-luminous compared to other classical SgXBs. In particular, 
their average X-ray luminosity was measured to be a factor of ~100 lower than that expected from a classical SgXB with 
a similar orbital period \citep{Romano2011,Romano2014b}. In IGR~J16479-4514, the SFXT with the shortest orbital period ($\sim$3 days) and, 
very likely, a circular orbit, a Suzaku observation allowed to almost continuously and deeply cover 80\% of the 
orbit at X-rays \citep{Sidoli2013b}. In this source, the ratio between the 
scattered X-rays observed during the X-ray eclipse and the level of
the uneclipsed emission permitted to derive an estimate of the wind density at the neutron star distance 
of 7$\times$10$^{-14}$~g~cm$^{-3}$ \citep{Sidoli2013b}. Assuming a spherical out-flowing supergiant wind, 
this density implies a ratio between the wind mass loss rate
and the wind terminal velocity  $\dot{M}_{w}/ \upsilon_{\infty}$ = $7\times$10$^{-17}$~M$_{\odot}$/km.
Assuming terminal wind velocities from 500 to 3000~km~s$^{-1}$ (see Sect.~\ref{basicsldw}), 
this ratio would imply an X-ray luminosity of L$_{\rm X}$=3-15$\times$10$^{36}$~erg~s$^{-1}$ according to 
the standard wind accretion scenario (see Eq.~\ref{eq:lx}). This is more than two orders of magnitude 
higher than the observed luminosity. The conclusion of all these studies is thus that, besides clumps and orbital 
elongations (only in some cases), a mechanism should be at work in all SFXTs to substantially reduce the mass 
accretion rate and the average X-ray luminosity. 

A first attempt to address these issues by combining wind accretion models with hydrodynamic 
simulations of massive star winds was carried out by \citet{osk2012}, who derived the implied X-ray 
variability of the system by assuming a direct accretion scenario and circular orbits\footnote{In their calculations 
the neutron star is not properly orbiting the companion but is assumed to be fixed at a certain distance while the variable and 
evolving wind is blown in its direction. The feedback of the X-ray radiation onto the 
stellar wind is also not included.} (see also Sect.~\ref{xraypars:clumps}). They obtained an extreme and continuous 
X-ray variability at any orbital phase of the 
system achieving up to eight orders of magnitude in luminosity. Not only this variability is significantly larger than that 
observed in the SFXTs, but also the sub-luminosity issue of these sources could not be addressed. 
On one hand, the large variability is related to the fact that the authors employed 1D hydrodynamic simulations 
of the stellar wind, in which clumps are known to grow to large sizes and densities (i.e., much larger than those expected 
from multi-dimensional simulations, see Sect.~\ref{sec:smallscale}). On the other hand, this work highlighted again the need 
for an additional mechanism to reduce the average mass accretion rate onto the neutron star, independently from the 
properties of the stellar wind. 

As illustrated in Sect.~\ref{sec:regimes}, a viable solution to reduce the mass accretion rate onto the neutron 
star is to assume that either centrifugal/magnetic gates are at work, or a settling accretion regime sets in. 
The possibility of having a centrifugal inhibition of accretion in SFXTs was first proposed by 
\citet{Grebenev2007}. If the neutron stars in SFXTs are spending most of their time in the supersonic propeller regime, 
a reduction of the X-ray luminosity by a factor of 100-1000 can be achieved compared to the standard direct accretion 
regime by assuming the neutron star is endowed with a standard magnetic field ($\mu_{30}$$\simeq$0.1-1) and a spin period in the range  
$\simeq$10-100~s (considering values for the stellar wind parameters in agreement with those expected for O-B supergiants, 
see Eq.~\ref{eq:rco} and \ref{eq:rmsuper}). Combining the centrifugal barrier with the presence of a moderately 
clumpy wind, one could argue that the sporadic encounter of a dense clump with the neutron star could provide 
the required increase in the local density to switch from the supersonic propeller to the direct accretion 
regime (see Eq.~ \ref{eq:rholim}) and produce a bright short flare as observed in the SFXTs. If a reasonable 
spread in density between different clumps is assumed, the X-ray dynamic range of the supersonic propeller could 
be combined with that implied by the simple direct accretion from a clumpy wind (a factor $\sim$10-100, 
see Sect.~\ref{xraypars:clumps}) to nearly explain the extreme variability of the SFXTs. 
The problem with the sole adoption of the centrifugal gate and the clumpy wind is that such combination of mechanisms 
would not be able to explain why only the SFXTs and not the classical SgXBs achieve an extreme X-ray variability: 
the neutron stars hosted in these systems are characterized by similar spin periods and in some cases 
these can be significantly longer than 10-100~s (Table~\ref{tab:windfed}).   

An alternative possibility would be to combine the presence of a magnetic gate with a clumpy wind model. 
As it was shown in Sect.~\ref{sec:regimes}, the magnetic gate can easily lead to variations in the system 
luminosity by a factor of 10$^{5}$ and thus it would be better suited to explain the SFXT X-ray dynamic range 
(see Eq.~\ref{eq:lshock}, \ref{eq:lsd1}, and \ref{eq:lx}).  
The advantage of the magnetic gate, compared to the centrifugal gate, is that it would require extremely 
large neutron star magnetic fields to be at work ($\mu_{30}$$\simeq$10-100) and particularly long 
spin periods ($>$1000~s). These values might not be unreasonable for a young neutron star in an HMXB  
and could provide a way to differentiate SFXTs from classical SgXBs \citep{Bozzo2008}.  
According to this idea, SFXTs should host neutron stars with systematically longer spin periods 
and more intense magnetic fields than classical SgXBs. 
However, to date, there are no consolidate measurements of SFXT spin periods 
(see Table~\ref{tab:windfed}), and the only SFXTs where a hint of the neutron star 
magnetic field value has been derived, shows a low field of a few 10$^{11}$~G\footnote{Note that this 
measurement has not been confirmed by \citet{Sguera2015}.} \citep{Sguera2010} or a 
few 10$^{12}$~G \citep{Bhalerao2015}. 

A different way to significantly reduce the accretion rate onto the neutron star with
respect to the direct accretion regime is the onset of a settling accretion regime.  
As we discussed in Sect.~\ref{sec:regimes}, this regime is believed to set in when the 
accreting material is not efficiently cooled down at the accretion radius (Eq.~\ref{eq:raccr}) 
and the Rayleigh-Taylor instability is only able to drive a reduced mass flow rate through the 
neutron star magnetosphere. Depending on the dominating cooling mechanism at the magnetospheric 
radius, it can be seen from Eq.~\ref{eq:sh03} and \ref{eq:sh01} that 
the accretion rate can be reduced by a factor ranging from 3 to 10 with respect to the direct 
accretion regime. 
On one hand, these reductions in the mass accretion rates are too small 
to reproduce the entire dynamic range of the SFXTs, and in particular to explain their lowest 
luminosity states ($L_{\rm X}$$\simeq$10$^{32}$~erg~s$^{-1}$). As both Eq.~\ref{eq:sh03} 
and \ref{eq:sh01} give the reduction in mass accretion rate as a function of the average 
source luminosity, it would be possible to achieve the low X-ray emission level of the 
SFXTs in quiescence within the settling accretion regime 
only by assuming that SFXTs are {\it a priori} significantly 
fainter in X-rays than the classical SgXBs. This can only be realized if the average mass-loss 
rate of the supergiant stars in the SFXTs is lower by a factor of 10-100 compared to 
those of supergiant stars in classical SgXBs. This reduction in mass-loss rate is, however, 
difficult to reconcile with the present information we have available on these stars, as the 
latter do not seem to be systematically different from supergiants in classical SgXBs 
(see Sect.~\ref{xraypars}).  
On the other hand, the high X-ray luminosity reached during the SFXTs brighter 
outbursts is not easily achieved within the settling accretion regime and would require 
the presence of very dense and massive clumps to increase the local mass accretion rates 
onto the neutron star by a factor of $\sim$100-1000. As such large clumps seem not realistic 
given our present knowledge of massive star winds (see Sects.~\ref{sec:smallscale}, \ref{sec:size} 
and \ref{xraypars:clumps}, but also
\ref{sec:largescale}), an alternative 
possibility was discussed by \citet{Shakura2014}. These authors suggested that the bright SFXT outbursts 
could be triggered by the collapse onto the neutron star of all (or part of) the hot shell that forms around 
the compact object in the settling accretion regime. The collapse of the shell might occur sporadically due to 
reconnection events between the supergiant magnetic field embedded in the accreting material and the 
neutron star magnetosphere\footnote{Note that this event is considered sporadic because it requires the two magnetic fields to be 
aligned along a preferred geometry and it is being assumed here that only the winds of supergiant stars in SFXTs 
are sufficiently magnetized to give rise to such effects as bright outbursts are not observed in classical SgXBs.}. 
The advantage of this idea is that the excess mass accretion rate 
required to reach the luminosities of the SFXTs during their outbursts is provided by accumulation 
within the hot shell around the neutron star rather than the presence of large clumps. 
\citet{Shakura2014} noted that the typical energy released in an SFXT bright flare (10$^{38}-10^{40}$~erg) 
would be consistent with the estimated mass of the entire hot shell around the compact object 
\citep[see also the discussion in][]{Drave2014}.  

\citet{PaizisSidoli2014} also proposed that the settling accretion model seems to be supported by the hard 
X-ray (20-100~keV) cumulative luminosity distributions of the SFXTs bright flares as observed with 
\emph{INTEGRAL} (IBIS/ISGRI) in the past 10 years. The main observational result of this study is that the 
cumulative luminosity distributions of the SFXTs flares are well described by a power-law and seem  
different from the one shown by classical persistent HMXBs (Vela~X-1 and 4U~1700-37, where 
the most frequent state is at high luminosity, around 1-2$\times$10$^{36}$). The latter are
more similar to log-normal distributions, as expected for a system accreting from a clumpy rather than from 
a smooth wind (see Sect.~\ref{xraypars:clumps}). 
\citet{PaizisSidoli2014} noted that the power-law fit of 
the SFXT cumulative luminosity distributions is reminiscent of Self-Organized Criticality 
(SOC) Systems \citep[see, e.g.,][]{Aschwanden2013}. 
A SOC \citep{Bak1987} is a system  naturally evolving into a critical state
where a minor event is able to start a chain reaction leading to a catastrophe, 
like the unpredictable ``avalanches'' in a sand-pile, when a certain instability threshold is reached.
Several phenomena are believed to behave like SOC systems: solar flares, 
lunar craters, earthquakes, landslides, forest fires, ...\citep{Newman2005}. 
The power-law scaling of SFXTs cumulative luminosity distributions was thus considered suggestive
of the fact that SFXTs flares can be associated to  ``avalanches'' in SOC systems, which
are triggered when a critical state is reached.
This would provide support in favour of the settling accretion model, where the hot matter stored in the 
shell can penetrate the magnetosphere only if it is able to cool down to a critical temperature 
(see discussion above and Sect.~\ref{sec:regimes}).
According to this interpretation, the observed power-law luminosity distributions of SFXTs flares might be 
coupled to the properties of the magnetized stellar wind and the physics of its interaction with the neutron 
star magnetosphere \citep{Shakura2014}, in analogy to what happens, e.g., in the case of the solar flares 
\citep[these are believed to be produced by the reconnection 
of the coronal magnetic field and show a power-law distribution;][]{Datlowe1974}. 

The results presented by \citet{PaizisSidoli2014} were later criticized by \citet{Bozzo2015}. 
These authors used the long-term monitoring observations of many SFXTs carried out with the XRT on-board \emph{Swift} 
to build the cumulative luminosity distributions of the SFXTs emission in the soft X-ray domain (0.5-10 keV).
Thanks to the higher sensitivity of the XRT compared to ISGRI for the low luminosity states  
of the SFXTs, it was possible to show that the approximated power-law shape of the cumulative luminosity 
distributions of these sources is no longer valid when the investigated emission range extends all the way down 
to $\simeq$10$^{33}$~erg~s$^{-1}$. From these results, it was concluded that the \emph{Swift/XRT} cumulative luminosity 
distributions of the SFXTs could have a similar functional shape as those of other SgXBs in the soft X-rays, 
but they are simply shifted to lower luminosities. This shift was interpreted as additional evidence of the strong inhibition 
of accretion that seems to affect all SFXTs \citep[see also][]{Romano2014b}. We note that the discrepancy 
between the soft and hard X-ray cumulative luminosity distributions of the SFXTs could be explained by the fact 
that INTEGRAL caught only the emission during the flares (so that the cumulative luminosity distributions reported 
by \citet{PaizisSidoli2014} are distributions of the X-ray flares peaks only), while the luminosity distributions reported by 
\citet{Bozzo2015} include data collected from all SFXT emission states. At present it is not possible 
to discriminate through the XRT data if the inhibition of accretion in SFXTs (compared to other SgXBs) is due to the presence 
of a centrifugal/magnetic gate or the onset of a settling accretion regime. As discussed by \citet{Bozzo2015} neither of the two possibilities 
seem able at present to satisfactorily explain the difference between classical SgXBs and SFXTs, as one would require large spin periods and 
magnetic fields that are not measured in the SFXTs and the other would imply a substantially lower mass-loss rate from the 
supergiant stars in SFXTs for which there is no clear supporting evidence. The issue of the peculiar SFXT X-ray variability 
thus remains highly debated.

Finally, it cannot be excluded that the class of SFXTs is inhomogeneous, including massive binaries 
where different mechanisms driving the X-ray flaring transient
behaviour are at work: sources where the flaring activity is sporadic and unpredictable are probably 
different from SFXTs where the outbursts are periodic, as it occurs for IGR~J11215-5952 
\citep{Sidoli2006,Sidoli2007}.
For this source, \citet{Sidoli2007} proposed that the periodic and short outbursts are produced 
when the neutron star crosses a denser and slower component in the supergiant wind, inclined with 
respect to the orbital plane. This putative second component of the supergiant wind could probably be 
due to the donor magnetic field that is able to steadily compress the
out-flowing wind onto the magnetic equatorial plane. 
The minimal supergiant magnetic field, B$_{\rm min}$, needed to compress the wind can be estimated 
from the ``wind magnetic confinement parameter'' $\eta$ \citep{ud-Doula2002}, assuming $\eta$=1:
B$_{\rm min} =  \sqrt{{\dot M_{\rm w}} \upsilon_{\infty}}/R_{*}$. 
Following \citet{Romano2009} and assuming the updated stellar parameters reported by \citet{Lorenzo2014}, 
(R$_{*}$=40~R$_\odot$, $\dot M_{\rm w}$=10$^{-6}$~M$_\odot$~yr$^{-1}$, $\upsilon_{\infty}$=1200~km~s$^{-1}$), 
one would get B$_{\rm min}$$\sim$30~G. So far, a measurement of the magnetic field in the supergiant companions of SFXTs 
has not been performed. If doable, it could help investigating the applicability of the proposed model for 
IGR~J11215-5952 in the near future. It should be noted, however, that no evidence for the second dense wind component 
around the supergiant star in IGR~J11215-5952 has been found to support the model proposed by \citet{Sidoli2007}, 
despite the intensive searches carried out by \citet{Negueruela2010}. 

A slightly different interpretation of the peculiar SFXT IGR~J11215-5952 was proposed by 
\citet{Lorenzo2014}. These authors obtained high-resolution spectra of the optical counterpart of 
IGR~J11215-5952, HD~306414, spanning three months and including one of its outbursts (in 2007 February). 
The supergiant star HD~306414 showed significant variability of the H$_{\alpha}$ line (shape and centroid of the
emission feature) simultaneously with the X-ray outburst, but similar variabilities 
were also observed seven weeks before the peak of the X-ray outburst (i.e. about 1/3 of the orbital period)
when no X-ray flaring activity was expected. It was thus suggested that 
IGR~J11215-5952 might have a large eccentricity (around 0.8) and that the neutron star in this system is 
able to accrete material from the wind of the companion only when it is close to its atmosphere during 
the periastron passage. In this configuration, the accretion would thus most likely take place from a transient 
tidal stream through the inner Lagrangian point, which could lead to the formation of a transient accretion disk, 
in turns triggering the periodic X-ray outbursts. 

The formation of short-lived accretion disks around the neutron stars in the SFXTs 
is also a possibility that has been discussed in several works to enhance the achievable 
X-ray luminosity in these systems. 
So far, no observational evidence of an accretion disk 
could be found during the observations of quiescent and out-bursting SFXTs, but \citet{Ducci2010} showed that 
an accretion disk could be formed especially in systems characterized by a short orbital period, a high X-ray luminosity, 
and a non negligible eccentricity. The idea is that in those cases in which the X-ray luminosity is high enough 
to significantly slow down  the stellar wind at the neutron star location and the compact object is close to the supergiant, 
the accretion radius becomes large enough that the angular momentum of the accreting material can not be neglected any longer. 
Although the formation and dissipation of these short-lived accretion disks is far from being understood, it is well known 
that an accretion disk can sustain a significantly higher mass accretion rate than the one of a wind-fed system and can thus lead 
to higher X-ray luminosities \citep[see, e.g.,][]{frank2002}. At least in one case, it has been convincingly 
shown that the peak luminosity reached during an SFXT outburst was far too high to be achievable through accretion from a  
supergiant stellar wind and it was suggested that a disk around the neutron star was formed slightly before the 
event \citep{Romano2015}. However, no significant changes in the X-ray emission from the source or other observational 
features could be found to firmly support this idea. Another extreme behaviour from a SFXT has been observed by INTEGRAL in the 
SFXT IGR~J18483-0311 \citep{Sguera2015}, with an unusually long outburst spanning 11 days (instead of a few hours), an interval 
of time lasting about 60\% of the orbit. The long-based observational data-set available now, after 10 years since the 
SFXTs' discovery, is allowing us to catch even the most extreme (and rare) behaviours from the members of the 
class.

\clearpage
\section{Conclusions}
\label{conclusions}

This review attempts to combine the efforts of two different scientific 
communities that have been independently working to improve our understanding of massive star 
winds by using either isolated objects or massive stars in binary systems. We  have largely 
reviewed the state-of-the-art of the research in the two fields and showed that the driving 
aim of these studies in both communities has been the investigation of the physical properties 
of dense structures populating the surroundings in isolated massive stars and 
feeding the compact objects in massive binaries. 
In the former, our relatively limited knowledge on the properties of these structures
lead to systematic uncertainties in the estimated mass loss rates from massive stars.
Consequently, our ability to reconstruct the long term evolution of these sources
and to properly evaluate their role influencing their environment is limited.
In the field of massive binaries, in turn, the lack of a proper understanding of the 
same structures represents one of our major limitations in the theoretical modelling 
of the wind accretion process onto the compact objects and thus on the ability to predict 
and reproduce their extreme variability in the X-ray domain 
(where the bulk of the energy from these systems is released).

The investigation of stellar winds in isolated massive stars has progressed significantly 
in the past years not only thanks to the availability of improved high sensitivity data in 
the optical, UV, and X-ray domain, but also to the development 
of more and more sophisticated numerical simulations. Although many advances have 
been made in different areas, we illustrated in the course of this review how a 
large number of fundamental issues are still far from being solved. Among these, various approaches 
are used by different groups in applying  micro and macro-clumping methods,  sometimes 
resulting in different derived mass loss rates and wind properties of massive stars.
The formation and origin of wind clumps is also still a matter of debate, as different mechanisms
have been proposed to produce these structures and drive their evolution in
shape, density, and velocity over time. 

In massive binaries, we particularly highlighted in the review how the degeneracy between the 
properties of the accretion flow regulated by the stellar wind (wind velocity and density, 
mass loss rates, ionization status) and the physical parameters of the compact object (mainly mass, 
radius, spin, and magnetic field strength) in all available models makes it very challenging 
to understand the details of the accretion processes and distinguish among the different proposed 
mechanisms to drive the release of the high energy radiations from these systems. 
The advent of the current generation of X-ray observatories, the main energy domain where 
observations of these systems are carried out, permitted to gain important insights into the 
accretion of stellar winds onto compact objects. This has been possible mainly through spectral 
and timing analyses performed by using integration times as short as the dynamical timescales 
of the relevant physical processes (heating/cooling of the accretion flow, free-fall motion).  
These results opened the interesting possibility of using the compact objects in these systems 
as probes {\it in situ} of the stellar wind and its structures. This could 
thus provide, in principle, independent measurements to be compared with those derived from the 
studies of isolated massive stars. The development of detailed theoretical models and 
advanced numerical simulations of wind accreting systems has shown, however, that it is very 
challenging to distinguish between different possible accretion regimes. The gravitational 
field of the compact object, as well as its rotational period and magnetic field 
strength, can greatly alter the geometry of the accretion flow. Thus the properties of the 
stellar wind and its structures inferred from measurements of the X-ray luminosity and 
spectral variations in the binary sources have to be taken with caution. The availability 
of more predictable properties for the stellar wind and its structures for a given massive 
star with a known spectral classification could allow us to remove some of the 
degeneracies and distinguish between different proposed accretion scenarios. 

It thus seems that combining observational, theoretical, and simulation efforts of massive star winds
and wind accretion in binary sources could certainly lead to significant simultaneous advantages for both 
fields of research, advancing our understanding of the accretion physics 
as well as of the galactic and cosmic evolution. For this reason we intend to pursue
the common interests of this combined community in the years to come by promoting: 
(i) the development of a new generation of hydrodynamic/magnetohydrodynamic simulations that can 
take into account the formation and evolution of stellar winds in presence of a compact object; (ii) 
the organization of multi-wavelength observational campaigns that take mutual advantage of 
the different diagnostic techniques exploited for isolated massive stars in the 
optical/UV/X-rays with those adopted by the binary community to the properties of 
stellar winds; (iii) advances in theoretical studies 
of the interaction between the X-ray radiation from accreting compact objects 
and stellar winds, especially in cases of highly inhomogeneous flows.

\section*{Acknowledgements}
 
 
We are grateful to the two anonymous referees for carefully 
reviewing such a large and detailed piece of work and for providing very
useful comments, which have allowed us to improve this paper.
 
SMN acknowledges the support of the Spanish Unemployment Agency, which allowed her to continue her scientific collaborations 
while being temporarily unemployed during the critical situation of the Spanish Research System until March 2016. 

SMN acknowledges support by research project ESP2016-76683-C3-1-R.

LS acknowledges the Italian Space Agency financial support INTEGRAL\linebreak[4] ASI/INAF 
agreement n. 2013-025.R.0, and the grant from PRIN-INAF 2014, \linebreak[4]
``Towards a unified picture of accretion in High Mass X-Ray Binaries'' (PI: Sidoli). 

AGG acknowledges support by Spanish MICINN under 
FPI Fellowship BES-2011-050874 and the Vicerectorat 
d'Investigaci\'o, Desenvolupament i Innovaci\'o de la Universitat d'Alacant under
 project GRE12-35.

IK, MK, and JW are supported by the  Bundesministerium f\"ur
Wirtschaft und Technologie under grant number 50OR1207
of the Deutsches Zentrum f\"ur Luft- und Raumfahrt. MK also
acknowledges support by the Bundesministerium f\"ur Wirtschaft und
Technologie under Deutsches Zentrum f\"ur Luft- und Raumfahrt grant
50OR1113.

AS is supported by the Deutsche Forschungsgemeinschaft (DFG) under grant
HA 1455/26.

JMT acknowledges research grants ESP2013-48637-C2-2P and ESP2014-53672-C3-3-P.

This publication was motivated by a team meeting sponsored by the International Space Science 
Institute at Bern, Switzerland.

\bibliographystyle{aps-nameyear}
\bibliography{bibfile_issi_v12}

\begin{thebibliography}{436}
\ifx \bisbn   \undefined \def \bisbn  #1{ISBN #1}\fi
\ifx \binits  \undefined \def \binits#1{#1} \fi
\ifx \bauthor  \undefined \def \bauthor#1{#1} \fi
\ifx \bjtitle  \undefined \def \bjtitle#1{\textrm{#1}}\fi
\ifx \batitle  \undefined \def \batitle#1{#1} \fi
\ifx \bctitle  \undefined \def \bctitle#1{#1} \fi
\ifx \bvolume  \undefined \def \bvolume#1{\textbf{#1}}\fi
\ifx \byear  \undefined \def \byear#1{#1} \fi
\ifx \bissue  \undefined \def \bissue#1{#1} \fi
\ifx \bfpage  \undefined \def \bfpage#1{#1} \fi
\ifx \blpage  \undefined \def \blpage #1{#1} \fi
\ifx \burl  \undefined \def \burl#1{#1} \fi
\ifx \doiurl  \undefined \def \doiurl#1{#1} \fi
\ifx \betal  \undefined \def \betal{et al.} \fi
\ifx \binstitute  \undefined \def \binstitute#1{#1} \fi
\ifx \beditor  \undefined \def \beditor#1{#1} \fi
\ifx \bpublisher  \undefined \def \bpublisher#1{#1} \fi
\ifx \bbtitle  \undefined \def \bbtitle#1{\textit{#1}} \fi
\ifx \bedition  \undefined \def \bedition#1{#1} \fi
\ifx \bseriesno  \undefined \def \bseriesno#1{#1} \fi
\ifx \blocation  \undefined \def \blocation#1{#1} \fi
\ifx \bsertitle  \undefined \def \bsertitle#1{#1} \fi
\ifx \bsnm \undefined \def \bsnm#1{#1} \fi
\ifx \bsuffix \undefined \def \bsuffix#1{#1} \fi
\ifx \bparticle \undefined \def \bparticle#1{#1} \fi
\ifx \barticle \undefined \def \barticle#1{#1} \fi
\ifx \botherref \undefined \def \botherref #1{#1} \fi
\ifx \url \undefined \def \url#1{#1} \fi
\ifx \bchapter \undefined \def \bchapter#1{#1} \fi
\ifx \bbook \undefined \def \bbook#1{#1} \fi
\ifx \bcomment \undefined \def \bcomment#1{#1} \fi
\ifx \oauthor \undefined \def \oauthor#1{#1} \fi
\ifx \citeauthoryear \undefined \def \citeauthoryear#1{#1} \fi
\ifx \texttildelow  \undefined \def \texttildelow{\symbol{126}} \fi
\def \endbibitem {}
\ifx \bconflocation  \undefined \def \bconflocation#1{#1} \fi

\bibitem[\protect\citeauthoryear{{Abbott}}{1980}]{Abbott80}
\begin{barticle}
\bauthor{\binits{D.C.} \bsnm{{Abbott}}},
\batitle{{The theory of radiatively driven stellar winds. I - A physical
  interpretation}}.
\bjtitle{\apj}
\bvolume{242},
\bfpage{1183}--\blpage{1207}
(\byear{1980}).
doi:\doiurl{10.1086/158550}
\end{barticle}
\endbibitem

\bibitem[\protect\citeauthoryear{{Abbott}}{1982}]{Abbott82}
\begin{barticle}
\bauthor{\binits{D.C.} \bsnm{{Abbott}}},
\batitle{{The theory of radiatively driven stellar winds. II - The line
  acceleration}}.
\bjtitle{\apj}
\bvolume{259},
\bfpage{282}--\blpage{301}
(\byear{1982})
\end{barticle}
\endbibitem

\bibitem[\protect\citeauthoryear{{Abbott} et~al.}{1980}]{Abbott1980}
\begin{barticle}
\bauthor{\binits{D.C.} \bsnm{{Abbott}}},
\bauthor{\binits{J.H.} \bsnm{{Bieging}}},
\bauthor{\binits{E.} \bsnm{{Churchwell}}},
\bauthor{\binits{J.P.} \bsnm{{Cassinelli}}},
\batitle{{VLA radio continuum measurements of mass loss from early-type
  stars}}.
\bjtitle{\apj}
\bvolume{238},
\bfpage{196}--\blpage{202}
(\byear{1980}).
doi:\doiurl{10.1086/157973}
\end{barticle}
\endbibitem

\bibitem[\protect\citeauthoryear{{Alecian} et~al.}{2013}]{Alecian13}
\begin{barticle}
\bauthor{\binits{E.} \bsnm{{Alecian}}},
\bauthor{\binits{G.A.} \bsnm{{Wade}}},
\bauthor{\binits{C.} \bsnm{{Catala}}},
\bauthor{\binits{J.H.} \bsnm{{Grunhut}}},
\bauthor{\binits{J.D.} \bsnm{{Landstreet}}},
\bauthor{\binits{S.} \bsnm{{Bagnulo}}},
\bauthor{\binits{T.} \bsnm{{B{\"o}hm}}},
\bauthor{\binits{C.P.} \bsnm{{Folsom}}},
\bauthor{\binits{S.} \bsnm{{Marsden}}},
\bauthor{\binits{I.} \bsnm{{Waite}}},
\batitle{{A high-resolution spectropolarimetric survey of Herbig Ae/Be stars -
  I. Observations and measurements}}.
\bjtitle{\mnras}
\bvolume{429},
\bfpage{1001}--\blpage{1026}
(\byear{2013}).
doi:\doiurl{10.1093/mnras/sts383}
\end{barticle}
\endbibitem

\bibitem[\protect\citeauthoryear{{Anders} and {Ebihara}}{1982}]{anders1982a}
\begin{barticle}
\bauthor{\binits{E.} \bsnm{{Anders}}},
\bauthor{\binits{M.} \bsnm{{Ebihara}}},
\batitle{Solar-system abundances of the elements}.
\bjtitle{Geochim.\ Cosmochim.\ Acta}
\bvolume{46},
\bfpage{2363}--\blpage{2380}
(\byear{1982})
\end{barticle}
\endbibitem

\bibitem[\protect\citeauthoryear{{Anders} and {Grevesse}}{1989}]{anders1989a}
\begin{barticle}
\bauthor{\binits{E.} \bsnm{{Anders}}},
\bauthor{\binits{N.} \bsnm{{Grevesse}}},
\batitle{Abundances of the elements - meteoritic and solar}.
\bjtitle{Geochim.\ Cosmochim.\ Acta}
\bvolume{53},
\bfpage{197}--\blpage{214}
(\byear{1989})
\end{barticle}
\endbibitem

\bibitem[\protect\citeauthoryear{{Ankay} et~al.}{2001}]{Ankay2001}
\begin{barticle}
\bauthor{\binits{A.} \bsnm{{Ankay}}},
\bauthor{\binits{L.} \bsnm{{Kaper}}},
\bauthor{\binits{J.H.J.} \bsnm{{de Bruijne}}},
\bauthor{\binits{J.} \bsnm{{Dewi}}},
\bauthor{\binits{R.} \bsnm{{Hoogerwerf}}},
\bauthor{\binits{G.J.} \bsnm{{Savonije}}},
\batitle{{The origin of the runaway high-mass X-ray binary HD
  153919/4U1700-37}}.
\bjtitle{\aap}
\bvolume{370},
\bfpage{170}--\blpage{175}
(\byear{2001}).
doi:\doiurl{10.1051/0004-6361:20010192}
\end{barticle}
\endbibitem

\bibitem[\protect\citeauthoryear{{Arons} and {Lea}}{1976}]{arons1976}
\begin{barticle}
\bauthor{\binits{J.} \bsnm{{Arons}}},
\bauthor{\binits{S.M.} \bsnm{{Lea}}},
\batitle{{Accretion onto magnetized neutron stars - Structure and interchange
  instability of a model magnetosphere}}.
\bjtitle{\apj}
\bvolume{207},
\bfpage{914}--\blpage{936}
(\byear{1976}).
doi:\doiurl{10.1086/154562}
\end{barticle}
\endbibitem

\bibitem[\protect\citeauthoryear{{Aschwanden}}{2013}]{Aschwanden2013}
\begin{botherref}
\oauthor{\binits{M.J.} \bsnm{{Aschwanden}}},
{A Macroscopic Description of Self-Organized Criticality Systems and
  Astrophysical Applications}.
ArXiv e-prints
(2013)
\end{botherref}
\endbibitem

\bibitem[\protect\citeauthoryear{{Asplund} et~al.}{2009}]{asplund2009a}
\begin{barticle}
\bauthor{\binits{M.} \bsnm{{Asplund}}},
\bauthor{\binits{N.} \bsnm{{Grevesse}}},
\bauthor{\binits{A.J.} \bsnm{{Sauval}}},
\bauthor{\binits{P.} \bsnm{{Scott}}},
\batitle{The chemical composition of the sun}.
\bjtitle{ARA\&A}
\bvolume{47},
\bfpage{481}--\blpage{522}
(\byear{2009})
\end{barticle}
\endbibitem

\bibitem[\protect\citeauthoryear{{Audley} et~al.}{2006}]{Audley2006}
\begin{barticle}
\bauthor{\binits{M.D.} \bsnm{{Audley}}},
\bauthor{\binits{F.} \bsnm{{Nagase}}},
\bauthor{\binits{K.} \bsnm{{Mitsuda}}},
\bauthor{\binits{L.} \bsnm{{Angelini}}},
\bauthor{\binits{R.L.} \bsnm{{Kelley}}},
\batitle{{ASCA observations of OAO 1657-415 and its dust-scattered X-ray
  halo}}.
\bjtitle{MNRAS}
\bvolume{367},
\bfpage{1147}--\blpage{1154}
(\byear{2006})
\end{barticle}
\endbibitem

\bibitem[\protect\citeauthoryear{{Babel}}{1995}]{Babel95}
\begin{barticle}
\bauthor{\binits{J.} \bsnm{{Babel}}},
\batitle{{Multi-component radiatively driven winds from A and B stars. I. The
  metallic wind of a main sequence A star.}}
\bjtitle{\aap}
\bvolume{301},
\bfpage{823}
(\byear{1995})
\end{barticle}
\endbibitem

\bibitem[\protect\citeauthoryear{{Bak} et~al.}{1987}]{Bak1987}
\begin{barticle}
\bauthor{\binits{P.} \bsnm{{Bak}}},
\bauthor{\binits{C.} \bsnm{{Tang}}},
\bauthor{\binits{K.} \bsnm{{Wiesenfeld}}},
\batitle{{Self-organized criticality - An explanation of 1/f noise}}.
\bjtitle{Physical Review Letters}
\bvolume{59},
\bfpage{381}--\blpage{384}
(\byear{1987}).
doi:\doiurl{10.1103/PhysRevLett.59.381}
\end{barticle}
\endbibitem

\bibitem[\protect\citeauthoryear{{Balucinska-Church} and
  {McCammon}}{1992}]{balucinska-church1992a}
\begin{barticle}
\bauthor{\binits{M.} \bsnm{{Balucinska-Church}}},
\bauthor{\binits{D.} \bsnm{{McCammon}}},
\batitle{Photoelectric absorption cross sections with variable abundances}.
\bjtitle{ApJ}
\bvolume{400},
\bfpage{699}
(\byear{1992})
\end{barticle}
\endbibitem

\bibitem[\protect\citeauthoryear{{Barba} et~al.}{2006}]{Barba2006}
\begin{barticle}
\bauthor{\binits{R.} \bsnm{{Barba}}},
\bauthor{\binits{R.} \bsnm{{Gamen}}},
\bauthor{\binits{N.} \bsnm{{Morrell}}},
\batitle{{HD 74194, a new binary supergiant fast X-ray transient?, possible
  optical counterpart of INTEGRAL hard X-ray source IGR J08408-4503}}.
\bjtitle{The Astronomer's Telegram}
\bvolume{819},
\bfpage{1}
(\byear{2006})
\end{barticle}
\endbibitem

\bibitem[\protect\citeauthoryear{{Basko} and {Sunyaev}}{1975}]{BaskoSunyaev:75}
\begin{barticle}
\bauthor{\binits{M.M.} \bsnm{{Basko}}},
\bauthor{\binits{R.A.} \bsnm{{Sunyaev}}},
\batitle{{Radiative transfer in a strong magnetic field and accreting X-ray
  pulsars}}.
\bjtitle{\aap}
\bvolume{42},
\bfpage{311}--\blpage{321}
(\byear{1975})
\end{barticle}
\endbibitem

\bibitem[\protect\citeauthoryear{{Bates} and
  {Gilheany}}{1990}]{BatesGilheany90}
\begin{barticle}
\bauthor{\binits{B.} \bsnm{{Bates}}},
\bauthor{\binits{S.} \bsnm{{Gilheany}}},
\batitle{{IUE observations of mass-loss spectral features in B5-B9
  supergiants}}.
\bjtitle{\mnras}
\bvolume{243},
\bfpage{320}--\blpage{329}
(\byear{1990})
\end{barticle}
\endbibitem

\bibitem[\protect\citeauthoryear{{Becker} and {Wolff}}{2007}]{becker2007a}
\begin{barticle}
\bauthor{\binits{P.A.} \bsnm{{Becker}}},
\bauthor{\binits{M.T.} \bsnm{{Wolff}}},
\batitle{Thermal and bulk comptonization in accretion-powered x-ray pulsars}.
\bjtitle{ApJ}
\bvolume{654},
\bfpage{435}--\blpage{457}
(\byear{2007})
\end{barticle}
\endbibitem

\bibitem[\protect\citeauthoryear{{Becker} et~al.}{2012}]{becker2012a}
\begin{barticle}
\bauthor{\binits{P.A.} \bsnm{{Becker}}},
\bauthor{\binits{D.} \bsnm{{Klochkov}}},
\bauthor{\binits{G.} \bsnm{{Sch{\"o}nherr}}},
\bauthor{\binits{O.} \bsnm{{Nishimura}}},
\bauthor{\binits{C.} \bsnm{{Ferrigno}}},
\bauthor{\binits{I.} \bsnm{{Caballero}}},
\bauthor{\binits{P.} \bsnm{{Kretschmar}}},
\bauthor{\binits{M.T.} \bsnm{{Wolff}}},
\bauthor{\binits{J.} \bsnm{{Wilms}}},
\bauthor{\binits{R.} \bsnm{{Staubert}}},
\batitle{Spectral formation in accreting x-ray pulsars: bimodal variation of
  the cyclotron energy with luminosity}.
\bjtitle{A\&A}
\bvolume{544},
\bfpage{123}
(\byear{2012})
\end{barticle}
\endbibitem

\bibitem[\protect\citeauthoryear{{Beloborodov}}{2002}]{beloborodov2002a}
\begin{barticle}
\bauthor{\binits{A.M.} \bsnm{{Beloborodov}}},
\batitle{Gravitational bending of light near compact objects}.
\bjtitle{ApJ}
\bvolume{566},
\bfpage{85}--\blpage{88}
(\byear{2002})
\end{barticle}
\endbibitem

\bibitem[\protect\citeauthoryear{Bernstein et~al.}{1958}]{bernstein57}
\begin{barticle}
\bauthor{\binits{I.B.} \bsnm{Bernstein}},
\bauthor{\binits{E.A.} \bsnm{Frieman}},
\bauthor{\binits{M.D.} \bsnm{Kruskal}},
\bauthor{\binits{R.M.} \bsnm{Kulsrud}},
\batitle{An energy principle for hydromagnetic stability problems}.
\bjtitle{Proceedings of the Royal Society of London. Series A, Mathematical and
  Physical Sciences}
\bvolume{244}(\bissue{1236}),
\bfpage{17}--\blpage{40}
(\byear{1958})
\end{barticle}
\endbibitem

\bibitem[\protect\citeauthoryear{{Bhalerao} et~al.}{2015}]{Bhalerao2015}
\begin{barticle}
\bauthor{\binits{V.} \bsnm{{Bhalerao}}},
\bauthor{\binits{P.} \bsnm{{Romano}}},
\bauthor{\binits{J.} \bsnm{{Tomsick}}},
\bauthor{\binits{L.} \bsnm{{Natalucci}}},
\bauthor{\binits{D.M.} \bsnm{{Smith}}},
\bauthor{\binits{E.} \bsnm{{Bellm}}},
\bauthor{\binits{S.E.} \bsnm{{Boggs}}},
\bauthor{\binits{D.} \bsnm{{Chakrabarty}}},
\bauthor{\binits{F.E.} \bsnm{{Christensen}}},
\bauthor{\binits{W.W.} \bsnm{{Craig}}},
\bauthor{\binits{F.} \bsnm{{Fuerst}}},
\bauthor{\binits{C.J.} \bsnm{{Hailey}}},
\bauthor{\binits{F.A.} \bsnm{{Harrison}}},
\bauthor{\binits{R.A.} \bsnm{{Krivonos}}},
\bauthor{\binits{T.-N.} \bsnm{{Lu}}},
\bauthor{\binits{K.} \bsnm{{Madsen}}},
\bauthor{\binits{D.} \bsnm{{Stern}}},
\bauthor{\binits{G.} \bsnm{{Younes}}},
\bauthor{\binits{W.} \bsnm{{Zhang}}},
\batitle{{NuSTAR detection of a cyclotron line in the supergiant fast X-ray
  transient IGR J17544-2619}}.
\bjtitle{\mnras}
\bvolume{447},
\bfpage{2274}--\blpage{2281}
(\byear{2015}).
doi:\doiurl{10.1093/mnras/stu2495}
\end{barticle}
\endbibitem

\bibitem[\protect\citeauthoryear{{Bird} et~al.}{2009}]{Bird2009}
\begin{barticle}
\bauthor{\binits{A.J.} \bsnm{{Bird}}},
\bauthor{\binits{A.} \bsnm{{Bazzano}}},
\bauthor{\binits{A.B.} \bsnm{{Hill}}},
\bauthor{\binits{V.A.} \bsnm{{McBride}}},
\bauthor{\binits{V.} \bsnm{{Sguera}}},
\bauthor{\binits{S.E.} \bsnm{{Shaw}}},
\bauthor{\binits{H.J.} \bsnm{{Watkins}}},
\batitle{{Discovery of a 30-d period in the supergiant fast X-ray transient SAX
  J1818.6-1703}}.
\bjtitle{\mnras}
\bvolume{393},
\bfpage{11}--\blpage{15}
(\byear{2009}).
doi:\doiurl{10.1111/j.1745-3933.2008.00583.x}
\end{barticle}
\endbibitem

\bibitem[\protect\citeauthoryear{{Blondin} and {Woo}}{1995}]{Blondin1995}
\begin{barticle}
\bauthor{\binits{J.M.} \bsnm{{Blondin}}},
\bauthor{\binits{J.W.} \bsnm{{Woo}}},
\batitle{{Wind dynamics in SMC X-1. 1: Hydrodynamic simulation}}.
\bjtitle{\apj}
\bvolume{445},
\bfpage{889}--\blpage{895}
(\byear{1995}).
doi:\doiurl{10.1086/175748}
\end{barticle}
\endbibitem

\bibitem[\protect\citeauthoryear{{Blondin} et~al.}{1990}]{Blondin1990}
\begin{barticle}
\bauthor{\binits{J.M.} \bsnm{{Blondin}}},
\bauthor{\binits{T.R.} \bsnm{{Kallman}}},
\bauthor{\binits{B.A.} \bsnm{{Fryxell}}},
\bauthor{\binits{R.E.} \bsnm{{Taam}}},
\batitle{{Hydrodynamic simulations of stellar wind disruption by a compact
  X-ray source}}.
\bjtitle{\apj}
\bvolume{356},
\bfpage{591}--\blpage{608}
(\byear{1990}).
doi:\doiurl{10.1086/168865}
\end{barticle}
\endbibitem

\bibitem[\protect\citeauthoryear{Blondin et~al.}{1991}]{Blondin1991}
\begin{barticle}
\bauthor{\binits{J.M.} \bsnm{Blondin}},
\bauthor{\binits{I.R.} \bsnm{Stevens}},
\bauthor{\binits{T.R.} \bsnm{Kallman}},
\batitle{Enhanced winds and tidal streams in massive x-ray binaries}.
\bjtitle{\apj}
\bvolume{371},
\bfpage{684}--\blpage{695}
(\byear{1991})
\end{barticle}
\endbibitem

\bibitem[\protect\citeauthoryear{{Bodaghee} et~al.}{2006}]{Bodaghee2006}
\begin{barticle}
\bauthor{\binits{A.} \bsnm{{Bodaghee}}},
\bauthor{\binits{R.} \bsnm{{Walter}}},
\bauthor{\binits{J.A.} \bsnm{{Zurita Heras}}},
\bauthor{\binits{A.J.} \bsnm{{Bird}}},
\bauthor{\binits{T.J.-L.} \bsnm{{Courvoisier}}},
\bauthor{\binits{A.} \bsnm{{Malizia}}},
\bauthor{\binits{R.} \bsnm{{Terrier}}},
\bauthor{\binits{P.} \bsnm{{Ubertini}}},
\batitle{{IGR J16393-4643: a new heavily-obscured X-ray pulsar}}.
\bjtitle{\aap}
\bvolume{447},
\bfpage{1027}--\blpage{1034}
(\byear{2006}).
doi:\doiurl{10.1051/0004-6361:20053809}
\end{barticle}
\endbibitem

\bibitem[\protect\citeauthoryear{{Bodaghee} et~al.}{2011}]{Bodaghee2011}
\begin{barticle}
\bauthor{\binits{A.} \bsnm{{Bodaghee}}},
\bauthor{\binits{J.A.} \bsnm{{Tomsick}}},
\bauthor{\binits{J.} \bsnm{{Rodriguez}}},
\bauthor{\binits{S.} \bsnm{{Chaty}}},
\bauthor{\binits{K.} \bsnm{{Pottschmidt}}},
\bauthor{\binits{R.} \bsnm{{Walter}}},
\bauthor{\binits{P.} \bsnm{{Romano}}},
\batitle{{Suzaku Observes Weak Flares from IGR J17391-3021 Representing a
  Common Low-activity State in this Supergiant Fast X-ray Transient}}.
\bjtitle{\apj}
\bvolume{727},
\bfpage{59}
(\byear{2011}).
doi:\doiurl{10.1088/0004-637X/727/1/59}
\end{barticle}
\endbibitem

\bibitem[\protect\citeauthoryear{{Bodaghee} et~al.}{2012}]{Bodaghee2012}
\begin{barticle}
\bauthor{\binits{A.} \bsnm{{Bodaghee}}},
\bauthor{\binits{F.} \bsnm{{Rahoui}}},
\bauthor{\binits{J.A.} \bsnm{{Tomsick}}},
\bauthor{\binits{J.} \bsnm{{Rodriguez}}},
\batitle{{Chandra Observations of Five INTEGRAL Sources: New X-Ray Positions
  for IGR J16393-4643 and IGR J17091-3624}}.
\bjtitle{\apj}
\bvolume{751},
\bfpage{113}
(\byear{2012}).
doi:\doiurl{10.1088/0004-637X/751/2/113}
\end{barticle}
\endbibitem

\bibitem[\protect\citeauthoryear{Bondi and Hoyle}{1944}]{BondiHoyle1944}
\begin{barticle}
\bauthor{\binits{H.} \bsnm{Bondi}},
\bauthor{\binits{F.} \bsnm{Hoyle}},
\batitle{On the mechanism of accretion by stars}.
\bjtitle{MNRAS}
\bvolume{104},
\bfpage{273}--\blpage{282}
(\byear{1944})
\end{barticle}
\endbibitem

\bibitem[\protect\citeauthoryear{{Boroson} et~al.}{2003}]{Boroson2003}
\begin{barticle}
\bauthor{\binits{B.} \bsnm{{Boroson}}},
\bauthor{\binits{S.D.} \bsnm{{Vrtilek}}},
\bauthor{\binits{T.} \bsnm{{Kallman}}},
\bauthor{\binits{M.} \bsnm{{Corcoran}}},
\batitle{{Chandra Grating Spectroscopy of the X-Ray Binary 4U 1700-37 in a
  Flaring State}}.
\bjtitle{\apj}
\bvolume{592},
\bfpage{516}--\blpage{531}
(\byear{2003}).
doi:\doiurl{10.1086/375636}
\end{barticle}
\endbibitem

\bibitem[\protect\citeauthoryear{{Bouret} et~al.}{2005}]{bouret2005}
\begin{barticle}
\bauthor{\binits{J.-C.} \bsnm{{Bouret}}},
\bauthor{\binits{T.} \bsnm{{Lanz}}},
\bauthor{\binits{D.J.} \bsnm{{Hillier}}},
\batitle{{Lower mass loss rates in O-type stars: Spectral signatures of dense
  clumps in the wind of two Galactic O4 stars}}.
\bjtitle{\aap}
\bvolume{438},
\bfpage{301}--\blpage{316}
(\byear{2005}).
doi:\doiurl{10.1051/0004-6361:20042531}
\end{barticle}
\endbibitem

\bibitem[\protect\citeauthoryear{{Bouret} et~al.}{2012}]{bouret2012}
\begin{barticle}
\bauthor{\binits{J.-C.} \bsnm{{Bouret}}},
\bauthor{\binits{D.J.} \bsnm{{Hillier}}},
\bauthor{\binits{T.} \bsnm{{Lanz}}},
\bauthor{\binits{A.W.} \bsnm{{Fullerton}}},
\batitle{{Properties of Galactic early-type O-supergiants. A combined FUV-UV
  and optical analysis}}.
\bjtitle{\aap}
\bvolume{544},
\bfpage{67}
(\byear{2012}).
doi:\doiurl{10.1051/0004-6361/201118594}
\end{barticle}
\endbibitem

\bibitem[\protect\citeauthoryear{{Bozzo} et~al.}{2008}]{Bozzo2008}
\begin{barticle}
\bauthor{\binits{E.} \bsnm{{Bozzo}}},
\bauthor{\binits{M.} \bsnm{{Falanga}}},
\bauthor{\binits{L.} \bsnm{{Stella}}},
\batitle{{Are There Magnetars in High-Mass X-Ray Binaries? The Case of
  Supergiant Fast X-Ray Transients}}.
\bjtitle{\apj}
\bvolume{683},
\bfpage{1031}--\blpage{1044}
(\byear{2008}).
doi:\doiurl{10.1086/589990}
\end{barticle}
\endbibitem

\bibitem[\protect\citeauthoryear{{Bozzo} et~al.}{2010}]{Bozzo2010}
\begin{barticle}
\bauthor{\binits{E.} \bsnm{{Bozzo}}},
\bauthor{\binits{L.} \bsnm{{Stella}}},
\bauthor{\binits{C.} \bsnm{{Ferrigno}}},
\bauthor{\binits{A.} \bsnm{{Giunta}}},
\bauthor{\binits{M.} \bsnm{{Falanga}}},
\bauthor{\binits{S.} \bsnm{{Campana}}},
\bauthor{\binits{G.} \bsnm{{Israel}}},
\bauthor{\binits{J.C.} \bsnm{{Leyder}}},
\batitle{{The supergiant fast X-ray transients XTE J1739-302 and IGR
  J08408-4503 in quiescence with XMM-Newton}}.
\bjtitle{\aap}
\bvolume{519},
\bfpage{6}
(\byear{2010}).
doi:\doiurl{10.1051/0004-6361/201014095}
\end{barticle}
\endbibitem

\bibitem[\protect\citeauthoryear{{Bozzo} et~al.}{2011}]{Bozzo2011}
\begin{barticle}
\bauthor{\binits{E.} \bsnm{{Bozzo}}},
\bauthor{\binits{A.} \bsnm{{Giunta}}},
\bauthor{\binits{G.} \bsnm{{Cusumano}}},
\bauthor{\binits{C.} \bsnm{{Ferrigno}}},
\bauthor{\binits{R.} \bsnm{{Walter}}},
\bauthor{\binits{S.} \bsnm{{Campana}}},
\bauthor{\binits{M.} \bsnm{{Falanga}}},
\bauthor{\binits{G.} \bsnm{{Israel}}},
\bauthor{\binits{L.} \bsnm{{Stella}}},
\batitle{{XMM-Newton observations of IGR J18410-0535: the ingestion of a clump
  by a supergiant fast X-ray transient}}.
\bjtitle{\aap}
\bvolume{531},
\bfpage{130}
(\byear{2011}).
doi:\doiurl{10.1051/0004-6361/201116726}
\end{barticle}
\endbibitem

\bibitem[\protect\citeauthoryear{{Bozzo} et~al.}{2012}]{Bozzo2012}
\begin{barticle}
\bauthor{\binits{E.} \bsnm{{Bozzo}}},
\bauthor{\binits{L.} \bsnm{{Pavan}}},
\bauthor{\binits{C.} \bsnm{{Ferrigno}}},
\bauthor{\binits{M.} \bsnm{{Falanga}}},
\bauthor{\binits{S.} \bsnm{{Campana}}},
\bauthor{\binits{S.} \bsnm{{Paltani}}},
\bauthor{\binits{L.} \bsnm{{Stella}}},
\bauthor{\binits{R.} \bsnm{{Walter}}},
\batitle{{XMM-Newton observations of four high mass X-ray binaries and IGR
  J17348-2045}}.
\bjtitle{\aap}
\bvolume{544},
\bfpage{118}
(\byear{2012}).
doi:\doiurl{10.1051/0004-6361/201218900}
\end{barticle}
\endbibitem

\bibitem[\protect\citeauthoryear{{Bozzo} et~al.}{2015}]{Bozzo2015}
\begin{barticle}
\bauthor{\binits{E.} \bsnm{{Bozzo}}},
\bauthor{\binits{P.} \bsnm{{Romano}}},
\bauthor{\binits{L.} \bsnm{{Ducci}}},
\bauthor{\binits{F.} \bsnm{{Bernardini}}},
\bauthor{\binits{M.} \bsnm{{Falanga}}},
\batitle{{Supergiant fast X-ray transients as an under-luminous class of
  supergiant X-ray binaries}}.
\bjtitle{Advances in Space Research}
\bvolume{55},
\bfpage{1255}--\blpage{1263}
(\byear{2015}).
doi:\doiurl{10.1016/j.asr.2014.11.012}
\end{barticle}
\endbibitem

\bibitem[\protect\citeauthoryear{{Bozzo} et~al.}{2016}]{Bozzo2016}
\begin{botherref}
\oauthor{\binits{E.} \bsnm{{Bozzo}}},
\oauthor{\binits{V.} \bsnm{{Bhalerao}}},
\oauthor{\binits{P.} \bsnm{{Pradhan}}},
\oauthor{\binits{J.} \bsnm{{Tomsick}}},
\oauthor{\binits{P.} \bsnm{{Romano}}},
\oauthor{\binits{C.} \bsnm{{Ferrigno}}},
\oauthor{\binits{S.} \bsnm{{Chaty}}},
\oauthor{\binits{L.} \bsnm{{Oskinova}}},
\oauthor{\binits{A.} \bsnm{{Manousakis}}},
\oauthor{\binits{R.} \bsnm{{Walter}}},
\oauthor{\binits{M.} \bsnm{{Falanga}}},
\oauthor{\binits{S.} \bsnm{{Campana}}},
\oauthor{\binits{L.} \bsnm{{Stella}}},
\oauthor{\binits{M.} \bsnm{{Ramolla}}},
\oauthor{\binits{R.} \bsnm{{Chini}}},
{Multi-wavelength observations of IGR J17544-2619 from quiescence to outburst}.
ArXiv e-prints
(2016)
\end{botherref}
\endbibitem

\bibitem[\protect\citeauthoryear{{Burnard} et~al.}{1983}]{burnard83}
\begin{barticle}
\bauthor{\binits{D.J.} \bsnm{{Burnard}}},
\bauthor{\binits{J.} \bsnm{{Arons}}},
\bauthor{\binits{S.M.} \bsnm{{Lea}}},
\batitle{{Accretion onto magnetized neutron stars - X-ray pulsars with
  intermediate rotation rates}}.
\bjtitle{\apj}
\bvolume{266},
\bfpage{175}--\blpage{187}
(\byear{1983}).
doi:\doiurl{10.1086/160768}
\end{barticle}
\endbibitem

\bibitem[\protect\citeauthoryear{{Butler} and
  {Giddings}}{1985}]{ButlerGiddings85}
\begin{botherref}
\oauthor{\binits{K.} \bsnm{{Butler}}},
\oauthor{\binits{J.R.} \bsnm{{Giddings}}},
{notitle}.
Newsl. Anal. Astron. Spectra
\textbf{9}
(1985)
\end{botherref}
\endbibitem

\bibitem[\protect\citeauthoryear{{Cassinelli} et~al.}{2001}]{cas2001}
\begin{barticle}
\bauthor{\binits{J.P.} \bsnm{{Cassinelli}}},
\bauthor{\binits{N.A.} \bsnm{{Miller}}},
\bauthor{\binits{W.L.} \bsnm{{Waldron}}},
\bauthor{\binits{J.J.} \bsnm{{MacFarlane}}},
\bauthor{\binits{D.H.} \bsnm{{Cohen}}},
\batitle{{Chandra Detection of Doppler-shifted X-Ray Line Profiles from the
  Wind of {$\zeta$} Puppis (O4 F)}}.
\bjtitle{\apjl}
\bvolume{554},
\bfpage{55}--\blpage{58}
(\byear{2001}).
doi:\doiurl{10.1086/320916}
\end{barticle}
\endbibitem

\bibitem[\protect\citeauthoryear{{Castor} et~al.}{1975}]{CAK}
\begin{barticle}
\bauthor{\binits{J.I.} \bsnm{{Castor}}},
\bauthor{\binits{D.C.} \bsnm{{Abbott}}},
\bauthor{\binits{R.I.} \bsnm{{Klein}}},
\batitle{{Radiation-driven winds in Of stars}}.
\bjtitle{\apj}
\bvolume{195},
\bfpage{157}--\blpage{174}
(\byear{1975})
\end{barticle}
\endbibitem

\bibitem[\protect\citeauthoryear{{Chaty} et~al.}{2008}]{Chaty2008}
\begin{barticle}
\bauthor{\binits{S.} \bsnm{{Chaty}}},
\bauthor{\binits{F.} \bsnm{{Rahoui}}},
\bauthor{\binits{C.} \bsnm{{Foellmi}}},
\bauthor{\binits{J.A.} \bsnm{{Tomsick}}},
\bauthor{\binits{J.} \bsnm{{Rodriguez}}},
\bauthor{\binits{R.} \bsnm{{Walter}}},
\batitle{{Multi-wavelength observations of Galactic hard X-ray sources
  discovered by INTEGRAL. I. The nature of the companion star}}.
\bjtitle{\aap}
\bvolume{484},
\bfpage{783}--\blpage{800}
(\byear{2008}).
doi:\doiurl{10.1051/0004-6361:20078768}
\end{barticle}
\endbibitem

\bibitem[\protect\citeauthoryear{{Clark} et~al.}{2009}]{Clark2009}
\begin{barticle}
\bauthor{\binits{D.J.} \bsnm{{Clark}}},
\bauthor{\binits{A.B.} \bsnm{{Hill}}},
\bauthor{\binits{A.J.} \bsnm{{Bird}}},
\bauthor{\binits{V.A.} \bsnm{{McBride}}},
\bauthor{\binits{S.} \bsnm{{Scaringi}}},
\bauthor{\binits{A.J.} \bsnm{{Dean}}},
\batitle{{Discovery of the orbital period in the supergiant fast X-ray
  transient IGR J17544-2619}}.
\bjtitle{\mnras}
\bvolume{399},
\bfpage{113}--\blpage{117}
(\byear{2009}).
doi:\doiurl{10.1111/j.1745-3933.2009.00737.x}
\end{barticle}
\endbibitem

\bibitem[\protect\citeauthoryear{{Clark} et~al.}{2010}]{Clark2010}
\begin{barticle}
\bauthor{\binits{D.J.} \bsnm{{Clark}}},
\bauthor{\binits{V.} \bsnm{{Sguera}}},
\bauthor{\binits{A.J.} \bsnm{{Bird}}},
\bauthor{\binits{V.A.} \bsnm{{McBride}}},
\bauthor{\binits{A.B.} \bsnm{{Hill}}},
\bauthor{\binits{S.} \bsnm{{Scaringi}}},
\bauthor{\binits{S.} \bsnm{{Drave}}},
\bauthor{\binits{A.} \bsnm{{Bazzano}}},
\bauthor{\binits{A.J.} \bsnm{{Dean}}},
\batitle{{The orbital period in the supergiant fast X-ray transient IGR
  J16465-4507}}.
\bjtitle{\mnras}
\bvolume{406},
\bfpage{75}--\blpage{79}
(\byear{2010}).
doi:\doiurl{10.1111/j.1745-3933.2010.00885.x}
\end{barticle}
\endbibitem

\bibitem[\protect\citeauthoryear{{Clark}}{2000}]{Clark2000}
\begin{barticle}
\bauthor{\binits{G.W.} \bsnm{{Clark}}},
\batitle{{The Orbit of the Binary X-Ray Pulsar 4U 1538-52 from Rossi X-Ray
  Timing Explorer Observations}}.
\bjtitle{\apjl}
\bvolume{542},
\bfpage{131}--\blpage{133}
(\byear{2000}).
doi:\doiurl{10.1086/312926}
\end{barticle}
\endbibitem

\bibitem[\protect\citeauthoryear{{Clark} et~al.}{1994}]{Clark1994}
\begin{barticle}
\bauthor{\binits{G.W.} \bsnm{{Clark}}},
\bauthor{\binits{J.W.} \bsnm{{Woo}}},
\bauthor{\binits{F.} \bsnm{{Nagase}}},
\batitle{{Properties of a B0 I stellar wind and interstellar grains derived
  from GINGA observations of the binary X-ray pulsar 4U 1538-52}}.
\bjtitle{\apj}
\bvolume{422},
\bfpage{336}--\blpage{350}
(\byear{1994}).
doi:\doiurl{10.1086/173729}
\end{barticle}
\endbibitem

\bibitem[\protect\citeauthoryear{{Coe} et~al.}{1996}]{Coe1996}
\begin{barticle}
\bauthor{\binits{M.J.} \bsnm{{Coe}}},
\bauthor{\binits{J.} \bsnm{{Fabregat}}},
\bauthor{\binits{I.} \bsnm{{Negueruela}}},
\bauthor{\binits{P.} \bsnm{{Roche}}},
\bauthor{\binits{I.A.} \bsnm{{Steele}}},
\batitle{{Discovery of the optical counterpart to the ASCA transient AX
  1845.0-0433}}.
\bjtitle{\mnras}
\bvolume{281},
\bfpage{333}--\blpage{338}
(\byear{1996})
\end{barticle}
\endbibitem

\bibitem[\protect\citeauthoryear{{Cohen} et~al.}{2010a}]{cohen2010}
\begin{barticle}
\bauthor{\binits{D.H.} \bsnm{{Cohen}}},
\bauthor{\binits{M.A.} \bsnm{{Leutenegger}}},
\bauthor{\binits{E.E.} \bsnm{{Wollman}}},
\bauthor{\binits{J.} \bsnm{{Zsarg{\'o}}}},
\bauthor{\binits{D.J.} \bsnm{{Hillier}}},
\bauthor{\binits{R.H.D.} \bsnm{{Townsend}}},
\bauthor{\binits{S.P.} \bsnm{{Owocki}}},
\batitle{{A mass-loss rate determination for {$\zeta$} Puppis from the
  quantitative analysis of X-ray emission-line profiles}}.
\bjtitle{\mnras}
\bvolume{405},
\bfpage{2391}--\blpage{2405}
(\byear{2010}a).
doi:\doiurl{10.1111/j.1365-2966.2010.16606.x}
\end{barticle}
\endbibitem

\bibitem[\protect\citeauthoryear{{Cohen} et~al.}{2010b}]{coh2010}
\begin{barticle}
\bauthor{\binits{D.H.} \bsnm{{Cohen}}},
\bauthor{\binits{M.A.} \bsnm{{Leutenegger}}},
\bauthor{\binits{E.E.} \bsnm{{Wollman}}},
\bauthor{\binits{J.} \bsnm{{Zsarg{\'o}}}},
\bauthor{\binits{D.J.} \bsnm{{Hillier}}},
\bauthor{\binits{R.H.D.} \bsnm{{Townsend}}},
\bauthor{\binits{S.P.} \bsnm{{Owocki}}},
\batitle{{A mass-loss rate determination for {$\zeta$} Puppis from the
  quantitative analysis of X-ray emission-line profiles}}.
\bjtitle{\mnras}
\bvolume{405},
\bfpage{2391}--\blpage{2405}
(\byear{2010}b).
doi:\doiurl{10.1111/j.1365-2966.2010.16606.x}
\end{barticle}
\endbibitem

\bibitem[\protect\citeauthoryear{{Cohen} et~al.}{2014}]{coh2014}
\begin{barticle}
\bauthor{\binits{D.H.} \bsnm{{Cohen}}},
\bauthor{\binits{E.E.} \bsnm{{Wollman}}},
\bauthor{\binits{M.A.} \bsnm{{Leutenegger}}},
\bauthor{\binits{J.O.} \bsnm{{Sundqvist}}},
\bauthor{\binits{A.W.} \bsnm{{Fullerton}}},
\bauthor{\binits{J.} \bsnm{{Zsarg{\'o}}}},
\bauthor{\binits{S.P.} \bsnm{{Owocki}}},
\batitle{{Measuring mass-loss rates and constraining shock physics using X-ray
  line profiles of O stars from the Chandra archive}}.
\bjtitle{\mnras}
\bvolume{439},
\bfpage{908}--\blpage{923}
(\byear{2014}).
doi:\doiurl{10.1093/mnras/stu008}
\end{barticle}
\endbibitem

\bibitem[\protect\citeauthoryear{{Coleiro} et~al.}{2013}]{Coleiro2013}
\begin{barticle}
\bauthor{\binits{A.} \bsnm{{Coleiro}}},
\bauthor{\binits{S.} \bsnm{{Chaty}}},
\bauthor{\binits{J.A.} \bsnm{{Zurita Heras}}},
\bauthor{\binits{F.} \bsnm{{Rahoui}}},
\bauthor{\binits{J.A.} \bsnm{{Tomsick}}},
\batitle{{Infrared identification of high-mass X-ray binaries discovered by
  INTEGRAL}}.
\bjtitle{\aap}
\bvolume{560},
\bfpage{108}
(\byear{2013}).
doi:\doiurl{10.1051/0004-6361/201322382}
\end{barticle}
\endbibitem

\bibitem[\protect\citeauthoryear{{Corbet} and {Krimm}}{2013}]{Corbet2013}
\begin{barticle}
\bauthor{\binits{R.H.D.} \bsnm{{Corbet}}},
\bauthor{\binits{H.A.} \bsnm{{Krimm}}},
\batitle{{Superorbital Periodic Modulation in Wind-accretion High-mass X-Ray
  Binaries from Swift Burst Alert Telescope Observations}}.
\bjtitle{\apj}
\bvolume{778},
\bfpage{45}
(\byear{2013}).
doi:\doiurl{10.1088/0004-637X/778/1/45}
\end{barticle}
\endbibitem

\bibitem[\protect\citeauthoryear{{Corbet} and {Mukai}}{2002}]{Corbet2002}
\begin{barticle}
\bauthor{\binits{R.H.D.} \bsnm{{Corbet}}},
\bauthor{\binits{K.} \bsnm{{Mukai}}},
\batitle{{The Orbit and Position of the X-Ray Pulsar XTE J1855-026: an
  Eclipsing Supergiant System}}.
\bjtitle{\apj}
\bvolume{577},
\bfpage{923}--\blpage{928}
(\byear{2002}).
doi:\doiurl{10.1086/342244}
\end{barticle}
\endbibitem

\bibitem[\protect\citeauthoryear{{Corbet} et~al.}{2004}]{Corbet2004}
\begin{barticle}
\bauthor{\binits{R.H.D.} \bsnm{{Corbet}}},
\bauthor{\binits{D.C.} \bsnm{{Hannikainen}}},
\bauthor{\binits{R.} \bsnm{{Remillard}}},
\batitle{{The Orbital Period of IGR J19140+098}}.
\bjtitle{The Astronomer's Telegram}
\bvolume{269},
\bfpage{1}
(\byear{2004})
\end{barticle}
\endbibitem

\bibitem[\protect\citeauthoryear{{Corbet} et~al.}{2010}]{Corbet2010a}
\begin{barticle}
\bauthor{\binits{R.H.D.} \bsnm{{Corbet}}},
\bauthor{\binits{H.A.} \bsnm{{Krimm}}},
\bauthor{\binits{G.K.} \bsnm{{Skinner}}},
\batitle{{A 44 Day Period in AX J1700.2-4220 from Swift/BAT Observations}}.
\bjtitle{The Astronomer's Telegram}
\bvolume{2559},
\bfpage{1}
(\byear{2010})
\end{barticle}
\endbibitem

\bibitem[\protect\citeauthoryear{{Corbet} et~al.}{1999}]{Corbet1999}
\begin{barticle}
\bauthor{\binits{R.H.D.} \bsnm{{Corbet}}},
\bauthor{\binits{F.E.} \bsnm{{Marshall}}},
\bauthor{\binits{A.G.} \bsnm{{Peele}}},
\bauthor{\binits{T.} \bsnm{{Takeshima}}},
\batitle{{Rossi X-Ray Timing Explorer Observations of the X-Ray Pulsar XTE
  J1855-026: A Possible New Supergiant System}}.
\bjtitle{\apj}
\bvolume{517},
\bfpage{956}--\blpage{963}
(\byear{1999}).
doi:\doiurl{10.1086/307235}
\end{barticle}
\endbibitem

\bibitem[\protect\citeauthoryear{{Corbet} et~al.}{2010a}]{Corbet2010}
\begin{barticle}
\bauthor{\binits{R.H.D.} \bsnm{{Corbet}}},
\bauthor{\binits{S.D.} \bsnm{{Barthelmy}}},
\bauthor{\binits{W.H.} \bsnm{{Baumgartner}}},
\bauthor{\binits{H.A.} \bsnm{{Krimm}}},
\bauthor{\binits{C.B.} \bsnm{{Markwardt}}},
\bauthor{\binits{G.K.} \bsnm{{Skinner}}},
\bauthor{\binits{J.} \bsnm{{Tueller}}},
\batitle{{A 10 Day Period in IGR J16328-4726 from Swift/BAT Observations}}.
\bjtitle{The Astronomer's Telegram}
\bvolume{2588},
\bfpage{1}
(\byear{2010}a)
\end{barticle}
\endbibitem

\bibitem[\protect\citeauthoryear{{Corbet} et~al.}{2010b}]{Corbet2010b}
\begin{barticle}
\bauthor{\binits{R.H.D.} \bsnm{{Corbet}}},
\bauthor{\binits{S.D.} \bsnm{{Barthelmy}}},
\bauthor{\binits{W.H.} \bsnm{{Baumgartner}}},
\bauthor{\binits{H.A.} \bsnm{{Krimm}}},
\bauthor{\binits{C.B.} \bsnm{{Markwardt}}},
\bauthor{\binits{G.K.} \bsnm{{Skinner}}},
\bauthor{\binits{J.} \bsnm{{Tueller}}},
\batitle{{A 6.8 Day Period in IGR J16493-4348 from Swift/BAT and RXTE/PCA
  Observations}}.
\bjtitle{The Astronomer's Telegram}
\bvolume{2599},
\bfpage{1}
(\byear{2010}b)
\end{barticle}
\endbibitem

\bibitem[\protect\citeauthoryear{{Corbet} et~al.}{2005}]{Corbet2005}
\begin{barticle}
\bauthor{\binits{R.} \bsnm{{Corbet}}},
\bauthor{\binits{L.} \bsnm{{Barbier}}},
\bauthor{\binits{S.} \bsnm{{Barthelmy}}},
\bauthor{\binits{J.} \bsnm{{Cummings}}},
\bauthor{\binits{E.} \bsnm{{Fenimore}}},
\bauthor{\binits{N.} \bsnm{{Gehrels}}},
\bauthor{\binits{D.} \bsnm{{Hullinger}}},
\bauthor{\binits{H.} \bsnm{{Krimm}}},
\bauthor{\binits{C.} \bsnm{{Markwardt}}},
\bauthor{\binits{D.} \bsnm{{Palmer}}},
\bauthor{\binits{A.} \bsnm{{Parsons}}},
\bauthor{\binits{T.} \bsnm{{Sakamoto}}},
\bauthor{\binits{G.} \bsnm{{Sato}}},
\bauthor{\binits{J.} \bsnm{{Tueller}}},
\bauthor{\bsnm{{Swift-Survey Team}}},
\batitle{{Swift/BAT Discovery of the Orbital Period of IGR J16320-4751}}.
\bjtitle{The Astronomer's Telegram}
\bvolume{649},
\bfpage{1}
(\byear{2005})
\end{barticle}
\endbibitem

\bibitem[\protect\citeauthoryear{{Corbet} et~al.}{2006}]{Corbet2006}
\begin{barticle}
\bauthor{\binits{R.} \bsnm{{Corbet}}},
\bauthor{\binits{L.} \bsnm{{Barbier}}},
\bauthor{\binits{S.} \bsnm{{Barthelmy}}},
\bauthor{\binits{J.} \bsnm{{Cummings}}},
\bauthor{\binits{E.} \bsnm{{Fenimore}}},
\bauthor{\binits{N.} \bsnm{{Gehrels}}},
\bauthor{\binits{D.} \bsnm{{Hullinger}}},
\bauthor{\binits{H.} \bsnm{{Krimm}}},
\bauthor{\binits{C.} \bsnm{{Markwardt}}},
\bauthor{\binits{D.} \bsnm{{Palmer}}},
\bauthor{\binits{A.} \bsnm{{Parsons}}},
\bauthor{\binits{T.} \bsnm{{Sakamoto}}},
\bauthor{\binits{G.} \bsnm{{Sato}}},
\bauthor{\binits{J.} \bsnm{{Tueller}}},
\bauthor{\binits{R.} \bsnm{{Remillard}}},
\batitle{{Swift/BAT and RXTE/ASM Discovery of the Orbital Period of IGR
  J16418-4532}}.
\bjtitle{The Astronomer's Telegram}
\bvolume{779},
\bfpage{1}
(\byear{2006})
\end{barticle}
\endbibitem

\bibitem[\protect\citeauthoryear{{Cox} et~al.}{2005}]{Cox2005}
\begin{barticle}
\bauthor{\binits{N.L.J.} \bsnm{{Cox}}},
\bauthor{\binits{L.} \bsnm{{Kaper}}},
\bauthor{\binits{M.R.} \bsnm{{Mokiem}}},
\batitle{{VLT/UVES spectroscopy of the O supergiant companion to <ASTROBJ>4U
  1907+09</ASTROBJ>(7)}}.
\bjtitle{\aap}
\bvolume{436},
\bfpage{661}--\blpage{669}
(\byear{2005}).
doi:\doiurl{10.1051/0004-6361:20040511}
\end{barticle}
\endbibitem

\bibitem[\protect\citeauthoryear{{Crampton} et~al.}{1985}]{Crampton1985}
\begin{barticle}
\bauthor{\binits{D.} \bsnm{{Crampton}}},
\bauthor{\binits{J.B.} \bsnm{{Hutchings}}},
\bauthor{\binits{A.P.} \bsnm{{Cowley}}},
\batitle{{The supergiant X-ray binary system 2S 0114 + 650}}.
\bjtitle{\apj}
\bvolume{299},
\bfpage{839}--\blpage{844}
(\byear{1985}).
doi:\doiurl{10.1086/163750}
\end{barticle}
\endbibitem

\bibitem[\protect\citeauthoryear{{Cranmer} and {Owocki}}{1995}]{CO95}
\begin{barticle}
\bauthor{\binits{S.R.} \bsnm{{Cranmer}}},
\bauthor{\binits{S.P.} \bsnm{{Owocki}}},
\batitle{{The effect of oblateness and gravity darkening on the radiation
  driving in winds from rapidly rotating B stars}}.
\bjtitle{\apj}
\bvolume{440},
\bfpage{308}--\blpage{321}
(\byear{1995})
\end{barticle}
\endbibitem

\bibitem[\protect\citeauthoryear{{Cranmer} and {Owocki}}{1996}]{CO96}
\begin{barticle}
\bauthor{\binits{S.R.} \bsnm{{Cranmer}}},
\bauthor{\binits{S.P.} \bsnm{{Owocki}}},
\batitle{{Hydrodynamical Simulations of Corotating Interaction Regions and
  Discrete Absorption Components in Rotating O-Star Winds}}.
\bjtitle{\apj}
\bvolume{462},
\bfpage{469}
(\byear{1996})
\end{barticle}
\endbibitem

\bibitem[\protect\citeauthoryear{{Crowther} et~al.}{2006}]{Crowther06}
\begin{barticle}
\bauthor{\binits{P.A.} \bsnm{{Crowther}}},
\bauthor{\binits{D.J.} \bsnm{{Lennon}}},
\bauthor{\binits{N.R.} \bsnm{{Walborn}}},
\batitle{{Physical parameters and wind properties of galactic early B
  supergiants}}.
\bjtitle{\aap}
\bvolume{446},
\bfpage{279}--\blpage{293}
(\byear{2006})
\end{barticle}
\endbibitem

\bibitem[\protect\citeauthoryear{{Datlowe} et~al.}{1974}]{Datlowe1974}
\begin{barticle}
\bauthor{\binits{D.W.} \bsnm{{Datlowe}}},
\bauthor{\binits{M.J.} \bsnm{{Elcan}}},
\bauthor{\binits{H.S.} \bsnm{{Hudson}}},
\batitle{{OSO-7 observations of solar X-rays in the energy range 10-100 keV}}.
\bjtitle{\solphys}
\bvolume{39},
\bfpage{155}--\blpage{174}
(\byear{1974}).
doi:\doiurl{10.1007/BF00154978}
\end{barticle}
\endbibitem

\bibitem[\protect\citeauthoryear{Davidson and
  Ostriker}{1973}]{DavidsonOstriker73}
\begin{barticle}
\bauthor{\binits{K.} \bsnm{Davidson}},
\bauthor{\binits{J.P.} \bsnm{Ostriker}},
\batitle{Neutron-star accretion in a stellar wind: Model for a pulsed x-ray
  source}.
\bjtitle{ApJ}
\bvolume{179},
\bfpage{585}--\blpage{598}
(\byear{1973})
\end{barticle}
\endbibitem

\bibitem[\protect\citeauthoryear{{Davies} and {Pringle}}{1981}]{davies1981}
\begin{barticle}
\bauthor{\binits{R.E.} \bsnm{{Davies}}},
\bauthor{\binits{J.E.} \bsnm{{Pringle}}},
\batitle{{Spindown of neutron stars in close binary systems. II}}.
\bjtitle{\mnras}
\bvolume{196},
\bfpage{209}--\blpage{224}
(\byear{1981})
\end{barticle}
\endbibitem

\bibitem[\protect\citeauthoryear{{Davies} et~al.}{1979}]{davies79}
\begin{barticle}
\bauthor{\binits{R.E.} \bsnm{{Davies}}},
\bauthor{\binits{A.C.} \bsnm{{Fabian}}},
\bauthor{\binits{J.E.} \bsnm{{Pringle}}},
\batitle{{Spindown of neutron stars in close binary systems}}.
\bjtitle{\mnras}
\bvolume{186},
\bfpage{779}--\blpage{782}
(\byear{1979})
\end{barticle}
\endbibitem

\bibitem[\protect\citeauthoryear{{Densham} and {Charles}}{1982}]{Densham1982}
\begin{barticle}
\bauthor{\binits{R.H.} \bsnm{{Densham}}},
\bauthor{\binits{P.A.} \bsnm{{Charles}}},
\batitle{{Optical photometry and spectroscopy of the X-ray pulsar 1E
  1145.1-6141}}.
\bjtitle{\mnras}
\bvolume{201},
\bfpage{171}--\blpage{178}
(\byear{1982})
\end{barticle}
\endbibitem

\bibitem[\protect\citeauthoryear{{Dessart} and {Owocki}}{2003}]{Dessart03}
\begin{barticle}
\bauthor{\binits{L.} \bsnm{{Dessart}}},
\bauthor{\binits{S.P.} \bsnm{{Owocki}}},
\batitle{{Two-dimensional simulations of the line-driven instability in
  hot-star winds}}.
\bjtitle{\aap}
\bvolume{406},
\bfpage{1}--\blpage{4}
(\byear{2003}).
doi:\doiurl{10.1051/0004-6361:20030810}
\end{barticle}
\endbibitem

\bibitem[\protect\citeauthoryear{{Dessart} and {Owocki}}{2005}]{Dessart05}
\begin{barticle}
\bauthor{\binits{L.} \bsnm{{Dessart}}},
\bauthor{\binits{S.P.} \bsnm{{Owocki}}},
\batitle{{2D simulations of the line-driven instability in hot-star winds. II.
  Approximations for the 2D radiation force}}.
\bjtitle{\aap}
\bvolume{437},
\bfpage{657}--\blpage{666}
(\byear{2005}).
doi:\doiurl{10.1051/0004-6361:20052778}
\end{barticle}
\endbibitem

\bibitem[\protect\citeauthoryear{Dickey and Lockman}{1990}]{dickey1990a}
\begin{barticle}
\bauthor{\binits{J.M.} \bsnm{Dickey}},
\bauthor{\binits{F.J.} \bsnm{Lockman}},
\batitle{{H I} in the galaxy}.
\bjtitle{ARA\&A}
\bvolume{28},
\bfpage{215}
(\byear{1990})
\end{barticle}
\endbibitem

\bibitem[\protect\citeauthoryear{{Domiciano de Souza} et~al.}{2003}]{deSouza03}
\begin{barticle}
\bauthor{\binits{A.} \bsnm{{Domiciano de Souza}}},
\bauthor{\binits{P.} \bsnm{{Kervella}}},
\bauthor{\binits{S.} \bsnm{{Jankov}}},
\bauthor{\binits{L.} \bsnm{{Abe}}},
\bauthor{\binits{F.} \bsnm{{Vakili}}},
\bauthor{\binits{E.} \bsnm{{di Folco}}},
\bauthor{\binits{F.} \bsnm{{Paresce}}},
\batitle{{The spinning-top Be star Achernar from VLTI-VINCI}}.
\bjtitle{\aap}
\bvolume{407},
\bfpage{47}--\blpage{50}
(\byear{2003})
\end{barticle}
\endbibitem

\bibitem[\protect\citeauthoryear{{Doroshenko} et~al.}{2011}]{Doroshenko:2011}
\begin{barticle}
\bauthor{\binits{V.} \bsnm{{Doroshenko}}},
\bauthor{\binits{A.} \bsnm{{Santangelo}}},
\bauthor{\binits{V.} \bsnm{{Suleimanov}}},
\batitle{{Witnessing the magnetospheric boundary at work in Vela X-1}}.
\bjtitle{\aap}
\bvolume{529},
\bfpage{52}
(\byear{2011}).
doi:\doiurl{10.1051/0004-6361/201116482}
\end{barticle}
\endbibitem

\bibitem[\protect\citeauthoryear{{Doroshenko} et~al.}{2012}]{Doroshenko:2012}
\begin{barticle}
\bauthor{\binits{V.} \bsnm{{Doroshenko}}},
\bauthor{\binits{A.} \bsnm{{Santangelo}}},
\bauthor{\binits{L.} \bsnm{{Ducci}}},
\bauthor{\binits{D.} \bsnm{{Klochkov}}},
\batitle{{Supergiant, fast, but not so transient 4U 1907+09}}.
\bjtitle{\aap}
\bvolume{548},
\bfpage{19}
(\byear{2012}).
doi:\doiurl{10.1051/0004-6361/201220085}
\end{barticle}
\endbibitem

\bibitem[\protect\citeauthoryear{{Drave} et~al.}{2010}]{Drave2010}
\begin{barticle}
\bauthor{\binits{S.P.} \bsnm{{Drave}}},
\bauthor{\binits{D.J.} \bsnm{{Clark}}},
\bauthor{\binits{A.J.} \bsnm{{Bird}}},
\bauthor{\binits{V.A.} \bsnm{{McBride}}},
\bauthor{\binits{A.B.} \bsnm{{Hill}}},
\bauthor{\binits{V.} \bsnm{{Sguera}}},
\bauthor{\binits{S.} \bsnm{{Scaringi}}},
\bauthor{\binits{A.} \bsnm{{Bazzano}}},
\batitle{{Discovery of the 51.47-d orbital period in the supergiant fast X-ray
  transient XTE J1739-302 with INTEGRAL}}.
\bjtitle{\mnras}
\bvolume{409},
\bfpage{1220}--\blpage{1226}
(\byear{2010}).
doi:\doiurl{10.1111/j.1365-2966.2010.17383.x}
\end{barticle}
\endbibitem

\bibitem[\protect\citeauthoryear{{Drave} et~al.}{2012}]{Drave2012}
\begin{barticle}
\bauthor{\binits{S.P.} \bsnm{{Drave}}},
\bauthor{\binits{A.J.} \bsnm{{Bird}}},
\bauthor{\binits{L.J.} \bsnm{{Townsend}}},
\bauthor{\binits{A.B.} \bsnm{{Hill}}},
\bauthor{\binits{V.A.} \bsnm{{McBride}}},
\bauthor{\binits{V.} \bsnm{{Sguera}}},
\bauthor{\binits{A.} \bsnm{{Bazzano}}},
\bauthor{\binits{D.J.} \bsnm{{Clark}}},
\batitle{{X-ray pulsations from the region of the supergiant fast X-ray
  transient IGR J17544-2619}}.
\bjtitle{\aap}
\bvolume{539},
\bfpage{21}
(\byear{2012}).
doi:\doiurl{10.1051/0004-6361/201117947}
\end{barticle}
\endbibitem

\bibitem[\protect\citeauthoryear{{Drave} et~al.}{2013a}]{Drave2013Atel}
\begin{barticle}
\bauthor{\binits{S.P.} \bsnm{{Drave}}},
\bauthor{\binits{A.J.} \bsnm{{Bird}}},
\bauthor{\binits{M.E.} \bsnm{{Goossens}}},
\bauthor{\binits{L.} \bsnm{{Sidoli}}},
\bauthor{\binits{V.} \bsnm{{Sguera}}},
\bauthor{\binits{M.} \bsnm{{Fiocchi}}},
\bauthor{\binits{A.} \bsnm{{Bazzano}}},
\batitle{{Confirmation of the superorbital modulation of the high mass X-ray
  binaries 4U 1909+07, IGR J16479-4514 and IGR J16418-4532 with
  INTEGRAL/IBIS}}.
\bjtitle{The Astronomer's Telegram}
\bvolume{5131},
\bfpage{1}
(\byear{2013}a)
\end{barticle}
\endbibitem

\bibitem[\protect\citeauthoryear{{Drave} et~al.}{2013b}]{Drave2013}
\begin{barticle}
\bauthor{\binits{S.P.} \bsnm{{Drave}}},
\bauthor{\binits{A.J.} \bsnm{{Bird}}},
\bauthor{\binits{L.} \bsnm{{Sidoli}}},
\bauthor{\binits{V.} \bsnm{{Sguera}}},
\bauthor{\binits{V.A.} \bsnm{{McBride}}},
\bauthor{\binits{A.B.} \bsnm{{Hill}}},
\bauthor{\binits{A.} \bsnm{{Bazzano}}},
\bauthor{\binits{M.E.} \bsnm{{Goossens}}},
\batitle{{INTEGRAL and XMM-Newton observations of IGR J16418-4532: evidence of
  accretion regime transitions in a supergiant fast X-ray transient}}.
\bjtitle{\mnras}
\bvolume{433},
\bfpage{528}--\blpage{542}
(\byear{2013}b).
doi:\doiurl{10.1093/mnras/stt754}
\end{barticle}
\endbibitem

\bibitem[\protect\citeauthoryear{{Drave} et~al.}{2014}]{Drave2014}
\begin{barticle}
\bauthor{\binits{S.P.} \bsnm{{Drave}}},
\bauthor{\binits{A.J.} \bsnm{{Bird}}},
\bauthor{\binits{L.} \bsnm{{Sidoli}}},
\bauthor{\binits{V.} \bsnm{{Sguera}}},
\bauthor{\binits{A.} \bsnm{{Bazzano}}},
\bauthor{\binits{A.B.} \bsnm{{Hill}}},
\bauthor{\binits{M.E.} \bsnm{{Goossens}}},
\batitle{{New insights on accretion in supergiant fast X-ray transients from
  XMM-Newton and INTEGRAL observations of IGR J17544-2619}}.
\bjtitle{\mnras}
\bvolume{439},
\bfpage{2175}--\blpage{2185}
(\byear{2014}).
doi:\doiurl{10.1093/mnras/stu110}
\end{barticle}
\endbibitem

\bibitem[\protect\citeauthoryear{{Ducci} et~al.}{2010}]{Ducci2010}
\begin{barticle}
\bauthor{\binits{L.} \bsnm{{Ducci}}},
\bauthor{\binits{L.} \bsnm{{Sidoli}}},
\bauthor{\binits{A.} \bsnm{{Paizis}}},
\batitle{{INTEGRAL results on supergiant fast X-ray transients and accretion
  mechanism interpretation: ionization effect and formation of transient
  accretion discs}}.
\bjtitle{\mnras}
\bvolume{408},
\bfpage{1540}--\blpage{1550}
(\byear{2010}).
doi:\doiurl{10.1111/j.1365-2966.2010.17216.x}
\end{barticle}
\endbibitem

\bibitem[\protect\citeauthoryear{{Ducci} et~al.}{2009}]{Ducci2009}
\begin{barticle}
\bauthor{\binits{L.} \bsnm{{Ducci}}},
\bauthor{\binits{L.} \bsnm{{Sidoli}}},
\bauthor{\binits{S.} \bsnm{{Mereghetti}}},
\bauthor{\binits{A.} \bsnm{{Paizis}}},
\bauthor{\binits{P.} \bsnm{{Romano}}},
\batitle{{The structure of blue supergiant winds and the accretion in
  supergiant high-mass X-ray binaries}}.
\bjtitle{\mnras}
\bvolume{398},
\bfpage{2152}--\blpage{2165}
(\byear{2009}).
doi:\doiurl{10.1111/j.1365-2966.2009.15265.x}
\end{barticle}
\endbibitem

\bibitem[\protect\citeauthoryear{{Ducci} et~al.}{2013}]{Ducci2013}
\begin{barticle}
\bauthor{\binits{L.} \bsnm{{Ducci}}},
\bauthor{\binits{V.} \bsnm{{Doroshenko}}},
\bauthor{\binits{M.} \bsnm{{Sasaki}}},
\bauthor{\binits{A.} \bsnm{{Santangelo}}},
\bauthor{\binits{P.} \bsnm{{Esposito}}},
\bauthor{\binits{P.} \bsnm{{Romano}}},
\bauthor{\binits{S.} \bsnm{{Vercellone}}},
\batitle{{Spectral and temporal properties of the supergiant fast X-ray
  transient IGR J18483-0311 observed by INTEGRAL}}.
\bjtitle{\aap}
\bvolume{559},
\bfpage{135}
(\byear{2013}).
doi:\doiurl{10.1051/0004-6361/201322299}
\end{barticle}
\endbibitem

\bibitem[\protect\citeauthoryear{{Dufton} et~al.}{2011}]{Dufton11}
\begin{barticle}
\bauthor{\binits{P.L.} \bsnm{{Dufton}}},
\bauthor{\binits{P.R.} \bsnm{{Dunstall}}},
\bauthor{\binits{C.J.} \bsnm{{Evans}}},
\bauthor{\binits{I.} \bsnm{{Brott}}},
\bauthor{\binits{M.} \bsnm{{Cantiello}}},
\bauthor{\binits{A.} \bsnm{{de Koter}}},
\bauthor{\binits{S.E.} \bsnm{{de Mink}}},
\bauthor{\binits{M.} \bsnm{{Fraser}}},
\bauthor{\binits{V.} \bsnm{{H{\'e}nault-Brunet}}},
\bauthor{\binits{I.D.} \bsnm{{Howarth}}},
\bauthor{\binits{N.} \bsnm{{Langer}}},
\bauthor{\binits{D.J.} \bsnm{{Lennon}}},
\bauthor{\binits{N.} \bsnm{{Markova}}},
\bauthor{\binits{H.} \bsnm{{Sana}}},
\bauthor{\binits{W.D.} \bsnm{{Taylor}}},
\batitle{{The VLT-FLAMES Tarantula Survey: The Fastest Rotating O-type Star and
  Shortest Period LMC Pulsar -- Remnants of a Supernova Disrupted Binary?}}
\bjtitle{\apjl}
\bvolume{743},
\bfpage{22}
(\byear{2011}).
doi:\doiurl{10.1088/2041-8205/743/1/L22}
\end{barticle}
\endbibitem

\bibitem[\protect\citeauthoryear{Edgar}{2004}]{Edgar2004}
\begin{barticle}
\bauthor{\binits{R.} \bsnm{Edgar}},
\batitle{A review of bondi--hoyle--lyttleton accretion}.
\bjtitle{New Astronomy Reviews}
\bvolume{48}(\bissue{10}),
\bfpage{843}--\blpage{859}
(\byear{2004}).
doi:\doiurl{http://dx.doi.org/10.1016/j.newar.2004.06.001}
\end{barticle}
\endbibitem

\bibitem[\protect\citeauthoryear{{Eikmann} et~al.}{2012}]{eikmann2012a}
\begin{bchapter}
\bauthor{\binits{W.} \bsnm{{Eikmann}}},
\bauthor{\binits{J.} \bsnm{{Wilms}}},
\bauthor{\binits{J.} \bsnm{{Lee}}},
\bctitle{Monte Carlo simulations of X-ray absorption in the interstellar
  medium},
in \bbtitle{An INTEGRAL view of the high-energy sky (the first 10 years)}.
\bsertitle{PoS},
\byear{2012}
\end{bchapter}
\endbibitem

\bibitem[\protect\citeauthoryear{{Eikmann} et~al.}{2014}]{eikmann2014a}
\begin{barticle}
\bauthor{\binits{W.} \bsnm{{Eikmann}}},
\bauthor{\binits{J.} \bsnm{{Wilms}}},
\bauthor{\binits{R.K.} \bsnm{{Smith}}},
\bauthor{\binits{J.C.} \bsnm{{Lee}}},
\batitle{X-rray transmission and reflection through a compton-thick medium via
  monte-carlo simulations}.
\bjtitle{Ac. Polytechnica}
\bvolume{54},
\bfpage{177}--\blpage{182}
(\byear{2014})
\end{barticle}
\endbibitem

\bibitem[\protect\citeauthoryear{{Elsner} and {Lamb}}{1977}]{elsner1977}
\begin{barticle}
\bauthor{\binits{R.F.} \bsnm{{Elsner}}},
\bauthor{\binits{F.K.} \bsnm{{Lamb}}},
\batitle{{Accretion by magnetic neutron stars. I - Magnetospheric structure and
  stability}}.
\bjtitle{\apj}
\bvolume{215},
\bfpage{897}--\blpage{913}
(\byear{1977}).
doi:\doiurl{10.1086/155427}
\end{barticle}
\endbibitem

\bibitem[\protect\citeauthoryear{{Endo} et~al.}{2002}]{Endo2002}
\begin{barticle}
\bauthor{\binits{T.} \bsnm{{Endo}}},
\bauthor{\binits{M.} \bsnm{{Ishida}}},
\bauthor{\binits{K.} \bsnm{{Masai}}},
\bauthor{\binits{H.} \bsnm{{Kunieda}}},
\bauthor{\binits{H.} \bsnm{{Inoue}}},
\bauthor{\binits{F.} \bsnm{{Nagase}}},
\batitle{{Broadening of Nearly Neutral Iron Emission Line of GX 301-2 Observed
  with ASCA}}.
\bjtitle{ApJ}
\bvolume{574},
\bfpage{879}--\blpage{898}
(\byear{2002})
\end{barticle}
\endbibitem

\bibitem[\protect\citeauthoryear{{Eversberg} et~al.}{1998}]{ev1998}
\begin{barticle}
\bauthor{\binits{T.} \bsnm{{Eversberg}}},
\bauthor{\binits{S.} \bsnm{{L{\'e}pine}}},
\bauthor{\binits{A.F.J.} \bsnm{{Moffat}}},
\batitle{{Outmoving Clumps in the Wind of the Hot O Supergiant {$\zeta$}
  Puppis}}.
\bjtitle{\apj}
\bvolume{494},
\bfpage{799}--\blpage{805}
(\byear{1998}).
doi:\doiurl{10.1086/305218}
\end{barticle}
\endbibitem

\bibitem[\protect\citeauthoryear{{Falanga} et~al.}{2015}]{Falanga:2015X}
\begin{botherref}
\oauthor{\binits{M.} \bsnm{{Falanga}}},
\oauthor{\binits{E.} \bsnm{{Bozzo}}},
\oauthor{\binits{A.} \bsnm{{Lutovinov}}},
\oauthor{\binits{J.M.} \bsnm{{Bonnet-Bidaud}}},
\oauthor{\binits{Y.} \bsnm{{Fetisova}}},
\oauthor{\binits{J.} \bsnm{{Puls}}},
{The ephemeris, orbital decay, and masses of 10 eclipsing HMXBs}.
ArXiv e-prints
(2015)
\end{botherref}
\endbibitem

\bibitem[\protect\citeauthoryear{{Farinelli} et~al.}{2012a}]{Farinelli2012b}
\begin{barticle}
\bauthor{\binits{R.} \bsnm{{Farinelli}}},
\bauthor{\binits{C.} \bsnm{{Ceccobello}}},
\bauthor{\binits{P.} \bsnm{{Romano}}},
\bauthor{\binits{L.} \bsnm{{Titarchuk}}},
\batitle{{Numerical solution of the radiative transfer equation: X-ray spectral
  formation from cylindrical accretion onto a magnetized neutron star}}.
\bjtitle{\aap}
\bvolume{538},
\bfpage{67}
(\byear{2012}a).
doi:\doiurl{10.1051/0004-6361/201118008}
\end{barticle}
\endbibitem

\bibitem[\protect\citeauthoryear{{Farinelli} et~al.}{2012b}]{Farinelli2012}
\begin{barticle}
\bauthor{\binits{R.} \bsnm{{Farinelli}}},
\bauthor{\binits{P.} \bsnm{{Romano}}},
\bauthor{\binits{V.} \bsnm{{Mangano}}},
\bauthor{\binits{C.} \bsnm{{Ceccobello}}},
\bauthor{\binits{L.} \bsnm{{Ducci}}},
\bauthor{\binits{S.} \bsnm{{Vercellone}}},
\bauthor{\binits{P.} \bsnm{{Esposito}}},
\bauthor{\binits{J.A.} \bsnm{{Kennea}}},
\bauthor{\binits{D.N.} \bsnm{{Burrows}}},
\batitle{{Swift observations of two supergiant fast X-ray transient prototypes
  in outburst}}.
\bjtitle{\mnras}
\bvolume{424},
\bfpage{2854}--\blpage{2863}
(\byear{2012}b).
doi:\doiurl{10.1111/j.1365-2966.2012.21422.x}
\end{barticle}
\endbibitem

\bibitem[\protect\citeauthoryear{{Farrell} et~al.}{2006}]{Farrell2006}
\begin{barticle}
\bauthor{\binits{S.A.} \bsnm{{Farrell}}},
\bauthor{\binits{R.K.} \bsnm{{Sood}}},
\bauthor{\binits{P.M.} \bsnm{{O'Neill}}},
\batitle{{Super-orbital period in the high-mass X-ray binary 2S 0114+650}}.
\bjtitle{\mnras}
\bvolume{367},
\bfpage{1457}--\blpage{1462}
(\byear{2006}).
doi:\doiurl{10.1111/j.1365-2966.2006.10150.x}
\end{barticle}
\endbibitem

\bibitem[\protect\citeauthoryear{{Farrell} et~al.}{2008}]{Farrell2008}
\begin{barticle}
\bauthor{\binits{S.A.} \bsnm{{Farrell}}},
\bauthor{\binits{R.K.} \bsnm{{Sood}}},
\bauthor{\binits{P.M.} \bsnm{{O'Neill}}},
\bauthor{\binits{S.} \bsnm{{Dieters}}},
\batitle{{A detailed study of 2S 0114+650 with the Rossi X-ray Timing
  Explorer}}.
\bjtitle{\mnras}
\bvolume{389},
\bfpage{608}--\blpage{628}
(\byear{2008}).
doi:\doiurl{10.1111/j.1365-2966.2008.13588.x}
\end{barticle}
\endbibitem

\bibitem[\protect\citeauthoryear{{Feldman}}{1992}]{feldman1992a}
\begin{barticle}
\bauthor{\binits{U.} \bsnm{{Feldman}}},
\batitle{Elemental abundances in the upper solar atmosphere.}
\bjtitle{Physica Scripta}
\bvolume{46},
\bfpage{202}--\blpage{220}
(\byear{1992})
\end{barticle}
\endbibitem

\bibitem[\protect\citeauthoryear{{Feldmeier}}{1995}]{Feldmeier95}
\begin{barticle}
\bauthor{\binits{A.} \bsnm{{Feldmeier}}},
\batitle{{Time-dependent structure and energy transfer in hot star winds.}}
\bjtitle{\aap}
\bvolume{299},
\bfpage{523}
(\byear{1995})
\end{barticle}
\endbibitem

\bibitem[\protect\citeauthoryear{{Feldmeier} and
  {Shlosman}}{2000}]{FeldmeierShlosman00}
\begin{barticle}
\bauthor{\binits{A.} \bsnm{{Feldmeier}}},
\bauthor{\binits{I.} \bsnm{{Shlosman}}},
\batitle{{Runaway of Line-driven Winds toward Critical and Overloaded
  Solutions}}.
\bjtitle{\apjl}
\bvolume{532},
\bfpage{125}--\blpage{128}
(\byear{2000})
\end{barticle}
\endbibitem

\bibitem[\protect\citeauthoryear{{Feldmeier} and
  {Shlosman}}{2002}]{FeldmeierShlosman02}
\begin{barticle}
\bauthor{\binits{A.} \bsnm{{Feldmeier}}},
\bauthor{\binits{I.} \bsnm{{Shlosman}}},
\batitle{{Abbott Wave-triggered Runaway in Line-driven Winds from Stars and
  Accretion Disks}}.
\bjtitle{\apj}
\bvolume{564},
\bfpage{385}--\blpage{394}
(\byear{2002})
\end{barticle}
\endbibitem

\bibitem[\protect\citeauthoryear{{Feldmeier} et~al.}{2003}]{feld2003}
\begin{barticle}
\bauthor{\binits{A.} \bsnm{{Feldmeier}}},
\bauthor{\binits{L.} \bsnm{{Oskinova}}},
\bauthor{\binits{W.-R.} \bsnm{{Hamann}}},
\batitle{{X-ray line emission from a fragmented stellar wind}}.
\bjtitle{\aap}
\bvolume{403},
\bfpage{217}--\blpage{224}
(\byear{2003}).
doi:\doiurl{10.1051/0004-6361:20030231}
\end{barticle}
\endbibitem

\bibitem[\protect\citeauthoryear{{Feldmeier} et~al.}{1997a}]{Feldmeier97b}
\begin{barticle}
\bauthor{\binits{A.} \bsnm{{Feldmeier}}},
\bauthor{\binits{J.} \bsnm{{Puls}}},
\bauthor{\binits{A.W.A.} \bsnm{{Pauldrach}}},
\batitle{{A possible origin for X-rays from O stars.}}
\bjtitle{\aap}
\bvolume{322},
\bfpage{878}--\blpage{895}
(\byear{1997}a)
\end{barticle}
\endbibitem

\bibitem[\protect\citeauthoryear{{Feldmeier} et~al.}{1997b}]{Feldmeier97a}
\begin{barticle}
\bauthor{\binits{A.} \bsnm{{Feldmeier}}},
\bauthor{\binits{R.} \bsnm{{Kudritzki}}},
\bauthor{\binits{R.} \bsnm{{Palsa}}},
\bauthor{\binits{A.W.A.} \bsnm{{Pauldrach}}},
\bauthor{\binits{J.} \bsnm{{Puls}}},
\batitle{{The X-ray emission from shock cooling zones in O star winds.}}
\bjtitle{\aap}
\bvolume{320},
\bfpage{899}--\blpage{912}
(\byear{1997}b)
\end{barticle}
\endbibitem

\bibitem[\protect\citeauthoryear{{Filliatre} and {Chaty}}{2004}]{Filliatre2004}
\begin{barticle}
\bauthor{\binits{P.} \bsnm{{Filliatre}}},
\bauthor{\binits{S.} \bsnm{{Chaty}}},
\batitle{{The Optical/Near-Infrared Counterpart of the INTEGRAL Obscured Source
  IGR J16318-4848: An sgB[e] in a High-Mass X-Ray Binary?}}
\bjtitle{\apj}
\bvolume{616},
\bfpage{469}--\blpage{484}
(\byear{2004}).
doi:\doiurl{10.1086/424869}
\end{barticle}
\endbibitem

\bibitem[\protect\citeauthoryear{{Fiocchi} et~al.}{2010}]{Fiocchi2010}
\begin{barticle}
\bauthor{\binits{M.} \bsnm{{Fiocchi}}},
\bauthor{\binits{V.} \bsnm{{Sguera}}},
\bauthor{\binits{A.} \bsnm{{Bazzano}}},
\bauthor{\binits{L.} \bsnm{{Bassani}}},
\bauthor{\binits{A.J.} \bsnm{{Bird}}},
\bauthor{\binits{L.} \bsnm{{Natalucci}}},
\bauthor{\binits{P.} \bsnm{{Ubertini}}},
\batitle{{IGR J16328-4726: A New Candidate Supergiant Fast X-ray Transient}}.
\bjtitle{\apjl}
\bvolume{725},
\bfpage{68}--\blpage{72}
(\byear{2010}).
doi:\doiurl{10.1088/2041-8205/725/1/L68}
\end{barticle}
\endbibitem

\bibitem[\protect\citeauthoryear{{Fiocchi} et~al.}{2013}]{Fiocchi2013}
\begin{barticle}
\bauthor{\binits{M.} \bsnm{{Fiocchi}}},
\bauthor{\binits{A.} \bsnm{{Bazzano}}},
\bauthor{\binits{A.J.} \bsnm{{Bird}}},
\bauthor{\binits{S.P.} \bsnm{{Drave}}},
\bauthor{\binits{L.} \bsnm{{Natalucci}}},
\bauthor{\binits{P.} \bsnm{{Persi}}},
\bauthor{\binits{L.} \bsnm{{Piro}}},
\bauthor{\binits{P.} \bsnm{{Ubertini}}},
\batitle{{The INTEGRAL Source IGR J16328-4726: A High-mass X-Ray Binary from
  the BeppoSAX Era}}.
\bjtitle{\apj}
\bvolume{762},
\bfpage{19}
(\byear{2013}).
doi:\doiurl{10.1088/0004-637X/762/1/19}
\end{barticle}
\endbibitem

\bibitem[\protect\citeauthoryear{{Frank} et~al.}{2002}]{frank2002}
\begin{bbook}
\bauthor{\binits{J.} \bsnm{{Frank}}},
\bauthor{\binits{A.} \bsnm{{King}}},
\bauthor{\binits{D.J.} \bsnm{{Raine}}},
\bbtitle{{Accretion Power in Astrophysics: Third Edition}}
\byear{2002}
\end{bbook}
\endbibitem

\bibitem[\protect\citeauthoryear{{Fransson} and {Fabian}}{1980}]{Fransson1980}
\begin{barticle}
\bauthor{\binits{C.} \bsnm{{Fransson}}},
\bauthor{\binits{A.C.} \bsnm{{Fabian}}},
\batitle{{X-ray induced shocks in stellar winds}}.
\bjtitle{\aap}
\bvolume{87},
\bfpage{102}--\blpage{108}
(\byear{1980})
\end{barticle}
\endbibitem

\bibitem[\protect\citeauthoryear{{Friend} and {Abbott}}{1986}]{FA86}
\begin{barticle}
\bauthor{\binits{D.B.} \bsnm{{Friend}}},
\bauthor{\binits{D.C.} \bsnm{{Abbott}}},
\batitle{{The theory of radiatively driven stellar winds. III - Wind models
  with finite disk correction and rotation}}.
\bjtitle{\apj}
\bvolume{311},
\bfpage{701}--\blpage{707}
(\byear{1986})
\end{barticle}
\endbibitem

\bibitem[\protect\citeauthoryear{{Friend} and {Castor}}{1982}]{Friend1982}
\begin{barticle}
\bauthor{\binits{D.B.} \bsnm{{Friend}}},
\bauthor{\binits{J.I.} \bsnm{{Castor}}},
\batitle{{Radiation-driven winds in X-ray binaries}}.
\bjtitle{\apj}
\bvolume{261},
\bfpage{293}--\blpage{300}
(\byear{1982}).
doi:\doiurl{10.1086/160340}
\end{barticle}
\endbibitem

\bibitem[\protect\citeauthoryear{{Friend} and {Castor}}{1983}]{FC83}
\begin{barticle}
\bauthor{\binits{D.B.} \bsnm{{Friend}}},
\bauthor{\binits{J.I.} \bsnm{{Castor}}},
\batitle{{Stellar winds driven by multiline scattering}}.
\bjtitle{\apj}
\bvolume{272},
\bfpage{259}--\blpage{272}
(\byear{1983})
\end{barticle}
\endbibitem

\bibitem[\protect\citeauthoryear{{Fritz} et~al.}{2006}]{Fritz2006}
\begin{barticle}
\bauthor{\binits{S.} \bsnm{{Fritz}}},
\bauthor{\binits{I.} \bsnm{{Kreykenbohm}}},
\bauthor{\binits{J.} \bsnm{{Wilms}}},
\bauthor{\binits{R.} \bsnm{{Staubert}}},
\bauthor{\binits{F.} \bsnm{{Bayazit}}},
\bauthor{\binits{K.} \bsnm{{Pottschmidt}}},
\bauthor{\binits{J.} \bsnm{{Rodriguez}}},
\bauthor{\binits{A.} \bsnm{{Santangelo}}},
\batitle{{A torque reversal of 4U 1907+09}}.
\bjtitle{\aap}
\bvolume{458},
\bfpage{885}--\blpage{893}
(\byear{2006}).
doi:\doiurl{10.1051/0004-6361:20065557}
\end{barticle}
\endbibitem

\bibitem[\protect\citeauthoryear{{Fullerton} and {Owocki}}{1992}]{FO92}
\begin{bchapter}
\bauthor{\binits{A.W.} \bsnm{{Fullerton}}},
\bauthor{\binits{S.P.} \bsnm{{Owocki}}},
\bctitle{{Can Nonstationary Velocity Plateaus Account for Slowly Moving
  Discrete Absorption Components? (Contributed Poster)}},
in \bbtitle{Nonisotropic and Variable Outflows from Stars},
ed. by \beditor{\binits{L.} \bsnm{{Drissen}}},
\beditor{\binits{C.} \bsnm{{Leitherer}}},
\beditor{\binits{A.} \bsnm{{Nota}}}
\bsertitle{Astronomical Society of the Pacific Conference Series},
vol. \bseriesno{22},
\byear{1992},
p. \bfpage{177}
\end{bchapter}
\endbibitem

\bibitem[\protect\citeauthoryear{{Fullerton} et~al.}{2006}]{Fullerton2006}
\begin{barticle}
\bauthor{\binits{A.W.} \bsnm{{Fullerton}}},
\bauthor{\binits{D.L.} \bsnm{{Massa}}},
\bauthor{\binits{R.K.} \bsnm{{Prinja}}},
\batitle{{The Discordance of Mass-Loss Estimates for Galactic O-Type Stars}}.
\bjtitle{\apj}
\bvolume{637},
\bfpage{1025}--\blpage{1039}
(\byear{2006}).
doi:\doiurl{10.1086/498560}
\end{barticle}
\endbibitem

\bibitem[\protect\citeauthoryear{{Fullerton} et~al.}{1997}]{Fullerton97}
\begin{barticle}
\bauthor{\binits{A.W.} \bsnm{{Fullerton}}},
\bauthor{\binits{D.L.} \bsnm{{Massa}}},
\bauthor{\binits{R.K.} \bsnm{{Prinja}}},
\bauthor{\binits{S.P.} \bsnm{{Owocki}}},
\bauthor{\binits{S.R.} \bsnm{{Cranmer}}},
\batitle{{Wind variability of B supergiants. III. Corotating spiral structures
  in the stellar wind of HD 64760.}}
\bjtitle{\aap}
\bvolume{327},
\bfpage{699}--\blpage{720}
(\byear{1997})
\end{barticle}
\endbibitem

\bibitem[\protect\citeauthoryear{{F{\"u}rst} et~al.}{2010}]{Furst2010}
\begin{barticle}
\bauthor{\binits{F.} \bsnm{{F{\"u}rst}}},
\bauthor{\binits{I.} \bsnm{{Kreykenbohm}}},
\bauthor{\binits{K.} \bsnm{{Pottschmidt}}},
\bauthor{\binits{J.} \bsnm{{Wilms}}},
\bauthor{\binits{M.} \bsnm{{Hanke}}},
\bauthor{\binits{R.E.} \bsnm{{Rothschild}}},
\bauthor{\binits{P.} \bsnm{{Kretschmar}}},
\bauthor{\binits{N.S.} \bsnm{{Schulz}}},
\bauthor{\binits{D.P.} \bsnm{{Huenemoerder}}},
\bauthor{\binits{D.} \bsnm{{Klochkov}}},
\bauthor{\binits{R.} \bsnm{{Staubert}}},
\batitle{{X-ray variation statistics and wind clumping in Vela X-1}}.
\bjtitle{\aap}
\bvolume{519},
\bfpage{37}
(\byear{2010}).
doi:\doiurl{10.1051/0004-6361/200913981}
\end{barticle}
\endbibitem

\bibitem[\protect\citeauthoryear{{F{\"u}rst} et~al.}{2011}]{Fuerst2011}
\begin{barticle}
\bauthor{\binits{F.} \bsnm{{F{\"u}rst}}},
\bauthor{\binits{S.} \bsnm{{Suchy}}},
\bauthor{\binits{I.} \bsnm{{Kreykenbohm}}},
\bauthor{\binits{L.} \bsnm{{Barrag{\'a}n}}},
\bauthor{\binits{J.} \bsnm{{Wilms}}},
\bauthor{\binits{K.} \bsnm{{Pottschmidt}}},
\bauthor{\binits{I.} \bsnm{{Caballero}}},
\bauthor{\binits{P.} \bsnm{{Kretschmar}}},
\bauthor{\binits{C.} \bsnm{{Ferrigno}}},
\bauthor{\binits{R.E.} \bsnm{{Rothschild}}},
\batitle{{Study of the many fluorescent lines and the absorption variability in
  GX 301-2 with XMM-Newton}}.
\bjtitle{\aap}
\bvolume{535},
\bfpage{9}
(\byear{2011}).
doi:\doiurl{10.1051/0004-6361/201117665}
\end{barticle}
\endbibitem

\bibitem[\protect\citeauthoryear{{F{\"u}rst} et~al.}{2014}]{Fuerst2014}
\begin{barticle}
\bauthor{\binits{F.} \bsnm{{F{\"u}rst}}},
\bauthor{\binits{K.} \bsnm{{Pottschmidt}}},
\bauthor{\binits{J.} \bsnm{{Wilms}}},
\bauthor{\binits{J.A.} \bsnm{{Tomsick}}},
\bauthor{\binits{M.} \bsnm{{Bachetti}}},
\bauthor{\binits{S.E.} \bsnm{{Boggs}}},
\bauthor{\binits{F.E.} \bsnm{{Christensen}}},
\bauthor{\binits{W.W.} \bsnm{{Craig}}},
\bauthor{\binits{B.W.} \bsnm{{Grefenstette}}},
\bauthor{\binits{C.J.} \bsnm{{Hailey}}},
\bauthor{\binits{F.} \bsnm{{Harrison}}},
\bauthor{\binits{K.K.} \bsnm{{Madsen}}},
\bauthor{\binits{J.M.} \bsnm{{Miller}}},
\bauthor{\binits{D.} \bsnm{{Stern}}},
\bauthor{\binits{D.J.} \bsnm{{Walton}}},
\bauthor{\binits{W.} \bsnm{{Zhang}}},
\batitle{{NuSTAR Discovery of a Luminosity Dependent Cyclotron Line Energy in
  Vela X-1}}.
\bjtitle{\apj}
\bvolume{780},
\bfpage{133}
(\byear{2014}).
doi:\doiurl{10.1088/0004-637X/780/2/133}
\end{barticle}
\endbibitem

\bibitem[\protect\citeauthoryear{{Gabler} et~al.}{1989}]{Gabler89}
\begin{barticle}
\bauthor{\binits{R.} \bsnm{{Gabler}}},
\bauthor{\binits{A.} \bsnm{{Gabler}}},
\bauthor{\binits{R.P.} \bsnm{{Kudritzki}}},
\bauthor{\binits{J.} \bsnm{{Puls}}},
\bauthor{\binits{A.} \bsnm{{Pauldrach}}},
\batitle{{Unified NLTE model atmospheres including spherical extension and
  stellar winds - Method and first results}}.
\bjtitle{\aap}
\bvolume{226},
\bfpage{162}--\blpage{182}
(\byear{1989})
\end{barticle}
\endbibitem

\bibitem[\protect\citeauthoryear{{Gamen} et~al.}{2015}]{Gamen2015}
\begin{barticle}
\bauthor{\binits{R.} \bsnm{{Gamen}}},
\bauthor{\binits{R.H.} \bsnm{{Barb{\`a}}}},
\bauthor{\binits{N.R.} \bsnm{{Walborn}}},
\bauthor{\binits{N.I.} \bsnm{{Morrell}}},
\bauthor{\binits{J.I.} \bsnm{{Arias}}},
\bauthor{\binits{J.} \bsnm{{Ma{\'{\i}}z Apell{\'a}niz}}},
\bauthor{\binits{A.} \bsnm{{Sota}}},
\bauthor{\binits{E.J.} \bsnm{{Alfaro}}},
\batitle{{The eccentric short-period orbit of the supergiant fast X-ray
  transient HD 74194 (=LM Vel)}}.
\bjtitle{\aap}
\bvolume{583},
\bfpage{4}
(\byear{2015}).
doi:\doiurl{10.1051/0004-6361/201527140}
\end{barticle}
\endbibitem

\bibitem[\protect\citeauthoryear{{Gayley} et~al.}{1995}]{Gayley95}
\begin{barticle}
\bauthor{\binits{K.G.} \bsnm{{Gayley}}},
\bauthor{\binits{S.P.} \bsnm{{Owocki}}},
\bauthor{\binits{S.R.} \bsnm{{Cranmer}}},
\batitle{{Momentum deposition on Wolf-Rayet winds: Nonisotropic diffusion with
  effective gray opacity}}.
\bjtitle{\apj}
\bvolume{442},
\bfpage{296}--\blpage{310}
(\byear{1995})
\end{barticle}
\endbibitem

\bibitem[\protect\citeauthoryear{{Giddings}}{1981}]{Giddings81}
\begin{botherref}
\oauthor{\binits{J.R.} \bsnm{{Giddings}}},
PhD thesis,
University of London,
1981
\end{botherref}
\endbibitem

\bibitem[\protect\citeauthoryear{{Gim{\'e}nez-Garc{\'{\i}}a}
  et~al.}{2015}]{Gimenez2015}
\begin{barticle}
\bauthor{\binits{A.} \bsnm{{Gim{\'e}nez-Garc{\'{\i}}a}}},
\bauthor{\binits{J.M.} \bsnm{{Torrej{\'o}n}}},
\bauthor{\binits{W.} \bsnm{{Eikmann}}},
\bauthor{\binits{S.} \bsnm{{Mart{\'{\i}}nez-N{\'u}{\~n}ez}}},
\bauthor{\binits{L.M.} \bsnm{{Oskinova}}},
\bauthor{\binits{J.J.} \bsnm{{Rodes-Roca}}},
\bauthor{\binits{G.} \bsnm{{Bernab{\'e}u}}},
\batitle{{An XMM-Newton view of FeK{$\alpha$} in high-mass X-ray binaries}}.
\bjtitle{\aap}
\bvolume{576},
\bfpage{108}
(\byear{2015}).
doi:\doiurl{10.1051/0004-6361/201425004}
\end{barticle}
\endbibitem

\bibitem[\protect\citeauthoryear{{Gim{\'e}nez-Garc{\'{\i}}a}
  et~al.}{2016}]{Gimenez2016}
\begin{barticle}
\bauthor{\binits{A.} \bsnm{{Gim{\'e}nez-Garc{\'{\i}}a}}},
\bauthor{\binits{T.} \bsnm{{Shenar}}},
\bauthor{\binits{J.M.} \bsnm{{Torrej{\'o}n}}},
\bauthor{\binits{L.} \bsnm{{Oskinova}}},
\bauthor{\binits{S.} \bsnm{{Mart{\'{\i}}nez-N{\'u}{\~n}ez}}},
\bauthor{\binits{W.-R.} \bsnm{{Hamann}}},
\bauthor{\binits{J.J.} \bsnm{{Rodes-Roca}}},
\bauthor{\binits{A.} \bsnm{{Gonz{\'a}lez-Gal{\'a}n}}},
\bauthor{\binits{J.} \bsnm{{Alonso-Santiago}}},
\bauthor{\binits{C.} \bsnm{{Gonz{\'a}lez-Fern{\'a}ndez}}},
\bauthor{\binits{G.} \bsnm{{Bernabeu}}},
\bauthor{\binits{A.} \bsnm{{Sander}}},
\batitle{{Measuring the stellar wind parameters in IGR J17544-2619 and Vela X-1
  constrains the accretion physics in supergiant fast X-ray transient and
  classical supergiant X-ray binaries}}.
\bjtitle{\aap}
\bvolume{591},
\bfpage{26}
(\byear{2016}).
doi:\doiurl{10.1051/0004-6361/201527551}
\end{barticle}
\endbibitem

\bibitem[\protect\citeauthoryear{{Gonz{\'a}lez-Riestra}
  et~al.}{2004}]{Gonzalez-Riestra2004}
\begin{barticle}
\bauthor{\binits{R.} \bsnm{{Gonz{\'a}lez-Riestra}}},
\bauthor{\binits{T.} \bsnm{{Oosterbroek}}},
\bauthor{\binits{E.} \bsnm{{Kuulkers}}},
\bauthor{\binits{A.} \bsnm{{Orr}}},
\bauthor{\binits{A.N.} \bsnm{{Parmar}}},
\batitle{{XMM-Newton observations of the INTEGRAL X-ray transient IGR
  J17544-2619}}.
\bjtitle{\aap}
\bvolume{420},
\bfpage{589}--\blpage{594}
(\byear{2004}).
doi:\doiurl{10.1051/0004-6361:20035940}
\end{barticle}
\endbibitem

\bibitem[\protect\citeauthoryear{{Goossens} et~al.}{2013}]{Goossens2013}
\begin{barticle}
\bauthor{\binits{M.E.} \bsnm{{Goossens}}},
\bauthor{\binits{A.J.} \bsnm{{Bird}}},
\bauthor{\binits{S.P.} \bsnm{{Drave}}},
\bauthor{\binits{A.} \bsnm{{Bazzano}}},
\bauthor{\binits{A.B.} \bsnm{{Hill}}},
\bauthor{\binits{V.A.} \bsnm{{McBride}}},
\bauthor{\binits{V.} \bsnm{{Sguera}}},
\bauthor{\binits{L.} \bsnm{{Sidoli}}},
\batitle{{Discovering a 5.72-d period in the supergiant fast X-ray transient AX
  J1845.0-0433}}.
\bjtitle{\mnras}
\bvolume{434},
\bfpage{2182}--\blpage{2187}
(\byear{2013}).
doi:\doiurl{10.1093/mnras/stt1166}
\end{barticle}
\endbibitem

\bibitem[\protect\citeauthoryear{{G{\"o}{\u g}{\"u}{\c s}}
  et~al.}{2011}]{Gogus:2011}
\begin{barticle}
\bauthor{\binits{E.} \bsnm{{G{\"o}{\u g}{\"u}{\c s}}}},
\bauthor{\binits{I.} \bsnm{{Kreykenbohm}}},
\bauthor{\binits{T.M.} \bsnm{{Belloni}}},
\batitle{{Discovery of a peculiar dip from GX 301-2}}.
\bjtitle{\aap}
\bvolume{525},
\bfpage{6}
(\byear{2011}).
doi:\doiurl{10.1051/0004-6361/201015905}
\end{barticle}
\endbibitem

\bibitem[\protect\citeauthoryear{{Gr{\"a}fener} and {Hamann}}{2005}]{GH05}
\begin{barticle}
\bauthor{\binits{G.} \bsnm{{Gr{\"a}fener}}},
\bauthor{\binits{W.-R.} \bsnm{{Hamann}}},
\batitle{{Hydrodynamic model atmospheres for WR stars. Self-consistent modeling
  of a WC star wind}}.
\bjtitle{\aap}
\bvolume{432},
\bfpage{633}--\blpage{645}
(\byear{2005}).
doi:\doiurl{10.1051/0004-6361:20041732}
\end{barticle}
\endbibitem

\bibitem[\protect\citeauthoryear{{Gr{\"a}fener} and {Hamann}}{2006}]{GH06}
\begin{bchapter}
\bauthor{\binits{G.} \bsnm{{Gr{\"a}fener}}},
\bauthor{\binits{W.-R.} \bsnm{{Hamann}}},
\bctitle{{The metallicity dependence of WR wind models}},
in \bbtitle{Stellar Evolution at Low Metallicity: Mass Loss, Explosions,
  Cosmology},
ed. by \beditor{\binits{H.J.G.L.M.} \bsnm{{Lamers}}},
\beditor{\binits{N.} \bsnm{{Langer}}},
\beditor{\binits{T.} \bsnm{{Nugis}}},
\beditor{\binits{K.} \bsnm{{Annuk}}}
\bsertitle{Astronomical Society of the Pacific Conference Series},
vol. \bseriesno{353},
\byear{2006},
p. \bfpage{171}
\end{bchapter}
\endbibitem

\bibitem[\protect\citeauthoryear{{Gr{\"a}fener} and {Hamann}}{2007}]{GH07}
\begin{bchapter}
\bauthor{\binits{G.} \bsnm{{Gr{\"a}fener}}},
\bauthor{\binits{W.-R.} \bsnm{{Hamann}}},
\bctitle{{Hydrodynamic Model Atmospheres for WR Stars: First Results and Their
  Consequences for Interacting Winds in Massive Binary Systems}},
in \bbtitle{Massive Stars in Interactive Binaries},
ed. by \beditor{\binits{N.} \bsnm{{St.-Louis}}},
\beditor{\binits{A.F.J.} \bsnm{{Moffat}}}
\bsertitle{Astronomical Society of the Pacific Conference Series},
vol. \bseriesno{367},
\byear{2007},
p. \bfpage{131}
\end{bchapter}
\endbibitem

\bibitem[\protect\citeauthoryear{{Gr{\"a}fener} et~al.}{2002}]{Graefener02}
\begin{barticle}
\bauthor{\binits{G.} \bsnm{{Gr{\"a}fener}}},
\bauthor{\binits{L.} \bsnm{{Koesterke}}},
\bauthor{\binits{W.-R.} \bsnm{{Hamann}}},
\batitle{{Line-blanketed model atmospheres for WR stars}}.
\bjtitle{\aap}
\bvolume{387},
\bfpage{244}--\blpage{257}
(\byear{2002}).
doi:\doiurl{10.1051/0004-6361:20020269}
\end{barticle}
\endbibitem

\bibitem[\protect\citeauthoryear{{Grebenev} and {Sunyaev}}{2007}]{Grebenev2007}
\begin{barticle}
\bauthor{\binits{S.A.} \bsnm{{Grebenev}}},
\bauthor{\binits{R.A.} \bsnm{{Sunyaev}}},
\batitle{{The first observation of AX J1749.1-2733 in a bright X-ray state.
  Another fast transient revealed by INTEGRAL}}.
\bjtitle{Astronomy Letters}
\bvolume{33},
\bfpage{149}--\blpage{158}
(\byear{2007}).
doi:\doiurl{10.1134/S1063773707030024}
\end{barticle}
\endbibitem

\bibitem[\protect\citeauthoryear{{Grevesse} and {Sauval}}{1998}]{grevesse1998a}
\begin{barticle}
\bauthor{\binits{N.} \bsnm{{Grevesse}}},
\bauthor{\binits{A.J.} \bsnm{{Sauval}}},
\batitle{Standard solar composition}.
\bjtitle{Space Sci.\ Rev.}
\bvolume{85},
\bfpage{161}--\blpage{174}
(\byear{1998})
\end{barticle}
\endbibitem

\bibitem[\protect\citeauthoryear{{Grinberg} et~al.}{2015}]{grin2015}
\begin{barticle}
\bauthor{\binits{V.} \bsnm{{Grinberg}}},
\bauthor{\binits{M.A.} \bsnm{{Leutenegger}}},
\bauthor{\binits{N.} \bsnm{{Hell}}},
\bauthor{\binits{K.} \bsnm{{Pottschmidt}}},
\bauthor{\binits{M.} \bsnm{{B{\"o}ck}}},
\bauthor{\binits{J.A.} \bsnm{{Garc{\'{\i}}a}}},
\bauthor{\binits{M.} \bsnm{{Hanke}}},
\bauthor{\binits{M.A.} \bsnm{{Nowak}}},
\bauthor{\binits{J.O.} \bsnm{{Sundqvist}}},
\bauthor{\binits{R.H.D.} \bsnm{{Townsend}}},
\bauthor{\binits{J.} \bsnm{{Wilms}}},
\batitle{{Long term variability of Cygnus X-1. VII. Orbital variability of the
  focussed wind in Cyg X-1/HDE 226868 system}}.
\bjtitle{\aap}
\bvolume{576},
\bfpage{117}
(\byear{2015}).
doi:\doiurl{10.1051/0004-6361/201425418}
\end{barticle}
\endbibitem

\bibitem[\protect\citeauthoryear{{Groenewegen} et~al.}{1989}]{Groenewegen89}
\begin{barticle}
\bauthor{\binits{M.A.T.} \bsnm{{Groenewegen}}},
\bauthor{\binits{H.J.G.L.M.} \bsnm{{Lamers}}},
\bauthor{\binits{A.W.A.} \bsnm{{Pauldrach}}},
\batitle{{The winds of O-stars. II - The terminal velocities of stellar winds
  of O-type stars}}.
\bjtitle{\aap}
\bvolume{221},
\bfpage{78}--\blpage{88}
(\byear{1989})
\end{barticle}
\endbibitem

\bibitem[\protect\citeauthoryear{{Groh} et~al.}{2014}]{groh2014}
\begin{barticle}
\bauthor{\binits{J.H.} \bsnm{{Groh}}},
\bauthor{\binits{G.} \bsnm{{Meynet}}},
\bauthor{\binits{S.} \bsnm{{Ekstr{\"o}m}}},
\bauthor{\binits{C.} \bsnm{{Georgy}}},
\batitle{{The evolution of massive stars and their spectra. I. A non-rotating
  60 M$_{\odot}$ star from the zero-age main sequence to the pre-supernova
  stage}}.
\bjtitle{\aap}
\bvolume{564},
\bfpage{30}
(\byear{2014}).
doi:\doiurl{10.1051/0004-6361/201322573}
\end{barticle}
\endbibitem

\bibitem[\protect\citeauthoryear{{Guver} et~al.}{2015}]{Guever15}
\begin{botherref}
\oauthor{\binits{T.} \bsnm{{Guver}}},
\oauthor{\binits{F.} \bsnm{{Ozel}}},
\oauthor{\binits{H.} \bsnm{{Marshall}}},
\oauthor{\binits{D.} \bsnm{{Psaltis}}},
\oauthor{\binits{M.} \bsnm{{Guainazzi}}},
\oauthor{\binits{M.} \bsnm{{Diaz-Trigo}}},
{Systematic Uncertainties in the Spectroscopic Measurements of Neutron-Star
  Masses and Radii from Thermonuclear X-ray Bursts. III. Absolute Flux
  Calibration}.
ArXiv e-prints
(2015)
\end{botherref}
\endbibitem

\bibitem[\protect\citeauthoryear{{Gvaramadze} et~al.}{2012}]{Gvar2012}
\begin{barticle}
\bauthor{\binits{V.V.} \bsnm{{Gvaramadze}}},
\bauthor{\binits{N.} \bsnm{{Langer}}},
\bauthor{\binits{J.} \bsnm{{Mackey}}},
\batitle{{{$\zeta$} Oph and the weak-wind problem}}.
\bjtitle{\mnras}
\bvolume{427},
\bfpage{50}--\blpage{54}
(\byear{2012}).
doi:\doiurl{10.1111/j.1745-3933.2012.01343.x}
\end{barticle}
\endbibitem

\bibitem[\protect\citeauthoryear{{Hainich} et~al.}{2014}]{Hainich2014}
\begin{barticle}
\bauthor{\binits{R.} \bsnm{{Hainich}}},
\bauthor{\binits{U.} \bsnm{{R{\"u}hling}}},
\bauthor{\binits{H.} \bsnm{{Todt}}},
\bauthor{\binits{L.M.} \bsnm{{Oskinova}}},
\bauthor{\binits{A.} \bsnm{{Liermann}}},
\bauthor{\binits{G.} \bsnm{{Gr{\"a}fener}}},
\bauthor{\binits{C.} \bsnm{{Foellmi}}},
\bauthor{\binits{O.} \bsnm{{Schnurr}}},
\bauthor{\binits{W.-R.} \bsnm{{Hamann}}},
\batitle{{The Wolf-Rayet stars in the Large Magellanic Cloud. A comprehensive
  analysis of the WN class}}.
\bjtitle{\aap}
\bvolume{565},
\bfpage{27}
(\byear{2014}).
doi:\doiurl{10.1051/0004-6361/201322696}
\end{barticle}
\endbibitem

\bibitem[\protect\citeauthoryear{{Hamann}}{1981}]{Hamann1981}
\begin{barticle}
\bauthor{\binits{W.-R.} \bsnm{{Hamann}}},
\batitle{{Line formation in expanding atmospheres - On the validity of the
  Sobolev approximation}}.
\bjtitle{\aap}
\bvolume{93},
\bfpage{353}--\blpage{361}
(\byear{1981})
\end{barticle}
\endbibitem

\bibitem[\protect\citeauthoryear{{Hamann} and {Gr{\"a}fener}}{2004}]{Hamann04}
\begin{barticle}
\bauthor{\binits{W.-R.} \bsnm{{Hamann}}},
\bauthor{\binits{G.} \bsnm{{Gr{\"a}fener}}},
\batitle{{Grids of model spectra for WN stars, ready for use}}.
\bjtitle{\aap}
\bvolume{427},
\bfpage{697}--\blpage{704}
(\byear{2004}).
doi:\doiurl{10.1051/0004-6361:20040506}
\end{barticle}
\endbibitem

\bibitem[\protect\citeauthoryear{{Hamann} and {Koesterke}}{1998}]{ham1998}
\begin{barticle}
\bauthor{\binits{W.-R.} \bsnm{{Hamann}}},
\bauthor{\binits{L.} \bsnm{{Koesterke}}},
\batitle{{Spectrum formation in clumped stellar winds: consequences for the
  analyses of Wolf-Rayet spectra}}.
\bjtitle{\aap}
\bvolume{335},
\bfpage{1003}--\blpage{1008}
(\byear{1998})
\end{barticle}
\endbibitem

\bibitem[\protect\citeauthoryear{{Hamann} et~al.}{2008}]{ham2008}
\begin{bbook}
\beditor{\binits{W.-R.} \bsnm{{Hamann}}},
\beditor{\binits{A.} \bsnm{{Feldmeier}}},
\beditor{\binits{L.M.} \bsnm{{Oskinova}}} (eds.),
\bctitle{{Clumping in hot-star winds}},
in \bbtitle{Clumping in Hot-Star Winds}
\byear{2008}
\end{bbook}
\endbibitem

\bibitem[\protect\citeauthoryear{{Hamann} et~al.}{2001}]{Hamann01}
\begin{barticle}
\bauthor{\binits{W.-R.} \bsnm{{Hamann}}},
\bauthor{\binits{J.C.} \bsnm{{Brown}}},
\bauthor{\binits{A.} \bsnm{{Feldmeier}}},
\bauthor{\binits{L.M.} \bsnm{{Oskinova}}},
\batitle{{On the wavelength drift of spectral features from structured hot star
  winds}}.
\bjtitle{\aap}
\bvolume{378},
\bfpage{946}--\blpage{953}
(\byear{2001})
\end{barticle}
\endbibitem

\bibitem[\protect\citeauthoryear{{Hanke}}{2011}]{hanke2011a}
\begin{botherref}
\oauthor{\binits{M.} \bsnm{{Hanke}}},
Probing the Environment of Accreting Compact Objects,
PhD thesis,
Dr.~Karl Remeis-Sternwarte, Astronomisches Institut der Universit{\"a}t
  Erlangen-N{\"u}rnberg, Sternwartstr.~7, 96049 Bamberg, Germany,
2011
\end{botherref}
\endbibitem

\bibitem[\protect\citeauthoryear{{Hanke} et~al.}{2009}]{Hanke2009}
\begin{barticle}
\bauthor{\binits{M.} \bsnm{{Hanke}}},
\bauthor{\binits{J.} \bsnm{{Wilms}}},
\bauthor{\binits{M.A.} \bsnm{{Nowak}}},
\bauthor{\binits{K.} \bsnm{{Pottschmidt}}},
\bauthor{\binits{N.S.} \bsnm{{Schulz}}},
\bauthor{\binits{J.C.} \bsnm{{Lee}}},
\batitle{{Chandra X-Ray Spectroscopy of the Focused Wind in the Cygnus X-1
  System. I. The Nondip Spectrum in the Low/Hard State}}.
\bjtitle{ApJ}
\bvolume{690},
\bfpage{330}--\blpage{346}
(\byear{2009})
\end{barticle}
\endbibitem

\bibitem[\protect\citeauthoryear{{Hannikainen} et~al.}{2007}]{Hannikainen2007}
\begin{barticle}
\bauthor{\binits{D.C.} \bsnm{{Hannikainen}}},
\bauthor{\binits{M.G.} \bsnm{{Rawlings}}},
\bauthor{\binits{P.} \bsnm{{Muhli}}},
\bauthor{\binits{O.} \bsnm{{Vilhu}}},
\bauthor{\binits{J.} \bsnm{{Schultz}}},
\bauthor{\binits{J.} \bsnm{{Rodriguez}}},
\batitle{{The nature of the infrared counterpart of IGR J19140+0951}}.
\bjtitle{\mnras}
\bvolume{380},
\bfpage{665}--\blpage{668}
(\byear{2007}).
doi:\doiurl{10.1111/j.1365-2966.2007.12092.x}
\end{barticle}
\endbibitem

\bibitem[\protect\citeauthoryear{{Hanson} et~al.}{1996}]{Hanson1996}
\begin{barticle}
\bauthor{\binits{M.M.} \bsnm{{Hanson}}},
\bauthor{\binits{P.S.} \bsnm{{Conti}}},
\bauthor{\binits{M.J.} \bsnm{{Rieke}}},
\batitle{{A Spectral Atlas of Hot, Luminous Stars at 2 Microns}}.
\bjtitle{\apjs}
\bvolume{107},
\bfpage{281}
(\byear{1996}).
doi:\doiurl{10.1086/192366}
\end{barticle}
\endbibitem

\bibitem[\protect\citeauthoryear{{Harding}}{1994}]{Harding1994}
\begin{bchapter}
\bauthor{\binits{A.K.} \bsnm{{Harding}}},
\bctitle{{Emission Processes in X-ray Pulsars}},
in \bbtitle{The Evolution of X-ray Binariese},
ed. by \beditor{\binits{S.} \bsnm{{Holt}}},
\beditor{\binits{C.S.} \bsnm{{Day}}}
\bsertitle{American Institute of Physics Conference Series},
vol. \bseriesno{308},
\byear{1994},
p. \bfpage{429}
\end{bchapter}
\endbibitem

\bibitem[\protect\citeauthoryear{{Harding} and {Leventhal}}{1992}]{harding1992}
\begin{barticle}
\bauthor{\binits{A.K.} \bsnm{{Harding}}},
\bauthor{\binits{M.} \bsnm{{Leventhal}}},
\batitle{{Can accretion onto isolated neutron stars produce gamma-ray bursts?}}
\bjtitle{\nat}
\bvolume{357},
\bfpage{388}
(\byear{1992}).
doi:\doiurl{10.1038/357388a0}
\end{barticle}
\endbibitem

\bibitem[\protect\citeauthoryear{{Hatchett} and {McCray}}{1977}]{hatchett77}
\begin{barticle}
\bauthor{\binits{S.} \bsnm{{Hatchett}}},
\bauthor{\binits{R.} \bsnm{{McCray}}},
\batitle{{X-ray sources in stellar winds}}.
\bjtitle{\apj}
\bvolume{211},
\bfpage{552}--\blpage{561}
(\byear{1977}).
doi:\doiurl{10.1086/154962}
\end{barticle}
\endbibitem

\bibitem[\protect\citeauthoryear{{Hauschildt}}{1992}]{Hauschildt92}
\begin{barticle}
\bauthor{\binits{P.H.} \bsnm{{Hauschildt}}},
\batitle{{A fast operator perturbation method for the solution of the special
  relativistic equation of radiative transfer in spherical symmetry}}.
\bjtitle{Journal of Quantitative Spectroscopy and Radiative Transfer}
\bvolume{47},
\bfpage{433}--\blpage{453}
(\byear{1992})
\end{barticle}
\endbibitem

\bibitem[\protect\citeauthoryear{Hayakawa}{1984}]{Hayakawa:84}
\begin{bchapter}
\bauthor{\binits{S.} \bsnm{Hayakawa}},
\bctitle{Sudden disappearance of X-ray emission from \mbox{Vela X-1} -- Can
  this be due to eclipse by a planet of HD77581?},
in \bbtitle{Proceedings of the Symposium on High-Energy Astrophysics and
  Cosmology, Pamporovo, Bulgaria, July 18-23, 1983}.
\bsertitle{Adv.\ Space Res.},
vol. \bseriesno{3},
\byear{1984},
pp. \bfpage{35}--\blpage{38}
\end{bchapter}
\endbibitem

\bibitem[\protect\citeauthoryear{{Henrichs} et~al.}{1988}]{Henrichs88}
\begin{bchapter}
\bauthor{\binits{H.F.} \bsnm{{Henrichs}}},
\bauthor{\binits{L.} \bsnm{{Kaper}}},
\bauthor{\binits{G.A.A.} \bsnm{{Zwarthoed}}},
\bctitle{{Rapid variability in O star winds}},
in \bbtitle{A Decade of UV Astronomy with the IUE Satellite, Volume 2},
vol. \bseriesno{2},
ed. by \beditor{\binits{E.J.} \bsnm{{Rolfe}}},
\byear{1988},
pp. \bfpage{145}--\blpage{149}
\end{bchapter}
\endbibitem

\bibitem[\protect\citeauthoryear{{Herv{\'e}} et~al.}{2013}]{herve2013}
\begin{barticle}
\bauthor{\binits{A.} \bsnm{{Herv{\'e}}}},
\bauthor{\binits{G.} \bsnm{{Rauw}}},
\bauthor{\binits{Y.} \bsnm{{Naz{\'e}}}},
\batitle{{A detailed X-ray investigation of {$\zeta$} Puppis. III. Spectral
  analysis of the whole RGS spectrum}}.
\bjtitle{\aap}
\bvolume{551},
\bfpage{83}
(\byear{2013}).
doi:\doiurl{10.1051/0004-6361/201219734}
\end{barticle}
\endbibitem

\bibitem[\protect\citeauthoryear{{Hillier}}{1991}]{hil1991}
\begin{barticle}
\bauthor{\binits{D.J.} \bsnm{{Hillier}}},
\batitle{{The effects of electron scattering and wind clumping for early
  emission line stars}}.
\bjtitle{\aap}
\bvolume{247},
\bfpage{455}--\blpage{468}
(\byear{1991})
\end{barticle}
\endbibitem

\bibitem[\protect\citeauthoryear{{Hillier} and
  {Miller}}{1998}]{HillierMiller98}
\begin{barticle}
\bauthor{\binits{D.J.} \bsnm{{Hillier}}},
\bauthor{\binits{D.L.} \bsnm{{Miller}}},
\batitle{{The Treatment of Non-LTE Line Blanketing in Spherically Expanding
  Outflows}}.
\bjtitle{\apj}
\bvolume{496},
\bfpage{407}
(\byear{1998})
\end{barticle}
\endbibitem

\bibitem[\protect\citeauthoryear{{Hillier} et~al.}{1993}]{hil1993}
\begin{barticle}
\bauthor{\binits{D.J.} \bsnm{{Hillier}}},
\bauthor{\binits{R.P.} \bsnm{{Kudritzki}}},
\bauthor{\binits{A.W.} \bsnm{{Pauldrach}}},
\bauthor{\binits{D.} \bsnm{{Baade}}},
\bauthor{\binits{J.P.} \bsnm{{Cassinelli}}},
\bauthor{\binits{J.} \bsnm{{Puls}}},
\bauthor{\binits{J.H.M.M.} \bsnm{{Schmitt}}},
\batitle{{The 0.1-2.5-KEV X-Ray Spectrum of the O4F-STAR Zeta-Puppis}}.
\bjtitle{\aap}
\bvolume{276},
\bfpage{117}
(\byear{1993})
\end{barticle}
\endbibitem

\bibitem[\protect\citeauthoryear{{Ho} and {Arons}}{1987}]{Ho1987}
\begin{barticle}
\bauthor{\binits{C.} \bsnm{{Ho}}},
\bauthor{\binits{J.} \bsnm{{Arons}}},
\batitle{{High-luminosity accretion in wind-driven binary X-ray sources}}.
\bjtitle{\apj}
\bvolume{316},
\bfpage{283}--\blpage{293}
(\byear{1987}).
doi:\doiurl{10.1086/165200}
\end{barticle}
\endbibitem

\bibitem[\protect\citeauthoryear{{Ho\-warth} and {Prinja}}{1989}]{HP89}
\begin{barticle}
\bauthor{\binits{I.D.} \bsnm{{Ho\-warth}}},
\bauthor{\binits{R.K.} \bsnm{{Prinja}}},
\batitle{{The stellar winds of 203 Galactic O stars - A quantitative
  ultraviolet survey}}.
\bjtitle{\apjs}
\bvolume{69},
\bfpage{527}--\blpage{592}
(\byear{1989})
\end{barticle}
\endbibitem

\bibitem[\protect\citeauthoryear{{Howarth}}{1992}]{Howarth92}
\begin{bchapter}
\bauthor{\binits{I.D.} \bsnm{{Howarth}}},
\bctitle{{Intrinsic Stellar-Wind Variability (Invited Paper)}},
in \bbtitle{Nonisotropic and Variable Outflows from Stars},
ed. by \beditor{\binits{L.} \bsnm{{Drissen}}},
\beditor{\binits{C.} \bsnm{{Leitherer}}},
\beditor{\binits{A.} \bsnm{{Nota}}}
\bsertitle{Astronomical Society of the Pacific Conference Series},
vol. \bseriesno{22},
\byear{1992},
p. \bfpage{155}
\end{bchapter}
\endbibitem

\bibitem[\protect\citeauthoryear{{Howarth} et~al.}{1995}]{Howarth95}
\begin{barticle}
\bauthor{\binits{I.D.} \bsnm{{Howarth}}},
\bauthor{\binits{R.K.} \bsnm{{Prinja}}},
\bauthor{\binits{D.} \bsnm{{Massa}}},
\batitle{{The IUE MEGA Campaign: The Rotationally Modulated Wind of zeta
  Puppis}}.
\bjtitle{\apjl}
\bvolume{452},
\bfpage{65}
(\byear{1995})
\end{barticle}
\endbibitem

\bibitem[\protect\citeauthoryear{Hoyle and
  Lyttleton}{1939}]{HoyleLyttleton1939}
\begin{barticle}
\bauthor{\binits{F.} \bsnm{Hoyle}},
\bauthor{\binits{R.A.} \bsnm{Lyttleton}},
\batitle{The effect of interstellar matter on climatic variation}.
\bjtitle{Mathematical Proceedings of the Cambridge Philosophical Society}
\bvolume{35},
\bfpage{405}--\blpage{415}
(\byear{1939}).
doi:\doiurl{10.1017/S0305004100021150}
\end{barticle}
\endbibitem

\bibitem[\protect\citeauthoryear{{Hubeny}}{1998}]{Hubeny98}
\begin{bchapter}
\bauthor{\binits{I.} \bsnm{{Hubeny}}},
\bctitle{{Non-LTE line-blanketed model atmospheres of hot stars}},
in \bbtitle{1997 Pacific Rim Conference on Stellar Astrophysics},
ed. by \beditor{\binits{K.L.} \bsnm{{Chan}}},
\beditor{\binits{K.S.} \bsnm{{Cheng}}},
\beditor{\binits{H.P.} \bsnm{{Singh}}}
\bsertitle{Astronomical Society of the Pacific Conference Series},
vol. \bseriesno{138},
\byear{1998},
p. \bfpage{139}
\end{bchapter}
\endbibitem

\bibitem[\protect\citeauthoryear{{Hubrig} et~al.}{2011}]{Hub2011}
\begin{barticle}
\bauthor{\binits{S.} \bsnm{{Hubrig}}},
\bauthor{\binits{L.M.} \bsnm{{Oskinova}}},
\bauthor{\binits{M.} \bsnm{{Sch{\"o}ller}}},
\batitle{{First detection of a magnetic field in the fast rotating runaway Oe
  star {$\zeta$} Ophiuchi}}.
\bjtitle{Astronomische Nachrichten}
\bvolume{332},
\bfpage{147}
(\byear{2011}).
doi:\doiurl{10.1002/asna.201111516}
\end{barticle}
\endbibitem

\bibitem[\protect\citeauthoryear{{Hubrig} et~al.}{2013}]{Hubrig13}
\begin{barticle}
\bauthor{\binits{S.} \bsnm{{Hubrig}}},
\bauthor{\binits{M.} \bsnm{{Sch{\"o}ller}}},
\bauthor{\binits{I.} \bsnm{{Ilyin}}},
\bauthor{\binits{N.V.} \bsnm{{Kharchenko}}},
\bauthor{\binits{L.M.} \bsnm{{Oskinova}}},
\bauthor{\binits{N.} \bsnm{{Langer}}},
\bauthor{\binits{J.F.} \bsnm{{Gonz{\'a}lez}}},
\bauthor{\binits{A.F.} \bsnm{{Kholtygin}}},
\bauthor{\binits{M.} \bsnm{{Briquet}}},
\bauthor{\bsnm{{Magori Collaboration}}},
\batitle{{Exploring the origin of magnetic fields in massive stars. II. New
  magnetic field measurements in cluster and field stars}}.
\bjtitle{\aap}
\bvolume{551},
\bfpage{33}
(\byear{2013}).
doi:\doiurl{10.1051/0004-6361/201220721}
\end{barticle}
\endbibitem

\bibitem[\protect\citeauthoryear{{Huenemoerder} et~al.}{2012}]{Huen2012}
\begin{barticle}
\bauthor{\binits{D.P.} \bsnm{{Huenemoerder}}},
\bauthor{\binits{L.M.} \bsnm{{Oskinova}}},
\bauthor{\binits{R.} \bsnm{{Ignace}}},
\bauthor{\binits{W.L.} \bsnm{{Waldron}}},
\bauthor{\binits{H.} \bsnm{{Todt}}},
\bauthor{\binits{K.} \bsnm{{Hamaguchi}}},
\bauthor{\binits{S.} \bsnm{{Kitamoto}}},
\batitle{{On the Weak-wind Problem in Massive Stars: X-Ray Spectra Reveal a
  Massive Hot Wind in {$\mu$} Columbae}}.
\bjtitle{\apjl}
\bvolume{756},
\bfpage{34}
(\byear{2012}).
doi:\doiurl{10.1088/2041-8205/756/2/L34}
\end{barticle}
\endbibitem

\bibitem[\protect\citeauthoryear{{Ibarra} et~al.}{2007}]{Ibarra2007}
\begin{barticle}
\bauthor{\binits{A.} \bsnm{{Ibarra}}},
\bauthor{\binits{G.} \bsnm{{Matt}}},
\bauthor{\binits{M.} \bsnm{{Guainazzi}}},
\bauthor{\binits{E.} \bsnm{{Kuulkers}}},
\bauthor{\binits{E.} \bsnm{{Jim{\'e}nez-Bail{\'o}n}}},
\bauthor{\binits{J.} \bsnm{{Rodriguez}}},
\bauthor{\binits{F.} \bsnm{{Nicastro}}},
\bauthor{\binits{R.} \bsnm{{Walter}}},
\batitle{{The XMM-Newton/INTEGRAL monitoring campaign of IGR J16318-4848}}.
\bjtitle{\aap}
\bvolume{465},
\bfpage{501}--\blpage{507}
(\byear{2007}).
doi:\doiurl{10.1051/0004-6361:20066225}
\end{barticle}
\endbibitem

\bibitem[\protect\citeauthoryear{{Ignace}}{2001}]{ign2001}
\begin{barticle}
\bauthor{\binits{R.} \bsnm{{Ignace}}},
\batitle{{Theoretical Profile Shapes for Optically Thin X-Ray Emission Lines
  from Spherical Stellar Winds}}.
\bjtitle{\apjl}
\bvolume{549},
\bfpage{119}--\blpage{123}
(\byear{2001}).
doi:\doiurl{10.1086/319141}
\end{barticle}
\endbibitem

\bibitem[\protect\citeauthoryear{{Ikhsanov}}{2001a}]{ikhsanov01c}
\begin{barticle}
\bauthor{\binits{N.R.} \bsnm{{Ikhsanov}}},
\batitle{{On the duration of the subsonic propeller state of neutron stars in
  wind-fed mass-exchange close binary systems}}.
\bjtitle{\aap}
\bvolume{368},
\bfpage{5}--\blpage{7}
(\byear{2001}a).
doi:\doiurl{10.1051/0004-6361:20010140}
\end{barticle}
\endbibitem

\bibitem[\protect\citeauthoryear{{Ikhsanov}}{2001b}]{ikhsanov01a}
\begin{barticle}
\bauthor{\binits{N.R.} \bsnm{{Ikhsanov}}},
\batitle{{On the origin of quiescent X-ray emission from A0535+26}}.
\bjtitle{\aap}
\bvolume{367},
\bfpage{549}--\blpage{556}
(\byear{2001}b).
doi:\doiurl{10.1051/0004-6361:20000464}
\end{barticle}
\endbibitem

\bibitem[\protect\citeauthoryear{{Ikhsanov}}{2001c}]{ikhsanov01b}
\begin{barticle}
\bauthor{\binits{N.R.} \bsnm{{Ikhsanov}}},
\batitle{{On the state of low luminous accreting neutron stars}}.
\bjtitle{\aap}
\bvolume{375},
\bfpage{944}--\blpage{949}
(\byear{2001}c).
doi:\doiurl{10.1051/0004-6361:20010909}
\end{barticle}
\endbibitem

\bibitem[\protect\citeauthoryear{{Ikhsanov}}{2005}]{ikhsanov05}
\begin{barticle}
\bauthor{\binits{N.R.} \bsnm{{Ikhsanov}}},
\batitle{{Neutron Stars in the Subsonic Propeller Stage}}.
\bjtitle{Astrophysics}
\bvolume{48},
\bfpage{400}--\blpage{410}
(\byear{2005}).
doi:\doiurl{10.1007/s10511-005-0039-y}
\end{barticle}
\endbibitem

\bibitem[\protect\citeauthoryear{{Ikhsanov}}{2007}]{ikhsanov07}
\begin{barticle}
\bauthor{\binits{N.R.} \bsnm{{Ikhsanov}}},
\batitle{{The origin of long-period X-ray pulsars}}.
\bjtitle{\mnras}
\bvolume{375},
\bfpage{698}--\blpage{704}
(\byear{2007}).
doi:\doiurl{10.1111/j.1365-2966.2006.11331.x}
\end{barticle}
\endbibitem

\bibitem[\protect\citeauthoryear{{Ikhsanov} and
  {Pustil'nik}}{1996}]{ikhsanov96}
\begin{barticle}
\bauthor{\binits{N.R.} \bsnm{{Ikhsanov}}},
\bauthor{\binits{L.A.} \bsnm{{Pustil'nik}}},
\batitle{{Stability of the magnetospheric boundary of a neutron star undergoing
  spherical accretion.}}
\bjtitle{\aap}
\bvolume{312},
\bfpage{338}--\blpage{344}
(\byear{1996})
\end{barticle}
\endbibitem

\bibitem[\protect\citeauthoryear{Illarionov and
  Sunyaev}{1975}]{IllarionovSunyaev75}
\begin{barticle}
\bauthor{\binits{A.F.} \bsnm{Illarionov}},
\bauthor{\binits{R.A.} \bsnm{Sunyaev}},
\batitle{Why the number of galactic x-ray stars is so small?}
\bjtitle{A\&A}
\bvolume{39},
\bfpage{185}--\blpage{195}
(\byear{1975})
\end{barticle}
\endbibitem

\bibitem[\protect\citeauthoryear{Inoue et~al.}{1984}]{Inoue:84}
\begin{barticle}
\bauthor{\binits{H.} \bsnm{Inoue}},
\bauthor{\binits{Y.} \bsnm{Ogawara}},
\bauthor{\binits{T.} \bsnm{Ohashi}},
\bauthor{\binits{I.} \bsnm{Waki}},
\bauthor{\binits{S.} \bsnm{Hayakawa}},
\bauthor{\binits{H.} \bsnm{Kunieda}},
\bauthor{\binits{F.} \bsnm{Nagase}},
\bauthor{\binits{H.} \bsnm{Tsunemi}},
\batitle{Sudden disappearance of \mbox{Vela X-1} pulses}.
\bjtitle{\pasj}
\bvolume{36},
\bfpage{709}--\blpage{713}
(\byear{1984})
\end{barticle}
\endbibitem

\bibitem[\protect\citeauthoryear{{in't Zand}}{2005}]{Zand2005}
\begin{barticle}
\bauthor{\binits{J.J.M.} \bsnm{{in't Zand}}},
\batitle{{Chandra observation of the fast X-ray transient IGR J17544-2619:
  evidence for a neutron star?}}
\bjtitle{\aap}
\bvolume{441},
\bfpage{1}--\blpage{4}
(\byear{2005}).
doi:\doiurl{10.1051/0004-6361:200500162}
\end{barticle}
\endbibitem

\bibitem[\protect\citeauthoryear{{in't Zand} et~al.}{1998}]{Zand1998}
\begin{barticle}
\bauthor{\binits{J.J.M.} \bsnm{{in't Zand}}},
\bauthor{\binits{A.} \bsnm{{Baykal}}},
\bauthor{\binits{T.E.} \bsnm{{Strohmayer}}},
\batitle{{Recent X-Ray Measurements of the Accretion-powered Pulsar 4U
  1907+09}}.
\bjtitle{\apj}
\bvolume{496},
\bfpage{386}--\blpage{394}
(\byear{1998}).
doi:\doiurl{10.1086/305362}
\end{barticle}
\endbibitem

\bibitem[\protect\citeauthoryear{{in't Zand} et~al.}{1997}]{intZand:1997}
\begin{barticle}
\bauthor{\binits{J.J.M.} \bsnm{{in't Zand}}},
\bauthor{\binits{T.E.} \bsnm{{Strohmayer}}},
\bauthor{\binits{A.} \bsnm{{Baykal}}},
\batitle{{Dipping Activity in the X-Ray Pulsar 4U 1907+09}}.
\bjtitle{\apjl}
\bvolume{479},
\bfpage{47}--\blpage{50}
(\byear{1997}).
doi:\doiurl{10.1086/310570}
\end{barticle}
\endbibitem

\bibitem[\protect\citeauthoryear{{Jain} et~al.}{2009}]{Jain2009}
\begin{barticle}
\bauthor{\binits{C.} \bsnm{{Jain}}},
\bauthor{\binits{B.} \bsnm{{Paul}}},
\bauthor{\binits{A.} \bsnm{{Dutta}}},
\batitle{{Discovery of a short orbital period in the Supergiant Fast X-ray
  Transient IGR J16479-4514}}.
\bjtitle{\mnras}
\bvolume{397},
\bfpage{11}--\blpage{15}
(\byear{2009}).
doi:\doiurl{10.1111/j.1745-3933.2009.00668.x}
\end{barticle}
\endbibitem

\bibitem[\protect\citeauthoryear{{Jain} et~al.}{2011}]{Jain2011}
\begin{barticle}
\bauthor{\binits{C.} \bsnm{{Jain}}},
\bauthor{\binits{B.} \bsnm{{Paul}}},
\bauthor{\binits{C.} \bsnm{{Maitra}}},
\batitle{{Detection of periodic X-ray modulation of SFXT IGR J16207-5129 in the
  Swift-BAT light curve}}.
\bjtitle{The Astronomer's Telegram}
\bvolume{3785},
\bfpage{1}
(\byear{2011})
\end{barticle}
\endbibitem

\bibitem[\protect\citeauthoryear{{Jones} et~al.}{1973}]{Jones1973}
\begin{barticle}
\bauthor{\binits{C.} \bsnm{{Jones}}},
\bauthor{\binits{W.} \bsnm{{Forman}}},
\bauthor{\binits{H.} \bsnm{{Tananbaum}}},
\bauthor{\binits{E.} \bsnm{{Schreier}}},
\bauthor{\binits{H.} \bsnm{{Gursky}}},
\bauthor{\binits{E.} \bsnm{{Kellogg}}},
\bauthor{\binits{R.} \bsnm{{Giacconi}}},
\batitle{{Evidence for the Binary Nature of 2u 1700-37}}.
\bjtitle{\apjl}
\bvolume{181},
\bfpage{43}
(\byear{1973}).
doi:\doiurl{10.1086/181181}
\end{barticle}
\endbibitem

\bibitem[\protect\citeauthoryear{{Kaastra} and {Mewe}}{1993}]{kaastra93a}
\begin{barticle}
\bauthor{\binits{J.S.} \bsnm{{Kaastra}}},
\bauthor{\binits{R.} \bsnm{{Mewe}}},
\batitle{{X-ray emission from thin plasmas. I - Multiple Auger ionisation and
  fluorescence processes for Be to Zn}}.
\bjtitle{\aaps}
\bvolume{97},
\bfpage{443}--\blpage{482}
(\byear{1993})
\end{barticle}
\endbibitem

\bibitem[\protect\citeauthoryear{{Kahn} et~al.}{2001}]{kahn2001}
\begin{barticle}
\bauthor{\binits{S.M.} \bsnm{{Kahn}}},
\bauthor{\binits{M.A.} \bsnm{{Leutenegger}}},
\bauthor{\binits{J.} \bsnm{{Cottam}}},
\bauthor{\binits{G.} \bsnm{{Rauw}}},
\bauthor{\binits{J.-M.} \bsnm{{Vreux}}},
\bauthor{\binits{A.J.F.} \bsnm{{den Boggende}}},
\bauthor{\binits{R.} \bsnm{{Mewe}}},
\bauthor{\binits{M.} \bsnm{{G{\"u}del}}},
\batitle{{High resolution X-ray spectroscopy of zeta Puppis with the XMM-Newton
  reflection grating spectrometer}}.
\bjtitle{\aap}
\bvolume{365},
\bfpage{312}--\blpage{317}
(\byear{2001}).
doi:\doiurl{10.1051/0004-6361:20000093}
\end{barticle}
\endbibitem

\bibitem[\protect\citeauthoryear{{Kalberla} et~al.}{2005}]{kalberla2005a}
\begin{barticle}
\bauthor{\binits{P.M.W.} \bsnm{{Kalberla}}},
\bauthor{\binits{W.B.} \bsnm{{Burton}}},
\bauthor{\binits{D.} \bsnm{{Hartmann}}},
\bauthor{\binits{E.M.} \bsnm{{Arnal}}},
\bauthor{\binits{E.} \bsnm{{Bajaja}}},
\bauthor{\binits{R.} \bsnm{{Morras}}},
\bauthor{\binits{W.G.L.} \bsnm{{P{\"o}ppel}}},
\batitle{The leiden/argentine/bonn (lab) survey of galactic hi. final data
  release of the combined lds and iar surveys with improved stray-radiation
  corrections}.
\bjtitle{A\&A}
\bvolume{440},
\bfpage{775}--\blpage{782}
(\byear{2005})
\end{barticle}
\endbibitem

\bibitem[\protect\citeauthoryear{Kallman and White}{1982}]{KallmanWhite:82}
\begin{barticle}
\bauthor{\binits{T.R.} \bsnm{Kallman}},
\bauthor{\binits{N.E.} \bsnm{White}},
\batitle{The anomalous x-ray absorption spectrum of \mbox{Vela X-1}}.
\bjtitle{\apj}
\bvolume{261},
\bfpage{35}--\blpage{39}
(\byear{1982})
\end{barticle}
\endbibitem

\bibitem[\protect\citeauthoryear{{Kallman} et~al.}{2004}]{Kallman2004}
\begin{barticle}
\bauthor{\binits{T.R.} \bsnm{{Kallman}}},
\bauthor{\binits{P.} \bsnm{{Palmeri}}},
\bauthor{\binits{M.A.} \bsnm{{Bautista}}},
\bauthor{\binits{C.} \bsnm{{Mendoza}}},
\bauthor{\binits{J.H.} \bsnm{{Krolik}}},
\batitle{{Photoionization Modeling and the K Lines of Iron}}.
\bjtitle{ApJS}
\bvolume{155},
\bfpage{675}--\blpage{701}
(\byear{2004})
\end{barticle}
\endbibitem

\bibitem[\protect\citeauthoryear{{Kaper} et~al.}{1994}]{Kaper1994}
\begin{barticle}
\bauthor{\binits{L.} \bsnm{{Kaper}}},
\bauthor{\binits{G.} \bsnm{{Hammerschlag-Hensberge}}},
\bauthor{\binits{E.J.} \bsnm{{Zuiderwijk}}},
\batitle{{Spectroscopic evidence for photo-ionization wakes in VELA X-1 and 4U
  1700--37}}.
\bjtitle{\aap}
\bvolume{289},
\bfpage{846}--\blpage{854}
(\byear{1994})
\end{barticle}
\endbibitem

\bibitem[\protect\citeauthoryear{{Kaper} et~al.}{2006}]{Kaper2006}
\begin{barticle}
\bauthor{\binits{L.} \bsnm{{Kaper}}},
\bauthor{\binits{A.} \bsnm{{van der Meer}}},
\bauthor{\binits{F.} \bsnm{{Najarro}}},
\batitle{{VLT/UVES spectroscopy of Wray 977, the hypergiant companion to the
  X-ray pulsar <ASTROBJ>GX301-2</ASTROBJ>}}.
\bjtitle{\aap}
\bvolume{457},
\bfpage{595}--\blpage{610}
(\byear{2006}).
doi:\doiurl{10.1051/0004-6361:20065393}
\end{barticle}
\endbibitem

\bibitem[\protect\citeauthoryear{{Kaper} et~al.}{1999}]{Kaper99}
\begin{barticle}
\bauthor{\binits{L.} \bsnm{{Kaper}}},
\bauthor{\binits{H.F.} \bsnm{{Henrichs}}},
\bauthor{\binits{J.S.} \bsnm{{Nichols}}},
\bauthor{\binits{J.H.} \bsnm{{Telting}}},
\batitle{{Long- and short-term variability in O-star winds. II. Quantitative
  analysis of DAC behaviour}}.
\bjtitle{\aap}
\bvolume{344},
\bfpage{231}--\blpage{262}
(\byear{1999})
\end{barticle}
\endbibitem

\bibitem[\protect\citeauthoryear{{Karino}}{2014}]{karino14}
\begin{barticle}
\bauthor{\binits{S.} \bsnm{{Karino}}},
\batitle{{Bimodality of wind-fed accretion in high-mass X-ray binaries}}.
\bjtitle{\pasj}
\bvolume{66},
\bfpage{34}
(\byear{2014}).
doi:\doiurl{10.1093/pasj/psu018}
\end{barticle}
\endbibitem

\bibitem[\protect\citeauthoryear{{Kaufer} et~al.}{2006}]{Kaufer06}
\begin{barticle}
\bauthor{\binits{A.} \bsnm{{Kaufer}}},
\bauthor{\binits{O.} \bsnm{{Stahl}}},
\bauthor{\binits{R.K.} \bsnm{{Prinja}}},
\bauthor{\binits{D.} \bsnm{{Witherick}}},
\batitle{{Multi-periodic photospheric pulsations and connected wind structures
  in <ASTROBJ>HD 64760</ASTROBJ>}}.
\bjtitle{\aap}
\bvolume{447},
\bfpage{325}--\blpage{341}
(\byear{2006})
\end{barticle}
\endbibitem

\bibitem[\protect\citeauthoryear{{Keszthelyi}}{2015}]{Keszthelyi15}
\begin{bbook}
\bauthor{\binits{Z.} \bsnm{{Keszthelyi}}},
\bbtitle{{Master Thesis. The Impact of Mass Loss on the Early Evolution of
  Massive Stars}}
(\bpublisher{Ludwig-Maximilians-Universit{\"a}t M{\"u}nchen}, \blocation{???},
  \byear{2015})
\end{bbook}
\endbibitem

\bibitem[\protect\citeauthoryear{{Kobulnicky} et~al.}{2010}]{Kobulnicky2010}
\begin{barticle}
\bauthor{\binits{H.A.} \bsnm{{Kobulnicky}}},
\bauthor{\binits{I.J.} \bsnm{{Gilbert}}},
\bauthor{\binits{D.C.} \bsnm{{Kiminki}}},
\batitle{{OB Stars and Stellar Bow shocks in Cygnus-X: A Novel Laboratory
  Estimating Stellar Mass Loss Rates}}.
\bjtitle{\apj}
\bvolume{710},
\bfpage{549}--\blpage{566}
(\byear{2010}).
doi:\doiurl{10.1088/0004-637X/710/1/549}
\end{barticle}
\endbibitem

\bibitem[\protect\citeauthoryear{{Koh} et~al.}{1997}]{Koh1997}
\begin{barticle}
\bauthor{\binits{D.T.} \bsnm{{Koh}}},
\bauthor{\binits{L.} \bsnm{{Bildsten}}},
\bauthor{\binits{D.} \bsnm{{Chakrabarty}}},
\bauthor{\binits{R.W.} \bsnm{{Nelson}}},
\bauthor{\binits{T.A.} \bsnm{{Prince}}},
\bauthor{\binits{B.A.} \bsnm{{Vaughan}}},
\bauthor{\binits{M.H.} \bsnm{{Finger}}},
\bauthor{\binits{R.B.} \bsnm{{Wilson}}},
\bauthor{\binits{B.C.} \bsnm{{Rubin}}},
\batitle{{Rapid Spin-Up Episodes in the Wind-fed Accreting Pulsar GX 301-2}}.
\bjtitle{\apj}
\bvolume{479},
\bfpage{933}--\blpage{947}
(\byear{1997})
\end{barticle}
\endbibitem

\bibitem[\protect\citeauthoryear{{Kramer} et~al.}{2003}]{kramer2003}
\begin{barticle}
\bauthor{\binits{R.H.} \bsnm{{Kramer}}},
\bauthor{\binits{D.H.} \bsnm{{Cohen}}},
\bauthor{\binits{S.P.} \bsnm{{Owocki}}},
\batitle{{X-Ray Emission-Line Profile Modeling of O Stars: Fitting a
  Spherically Symmetric Analytic Wind-Shock Model to the Chandra Spectrum of
  {$\zeta$} Puppis}}.
\bjtitle{\apj}
\bvolume{592},
\bfpage{532}--\blpage{538}
(\byear{2003}).
doi:\doiurl{10.1086/375390}
\end{barticle}
\endbibitem

\bibitem[\protect\citeauthoryear{{Kraus}}{2001}]{Kraus2001}
\begin{barticle}
\bauthor{\binits{U.} \bsnm{{Kraus}}},
\batitle{{Hollow Accretion Columns on Neutron Stars and the Effects of
  Gravitational Light Bending}}.
\bjtitle{\apj}
\bvolume{563},
\bfpage{289}--\blpage{300}
(\byear{2001}).
doi:\doiurl{10.1086/323791}
\end{barticle}
\endbibitem

\bibitem[\protect\citeauthoryear{Kretschmar et~al.}{2000}]{Kretschmar:CGRO99}
\begin{bchapter}
\bauthor{\binits{P.} \bsnm{Kretschmar}},
\bauthor{\binits{I.} \bsnm{Kreykenbohm}},
\bauthor{\binits{J.} \bsnm{Wilms}},
\bauthor{\binits{R.} \bsnm{Staubert}},
\bauthor{\binits{W.} \bsnm{Heindl}},
\bauthor{\binits{D.} \bsnm{Gruber}},
\bauthor{\binits{R.} \bsnm{Rothschild}},
\bctitle{Disappearing Pulses in \mbox{Vela X-1}},
in \bbtitle{Proceedings of the Fifth Compton Symposium},
ed. by \beditor{\binits{M.L.} \bsnm{{McConnell}}},
\beditor{\binits{J.M.} \bsnm{{Ryan}}}
\bsertitle{American Institute of Physics Conference Series},
vol. \bseriesno{510},
\byear{2000},
pp. \bfpage{163}--\blpage{167}
\end{bchapter}
\endbibitem

\bibitem[\protect\citeauthoryear{{Kreykenbohm} et~al.}{2008}]{krey2008}
\begin{barticle}
\bauthor{\binits{I.} \bsnm{{Kreykenbohm}}},
\bauthor{\binits{J.} \bsnm{{Wilms}}},
\bauthor{\binits{P.} \bsnm{{Kretschmar}}},
\bauthor{\binits{J.M.} \bsnm{{Torrej{\'o}n}}},
\bauthor{\binits{K.} \bsnm{{Pottschmidt}}},
\bauthor{\binits{M.} \bsnm{{Hanke}}},
\bauthor{\binits{A.} \bsnm{{Santangelo}}},
\bauthor{\binits{C.} \bsnm{{Ferrigno}}},
\bauthor{\binits{R.} \bsnm{{Staubert}}},
\batitle{{High variability in Vela X-1: giant flares and off states}}.
\bjtitle{\aap}
\bvolume{492},
\bfpage{511}--\blpage{525}
(\byear{2008}).
doi:\doiurl{10.1051/0004-6361:200809956}
\end{barticle}
\endbibitem

\bibitem[\protect\citeauthoryear{Kreykenbohm et~al.}{1999}]{Kreykenbohm:99}
\begin{barticle}
\bauthor{\binits{I.} \bsnm{Kreykenbohm}},
\bauthor{\binits{P.} \bsnm{Kretschmar}},
\bauthor{\binits{J.} \bsnm{Wilms}},
\bauthor{\binits{R.} \bsnm{Staubert}},
\bauthor{\binits{E.} \bsnm{Kendziorra}},
\bauthor{\binits{D.E.} \bsnm{Gruber}},
\bauthor{\binits{W.A.} \bsnm{Heindl}},
\bauthor{\binits{R.R.} \bsnm{Rothschild}},
\batitle{Vela x-1 as seen by rxte}.
\bjtitle{\aap}
\bvolume{341},
\bfpage{141}--\blpage{150}
(\byear{1999})
\end{barticle}
\endbibitem

\bibitem[\protect\citeauthoryear{{Krolik} and {Kallman}}{1984}]{Krolik1984}
\begin{barticle}
\bauthor{\binits{J.H.} \bsnm{{Krolik}}},
\bauthor{\binits{T.R.} \bsnm{{Kallman}}},
\batitle{{Soft X-ray opacity in hot and photoionized gases}}.
\bjtitle{\apj}
\bvolume{286},
\bfpage{366}--\blpage{370}
(\byear{1984}).
doi:\doiurl{10.1086/162608}
\end{barticle}
\endbibitem

\bibitem[\protect\citeauthoryear{{Krticka} et~al.}{2015}]{Krticka2015}
\begin{botherref}
\oauthor{\binits{J.} \bsnm{{Krticka}}},
\oauthor{\binits{J.} \bsnm{{Kubat}}},
\oauthor{\binits{I.} \bsnm{{Krtickova}}},
{X-ray irradiation of the winds in binaries with massive components}.
ArXiv e-prints
(2015)
\end{botherref}
\endbibitem

\bibitem[\protect\citeauthoryear{{Krti{\v c}ka}}{2006}]{Krticka06}
\begin{botherref}
\oauthor{\binits{J.} \bsnm{{Krti{\v c}ka}}},
{NLTE models of line-driven stellar winds - II. O stars in the Small Magellanic
  Cloud}.
\mnras,
266
(2006)
\end{botherref}
\endbibitem

\bibitem[\protect\citeauthoryear{{Krti{\v c}ka} and {Kub{\'a}t}}{2000}]{KK00}
\begin{barticle}
\bauthor{\binits{J.} \bsnm{{Krti{\v c}ka}}},
\bauthor{\binits{J.} \bsnm{{Kub{\'a}t}}},
\batitle{{Isothermal two-component stellar wind of hot stars}}.
\bjtitle{\aap}
\bvolume{359},
\bfpage{983}--\blpage{990}
(\byear{2000})
\end{barticle}
\endbibitem

\bibitem[\protect\citeauthoryear{{Krti{\v c}ka} and {Kub{\'a}t}}{2001}]{KK01}
\begin{barticle}
\bauthor{\binits{J.} \bsnm{{Krti{\v c}ka}}},
\bauthor{\binits{J.} \bsnm{{Kub{\'a}t}}},
\batitle{{Multicomponent radiatively driven stellar winds. I. Nonisothermal
  three-component wind of hot B stars}}.
\bjtitle{\aap}
\bvolume{369},
\bfpage{222}--\blpage{238}
(\byear{2001}).
doi:\doiurl{10.1051/0004-6361:20010121}
\end{barticle}
\endbibitem

\bibitem[\protect\citeauthoryear{{Krti{\v c}ka} and {Kub{\'a}t}}{2004}]{KK04}
\begin{barticle}
\bauthor{\binits{J.} \bsnm{{Krti{\v c}ka}}},
\bauthor{\binits{J.} \bsnm{{Kub{\'a}t}}},
\batitle{{NLTE models of line-driven stellar winds. I. Method of calculation
  and first results for O stars}}.
\bjtitle{\aap}
\bvolume{417},
\bfpage{1003}--\blpage{1016}
(\byear{2004}).
doi:\doiurl{10.1051/0004-6361:20034030}
\end{barticle}
\endbibitem

\bibitem[\protect\citeauthoryear{{Krti{\v c}ka} et~al.}{2012}]{Krticka2012}
\begin{barticle}
\bauthor{\binits{J.} \bsnm{{Krti{\v c}ka}}},
\bauthor{\binits{J.} \bsnm{{Kub{\'a}t}}},
\bauthor{\binits{J.} \bsnm{{Skalick{\'y}}}},
\batitle{{X-Ray Photoionized Bubble in the Wind of Vela X-1 Pulsar Supergiant
  Companion}}.
\bjtitle{\apj}
\bvolume{757},
\bfpage{162}
(\byear{2012}).
doi:\doiurl{10.1088/0004-637X/757/2/162}
\end{barticle}
\endbibitem

\bibitem[\protect\citeauthoryear{{Krti{\v c}ka} et~al.}{2003}]{Krticka03}
\begin{barticle}
\bauthor{\binits{J.} \bsnm{{Krti{\v c}ka}}},
\bauthor{\binits{S.P.} \bsnm{{Owocki}}},
\bauthor{\binits{J.} \bsnm{{Kub{\'a}t}}},
\bauthor{\binits{R.K.} \bsnm{{Galloway}}},
\bauthor{\binits{J.C.} \bsnm{{Brown}}},
\batitle{{On multicomponent effects in stellar winds of stars at extremely low
  metallicity}}.
\bjtitle{\aap}
\bvolume{402},
\bfpage{713}--\blpage{718}
(\byear{2003})
\end{barticle}
\endbibitem

\bibitem[\protect\citeauthoryear{{Krti{\v c}ka} et~al.}{2009}]{krt2009}
\begin{barticle}
\bauthor{\binits{J.} \bsnm{{Krti{\v c}ka}}},
\bauthor{\binits{A.} \bsnm{{Feldmeier}}},
\bauthor{\binits{L.M.} \bsnm{{Oskinova}}},
\bauthor{\binits{J.} \bsnm{{Kub{\'a}t}}},
\bauthor{\binits{W.-R.} \bsnm{{Hamann}}},
\batitle{{X-ray emission from hydrodynamical simulations in non-LTE wind
  models}}.
\bjtitle{\aap}
\bvolume{508},
\bfpage{841}--\blpage{848}
(\byear{2009}).
doi:\doiurl{10.1051/0004-6361/200912642}
\end{barticle}
\endbibitem

\bibitem[\protect\citeauthoryear{{Kudritzki}}{2002}]{kud2002}
\begin{barticle}
\bauthor{\binits{R.P.} \bsnm{{Kudritzki}}},
\batitle{{Line-driven Winds, Ionizing Fluxes, and Ultraviolet Spectra of Hot
  Stars at Extremely Low Metallicity. I. Very Massive O Stars}}.
\bjtitle{\apj}
\bvolume{577},
\bfpage{389}--\blpage{408}
(\byear{2002}).
doi:\doiurl{10.1086/342178}
\end{barticle}
\endbibitem

\bibitem[\protect\citeauthoryear{{Kudritzki} and {Puls}}{2000}]{KP00}
\begin{barticle}
\bauthor{\binits{R.-P.} \bsnm{{Kudritzki}}},
\bauthor{\binits{J.} \bsnm{{Puls}}},
\batitle{{Winds from Hot Stars}}.
\bjtitle{\araa}
\bvolume{38},
\bfpage{613}--\blpage{666}
(\byear{2000})
\end{barticle}
\endbibitem

\bibitem[\protect\citeauthoryear{{Kudritzki} et~al.}{1995}]{Kud95}
\begin{bchapter}
\bauthor{\binits{R.-P.} \bsnm{{Kudritzki}}},
\bauthor{\binits{D.J.} \bsnm{{Lennon}}},
\bauthor{\binits{J.} \bsnm{{Puls}}},
\bctitle{{Quantitative Spectroscopy of Luminous Blue Stars in Distant
  Galaxies}},
in \bbtitle{Science with the VLT},
ed. by \beditor{\binits{J.R.} \bsnm{{Walsh}}},
\beditor{\binits{I.J.} \bsnm{{Danziger}}},
\byear{1995},
p. \bfpage{246}
\end{bchapter}
\endbibitem

\bibitem[\protect\citeauthoryear{{Kudritzki} et~al.}{1989}]{Kud89}
\begin{barticle}
\bauthor{\binits{R.-P.} \bsnm{{Kudritzki}}},
\bauthor{\binits{A.} \bsnm{{Pauldrach}}},
\bauthor{\binits{J.} \bsnm{{Puls}}},
\bauthor{\binits{D.C.} \bsnm{{Abbott}}},
\batitle{{Radiation-driven winds of hot stars. VI - Analytical solutions for
  wind models including the finite cone angle effect}}.
\bjtitle{\aap}
\bvolume{219},
\bfpage{205}--\blpage{218}
(\byear{1989})
\end{barticle}
\endbibitem

\bibitem[\protect\citeauthoryear{{Kudritzki} et~al.}{1999}]{Kud99}
\begin{barticle}
\bauthor{\binits{R.-P.} \bsnm{{Kudritzki}}},
\bauthor{\binits{J.} \bsnm{{Puls}}},
\bauthor{\binits{D.J.} \bsnm{{Lennon}}},
\bauthor{\binits{K.A.} \bsnm{{Venn}}},
\bauthor{\binits{J.} \bsnm{{Reetz}}},
\bauthor{\binits{F.} \bsnm{{Najarro}}},
\bauthor{\binits{J.K.} \bsnm{{McCarthy}}},
\bauthor{\binits{A.} \bsnm{{Herrero}}},
\batitle{{The wind momentum-luminosity relationship of galactic A- and
  B-supergiants}}.
\bjtitle{\aap}
\bvolume{350},
\bfpage{970}--\blpage{984}
(\byear{1999})
\end{barticle}
\endbibitem

\bibitem[\protect\citeauthoryear{{Kulkarni} and
  {Romanova}}{2008}]{Kulkarni2008}
\begin{barticle}
\bauthor{\binits{A.K.} \bsnm{{Kulkarni}}},
\bauthor{\binits{M.M.} \bsnm{{Romanova}}},
\batitle{{Accretion to magnetized stars through the Rayleigh-Taylor
  instability: global 3D simulations}}.
\bjtitle{\mnras}
\bvolume{386},
\bfpage{673}--\blpage{687}
(\byear{2008}).
doi:\doiurl{10.1111/j.1365-2966.2008.13094.x}
\end{barticle}
\endbibitem

\bibitem[\protect\citeauthoryear{{La Parola} et~al.}{2010}]{LaParola2010}
\begin{barticle}
\bauthor{\binits{V.} \bsnm{{La Parola}}},
\bauthor{\binits{G.} \bsnm{{Cusumano}}},
\bauthor{\binits{P.} \bsnm{{Romano}}},
\bauthor{\binits{A.} \bsnm{{Segreto}}},
\bauthor{\binits{S.} \bsnm{{Vercellone}}},
\bauthor{\binits{G.} \bsnm{{Chincarini}}},
\batitle{{Detection of an orbital period in the supergiant high-mass X-ray
  binary IGR J16465-4507 with Swift-BAT}}.
\bjtitle{\mnras}
\bvolume{405},
\bfpage{66}--\blpage{70}
(\byear{2010}).
doi:\doiurl{10.1111/j.1745-3933.2010.00860.x}
\end{barticle}
\endbibitem

\bibitem[\protect\citeauthoryear{{Lamers} and {Cassinelli}}{1999}]{Lamers1999}
\begin{bbook}
\bauthor{\binits{H.J.G.L.M.} \bsnm{{Lamers}}},
\bauthor{\binits{J.P.} \bsnm{{Cassinelli}}},
\bbtitle{{Introduction to Stellar Winds}}
\byear{1999}
\end{bbook}
\endbibitem

\bibitem[\protect\citeauthoryear{{Lamers} and {Morton}}{1976}]{Lamers1976}
\begin{barticle}
\bauthor{\binits{H.J.G.L.M.} \bsnm{{Lamers}}},
\bauthor{\binits{D.C.} \bsnm{{Morton}}},
\batitle{{Mass ejection from the O4f star Zeta Puppis}}.
\bjtitle{\apjs}
\bvolume{32},
\bfpage{715}--\blpage{736}
(\byear{1976}).
doi:\doiurl{10.1086/190413}
\end{barticle}
\endbibitem

\bibitem[\protect\citeauthoryear{{Lamers} et~al.}{1987}]{Lamers1987}
\begin{barticle}
\bauthor{\binits{H.J.G.L.M.} \bsnm{{Lamers}}},
\bauthor{\binits{M.} \bsnm{{Cerruti-Sola}}},
\bauthor{\binits{M.} \bsnm{{Perinotto}}},
\batitle{{The 'SEI' method for accurate and efficient calculations of line
  profiles in spherically symmetric stellar winds}}.
\bjtitle{\apj}
\bvolume{314},
\bfpage{726}--\blpage{738}
(\byear{1987}).
doi:\doiurl{10.1086/165100}
\end{barticle}
\endbibitem

\bibitem[\protect\citeauthoryear{{Lamers} et~al.}{1982}]{Lamersetal82}
\begin{barticle}
\bauthor{\binits{H.J.G.L.M.} \bsnm{{Lamers}}},
\bauthor{\binits{R.} \bsnm{{Gathier}}},
\bauthor{\binits{T.P.} \bsnm{{Snow}} \bsuffix{Jr.}},
\batitle{{Narrow components in the profiles of ultraviolet resonance lines -
  Evidence for a two-component stellar wind for O and B stars}}.
\bjtitle{\apj}
\bvolume{258},
\bfpage{186}--\blpage{200}
(\byear{1982})
\end{barticle}
\endbibitem

\bibitem[\protect\citeauthoryear{{Lanz} and {Hubeny}}{2003}]{LH2003}
\begin{barticle}
\bauthor{\binits{T.} \bsnm{{Lanz}}},
\bauthor{\binits{I.} \bsnm{{Hubeny}}},
\batitle{{A Grid of Non-LTE Line-blanketed Model Atmospheres of O-Type Stars}}.
\bjtitle{\apjs}
\bvolume{146},
\bfpage{417}--\blpage{441}
(\byear{2003}).
doi:\doiurl{10.1086/374373}
\end{barticle}
\endbibitem

\bibitem[\protect\citeauthoryear{Lapshov et~al.}{1992}]{Lapshov:92}
\begin{barticle}
\bauthor{\binits{I.Y.} \bsnm{Lapshov}},
\bauthor{\binits{R.A.} \bsnm{Sunyaev}},
\bauthor{\binits{M.A.} \bsnm{Chichkov}},
\bauthor{\binits{V.V.} \bsnm{Dremin}},
\bauthor{\binits{S.} \bsnm{Brandt}},
\bauthor{\binits{N.} \bsnm{Lund}},
\batitle{Two years of observation of the x-ray pulsar \mbox{Vela X-1} with the
  {\sl watch\/} instrument on the {\sl granat\/} observatory}.
\bjtitle{Sov.\ Astron.\ Lett.}
\bvolume{18},
\bfpage{16}--\blpage{19}
(\byear{1992})
\end{barticle}
\endbibitem

\bibitem[\protect\citeauthoryear{{Lefever} et~al.}{2010}]{Lefever10}
\begin{barticle}
\bauthor{\binits{K.} \bsnm{{Lefever}}},
\bauthor{\binits{J.} \bsnm{{Puls}}},
\bauthor{\binits{T.} \bsnm{{Morel}}},
\bauthor{\binits{C.} \bsnm{{Aerts}}},
\bauthor{\binits{L.} \bsnm{{Decin}}},
\bauthor{\binits{M.} \bsnm{{Briquet}}},
\batitle{{Spectroscopic determination of the fundamental parameters of 66
  B-type stars in the field-of-view of the CoRoT satellite}}.
\bjtitle{\aap}
\bvolume{515},
\bfpage{74}
(\byear{2010}).
doi:\doiurl{10.1051/0004-6361/200911956}
\end{barticle}
\endbibitem

\bibitem[\protect\citeauthoryear{{Leitherer} et~al.}{1992}]{Leitherer92}
\begin{barticle}
\bauthor{\binits{C.} \bsnm{{Leitherer}}},
\bauthor{\binits{C.} \bsnm{{Robert}}},
\bauthor{\binits{L.} \bsnm{{Drissen}}},
\batitle{{Deposition of mass, momentum, and energy by massive stars into the
  interstellar medium}}.
\bjtitle{\apj}
\bvolume{401},
\bfpage{596}--\blpage{617}
(\byear{1992}).
doi:\doiurl{10.1086/172089}
\end{barticle}
\endbibitem

\bibitem[\protect\citeauthoryear{{Leutenegger} et~al.}{2007}]{leu2007}
\begin{barticle}
\bauthor{\binits{M.A.} \bsnm{{Leutenegger}}},
\bauthor{\binits{S.P.} \bsnm{{Owocki}}},
\bauthor{\binits{S.M.} \bsnm{{Kahn}}},
\bauthor{\binits{F.B.S.} \bsnm{{Paerels}}},
\batitle{{Evidence for the Importance of Resonance Scattering in X-Ray Emission
  Line Profiles of the O Star {$\zeta$} Puppis}}.
\bjtitle{\apj}
\bvolume{659},
\bfpage{642}--\blpage{649}
(\byear{2007}).
doi:\doiurl{10.1086/512031}
\end{barticle}
\endbibitem

\bibitem[\protect\citeauthoryear{{Leutenegger} et~al.}{2013}]{leu2013}
\begin{barticle}
\bauthor{\binits{M.A.} \bsnm{{Leutenegger}}},
\bauthor{\binits{D.H.} \bsnm{{Cohen}}},
\bauthor{\binits{J.O.} \bsnm{{Sundqvist}}},
\bauthor{\binits{S.P.} \bsnm{{Owocki}}},
\batitle{{Constraints on Porosity and Mass Loss in O-star Winds from the
  Modeling of X-Ray Emission Line Profile Shapes}}.
\bjtitle{\apj}
\bvolume{770},
\bfpage{80}
(\byear{2013}).
doi:\doiurl{10.1088/0004-637X/770/1/80}
\end{barticle}
\endbibitem

\bibitem[\protect\citeauthoryear{{Levine} and {Corbet}}{2006}]{Levine2006}
\begin{barticle}
\bauthor{\binits{A.M.} \bsnm{{Levine}}},
\bauthor{\binits{R.} \bsnm{{Corbet}}},
\batitle{{Detection of Additional Periodicities in RXTE ASM Light Curves}}.
\bjtitle{The Astronomer's Telegram}
\bvolume{940},
\bfpage{1}
(\byear{2006})
\end{barticle}
\endbibitem

\bibitem[\protect\citeauthoryear{{Leyder} et~al.}{2007}]{Leyder2007}
\begin{barticle}
\bauthor{\binits{J.-C.} \bsnm{{Leyder}}},
\bauthor{\binits{R.} \bsnm{{Walter}}},
\bauthor{\binits{M.} \bsnm{{Lazos}}},
\bauthor{\binits{N.} \bsnm{{Masetti}}},
\bauthor{\binits{N.} \bsnm{{Produit}}},
\batitle{{Hard X-ray flares in <ASTROBJ>IGR J08408-4503</ASTROBJ> unveil clumpy
  stellar winds}}.
\bjtitle{\aap}
\bvolume{465},
\bfpage{35}--\blpage{38}
(\byear{2007}).
doi:\doiurl{10.1051/0004-6361:20066317}
\end{barticle}
\endbibitem

\bibitem[\protect\citeauthoryear{{Lobel}}{2013}]{Lobel13}
\begin{bchapter}
\bauthor{\binits{A.} \bsnm{{Lobel}}},
\bctitle{{3-D radiative transfer modeling of rotational modulations in the blue
  supergiant J Puppis}},
in \bbtitle{Massive Stars: From alpha to Omega},
\byear{2013}
\end{bchapter}
\endbibitem

\bibitem[\protect\citeauthoryear{{Lobel} and {Blomme}}{2008}]{LobelBlomme08}
\begin{barticle}
\bauthor{\binits{A.} \bsnm{{Lobel}}},
\bauthor{\binits{R.} \bsnm{{Blomme}}},
\batitle{{Modeling Ultraviolet Wind Line Variability in Massive Hot Stars}}.
\bjtitle{\apj}
\bvolume{678},
\bfpage{408}--\blpage{430}
(\byear{2008})
\end{barticle}
\endbibitem

\bibitem[\protect\citeauthoryear{{Lodders}}{2003}]{lodders2003a}
\begin{barticle}
\bauthor{\binits{K.} \bsnm{{Lodders}}},
\batitle{Solar system abundances and condensation temperatures of the
  elements}.
\bjtitle{ApJ}
\bvolume{591},
\bfpage{1220}--\blpage{1247}
(\byear{2003})
\end{barticle}
\endbibitem

\bibitem[\protect\citeauthoryear{{Lorenzo} et~al.}{2010}]{Lorenzo2010}
\begin{bchapter}
\bauthor{\binits{J.} \bsnm{{Lorenzo}}},
\bauthor{\binits{I.} \bsnm{{Negueruela}}},
\bauthor{\binits{A.J.} \bsnm{{Norton}}},
\bctitle{{HD 306414, the Optical Counterpart to the Peculiar X-Ray Transient
  IGR J11215-5952}},
in \bbtitle{High Energy Phenomena in Massive Stars},
ed. by \beditor{\binits{J.} \bsnm{{Mart{\'{\i}}}}},
\beditor{\binits{P.L.} \bsnm{{Luque-Escamilla}}},
\beditor{\binits{J.A.} \bsnm{{Combi}}}
\bsertitle{Astronomical Society of the Pacific Conference Series},
vol. \bseriesno{422},
\byear{2010},
p. \bfpage{259}
\end{bchapter}
\endbibitem

\bibitem[\protect\citeauthoryear{{Lorenzo} et~al.}{2014}]{Lorenzo2014}
\begin{barticle}
\bauthor{\binits{J.} \bsnm{{Lorenzo}}},
\bauthor{\binits{I.} \bsnm{{Negueruela}}},
\bauthor{\binits{N.} \bsnm{{Castro}}},
\bauthor{\binits{A.J.} \bsnm{{Norton}}},
\bauthor{\binits{F.} \bsnm{{Vilardell}}},
\bauthor{\binits{A.} \bsnm{{Herrero}}},
\batitle{{Astrophysical parameters of the peculiar X-ray transient IGR
  J11215-5952}}.
\bjtitle{\aap}
\bvolume{562},
\bfpage{18}
(\byear{2014}).
doi:\doiurl{10.1051/0004-6361/201321913}
\end{barticle}
\endbibitem

\bibitem[\protect\citeauthoryear{{Lucy}}{1984}]{Lucy84}
\begin{barticle}
\bauthor{\binits{L.B.} \bsnm{{Lucy}}},
\batitle{{Wave amplification in line-driven winds}}.
\bjtitle{\apj}
\bvolume{284},
\bfpage{351}--\blpage{356}
(\byear{1984}).
doi:\doiurl{10.1086/162413}
\end{barticle}
\endbibitem

\bibitem[\protect\citeauthoryear{{Lucy} and {Abbott}}{1993}]{LA93}
\begin{barticle}
\bauthor{\binits{L.B.} \bsnm{{Lucy}}},
\bauthor{\binits{D.C.} \bsnm{{Abbott}}},
\batitle{{Multiline transfer and the dynamics of Wolf-Rayet winds}}.
\bjtitle{\apj}
\bvolume{405},
\bfpage{738}--\blpage{746}
(\byear{1993})
\end{barticle}
\endbibitem

\bibitem[\protect\citeauthoryear{{Lucy} and {Solomon}}{1970}]{LS70}
\begin{barticle}
\bauthor{\binits{L.B.} \bsnm{{Lucy}}},
\bauthor{\binits{P.M.} \bsnm{{Solomon}}},
\batitle{{Mass Loss by Hot Stars}}.
\bjtitle{\apj}
\bvolume{159},
\bfpage{879}--\blpage{893}
(\byear{1970})
\end{barticle}
\endbibitem

\bibitem[\protect\citeauthoryear{{Lutovinov} et~al.}{2005}]{Lutovinov2005}
\begin{barticle}
\bauthor{\binits{A.} \bsnm{{Lutovinov}}},
\bauthor{\binits{J.} \bsnm{{Rodriguez}}},
\bauthor{\binits{M.} \bsnm{{Revnivtsev}}},
\bauthor{\binits{P.} \bsnm{{Shtykovskiy}}},
\batitle{{Discovery of X-ray pulsations from IGR J16320-4751 = AX
  J1631.9-4752}}.
\bjtitle{\aap}
\bvolume{433},
\bfpage{41}--\blpage{44}
(\byear{2005}).
doi:\doiurl{10.1051/0004-6361:200500092}
\end{barticle}
\endbibitem

\bibitem[\protect\citeauthoryear{{Maccarone} et~al.}{2014}]{Maccarone2014}
\begin{barticle}
\bauthor{\binits{T.J.} \bsnm{{Maccarone}}},
\bauthor{\binits{T.M.} \bsnm{{Girard}}},
\bauthor{\binits{D.I.} \bsnm{{Casetti-Dinescu}}},
\batitle{{High proper motion X-ray binaries from the Yale Southern Proper
  Motion Survey}}.
\bjtitle{\mnras}
\bvolume{440},
\bfpage{1626}--\blpage{1633}
(\byear{2014}).
doi:\doiurl{10.1093/mnras/stu320}
\end{barticle}
\endbibitem

\bibitem[\protect\citeauthoryear{{Macfarlane} et~al.}{1991}]{macf1991}
\begin{barticle}
\bauthor{\binits{J.J.} \bsnm{{Macfarlane}}},
\bauthor{\binits{J.P.} \bsnm{{Cassinelli}}},
\bauthor{\binits{B.Y.} \bsnm{{Welsh}}},
\bauthor{\binits{P.W.} \bsnm{{Vedder}}},
\bauthor{\binits{J.V.} \bsnm{{Vallerga}}},
\bauthor{\binits{W.L.} \bsnm{{Waldron}}},
\batitle{{Predicted extreme-ultraviolet line emission for nearby main-sequence
  B stars}}.
\bjtitle{\apj}
\bvolume{380},
\bfpage{564}--\blpage{574}
(\byear{1991}).
doi:\doiurl{10.1086/170614}
\end{barticle}
\endbibitem

\bibitem[\protect\citeauthoryear{{Manousakis} and
  {Walter}}{2011}]{Manousakis2011}
\begin{barticle}
\bauthor{\binits{A.} \bsnm{{Manousakis}}},
\bauthor{\binits{R.} \bsnm{{Walter}}},
\batitle{{X-ray wind tomography of the highly absorbed HMXB IGR J17252-3616}}.
\bjtitle{A\&A}
\bvolume{526},
\bfpage{62}
(\byear{2011})
\end{barticle}
\endbibitem

\bibitem[\protect\citeauthoryear{{Manousakis} and
  {Walter}}{2015}]{ManousakisWalter:2015}
\begin{barticle}
\bauthor{\binits{A.} \bsnm{{Manousakis}}},
\bauthor{\binits{R.} \bsnm{{Walter}}},
\batitle{{Origin of the X-ray off-states in Vela X-1}}.
\bjtitle{\aap}
\bvolume{575},
\bfpage{58}
(\byear{2015}).
doi:\doiurl{10.1051/0004-6361/201321414}
\end{barticle}
\endbibitem

\bibitem[\protect\citeauthoryear{{Marcolino} et~al.}{2009}]{Marcolino09}
\begin{barticle}
\bauthor{\binits{W.L.F.} \bsnm{{Marcolino}}},
\bauthor{\binits{J.} \bsnm{{Bouret}}},
\bauthor{\binits{F.} \bsnm{{Martins}}},
\bauthor{\binits{D.J.} \bsnm{{Hillier}}},
\bauthor{\binits{T.} \bsnm{{Lanz}}},
\bauthor{\binits{C.} \bsnm{{Escolano}}},
\batitle{{Analysis of Galactic late-type O dwarfs: more constraints on the weak
  wind problem}}.
\bjtitle{\aap}
\bvolume{498},
\bfpage{837}--\blpage{852}
(\byear{2009}).
doi:\doiurl{10.1051/0004-6361/200811289}
\end{barticle}
\endbibitem

\bibitem[\protect\citeauthoryear{{Markova}}{1986}]{Markova86}
\begin{barticle}
\bauthor{\binits{N.} \bsnm{{Markova}}},
\batitle{{The ejection of shells in the stellar wind of P CYG - The most
  plausible explanation of the Balmer-line radial velocity variations}}.
\bjtitle{\aap}
\bvolume{162},
\bfpage{3}--\blpage{5}
(\byear{1986})
\end{barticle}
\endbibitem

\bibitem[\protect\citeauthoryear{{Markova} and {Puls}}{2008}]{MP08}
\begin{barticle}
\bauthor{\binits{N.} \bsnm{{Markova}}},
\bauthor{\binits{J.} \bsnm{{Puls}}},
\batitle{{Bright OB stars in the Galaxy. IV. Stellar and wind parameters of
  early to late B supergiants}}.
\bjtitle{\aap}
\bvolume{478},
\bfpage{823}--\blpage{842}
(\byear{2008})
\end{barticle}
\endbibitem

\bibitem[\protect\citeauthoryear{{Markova} et~al.}{2014}]{Markova14}
\begin{barticle}
\bauthor{\binits{N.} \bsnm{{Markova}}},
\bauthor{\binits{J.} \bsnm{{Puls}}},
\bauthor{\binits{S.} \bsnm{{Sim{\'o}n-D{\'{\i}}az}}},
\bauthor{\binits{A.} \bsnm{{Herrero}}},
\bauthor{\binits{H.} \bsnm{{Markov}}},
\bauthor{\binits{N.} \bsnm{{Langer}}},
\batitle{{Spectroscopic and physical parameters of Galactic O-type stars. II.
  Observational constraints on projected rotational and extra broadening
  velocities as a function of fundamental parameters and stellar evolution}}.
\bjtitle{\aap}
\bvolume{562},
\bfpage{37}
(\byear{2014}).
doi:\doiurl{10.1051/0004-6361/201322661}
\end{barticle}
\endbibitem

\bibitem[\protect\citeauthoryear{{Mart{\'{\i}}nez-N{\'u}{\~n}ez}
  et~al.}{2014}]{martinez-nunez2014a}
\begin{barticle}
\bauthor{\binits{S.} \bsnm{{Mart{\'{\i}}nez-N{\'u}{\~n}ez}}},
\bauthor{\binits{J.M.} \bsnm{{Torrej{\'o}n}}},
\bauthor{\binits{M.} \bsnm{{K{\"u}hnel}}},
\bauthor{\binits{P.} \bsnm{{Kretschmar}}},
\bauthor{\binits{M.} \bsnm{{Stuhlinger}}},
\bauthor{\binits{J.J.} \bsnm{{Rodes-Roca}}},
\bauthor{\binits{F.} \bsnm{{F{\"u}rst}}},
\bauthor{\binits{I.} \bsnm{{Kreykenbohm}}},
\bauthor{\binits{A.} \bsnm{{Martin-Carrillo}}},
\bauthor{\binits{A.M.T.} \bsnm{{Pollock}}},
\bauthor{\binits{J.} \bsnm{{Wilms}}},
\batitle{The accretion environment in vela x-1 during a flaring period using
  xmm-newton}.
\bjtitle{A\&A}
\bvolume{563},
\bfpage{70}
(\byear{2014})
\end{barticle}
\endbibitem

\bibitem[\protect\citeauthoryear{{Mart{\'{\i}}nez-N{\'u}{\~n}ez}
  et~al.}{2015}]{martinez-nunez2015}
\begin{barticle}
\bauthor{\binits{S.} \bsnm{{Mart{\'{\i}}nez-N{\'u}{\~n}ez}}},
\bauthor{\binits{A.} \bsnm{{Sander}}},
\bauthor{\binits{A.} \bsnm{{G{\'{\i}}menez-Garc{\'{\i}}a}}},
\bauthor{\binits{A.} \bsnm{{G{\'o}nzalez-Gal{\'a}n}}},
\bauthor{\binits{J.M.} \bsnm{{Torrej{\'o}n}}},
\bauthor{\binits{C.} \bsnm{{G{\'o}nzalez-Fern{\'a}ndez}}},
\bauthor{\binits{W.-R.} \bsnm{{Hamann}}},
\batitle{{The donor star of the X-ray pulsar X1908+075}}.
\bjtitle{\aap}
\bvolume{578},
\bfpage{107}
(\byear{2015}).
doi:\doiurl{10.1051/0004-6361/201424823}
\end{barticle}
\endbibitem

\bibitem[\protect\citeauthoryear{{Martins} and {Hillier}}{2012}]{Martins12}
\begin{barticle}
\bauthor{\binits{F.} \bsnm{{Martins}}},
\bauthor{\binits{D.J.} \bsnm{{Hillier}}},
\batitle{{On the formation of C iii 4647-50-51 and C iii 5696 in O star
  atmospheres}}.
\bjtitle{\aap}
\bvolume{545},
\bfpage{95}
(\byear{2012}).
doi:\doiurl{10.1051/0004-6361/201219788}
\end{barticle}
\endbibitem

\bibitem[\protect\citeauthoryear{{Masetti} et~al.}{2006}]{Masetti2006}
\begin{barticle}
\bauthor{\binits{N.} \bsnm{{Masetti}}},
\bauthor{\binits{M.L.} \bsnm{{Pretorius}}},
\bauthor{\binits{E.} \bsnm{{Palazzi}}},
\bauthor{\binits{L.} \bsnm{{Bassani}}},
\bauthor{\binits{A.} \bsnm{{Bazzano}}},
\bauthor{\binits{A.J.} \bsnm{{Bird}}},
\bauthor{\binits{P.A.} \bsnm{{Charles}}},
\bauthor{\binits{A.J.} \bsnm{{Dean}}},
\bauthor{\binits{A.} \bsnm{{Malizia}}},
\bauthor{\binits{P.} \bsnm{{Nkundabakura}}},
\bauthor{\binits{J.B.} \bsnm{{Stephen}}},
\bauthor{\binits{P.} \bsnm{{Ubertini}}},
\batitle{{Unveiling the nature of INTEGRAL objects through optical
  spectroscopy. III. Observations of seven southern sources}}.
\bjtitle{\aap}
\bvolume{449},
\bfpage{1139}--\blpage{1149}
(\byear{2006}).
doi:\doiurl{10.1051/0004-6361:20054332}
\end{barticle}
\endbibitem

\bibitem[\protect\citeauthoryear{{Masetti} et~al.}{2008}]{Masetti2008}
\begin{barticle}
\bauthor{\binits{N.} \bsnm{{Masetti}}},
\bauthor{\binits{E.} \bsnm{{Mason}}},
\bauthor{\binits{L.} \bsnm{{Morelli}}},
\bauthor{\binits{S.A.} \bsnm{{Cellone}}},
\bauthor{\binits{V.A.} \bsnm{{McBride}}},
\bauthor{\binits{E.} \bsnm{{Palazzi}}},
\bauthor{\binits{L.} \bsnm{{Bassani}}},
\bauthor{\binits{A.} \bsnm{{Bazzano}}},
\bauthor{\binits{A.J.} \bsnm{{Bird}}},
\bauthor{\binits{P.A.} \bsnm{{Charles}}},
\bauthor{\binits{A.J.} \bsnm{{Dean}}},
\bauthor{\binits{G.} \bsnm{{Galaz}}},
\bauthor{\binits{N.} \bsnm{{Gehrels}}},
\bauthor{\binits{R.} \bsnm{{Landi}}},
\bauthor{\binits{A.} \bsnm{{Malizia}}},
\bauthor{\binits{D.} \bsnm{{Minniti}}},
\bauthor{\binits{F.} \bsnm{{Panessa}}},
\bauthor{\binits{G.E.} \bsnm{{Romero}}},
\bauthor{\binits{J.B.} \bsnm{{Stephen}}},
\bauthor{\binits{P.} \bsnm{{Ubertini}}},
\bauthor{\binits{R.} \bsnm{{Walter}}},
\batitle{{Unveiling the nature of INTEGRAL objects through optical
  spectroscopy. VI. A multi-observatory identification campaign}}.
\bjtitle{\aap}
\bvolume{482},
\bfpage{113}--\blpage{132}
(\byear{2008}).
doi:\doiurl{10.1051/0004-6361:20079332}
\end{barticle}
\endbibitem

\bibitem[\protect\citeauthoryear{{Mason} et~al.}{2009}]{Mason2009}
\begin{barticle}
\bauthor{\binits{A.B.} \bsnm{{Mason}}},
\bauthor{\binits{J.S.} \bsnm{{Clark}}},
\bauthor{\binits{A.J.} \bsnm{{Norton}}},
\bauthor{\binits{I.} \bsnm{{Negueruela}}},
\bauthor{\binits{P.} \bsnm{{Roche}}},
\batitle{{Spectral classification of the mass donors in the high-mass X-ray
  binaries EXO 1722-363 and OAO 1657-415}}.
\bjtitle{A\&A}
\bvolume{505},
\bfpage{281}--\blpage{286}
(\byear{2009})
\end{barticle}
\endbibitem

\bibitem[\protect\citeauthoryear{{Mason} et~al.}{2011}]{Mason2011}
\begin{barticle}
\bauthor{\binits{A.B.} \bsnm{{Mason}}},
\bauthor{\binits{A.J.} \bsnm{{Norton}}},
\bauthor{\binits{J.S.} \bsnm{{Clark}}},
\bauthor{\binits{I.} \bsnm{{Negueruela}}},
\bauthor{\binits{P.} \bsnm{{Roche}}},
\batitle{{The masses of the neutron and donor star in the high-mass X-ray
  binary IGR J18027-2016}}.
\bjtitle{\aap}
\bvolume{532},
\bfpage{124}
(\byear{2011}).
doi:\doiurl{10.1051/0004-6361/201117392}
\end{barticle}
\endbibitem

\bibitem[\protect\citeauthoryear{{Mason} et~al.}{2012}]{Mason2012}
\begin{barticle}
\bauthor{\binits{A.B.} \bsnm{{Mason}}},
\bauthor{\binits{J.S.} \bsnm{{Clark}}},
\bauthor{\binits{A.J.} \bsnm{{Norton}}},
\bauthor{\binits{P.A.} \bsnm{{Crowther}}},
\bauthor{\binits{T.M.} \bsnm{{Tauris}}},
\bauthor{\binits{N.} \bsnm{{Langer}}},
\bauthor{\binits{I.} \bsnm{{Negueruela}}},
\bauthor{\binits{P.} \bsnm{{Roche}}},
\batitle{{The evolution and masses of the neutron star and donor star in the
  high mass X-ray binary OAO 1657-415}}.
\bjtitle{\mnras}
\bvolume{422},
\bfpage{199}--\blpage{206}
(\byear{2012}).
doi:\doiurl{10.1111/j.1365-2966.2012.20596.x}
\end{barticle}
\endbibitem

\bibitem[\protect\citeauthoryear{{Massa} et~al.}{1995}]{Massaetal95}
\begin{barticle}
\bauthor{\binits{D.} \bsnm{{Massa}}},
\bauthor{\binits{A.W.} \bsnm{{Fullerton}}},
\bauthor{\binits{J.S.} \bsnm{{Nichols}}},
\bauthor{\binits{S.P.} \bsnm{{Owocki}}},
\bauthor{\binits{R.K.} \bsnm{{Prinja}}},
\bauthor{\bsnm{{and 28 co-authors}}},
\batitle{{The IUE MEGA Campaign: Wind Variability and Rotation in Early-Type
  Stars}}.
\bjtitle{\apjl}
\bvolume{452},
\bfpage{53}
(\byear{1995})
\end{barticle}
\endbibitem

\bibitem[\protect\citeauthoryear{{Massa} et~al.}{2003}]{Massa2003}
\begin{barticle}
\bauthor{\binits{D.} \bsnm{{Massa}}},
\bauthor{\binits{A.W.} \bsnm{{Fullerton}}},
\bauthor{\binits{G.} \bsnm{{Sonneborn}}},
\bauthor{\binits{J.B.} \bsnm{{Hutchings}}},
\batitle{{Constraints on the Ionization Balance of Hot-Star Winds from FUSE
  Observations of O Stars in the Large Magellanic Cloud}}.
\bjtitle{\apj}
\bvolume{586},
\bfpage{996}--\blpage{1018}
(\byear{2003}).
doi:\doiurl{10.1086/367786}
\end{barticle}
\endbibitem

\bibitem[\protect\citeauthoryear{{Massey} et~al.}{2013}]{Massey13}
\begin{barticle}
\bauthor{\binits{P.} \bsnm{{Massey}}},
\bauthor{\binits{K.F.} \bsnm{{Neugent}}},
\bauthor{\binits{D.J.} \bsnm{{Hillier}}},
\bauthor{\binits{J.} \bsnm{{Puls}}},
\batitle{{A Bake-off between CMFGEN and FASTWIND: Modeling the Physical
  Properties of SMC and LMC O-type Stars}}.
\bjtitle{\apj}
\bvolume{768},
\bfpage{6}
(\byear{2013}).
doi:\doiurl{10.1088/0004-637X/768/1/6}
\end{barticle}
\endbibitem

\bibitem[\protect\citeauthoryear{{Mathis}}{1990}]{mathis1990a}
\begin{barticle}
\bauthor{\binits{J.S.} \bsnm{{Mathis}}},
\batitle{Interstellar dust and extinction}.
\bjtitle{ARA\&A}
\bvolume{28},
\bfpage{37}--\blpage{70}
(\byear{1990})
\end{barticle}
\endbibitem

\bibitem[\protect\citeauthoryear{{Milne}}{1926}]{Milne26}
\begin{barticle}
\bauthor{\binits{E.A.} \bsnm{{Milne}}},
\batitle{{On the possibility of the emission of high-speed atoms from the sun
  and stars}}.
\bjtitle{\mnras}
\bvolume{86},
\bfpage{459}--\blpage{473}
(\byear{1926})
\end{barticle}
\endbibitem

\bibitem[\protect\citeauthoryear{{Mi{\v s}kovi{\v c}ov{\'a}}
  et~al.}{2011}]{misko2011a}
\begin{barticle}
\bauthor{\binits{I.} \bsnm{{Mi{\v s}kovi{\v c}ov{\'a}}}},
\bauthor{\binits{M.} \bsnm{{Hanke}}},
\bauthor{\binits{J.} \bsnm{{Wilms}}},
\bauthor{\binits{M.A.} \bsnm{{Nowak}}},
\bauthor{\binits{K.} \bsnm{{Pottschmidt}}},
\bauthor{\binits{N.S.} \bsnm{{Schulz}}},
\batitle{{Spectroscopy of the Stellar Wind in the Cygnus X-1 System}}.
\bjtitle{Acta Polytechnica}
\bvolume{51}(\bissue{2}),
\bfpage{020000}
(\byear{2011}).
doi:\doiurl{eprintid: arXiv:1103.2711}
\end{barticle}
\endbibitem

\bibitem[\protect\citeauthoryear{{Mokiem} et~al.}{2005}]{Mokiem05}
\begin{barticle}
\bauthor{\binits{M.R.} \bsnm{{Mokiem}}},
\bauthor{\binits{A.} \bsnm{{de Koter}}},
\bauthor{\binits{J.} \bsnm{{Puls}}},
\bauthor{\binits{A.} \bsnm{{Herrero}}},
\bauthor{\binits{F.} \bsnm{{Najarro}}},
\bauthor{\binits{M.R.} \bsnm{{Villamariz}}},
\batitle{{Spectral analysis of early-type stars using a genetic algorithm based
  fitting method}}.
\bjtitle{\aap}
\bvolume{441},
\bfpage{711}--\blpage{733}
(\byear{2005})
\end{barticle}
\endbibitem

\bibitem[\protect\citeauthoryear{{Mokiem} et~al.}{2007}]{Mokiem07}
\begin{barticle}
\bauthor{\binits{M.R.} \bsnm{{Mokiem}}},
\bauthor{\binits{A.} \bsnm{{de Koter}}},
\bauthor{\binits{J.S.} \bsnm{{Vink}}},
\bauthor{\binits{J.} \bsnm{{Puls}}},
\bauthor{\binits{C.J.} \bsnm{{Evans}}},
\bauthor{\binits{S.J.} \bsnm{{Smartt}}},
\bauthor{\binits{P.A.} \bsnm{{Crowther}}},
\bauthor{\binits{A.} \bsnm{{Herrero}}},
\bauthor{\binits{N.} \bsnm{{Langer}}},
\bauthor{\binits{D.J.} \bsnm{{Lennon}}},
\bauthor{\binits{F.} \bsnm{{Najarro}}},
\bauthor{\binits{M.R.} \bsnm{{Villamariz}}},
\batitle{{The empirical metallicity dependence of the mass-loss rate of O- and
  early B-type stars}}.
\bjtitle{\aap}
\bvolume{473},
\bfpage{603}--\blpage{614}
(\byear{2007})
\end{barticle}
\endbibitem

\bibitem[\protect\citeauthoryear{{Monnier} et~al.}{2007}]{Monier07}
\begin{barticle}
\bauthor{\binits{J.D.} \bsnm{{Monnier}}},
\bauthor{\binits{M.} \bsnm{{Zhao}}},
\bauthor{\binits{E.} \bsnm{{Pedretti}}},
\bauthor{\binits{N.} \bsnm{{Thureau}}},
\bauthor{\binits{M.} \bsnm{{Ireland}}},
\bauthor{\binits{P.} \bsnm{{Muirhead}}},
\bauthor{\binits{J.-P.} \bsnm{{Berger}}},
\bauthor{\binits{R.} \bsnm{{Millan-Gabet}}},
\bauthor{\binits{G.} \bsnm{{Van Belle}}},
\bauthor{\binits{T.} \bsnm{{ten Brummelaar}}},
\bauthor{\binits{H.} \bsnm{{McAlister}}},
\bauthor{\binits{S.} \bsnm{{Ridgway}}},
\bauthor{\binits{N.} \bsnm{{Turner}}},
\bauthor{\binits{L.} \bsnm{{Sturmann}}},
\bauthor{\binits{J.} \bsnm{{Sturmann}}},
\bauthor{\binits{D.} \bsnm{{Berger}}},
\batitle{{Imaging the Surface of Altair}}.
\bjtitle{Science}
\bvolume{317},
\bfpage{342}--\blpage{345}
(\byear{2007})
\end{barticle}
\endbibitem

\bibitem[\protect\citeauthoryear{{Mori} and {Ruderman}}{2003}]{mori2003}
\begin{barticle}
\bauthor{\binits{K.} \bsnm{{Mori}}},
\bauthor{\binits{M.A.} \bsnm{{Ruderman}}},
\batitle{{Isolated Magnetar Spin-Down, Soft X-Ray Emission, and RX
  J1856.5-3754}}.
\bjtitle{\apjl}
\bvolume{592},
\bfpage{75}--\blpage{78}
(\byear{2003}).
doi:\doiurl{10.1086/377705}
\end{barticle}
\endbibitem

\bibitem[\protect\citeauthoryear{{Muijres} et~al.}{2011}]{Mui2011}
\begin{barticle}
\bauthor{\binits{L.E.} \bsnm{{Muijres}}},
\bauthor{\binits{A.} \bsnm{{de Koter}}},
\bauthor{\binits{J.S.} \bsnm{{Vink}}},
\bauthor{\binits{J.} \bsnm{{Krti{\v c}ka}}},
\bauthor{\binits{J.} \bsnm{{Kub{\'a}t}}},
\bauthor{\binits{N.} \bsnm{{Langer}}},
\batitle{{Predictions of the effect of clumping on the wind properties of
  O-type stars}}.
\bjtitle{\aap}
\bvolume{526},
\bfpage{32}
(\byear{2011}).
doi:\doiurl{10.1051/0004-6361/201014290}
\end{barticle}
\endbibitem

\bibitem[\protect\citeauthoryear{{Mukherjee} et~al.}{2006}]{Mukherjee2006}
\begin{barticle}
\bauthor{\binits{U.} \bsnm{{Mukherjee}}},
\bauthor{\binits{H.} \bsnm{{Raichur}}},
\bauthor{\binits{B.} \bsnm{{Paul}}},
\bauthor{\binits{S.} \bsnm{{Naik}}},
\bauthor{\binits{N.} \bsnm{{Bhatt}}},
\batitle{{Orbital Evolution and Orbital Phase Resolved Spectroscopy of the HMXB
  Pulsar 4U 1538-52 with RXTE-PCA and BeppoSAX}}.
\bjtitle{Journal of Astrophysics and Astronomy}
\bvolume{27},
\bfpage{411}--\blpage{423}
(\byear{2006})
\end{barticle}
\endbibitem

\bibitem[\protect\citeauthoryear{{Mullan}}{1984}]{Mullan84}
\begin{barticle}
\bauthor{\binits{D.J.} \bsnm{{Mullan}}},
\batitle{{Corotating interaction regions in stellar winds}}.
\bjtitle{\apj}
\bvolume{283},
\bfpage{303}--\blpage{312}
(\byear{1984})
\end{barticle}
\endbibitem

\bibitem[\protect\citeauthoryear{{Mullan}}{1986}]{Mullan86}
\begin{barticle}
\bauthor{\binits{D.J.} \bsnm{{Mullan}}},
\batitle{{Displaced narrow absorption components in the spectra of mass-losing
  OB stars - Indications of corotating interaction regions?}}
\bjtitle{\aap}
\bvolume{165},
\bfpage{157}--\blpage{162}
(\byear{1986})
\end{barticle}
\endbibitem

\bibitem[\protect\citeauthoryear{{Nagase} et~al.}{1986}]{Nagase1986}
\begin{barticle}
\bauthor{\binits{F.} \bsnm{{Nagase}}},
\bauthor{\binits{S.} \bsnm{{Hayakawa}}},
\bauthor{\binits{N.} \bsnm{{Sato}}},
\bauthor{\binits{K.} \bsnm{{Masai}}},
\bauthor{\binits{H.} \bsnm{{Inoue}}},
\batitle{{Circumstellar matter in the VELA X-1/HD 77581 system}}.
\bjtitle{\pasj}
\bvolume{38},
\bfpage{547}--\blpage{569}
(\byear{1986})
\end{barticle}
\endbibitem

\bibitem[\protect\citeauthoryear{{Naz{\'e}} et~al.}{2013}]{Naze2013}
\begin{barticle}
\bauthor{\binits{Y.} \bsnm{{Naz{\'e}}}},
\bauthor{\binits{L.M.} \bsnm{{Oskinova}}},
\bauthor{\binits{E.} \bsnm{{Gosset}}},
\batitle{{A Detailed X-Ray Investigation of {$\zeta$} Puppis. II. The
  Variability on Short and Long Timescales}}.
\bjtitle{\apj}
\bvolume{763},
\bfpage{143}
(\byear{2013}).
doi:\doiurl{10.1088/0004-637X/763/2/143}
\end{barticle}
\endbibitem

\bibitem[\protect\citeauthoryear{{Negueruela} and
  {Smith}}{2006}]{Negueruela2006:aTel831}
\begin{barticle}
\bauthor{\binits{I.} \bsnm{{Negueruela}}},
\bauthor{\binits{D.M.} \bsnm{{Smith}}},
\batitle{{Optical counterpart to SAX J1818.6-1703}}.
\bjtitle{\ATel}
\bvolume{831},
\bfpage{1}
(\byear{2006})
\end{barticle}
\endbibitem

\bibitem[\protect\citeauthoryear{{Negueruela} et~al.}{2005}]{Negueruela2005hd}
\begin{barticle}
\bauthor{\binits{I.} \bsnm{{Negueruela}}},
\bauthor{\binits{D.M.} \bsnm{{Smith}}},
\bauthor{\binits{S.} \bsnm{{Chaty}}},
\batitle{{HD 306414 and IGR J11215-5952}}.
\bjtitle{\ATel}
\bvolume{470},
\bfpage{1}
(\byear{2005})
\end{barticle}
\endbibitem

\bibitem[\protect\citeauthoryear{{Negueruela} et~al.}{2008}]{Negueruela2008int}
\begin{bchapter}
\bauthor{\binits{I.} \bsnm{{Negueruela}}},
\bauthor{\binits{J.M.} \bsnm{{Torrej{\'o}n}}},
\bauthor{\binits{P.} \bsnm{{Reig}}},
\bctitle{{Optical and Infrared characterisation of High Mass X-ray Binaries
  discovered by INTEGRAL}},
in \bbtitle{Proceedings of the 7th INTEGRAL Workshop},
\byear{2008}
\end{bchapter}
\endbibitem

\bibitem[\protect\citeauthoryear{{Negueruela} et~al.}{2006a}]{Negueruela2005a}
\begin{bchapter}
\bauthor{\binits{I.} \bsnm{{Negueruela}}},
\bauthor{\binits{D.M.} \bsnm{{Smith}}},
\bauthor{\binits{P.} \bsnm{{Reig}}},
\bauthor{\binits{S.} \bsnm{{Chaty}}},
\bauthor{\binits{J.M.} \bsnm{{Torrej{\'o}n}}},
\bctitle{{Supergiant Fast X-ray Transients: a new class of high mass X-ray
  binaries unveiled by INTEGRAL}},
in \bbtitle{Proc. of the ``The X-ray Universe 2005'', 26-30 September 2005, El
  Escorial, Madrid, Spain. Ed. by A. Wilson. ESA SP-604, Volume 1, Noordwijk:
  ESA Pub. Division, ISBN 92-9092-915-4, 2006},
\byear{2006}a,
p. \bfpage{165}
\end{bchapter}
\endbibitem

\bibitem[\protect\citeauthoryear{{Negueruela} et~al.}{2006b}]{Negueruela2006}
\begin{barticle}
\bauthor{\binits{I.} \bsnm{{Negueruela}}},
\bauthor{\binits{D.M.} \bsnm{{Smith}}},
\bauthor{\binits{T.E.} \bsnm{{Harrison}}},
\bauthor{\binits{J.M.} \bsnm{{Torrej{\'o}n}}},
\batitle{{The Optical Counterpart to the Peculiar X-Ray Transient XTE
  J1739-302}}.
\bjtitle{\apj}
\bvolume{638},
\bfpage{982}--\blpage{986}
(\byear{2006}b).
doi:\doiurl{10.1086/498935}
\end{barticle}
\endbibitem

\bibitem[\protect\citeauthoryear{{Negueruela} et~al.}{2007}]{Negueruela2007}
\begin{bchapter}
\bauthor{\binits{I.} \bsnm{{Negueruela}}},
\bauthor{\binits{D.M.} \bsnm{{Smith}}},
\bauthor{\binits{J.M.} \bsnm{{Torrej{\'o}n}}},
\bauthor{\binits{P.} \bsnm{{Reig}}},
\bctitle{{Supergiant Fast X-Ray Transients: A Common Behaviour or a Class of
  Objects?}},
in \bbtitle{ESA Special Publication}.
\bsertitle{ESA Special Publication},
vol. \bseriesno{622},
\byear{2007},
p. \bfpage{255}
\end{bchapter}
\endbibitem

\bibitem[\protect\citeauthoryear{{Negueruela} et~al.}{2008}]{Negueruela2008a}
\begin{barticle}
\bauthor{\binits{I.} \bsnm{{Negueruela}}},
\bauthor{\binits{J.} \bsnm{{Casares}}},
\bauthor{\binits{F.} \bsnm{{Verrecchia}}},
\bauthor{\binits{P.} \bsnm{{Blay}}},
\bauthor{\binits{G.L.} \bsnm{{Israel}}},
\bauthor{\binits{S.} \bsnm{{Covino}}},
\batitle{{XTE J1855-026 is a supergiant X-ray binary}}.
\bjtitle{The Astronomer's Telegram}
\bvolume{1876},
\bfpage{1}
(\byear{2008})
\end{barticle}
\endbibitem

\bibitem[\protect\citeauthoryear{{Negueruela} et~al.}{2010}]{Negueruela2010}
\begin{bchapter}
\bauthor{\binits{I.} \bsnm{{Negueruela}}},
\bauthor{\binits{J.} \bsnm{{Lorenzo}}},
\bauthor{\binits{A.} \bsnm{{Herrero}}},
\bauthor{\binits{A.J.} \bsnm{{Norton}}},
\bctitle{{HD 306414: Optical Counterpart to the Peculiar X-ray Transient IGR
  J11215-5952}},
in \bbtitle{Binaries - Key to Comprehension of the Universe},
ed. by \beditor{\binits{A.} \bsnm{{Pr{\v s}a}}},
\beditor{\binits{M.} \bsnm{{Zejda}}}
\bsertitle{Astronomical Society of the Pacific Conference Series},
vol. \bseriesno{435},
\byear{2010},
p. \bfpage{407}
\end{bchapter}
\endbibitem

\bibitem[\protect\citeauthoryear{{Nespoli} et~al.}{2008}]{Nespoli2008}
\begin{barticle}
\bauthor{\binits{E.} \bsnm{{Nespoli}}},
\bauthor{\binits{J.} \bsnm{{Fabregat}}},
\bauthor{\binits{R.E.} \bsnm{{Mennickent}}},
\batitle{{Unveiling the nature of six HMXBs through IR spectroscopy}}.
\bjtitle{\aap}
\bvolume{486},
\bfpage{911}--\blpage{917}
(\byear{2008}).
doi:\doiurl{10.1051/0004-6361:200809645}
\end{barticle}
\endbibitem

\bibitem[\protect\citeauthoryear{{Nespoli} et~al.}{2010}]{Nespoli2010}
\begin{barticle}
\bauthor{\binits{E.} \bsnm{{Nespoli}}},
\bauthor{\binits{J.} \bsnm{{Fabregat}}},
\bauthor{\binits{R.E.} \bsnm{{Mennickent}}},
\batitle{{Unveiling the nature of IGR J16493-4348 with IR spectroscopy}}.
\bjtitle{\aap}
\bvolume{516},
\bfpage{106}
(\byear{2010}).
doi:\doiurl{10.1051/0004-6361/201014348}
\end{barticle}
\endbibitem

\bibitem[\protect\citeauthoryear{{Newman}}{2005}]{Newman2005}
\begin{barticle}
\bauthor{\binits{M.} \bsnm{{Newman}}},
\batitle{{Power laws, Pareto distributions and Zipf's law}}.
\bjtitle{Contemporary Physics}
\bvolume{46},
\bfpage{323}--\blpage{351}
(\byear{2005}).
doi:\doiurl{10.1080/00107510500052444}
\end{barticle}
\endbibitem

\bibitem[\protect\citeauthoryear{{Noebauer} and {Sim}}{2015}]{Noe2015}
\begin{barticle}
\bauthor{\binits{U.M.} \bsnm{{Noebauer}}},
\bauthor{\binits{S.A.} \bsnm{{Sim}}},
\batitle{{Self-consistent modelling of line-driven hot-star winds with Monte
  Carlo radiation hydrodynamics}}.
\bjtitle{\mnras}
\bvolume{453},
\bfpage{3120}--\blpage{3134}
(\byear{2015}).
doi:\doiurl{10.1093/mnras/stv1849}
\end{barticle}
\endbibitem

\bibitem[\protect\citeauthoryear{{Nugis} and {Lamers}}{2002}]{NL02}
\begin{barticle}
\bauthor{\binits{T.} \bsnm{{Nugis}}},
\bauthor{\binits{H.J.G.L.M.} \bsnm{{Lamers}}},
\batitle{The mass-loss rates of wolf-rayet stars explained by optically thick
  radiation driven wind models}.
\bjtitle{A\&A}
\bvolume{389},
\bfpage{162}--\blpage{179}
(\byear{2002})
\end{barticle}
\endbibitem

\bibitem[\protect\citeauthoryear{{Ogilvie} and {Dubus}}{2001}]{Ogilvie2001}
\begin{barticle}
\bauthor{\binits{G.I.} \bsnm{{Ogilvie}}},
\bauthor{\binits{G.} \bsnm{{Dubus}}},
\batitle{{Precessing warped accretion discs in X-ray binaries}}.
\bjtitle{\mnras}
\bvolume{320},
\bfpage{485}--\blpage{503}
(\byear{2001}).
doi:\doiurl{10.1046/j.1365-8711.2001.04011.x}
\end{barticle}
\endbibitem

\bibitem[\protect\citeauthoryear{{Oskinova} et~al.}{2004}]{osk2004}
\begin{barticle}
\bauthor{\binits{L.M.} \bsnm{{Oskinova}}},
\bauthor{\binits{A.} \bsnm{{Feldmeier}}},
\bauthor{\binits{W.-R.} \bsnm{{Hamann}}},
\batitle{{X-ray emission lines from inhomogeneous stellar winds}}.
\bjtitle{\aap}
\bvolume{422},
\bfpage{675}--\blpage{691}
(\byear{2004}).
doi:\doiurl{10.1051/0004-6361:20047187}
\end{barticle}
\endbibitem

\bibitem[\protect\citeauthoryear{{Oskinova} et~al.}{2006}]{osk2006}
\begin{barticle}
\bauthor{\binits{L.M.} \bsnm{{Oskinova}}},
\bauthor{\binits{A.} \bsnm{{Feldmeier}}},
\bauthor{\binits{W.} \bsnm{{Hamann}}},
\batitle{{High-resolution X-ray spectroscopy of bright O-type stars}}.
\bjtitle{\mnras}
\bvolume{372},
\bfpage{313}--\blpage{326}
(\byear{2006}).
doi:\doiurl{10.1111/j.1365-2966.2006.10858.x}
\end{barticle}
\endbibitem

\bibitem[\protect\citeauthoryear{{Oskinova} et~al.}{2012}]{osk2012}
\begin{barticle}
\bauthor{\binits{L.M.} \bsnm{{Oskinova}}},
\bauthor{\binits{A.} \bsnm{{Feldmeier}}},
\bauthor{\binits{P.} \bsnm{{Kretschmar}}},
\batitle{{Clumped stellar winds in supergiant high-mass X-ray binaries: X-ray
  variability and photoionization}}.
\bjtitle{\mnras}
\bvolume{421},
\bfpage{2820}--\blpage{2831}
(\byear{2012}).
doi:\doiurl{10.1111/j.1365-2966.2012.20507.x}
\end{barticle}
\endbibitem

\bibitem[\protect\citeauthoryear{{Oskinova} et~al.}{2007}]{osk2007}
\begin{barticle}
\bauthor{\binits{L.M.} \bsnm{{Oskinova}}},
\bauthor{\binits{W.} \bsnm{{Hamann}}},
\bauthor{\binits{A.} \bsnm{{Feldmeier}}},
\batitle{{Neglecting the porosity of hot-star winds can lead to underestimating
  mass-loss rates}}.
\bjtitle{\aap}
\bvolume{476},
\bfpage{1331}--\blpage{1340}
(\byear{2007}).
doi:\doiurl{10.1051/0004-6361:20066377}
\end{barticle}
\endbibitem

\bibitem[\protect\citeauthoryear{{Oskinova} et~al.}{2011}]{osk2011}
\begin{barticle}
\bauthor{\binits{L.} \bsnm{{Oskinova}}},
\bauthor{\binits{W.-R.} \bsnm{{Hamann}}},
\bauthor{\binits{R.} \bsnm{{Ignace}}},
\bauthor{\binits{A.} \bsnm{{Feldmeier}}},
\batitle{{X-rays, clumping and wind structures}}.
\bjtitle{Bulletin de la Societe Royale des Sciences de Liege}
\bvolume{80},
\bfpage{54}--\blpage{66}
(\byear{2011})
\end{barticle}
\endbibitem

\bibitem[\protect\citeauthoryear{{Owocki}}{1999}]{Owocki99}
\begin{bchapter}
\bauthor{\binits{S.P.} \bsnm{{Owocki}}},
\bctitle{{Co-Rotating Interaction Regions in 2D Hot-Star Wind Models with
  Line-Driven Instability}},
in \bbtitle{IAU Colloq. 169: Variable and Non-spherical Stellar Winds in
  Luminous Hot Stars},
ed. by \beditor{\binits{B.} \bsnm{{Wolf}}},
\beditor{\binits{O.} \bsnm{{Stahl}}},
\beditor{\binits{A.W.} \bsnm{{Fullerton}}}
\bsertitle{Lecture Notes in Physics, Berlin Springer Verlag},
vol. \bseriesno{523},
\byear{1999},
p. \bfpage{294}
\end{bchapter}
\endbibitem

\bibitem[\protect\citeauthoryear{{Owocki}}{2008}]{Owocki2008}
\begin{bchapter}
\bauthor{\binits{S.P.} \bsnm{{Owocki}}},
\bctitle{{Dynamical simulation of the ``velocity-porosity'' reduction in
  observed strength of stellar wind lines}},
in \bbtitle{Clumping in Hot-Star Winds},
ed. by \beditor{\binits{W.-R.} \bsnm{{Hamann}}},
\beditor{\binits{A.} \bsnm{{Feldmeier}}},
\beditor{\binits{L.M.} \bsnm{{Oskinova}}},
\byear{2008},
p. \bfpage{121}
\end{bchapter}
\endbibitem

\bibitem[\protect\citeauthoryear{{Owocki} and {Cohen}}{2001}]{oc2001}
\begin{barticle}
\bauthor{\binits{S.P.} \bsnm{{Owocki}}},
\bauthor{\binits{D.H.} \bsnm{{Cohen}}},
\batitle{{X-Ray Line Profiles from Parameterized Emission within an
  Accelerating Stellar Wind}}.
\bjtitle{\apj}
\bvolume{559},
\bfpage{1108}--\blpage{1116}
(\byear{2001}).
doi:\doiurl{10.1086/322413}
\end{barticle}
\endbibitem

\bibitem[\protect\citeauthoryear{{Owocki} and {Cohen}}{2006}]{oc2006}
\begin{barticle}
\bauthor{\binits{S.P.} \bsnm{{Owocki}}},
\bauthor{\binits{D.H.} \bsnm{{Cohen}}},
\batitle{{The Effect of Porosity on X-Ray Emission-Line Profiles from Hot-Star
  Winds}}.
\bjtitle{\apj}
\bvolume{648},
\bfpage{565}--\blpage{571}
(\byear{2006}).
doi:\doiurl{10.1086/505698}
\end{barticle}
\endbibitem

\bibitem[\protect\citeauthoryear{{Owocki} and {Puls}}{1996}]{Owocki96}
\begin{barticle}
\bauthor{\binits{S.P.} \bsnm{{Owocki}}},
\bauthor{\binits{J.} \bsnm{{Puls}}},
\batitle{{Nonlocal Escape-Integral Approximations for the Line Force in
  Structured Line-driven Stellar Winds}}.
\bjtitle{\apj}
\bvolume{462},
\bfpage{894}
(\byear{1996}).
doi:\doiurl{10.1086/177203}
\end{barticle}
\endbibitem

\bibitem[\protect\citeauthoryear{{Owocki} and {Puls}}{1999}]{OP99}
\begin{barticle}
\bauthor{\binits{S.P.} \bsnm{{Owocki}}},
\bauthor{\binits{J.} \bsnm{{Puls}}},
\batitle{{Line-driven Stellar Winds: The Dynamical Role of Diffuse Radiation
  Gradients and Limitations to the Sobolev Approach}}.
\bjtitle{\apj}
\bvolume{510},
\bfpage{355}--\blpage{368}
(\byear{1999})
\end{barticle}
\endbibitem

\bibitem[\protect\citeauthoryear{{Owocki} and {Puls}}{2002}]{OP02}
\begin{barticle}
\bauthor{\binits{S.P.} \bsnm{{Owocki}}},
\bauthor{\binits{J.} \bsnm{{Puls}}},
\batitle{{Ion Runaway Instability in Low-Density, Line-driven Stellar Winds}}.
\bjtitle{\apj}
\bvolume{568},
\bfpage{965}--\blpage{978}
(\byear{2002})
\end{barticle}
\endbibitem

\bibitem[\protect\citeauthoryear{{Owocki} and {Rybicki}}{1984}]{Owocki84}
\begin{barticle}
\bauthor{\binits{S.P.} \bsnm{{Owocki}}},
\bauthor{\binits{G.B.} \bsnm{{Rybicki}}},
\batitle{{Instabilities in line-driven stellar winds. I - Dependence on
  perturbation wavelength}}.
\bjtitle{\apj}
\bvolume{284},
\bfpage{337}--\blpage{350}
(\byear{1984}).
doi:\doiurl{10.1086/162412}
\end{barticle}
\endbibitem

\bibitem[\protect\citeauthoryear{{Owocki} and {Rybicki}}{1985}]{Owocki85}
\begin{barticle}
\bauthor{\binits{S.P.} \bsnm{{Owocki}}},
\bauthor{\binits{G.B.} \bsnm{{Rybicki}}},
\batitle{{Instabilities in line-driven stellar winds. II - Effect of
  scattering}}.
\bjtitle{\apj}
\bvolume{299},
\bfpage{265}--\blpage{276}
(\byear{1985}).
doi:\doiurl{10.1086/163697}
\end{barticle}
\endbibitem

\bibitem[\protect\citeauthoryear{{Owocki} et~al.}{1988}]{Owocki88}
\begin{barticle}
\bauthor{\binits{S.P.} \bsnm{{Owocki}}},
\bauthor{\binits{J.I.} \bsnm{{Castor}}},
\bauthor{\binits{G.B.} \bsnm{{Rybicki}}},
\batitle{{Time-dependent models of radiatively driven stellar winds. I -
  Nonlinear evolution of instabilities for a pure absorption model}}.
\bjtitle{\apj}
\bvolume{335},
\bfpage{914}--\blpage{930}
(\byear{1988}).
doi:\doiurl{10.1086/166977}
\end{barticle}
\endbibitem

\bibitem[\protect\citeauthoryear{{Owocki} et~al.}{1995}]{OCF95}
\begin{barticle}
\bauthor{\binits{S.P.} \bsnm{{Owocki}}},
\bauthor{\binits{S.R.} \bsnm{{Cranmer}}},
\bauthor{\binits{A.W.} \bsnm{{Fullerton}}},
\batitle{{Periodic Variations in Ultraviolet Spectral Lines of the B0.5 Ib Star
  HD 64760: Evidence for Corotating Wind Streams Rooted in Surface
  Variations}}.
\bjtitle{\apjl}
\bvolume{453},
\bfpage{37}
(\byear{1995})
\end{barticle}
\endbibitem

\bibitem[\protect\citeauthoryear{{Owocki} et~al.}{2004}]{OGS04}
\begin{barticle}
\bauthor{\binits{S.P.} \bsnm{{Owocki}}},
\bauthor{\binits{K.G.} \bsnm{{Gayley}}},
\bauthor{\binits{N.J.} \bsnm{{Shaviv}}},
\batitle{{A Porosity-Length Formalism for Photon-Tiring-limited Mass Loss from
  Stars above the Eddington Limit}}.
\bjtitle{\apj}
\bvolume{616},
\bfpage{525}--\blpage{541}
(\byear{2004})
\end{barticle}
\endbibitem

\bibitem[\protect\citeauthoryear{{Owocki} et~al.}{2013}]{Ow2013}
\begin{barticle}
\bauthor{\binits{S.P.} \bsnm{{Owocki}}},
\bauthor{\binits{J.O.} \bsnm{{Sundqvist}}},
\bauthor{\binits{D.H.} \bsnm{{Cohen}}},
\bauthor{\binits{K.G.} \bsnm{{Gayley}}},
\batitle{{Thin-shell mixing in radiative wind-shocks and the Lx-Lbol scaling of
  O-star X-rays}}.
\bjtitle{\mnras}
\bvolume{429},
\bfpage{3379}--\blpage{3389}
(\byear{2013}).
doi:\doiurl{10.1093/mnras/sts599}
\end{barticle}
\endbibitem

\bibitem[\protect\citeauthoryear{{Paizis} and
  {Sidoli}}{2014}]{PaizisSidoli2014}
\begin{barticle}
\bauthor{\binits{A.} \bsnm{{Paizis}}},
\bauthor{\binits{L.} \bsnm{{Sidoli}}},
\batitle{{Cumulative luminosity distributions of supergiant fast X-ray
  transients in hard X-rays}}.
\bjtitle{\mnras}
\bvolume{439},
\bfpage{3439}--\blpage{3452}
(\byear{2014}).
doi:\doiurl{10.1093/mnras/stu191}
\end{barticle}
\endbibitem

\bibitem[\protect\citeauthoryear{{Palmeri} et~al.}{2003}]{palmeri03a}
\begin{barticle}
\bauthor{\binits{P.} \bsnm{{Palmeri}}},
\bauthor{\binits{C.} \bsnm{{Mendoza}}},
\bauthor{\binits{T.R.} \bsnm{{Kallman}}},
\bauthor{\binits{M.A.} \bsnm{{Bautista}}},
\bauthor{\binits{M.} \bsnm{{Mel{\'e}ndez}}},
\batitle{{Modeling of iron K lines: Radiative and Auger decay data for Fe II-Fe
  IX}}.
\bjtitle{\aap}
\bvolume{410},
\bfpage{359}--\blpage{364}
(\byear{2003}).
doi:\doiurl{10.1051/0004-6361:20031262}
\end{barticle}
\endbibitem

\bibitem[\protect\citeauthoryear{{Pauldrach}}{1987}]{Pauldrach87}
\begin{barticle}
\bauthor{\binits{A.} \bsnm{{Pauldrach}}},
\batitle{{Radiation driven winds of hot luminous stars. III - Detailed
  statistical equilibrium calculations for hydrogen to zinc}}.
\bjtitle{\aap}
\bvolume{183},
\bfpage{295}--\blpage{313}
(\byear{1987})
\end{barticle}
\endbibitem

\bibitem[\protect\citeauthoryear{{Pauldrach} et~al.}{1986}]{PPK}
\begin{barticle}
\bauthor{\binits{A.} \bsnm{{Pauldrach}}},
\bauthor{\binits{J.} \bsnm{{Puls}}},
\bauthor{\binits{R.P.} \bsnm{{Kudritzki}}},
\batitle{{Radiation-driven winds of hot luminous stars - Improvements of the
  theory and first results}}.
\bjtitle{\aap}
\bvolume{164},
\bfpage{86}--\blpage{100}
(\byear{1986})
\end{barticle}
\endbibitem

\bibitem[\protect\citeauthoryear{{Pauldrach} and
  {Puls}}{1990}]{PauldrachPuls90}
\begin{barticle}
\bauthor{\binits{A.W.A.} \bsnm{{Pauldrach}}},
\bauthor{\binits{J.} \bsnm{{Puls}}},
\batitle{{Radiation-driven winds of hot stars. VIII - The bistable wind of the
  luminous blue variable P Cygni (B1 Ia/+/)}}.
\bjtitle{\aap}
\bvolume{237},
\bfpage{409}--\blpage{424}
(\byear{1990})
\end{barticle}
\endbibitem

\bibitem[\protect\citeauthoryear{{Pauldrach} et~al.}{2001}]{Pauldrach01}
\begin{barticle}
\bauthor{\binits{A.W.A.} \bsnm{{Pauldrach}}},
\bauthor{\binits{T.L.} \bsnm{{Hoffmann}}},
\bauthor{\binits{M.} \bsnm{{Lennon}}},
\batitle{{Radiation-driven winds of hot luminous stars. XIII. A description of
  NLTE line blocking and blanketing towards realistic models for expanding
  atmospheres}}.
\bjtitle{\aap}
\bvolume{375},
\bfpage{161}--\blpage{195}
(\byear{2001}).
doi:\doiurl{10.1051/0004-6361:20010805}
\end{barticle}
\endbibitem

\bibitem[\protect\citeauthoryear{{Pauldrach} et~al.}{2003}]{Pauldrach03}
\begin{bchapter}
\bauthor{\binits{A.W.A.} \bsnm{{Pauldrach}}},
\bauthor{\binits{T.L.} \bsnm{{Hoffmann}}},
\bauthor{\binits{R.H.} \bsnm{{M{\'e}ndez}}},
\bctitle{{Radiation Driven Atmospheres of O-type stars: Constraints on the
  Mass-Luminosity Relation of Central Stars of Planetary Nebulae (invited
  review)}},
in \bbtitle{Planetary Nebulae: Their Evolution and Role in the Universe},
ed. by \beditor{\binits{S.} \bsnm{{Kwok}}},
\beditor{\binits{M.} \bsnm{{Dopita}}},
\beditor{\binits{R.} \bsnm{{Sutherland}}}
\bsertitle{IAU Symposium},
vol. \bseriesno{209},
\byear{2003},
p. \bfpage{177}
\end{bchapter}
\endbibitem

\bibitem[\protect\citeauthoryear{{Pauldrach} et~al.}{1994}]{Pauldrach94}
\begin{barticle}
\bauthor{\binits{A.W.A.} \bsnm{{Pauldrach}}},
\bauthor{\binits{R.P.} \bsnm{{Kudritzki}}},
\bauthor{\binits{J.} \bsnm{{Puls}}},
\bauthor{\binits{K.} \bsnm{{Butler}}},
\bauthor{\binits{J.} \bsnm{{Hunsinger}}},
\batitle{{Radiation-driven winds of hot luminous stars. 12: A first step
  towards detailed UV-line diagnostics of O-stars}}.
\bjtitle{\aap}
\bvolume{283},
\bfpage{525}--\blpage{560}
(\byear{1994})
\end{barticle}
\endbibitem

\bibitem[\protect\citeauthoryear{{Pearlman} et~al.}{2013}]{Pearlman2013}
\begin{bchapter}
\bauthor{\binits{A.B.} \bsnm{{Pearlman}}},
\bauthor{\binits{R.} \bsnm{{Corbet}}},
\bauthor{\binits{K.} \bsnm{{Pottschmidt}}},
\bctitle{{Superorbital Modulation and Orbital Parameters of the Eclipsing
  High-Mass X-ray Pulsar IGR J16493-4348}},
in \bbtitle{American Astronomical Society Meeting Abstracts 221}.
\bsertitle{American Astronomical Society Meeting Abstracts},
vol. \bseriesno{221},
\byear{2013},
pp. \bfpage{142}--\blpage{38}
\end{bchapter}
\endbibitem

\bibitem[\protect\citeauthoryear{{Pearlman} et~al.}{2011}]{Perlman2011}
\begin{bchapter}
\bauthor{\binits{A.B.} \bsnm{{Pearlman}}},
\bauthor{\binits{R.H.D.} \bsnm{{Corbet}}},
\bauthor{\binits{K.} \bsnm{{Pottschmidt}}},
\bauthor{\binits{G.K.} \bsnm{{Skinner}}},
\bctitle{{The Orbital Parameters and Nature of the X-ray Pulsar IGR J16393-4643
  Using Pulse Timing Analysis}},
in \bbtitle{AAS/High Energy Astrophysics Division}.
\bsertitle{AAS/High Energy Astrophysics Division},
vol. \bseriesno{12},
\byear{2011},
pp. \bfpage{42}--\blpage{06}
\end{bchapter}
\endbibitem

\bibitem[\protect\citeauthoryear{{Petit} et~al.}{2013}]{Petit13}
\begin{barticle}
\bauthor{\binits{V.} \bsnm{{Petit}}},
\bauthor{\binits{S.P.} \bsnm{{Owocki}}},
\bauthor{\binits{G.A.} \bsnm{{Wade}}},
\bauthor{\binits{D.H.} \bsnm{{Cohen}}},
\bauthor{\binits{J.O.} \bsnm{{Sundqvist}}},
\bauthor{\binits{M.} \bsnm{{Gagn{\'e}}}},
\bauthor{\binits{J.} \bsnm{{Ma{\'{\i}}z Apell{\'a}niz}}},
\bauthor{\binits{M.E.} \bsnm{{Oksala}}},
\bauthor{\binits{D.A.} \bsnm{{Bohlender}}},
\bauthor{\binits{T.} \bsnm{{Rivinius}}},
\bauthor{\binits{H.F.} \bsnm{{Henrichs}}},
\bauthor{\binits{E.} \bsnm{{Alecian}}},
\bauthor{\binits{R.H.D.} \bsnm{{Townsend}}},
\bauthor{\binits{A.} \bsnm{{ud-Doula}}},
\bauthor{\bsnm{{MiMeS Collaboration}}},
\batitle{{A magnetic confinement versus rotation classification of massive-star
  magnetospheres}}.
\bjtitle{\mnras}
\bvolume{429},
\bfpage{398}--\blpage{422}
(\byear{2013}).
doi:\doiurl{10.1093/mnras/sts344}
\end{barticle}
\endbibitem

\bibitem[\protect\citeauthoryear{{Petrenz} and {Puls}}{1996}]{PetrenzPuls96}
\begin{barticle}
\bauthor{\binits{P.} \bsnm{{Petrenz}}},
\bauthor{\binits{J.} \bsnm{{Puls}}},
\batitle{{H{$\alpha$} line formation in hot star winds: the influence of
  rotation.}}
\bjtitle{\aap}
\bvolume{312},
\bfpage{195}--\blpage{220}
(\byear{1996})
\end{barticle}
\endbibitem

\bibitem[\protect\citeauthoryear{{Petrov} et~al.}{2014}]{petrov2014}
\begin{barticle}
\bauthor{\binits{B.} \bsnm{{Petrov}}},
\bauthor{\binits{J.S.} \bsnm{{Vink}}},
\bauthor{\binits{G.} \bsnm{{Gr{\"a}fener}}},
\batitle{{On the H{$\alpha$} behaviour of blue supergiants: rise and fall over
  the bi-stability jump}}.
\bjtitle{\aap}
\bvolume{565},
\bfpage{62}
(\byear{2014}).
doi:\doiurl{10.1051/0004-6361/201322754}
\end{barticle}
\endbibitem

\bibitem[\protect\citeauthoryear{{Prinja}}{1992}]{Prinja92}
\begin{bchapter}
\bauthor{\binits{B.K.} \bsnm{{Prinja}}},
\bctitle{{UV P Cygni Profile Variability in 0 Stars (Invited Paper)}},
in \bbtitle{Nonisotropic and Variable Outflows from Stars},
ed. by \beditor{\binits{L.} \bsnm{{Drissen}}},
\beditor{\binits{C.} \bsnm{{Leitherer}}},
\beditor{\binits{A.} \bsnm{{Nota}}}
\bsertitle{Astronomical Society of the Pacific Conference Series},
vol. \bseriesno{22},
\byear{1992},
p. \bfpage{167}
\end{bchapter}
\endbibitem

\bibitem[\protect\citeauthoryear{{Prinja}}{1988}]{Prinja88}
\begin{barticle}
\bauthor{\binits{R.K.} \bsnm{{Prinja}}},
\batitle{{Evidence for rotationally modulated variability in O star winds}}.
\bjtitle{\mnras}
\bvolume{231},
\bfpage{21}--\blpage{24}
(\byear{1988})
\end{barticle}
\endbibitem

\bibitem[\protect\citeauthoryear{{Prinja}}{1994}]{Prinja94}
\begin{bchapter}
\bauthor{\binits{R.K.} \bsnm{{Prinja}}},
\bctitle{{Time-Dependent Phenomena in OB Star Winds}},
in \bbtitle{Pulsation; Rotation; and Mass Loss in Early-Type Stars},
ed. by \beditor{\binits{L.A.} \bsnm{{Balona}}},
\beditor{\binits{H.F.} \bsnm{{Henrichs}}},
\beditor{\binits{J.M.} \bsnm{{Le Contel}}}
\bsertitle{IAU Symposium},
vol. \bseriesno{162},
\byear{1994},
p. \bfpage{507}
\end{bchapter}
\endbibitem

\bibitem[\protect\citeauthoryear{{Prinja} and {Massa}}{2010}]{pm2010}
\begin{barticle}
\bauthor{\binits{R.K.} \bsnm{{Prinja}}},
\bauthor{\binits{D.L.} \bsnm{{Massa}}},
\batitle{{Signature of wide-spread clumping in B supergiant winds}}.
\bjtitle{\aap}
\bvolume{521},
\bfpage{55}
(\byear{2010}).
doi:\doiurl{10.1051/0004-6361/201015252}
\end{barticle}
\endbibitem

\bibitem[\protect\citeauthoryear{{Prinja} and {Massa}}{2013}]{pm2013}
\begin{barticle}
\bauthor{\binits{R.K.} \bsnm{{Prinja}}},
\bauthor{\binits{D.L.} \bsnm{{Massa}}},
\batitle{{Ultraviolet diagnostic of porosity-free mass-loss estimates in B
  stars}}.
\bjtitle{\aap}
\bvolume{559},
\bfpage{15}
(\byear{2013}).
doi:\doiurl{10.1051/0004-6361/201321799}
\end{barticle}
\endbibitem

\bibitem[\protect\citeauthoryear{{Prinja} and {Smith}}{1992}]{PrinjaSmith92}
\begin{barticle}
\bauthor{\binits{R.K.} \bsnm{{Prinja}}},
\bauthor{\binits{L.J.} \bsnm{{Smith}}},
\batitle{{Migrating optical depth enhancements in the UV wind lines of the
  Wolf-Rayet star HD 93131}}.
\bjtitle{\aap}
\bvolume{266},
\bfpage{377}--\blpage{384}
(\byear{1992})
\end{barticle}
\endbibitem

\bibitem[\protect\citeauthoryear{{Prinja} et~al.}{1990}]{Prinjaetal90}
\begin{barticle}
\bauthor{\binits{R.K.} \bsnm{{Prinja}}},
\bauthor{\binits{M.J.} \bsnm{{Barlow}}},
\bauthor{\binits{I.D.} \bsnm{{Howarth}}},
\batitle{{Terminal velocities for a large sample of O stars, B supergiants, and
  Wolf-Rayet stars}}.
\bjtitle{\apj}
\bvolume{361},
\bfpage{607}--\blpage{620}
(\byear{1990})
\end{barticle}
\endbibitem

\bibitem[\protect\citeauthoryear{{Prinja} et~al.}{1995}]{Prinjaetal95}
\begin{barticle}
\bauthor{\binits{R.K.} \bsnm{{Prinja}}},
\bauthor{\binits{D.} \bsnm{{Massa}}},
\bauthor{\binits{A.W.} \bsnm{{Fullerton}}},
\batitle{{The IUE MEGA Campaign: Modulated Structure in the Wind of HD 64760
  (B0.5 Ib)}}.
\bjtitle{\apjl}
\bvolume{452},
\bfpage{61}
(\byear{1995})
\end{barticle}
\endbibitem

\bibitem[\protect\citeauthoryear{{Prinja} et~al.}{1992}]{Prinjaetal92}
\begin{barticle}
\bauthor{\binits{R.K.} \bsnm{{Prinja}}},
\bauthor{\binits{L.A.} \bsnm{{Balona}}},
\bauthor{\binits{C.T.} \bsnm{{Bolton}}},
\bauthor{\binits{R.A.} \bsnm{{Crowe}}},
\bauthor{\binits{M.S.} \bsnm{{Fieldus}}},
\bauthor{\binits{A.W.} \bsnm{{Fullerton}}},
\bauthor{\binits{D.R.} \bsnm{{Gies}}},
\bauthor{\binits{I.D.} \bsnm{{Howarth}}},
\bauthor{\binits{D.} \bsnm{{McDavid}}},
\bauthor{\binits{A.H.N.} \bsnm{{Reid}}},
\batitle{{Time series observations of O stars. I - IUE observations of
  variability in the stellar wind of Zeta Puppis}}.
\bjtitle{\apj}
\bvolume{390},
\bfpage{266}--\blpage{272}
(\byear{1992})
\end{barticle}
\endbibitem

\bibitem[\protect\citeauthoryear{{Przybilla} et~al.}{2008}]{Przybilla08}
\begin{barticle}
\bauthor{\binits{N.} \bsnm{{Przybilla}}},
\bauthor{\binits{M.} \bsnm{{Nieva}}},
\bauthor{\binits{K.} \bsnm{{Butler}}},
\batitle{{A Cosmic Abundance Standard: Chemical Homogeneity of the Solar
  Neighborhood and the ISM Dust-Phase Composition}}.
\bjtitle{\apjl}
\bvolume{688},
\bfpage{103}--\blpage{106}
(\byear{2008}).
doi:\doiurl{10.1086/595618}
\end{barticle}
\endbibitem

\bibitem[\protect\citeauthoryear{{Puls}}{1987}]{Puls87}
\begin{barticle}
\bauthor{\binits{J.} \bsnm{{Puls}}},
\batitle{{Radiation-driven winds of hot luminous stars. IV - The influence of
  multi-line effects}}.
\bjtitle{\aap}
\bvolume{184},
\bfpage{227}--\blpage{248}
(\byear{1987})
\end{barticle}
\endbibitem

\bibitem[\protect\citeauthoryear{{Puls} et~al.}{2000}]{Puls00}
\begin{barticle}
\bauthor{\binits{J.} \bsnm{{Puls}}},
\bauthor{\binits{U.} \bsnm{{Springmann}}},
\bauthor{\binits{M.} \bsnm{{Lennon}}},
\batitle{{Radiation driven winds of hot luminous stars. XIV. Line statistics
  and radiative driving}}.
\bjtitle{\aaps}
\bvolume{141},
\bfpage{23}--\blpage{64}
(\byear{2000})
\end{barticle}
\endbibitem

\bibitem[\protect\citeauthoryear{{Puls} et~al.}{2010}]{Puls10}
\begin{botherref}
\oauthor{\binits{J.} \bsnm{{Puls}}},
\oauthor{\binits{J.O.} \bsnm{{Sundqvist}}},
\oauthor{\binits{J.G.} \bsnm{{Rivero Gonz{\'a}lez}}},
{OB-stars as extreme condition test beds}.
ArXiv e-prints
(2010)
\end{botherref}
\endbibitem

\bibitem[\protect\citeauthoryear{{Puls} et~al.}{2008}]{Puls08}
\begin{barticle}
\bauthor{\binits{J.} \bsnm{{Puls}}},
\bauthor{\binits{J.S.} \bsnm{{Vink}}},
\bauthor{\binits{F.} \bsnm{{Najarro}}},
\batitle{{Mass loss from hot massive stars}}.
\bjtitle{\aapr}
\bvolume{16},
\bfpage{209}--\blpage{325}
(\byear{2008})
\end{barticle}
\endbibitem

\bibitem[\protect\citeauthoryear{{Puls} et~al.}{1996}]{Puls96}
\begin{barticle}
\bauthor{\binits{J.} \bsnm{{Puls}}},
\bauthor{\binits{R.-P.} \bsnm{{Kudritzki}}},
\bauthor{\binits{A.} \bsnm{{Herrero}}},
\bauthor{\binits{A.W.A.} \bsnm{{Pauldrach}}},
\bauthor{\binits{S.M.} \bsnm{{Haser}}},
\bauthor{\binits{D.J.} \bsnm{{Lennon}}},
\bauthor{\binits{R.} \bsnm{{Gabler}}},
\bauthor{\binits{S.A.} \bsnm{{Voels}}},
\bauthor{\binits{J.M.} \bsnm{{Vilchez}}},
\bauthor{\binits{S.} \bsnm{{Wachter}}},
\bauthor{\binits{A.} \bsnm{{Feldmeier}}},
\batitle{{O-star mass-loss and wind momentum rates in the Galaxy and the
  Magellanic Clouds Observations and theoretical predictions.}}
\bjtitle{\aap}
\bvolume{305},
\bfpage{171}
(\byear{1996})
\end{barticle}
\endbibitem

\bibitem[\protect\citeauthoryear{{Puls} et~al.}{2005}]{Puls05}
\begin{barticle}
\bauthor{\binits{J.} \bsnm{{Puls}}},
\bauthor{\binits{M.A.} \bsnm{{Urbaneja}}},
\bauthor{\binits{R.} \bsnm{{Venero}}},
\bauthor{\binits{T.} \bsnm{{Repolust}}},
\bauthor{\binits{U.} \bsnm{{Springmann}}},
\bauthor{\binits{A.} \bsnm{{Jokuthy}}},
\bauthor{\binits{M.R.} \bsnm{{Mokiem}}},
\batitle{{Atmospheric NLTE-models for the spectroscopic analysis of blue stars
  with winds. II. Line-blanketed models}}.
\bjtitle{\aap}
\bvolume{435},
\bfpage{669}--\blpage{698}
(\byear{2005})
\end{barticle}
\endbibitem

\bibitem[\protect\citeauthoryear{{Puls} et~al.}{2006}]{puls2006}
\begin{barticle}
\bauthor{\binits{J.} \bsnm{{Puls}}},
\bauthor{\binits{N.} \bsnm{{Markova}}},
\bauthor{\binits{S.} \bsnm{{Scuderi}}},
\bauthor{\binits{C.} \bsnm{{Stanghellini}}},
\bauthor{\binits{O.G.} \bsnm{{Taranova}}},
\bauthor{\binits{A.W.} \bsnm{{Burnley}}},
\bauthor{\binits{I.D.} \bsnm{{Howarth}}},
\batitle{{Bright OB stars in the Galaxy. III. Constraints on the radial
  stratification of the clumping factor in hot star winds from a combined
  H$_{α}$, IR and radio analysis}}.
\bjtitle{\aap}
\bvolume{454},
\bfpage{625}--\blpage{651}
(\byear{2006}).
doi:\doiurl{10.1051/0004-6361:20065073}
\end{barticle}
\endbibitem

\bibitem[\protect\citeauthoryear{{Quaintrell} et~al.}{2003}]{Quaintrell2003}
\begin{barticle}
\bauthor{\binits{H.} \bsnm{{Quaintrell}}},
\bauthor{\binits{A.J.} \bsnm{{Norton}}},
\bauthor{\binits{T.D.C.} \bsnm{{Ash}}},
\bauthor{\binits{P.} \bsnm{{Roche}}},
\bauthor{\binits{B.} \bsnm{{Willems}}},
\bauthor{\binits{T.R.} \bsnm{{Bedding}}},
\bauthor{\binits{I.K.} \bsnm{{Baldry}}},
\bauthor{\binits{R.P.} \bsnm{{Fender}}},
\batitle{{The mass of the neutron star in Vela X-1 and tidally induced
  non-radial oscillations in GP Vel}}.
\bjtitle{\aap}
\bvolume{401},
\bfpage{313}--\blpage{323}
(\byear{2003}).
doi:\doiurl{10.1051/0004-6361:20030120}
\end{barticle}
\endbibitem

\bibitem[\protect\citeauthoryear{{Rahoui} and {Chaty}}{2008}]{RahouiChaty2008}
\begin{barticle}
\bauthor{\binits{F.} \bsnm{{Rahoui}}},
\bauthor{\binits{S.} \bsnm{{Chaty}}},
\batitle{{IGR J18483-0311: a new intermediate supergiant fast X-ray
  transient}}.
\bjtitle{\aap}
\bvolume{492},
\bfpage{163}--\blpage{166}
(\byear{2008}).
doi:\doiurl{10.1051/0004-6361:200810695}
\end{barticle}
\endbibitem

\bibitem[\protect\citeauthoryear{{Rahoui} et~al.}{2008}]{Rahoui2008}
\begin{barticle}
\bauthor{\binits{F.} \bsnm{{Rahoui}}},
\bauthor{\binits{S.} \bsnm{{Chaty}}},
\bauthor{\binits{P.-O.} \bsnm{{Lagage}}},
\bauthor{\binits{E.} \bsnm{{Pantin}}},
\batitle{{Multi-wavelength observations of Galactic hard X-ray sources
  discovered by INTEGRAL. II. The environment of the companion star}}.
\bjtitle{\aap}
\bvolume{484},
\bfpage{801}--\blpage{813}
(\byear{2008}).
doi:\doiurl{10.1051/0004-6361:20078774}
\end{barticle}
\endbibitem

\bibitem[\protect\citeauthoryear{{Rampy} et~al.}{2009}]{Rampy2009}
\begin{barticle}
\bauthor{\binits{R.A.} \bsnm{{Rampy}}},
\bauthor{\binits{D.M.} \bsnm{{Smith}}},
\bauthor{\binits{I.} \bsnm{{Negueruela}}},
\batitle{{IGR J17544-2619 in Depth With Suzaku: Direct Evidence for Clumpy
  Winds in a Supergiant Fast X-ray Transient}}.
\bjtitle{\apj}
\bvolume{707},
\bfpage{243}--\blpage{249}
(\byear{2009}).
doi:\doiurl{10.1088/0004-637X/707/1/243}
\end{barticle}
\endbibitem

\bibitem[\protect\citeauthoryear{{Rauw} et~al.}{2015}]{rauw2015}
\begin{botherref}
\oauthor{\binits{G.} \bsnm{{Rauw}}},
\oauthor{\binits{A.} \bsnm{{Herve}}},
\oauthor{\binits{Y.} \bsnm{{Naze}}},
\oauthor{\binits{J.N.} \bsnm{{Gonzalez-Perez}}},
\oauthor{\binits{A.} \bsnm{{Hempelmann}}},
\oauthor{\binits{M.} \bsnm{{Mittag}}},
\oauthor{\binits{J.H.M.M.} \bsnm{{Schmitt}}},
\oauthor{\binits{K.-P.} \bsnm{{Schroeder}}},
\oauthor{\binits{E.} \bsnm{{Gosset}}},
\oauthor{\binits{P.} \bsnm{{Eenens}}},
\oauthor{\binits{J.M.} \bsnm{{Uuh-Sonda}}},
{Simultaneous X-ray and optical spectroscopy of the Oef supergiant lambda Cep}.
ArXiv e-prints
(2015)
\end{botherref}
\endbibitem

\bibitem[\protect\citeauthoryear{{Ray} and {Chakrabarty}}{2002}]{Ray2002}
\begin{barticle}
\bauthor{\binits{P.S.} \bsnm{{Ray}}},
\bauthor{\binits{D.} \bsnm{{Chakrabarty}}},
\batitle{{The Orbit of the High-Mass X-Ray Binary Pulsar 1E 1145.1-6141}}.
\bjtitle{\apj}
\bvolume{581},
\bfpage{1293}--\blpage{1296}
(\byear{2002}).
doi:\doiurl{10.1086/344300}
\end{barticle}
\endbibitem

\bibitem[\protect\citeauthoryear{{Rea} and {Esposito}}{2011}]{rea2011}
\begin{bchapter}
\bauthor{\binits{N.} \bsnm{{Rea}}},
\bauthor{\binits{P.} \bsnm{{Esposito}}},
\bctitle{{Magnetar outbursts: an observational review}},
in \bbtitle{High-Energy Emission from Pulsars and their Systems},
ed. by \beditor{\binits{D.F.} \bsnm{{Torres}}},
\beditor{\binits{N.} \bsnm{{Rea}}},
\byear{2011},
p. \bfpage{247}.
doi:\doiurl{10.1007/978-3-642-17251-9\_21}
\end{bchapter}
\endbibitem

\bibitem[\protect\citeauthoryear{{Reig} et~al.}{1996}]{Reig1996}
\begin{barticle}
\bauthor{\binits{P.} \bsnm{{Reig}}},
\bauthor{\binits{D.} \bsnm{{Chakrabarty}}},
\bauthor{\binits{M.J.} \bsnm{{Coe}}},
\bauthor{\binits{J.} \bsnm{{Fabregat}}},
\bauthor{\binits{I.} \bsnm{{Negueruela}}},
\bauthor{\binits{T.A.} \bsnm{{Prince}}},
\bauthor{\binits{P.} \bsnm{{Roche}}},
\bauthor{\binits{I.A.} \bsnm{{Steele}}},
\batitle{{Astrophysical parameters of the massive X-ray binary 2S 0114+650.}}
\bjtitle{\aap}
\bvolume{311},
\bfpage{879}--\blpage{888}
(\byear{1996})
\end{barticle}
\endbibitem

\bibitem[\protect\citeauthoryear{{Repolust} et~al.}{2004}]{Repolust04}
\begin{barticle}
\bauthor{\binits{T.} \bsnm{{Repolust}}},
\bauthor{\binits{J.} \bsnm{{Puls}}},
\bauthor{\binits{A.} \bsnm{{Herrero}}},
\batitle{{Stellar and wind parameters of Galactic O-stars. The influence of
  line-blocking/blanketing}}.
\bjtitle{\aap}
\bvolume{415},
\bfpage{349}--\blpage{376}
(\byear{2004})
\end{barticle}
\endbibitem

\bibitem[\protect\citeauthoryear{{Reynolds} et~al.}{1992}]{Reynolds1992}
\begin{barticle}
\bauthor{\binits{A.P.} \bsnm{{Reynolds}}},
\bauthor{\binits{S.A.} \bsnm{{Bell}}},
\bauthor{\binits{R.W.} \bsnm{{Hilditch}}},
\batitle{{Optical spectroscopy of the massive X-ray binary QV Nor (4U 1538 -
  52)}}.
\bjtitle{\mnras}
\bvolume{256},
\bfpage{631}--\blpage{640}
(\byear{1992})
\end{barticle}
\endbibitem

\bibitem[\protect\citeauthoryear{{Rivero Gonz{\'a}lez} et~al.}{2012}]{Rivero12}
\begin{barticle}
\bauthor{\binits{J.G.} \bsnm{{Rivero Gonz{\'a}lez}}},
\bauthor{\binits{J.} \bsnm{{Puls}}},
\bauthor{\binits{F.} \bsnm{{Najarro}}},
\bauthor{\binits{I.} \bsnm{{Brott}}},
\batitle{{Nitrogen line spectroscopy of O-stars. II. Surface nitrogen
  abundances for O-stars in the Large Magellanic Cloud}}.
\bjtitle{\aap}
\bvolume{537},
\bfpage{79}
(\byear{2012}).
doi:\doiurl{10.1051/0004-6361/201117790}
\end{barticle}
\endbibitem

\bibitem[\protect\citeauthoryear{{Rivers} et~al.}{2010}]{Rivers:2010}
\begin{barticle}
\bauthor{\binits{E.} \bsnm{{Rivers}}},
\bauthor{\binits{A.} \bsnm{{Markowitz}}},
\bauthor{\binits{K.} \bsnm{{Pottschmidt}}},
\bauthor{\binits{S.} \bsnm{{Roth}}},
\bauthor{\binits{L.} \bsnm{{Barrag{\'a}n}}},
\bauthor{\binits{F.} \bsnm{{F{\"u}rst}}},
\bauthor{\binits{S.} \bsnm{{Suchy}}},
\bauthor{\binits{I.} \bsnm{{Kreykenbohm}}},
\bauthor{\binits{J.} \bsnm{{Wilms}}},
\bauthor{\binits{R.} \bsnm{{Rothschild}}},
\batitle{{A Comprehensive Spectral Analysis of the X-Ray Pulsar 4U 1907+09 from
  Two Observations with the Suzaku X-ray Observatory}}.
\bjtitle{\apj}
\bvolume{709},
\bfpage{179}--\blpage{190}
(\byear{2010}).
doi:\doiurl{10.1088/0004-637X/709/1/179}.
\burl{http://cdsads.u-strasbg.fr/abs/2010ApJ...709..179R}
\end{barticle}
\endbibitem

\bibitem[\protect\citeauthoryear{{Roberts} et~al.}{2001}]{Roberts:2001}
\begin{barticle}
\bauthor{\binits{M.S.E.} \bsnm{{Roberts}}},
\bauthor{\binits{P.F.} \bsnm{{Michelson}}},
\bauthor{\binits{D.A.} \bsnm{{Leahy}}},
\bauthor{\binits{T.A.} \bsnm{{Hall}}},
\bauthor{\binits{J.P.} \bsnm{{Finley}}},
\bauthor{\binits{L.R.} \bsnm{{Cominsky}}},
\bauthor{\binits{R.} \bsnm{{Srinivasan}}},
\batitle{{Phase-dependent Spectral Variability in 4U 1907+09}}.
\bjtitle{\apj}
\bvolume{555},
\bfpage{967}--\blpage{977}
(\byear{2001}).
doi:\doiurl{10.1086/321487}
\end{barticle}
\endbibitem

\bibitem[\protect\citeauthoryear{{Rodriguez} et~al.}{2006}]{Rodriguez2006}
\begin{barticle}
\bauthor{\binits{J.} \bsnm{{Rodriguez}}},
\bauthor{\binits{A.} \bsnm{{Bodaghee}}},
\bauthor{\binits{P.} \bsnm{{Kaaret}}},
\bauthor{\binits{J.A.} \bsnm{{Tomsick}}},
\bauthor{\binits{E.} \bsnm{{Kuulkers}}},
\bauthor{\binits{G.} \bsnm{{Malaguti}}},
\bauthor{\binits{P.-O.} \bsnm{{Petrucci}}},
\bauthor{\binits{C.} \bsnm{{Cabanac}}},
\bauthor{\binits{M.} \bsnm{{Chernyakova}}},
\bauthor{\binits{S.} \bsnm{{Corbel}}},
\bauthor{\binits{S.} \bsnm{{Deluit}}},
\bauthor{\binits{G.} \bsnm{{Di Cocco}}},
\bauthor{\binits{K.} \bsnm{{Ebisawa}}},
\bauthor{\binits{A.} \bsnm{{Goldwurm}}},
\bauthor{\binits{G.} \bsnm{{Henri}}},
\bauthor{\binits{F.} \bsnm{{Lebrun}}},
\bauthor{\binits{A.} \bsnm{{Paizis}}},
\bauthor{\binits{R.} \bsnm{{Walter}}},
\bauthor{\binits{L.} \bsnm{{Foschini}}},
\batitle{{INTEGRAL and XMM-Newton observations of the X-ray pulsar IGR
  J16320-4751/AX J1631.9-4752}}.
\bjtitle{MNRAS}
\bvolume{366},
\bfpage{274}--\blpage{282}
(\byear{2006})
\end{barticle}
\endbibitem

\bibitem[\protect\citeauthoryear{{Romano} et~al.}{2009a}]{Romano2009b}
\begin{barticle}
\bauthor{\binits{P.} \bsnm{{Romano}}},
\bauthor{\binits{L.} \bsnm{{Sidoli}}},
\bauthor{\binits{G.} \bsnm{{Cusumano}}},
\bauthor{\binits{S.} \bsnm{{Vercellone}}},
\bauthor{\binits{V.} \bsnm{{Mangano}}},
\bauthor{\binits{H.A.} \bsnm{{Krimm}}},
\batitle{{Disentangling the System Geometry of the Supergiant Fast X-Ray
  Transient IGR J11215-5952 with Swift}}.
\bjtitle{\apj}
\bvolume{696},
\bfpage{2068}--\blpage{2074}
(\byear{2009}a).
doi:\doiurl{10.1088/0004-637X/696/2/2068}
\end{barticle}
\endbibitem

\bibitem[\protect\citeauthoryear{{Romano} et~al.}{2009b}]{Romano2009}
\begin{barticle}
\bauthor{\binits{P.} \bsnm{{Romano}}},
\bauthor{\binits{L.} \bsnm{{Sidoli}}},
\bauthor{\binits{G.} \bsnm{{Cusumano}}},
\bauthor{\binits{V.} \bsnm{{La Parola}}},
\bauthor{\binits{S.} \bsnm{{Vercellone}}},
\bauthor{\binits{C.} \bsnm{{Pagani}}},
\bauthor{\binits{L.} \bsnm{{Ducci}}},
\bauthor{\binits{V.} \bsnm{{Mangano}}},
\bauthor{\binits{J.} \bsnm{{Cummings}}},
\bauthor{\binits{H.A.} \bsnm{{Krimm}}},
\bauthor{\binits{C.} \bsnm{{Guidorzi}}},
\bauthor{\binits{J.A.} \bsnm{{Kennea}}},
\bauthor{\binits{E.A.} \bsnm{{Hoversten}}},
\bauthor{\binits{D.N.} \bsnm{{Burrows}}},
\bauthor{\binits{N.} \bsnm{{Gehrels}}},
\batitle{{Monitoring supergiant fast X-ray transients with Swift: results from
  the first year}}.
\bjtitle{\mnras}
\bvolume{399},
\bfpage{2021}--\blpage{2032}
(\byear{2009}b).
doi:\doiurl{10.1111/j.1365-2966.2009.15356.x}
\end{barticle}
\endbibitem

\bibitem[\protect\citeauthoryear{{Romano} et~al.}{2010}]{Romano2010}
\begin{barticle}
\bauthor{\binits{P.} \bsnm{{Romano}}},
\bauthor{\binits{L.} \bsnm{{Sidoli}}},
\bauthor{\binits{L.} \bsnm{{Ducci}}},
\bauthor{\binits{G.} \bsnm{{Cusumano}}},
\bauthor{\binits{V.} \bsnm{{La Parola}}},
\bauthor{\binits{C.} \bsnm{{Pagani}}},
\bauthor{\binits{K.L.} \bsnm{{Page}}},
\bauthor{\binits{J.A.} \bsnm{{Kennea}}},
\bauthor{\binits{D.N.} \bsnm{{Burrows}}},
\bauthor{\binits{N.} \bsnm{{Gehrels}}},
\bauthor{\binits{V.} \bsnm{{Sguera}}},
\bauthor{\binits{A.} \bsnm{{Bazzano}}},
\batitle{{Swift/XRT monitoring of the supergiant fast X-ray transient IGR
  J18483-0311 for an entire orbital period}}.
\bjtitle{\mnras}
\bvolume{401},
\bfpage{1564}--\blpage{1569}
(\byear{2010}).
doi:\doiurl{10.1111/j.1365-2966.2009.15789.x}
\end{barticle}
\endbibitem

\bibitem[\protect\citeauthoryear{{Romano} et~al.}{2011}]{Romano2011}
\begin{barticle}
\bauthor{\binits{P.} \bsnm{{Romano}}},
\bauthor{\binits{V.} \bsnm{{La Parola}}},
\bauthor{\binits{S.} \bsnm{{Vercellone}}},
\bauthor{\binits{G.} \bsnm{{Cusumano}}},
\bauthor{\binits{L.} \bsnm{{Sidoli}}},
\bauthor{\binits{H.A.} \bsnm{{Krimm}}},
\bauthor{\binits{C.} \bsnm{{Pagani}}},
\bauthor{\binits{P.} \bsnm{{Esposito}}},
\bauthor{\binits{E.A.} \bsnm{{Hoversten}}},
\bauthor{\binits{J.A.} \bsnm{{Kennea}}},
\bauthor{\binits{K.L.} \bsnm{{Page}}},
\bauthor{\binits{D.N.} \bsnm{{Burrows}}},
\bauthor{\binits{N.} \bsnm{{Gehrels}}},
\batitle{{Two years of monitoring supergiant fast X-ray transients with
  Swift}}.
\bjtitle{\mnras}
\bvolume{410},
\bfpage{1825}--\blpage{1836}
(\byear{2011}).
doi:\doiurl{10.1111/j.1365-2966.2010.17564.x}
\end{barticle}
\endbibitem

\bibitem[\protect\citeauthoryear{{Romano} et~al.}{2012}]{Romano2012}
\begin{barticle}
\bauthor{\binits{P.} \bsnm{{Romano}}},
\bauthor{\binits{V.} \bsnm{{Mangano}}},
\bauthor{\binits{L.} \bsnm{{Ducci}}},
\bauthor{\binits{P.} \bsnm{{Esposito}}},
\bauthor{\binits{P.A.} \bsnm{{Evans}}},
\bauthor{\binits{S.} \bsnm{{Vercellone}}},
\bauthor{\binits{J.A.} \bsnm{{Kennea}}},
\bauthor{\binits{D.N.} \bsnm{{Burrows}}},
\bauthor{\binits{N.} \bsnm{{Gehrels}}},
\batitle{{Swift/X-ray Telescope monitoring of the candidate supergiant fast
  X-ray transient IGR J16418-4532}}.
\bjtitle{\mnras}
\bvolume{419},
\bfpage{2695}--\blpage{2702}
(\byear{2012}).
doi:\doiurl{10.1111/j.1365-2966.2011.19916.x}
\end{barticle}
\endbibitem

\bibitem[\protect\citeauthoryear{{Romano} et~al.}{2014a}]{Romano2014b}
\begin{barticle}
\bauthor{\binits{P.} \bsnm{{Romano}}},
\bauthor{\binits{L.} \bsnm{{Ducci}}},
\bauthor{\binits{V.} \bsnm{{Mangano}}},
\bauthor{\binits{P.} \bsnm{{Esposito}}},
\bauthor{\binits{E.} \bsnm{{Bozzo}}},
\bauthor{\binits{S.} \bsnm{{Vercellone}}},
\batitle{{Soft X-ray characterisation of the long-term properties of supergiant
  fast X-ray transients}}.
\bjtitle{\aap}
\bvolume{568},
\bfpage{55}
(\byear{2014}a).
doi:\doiurl{10.1051/0004-6361/201423867}
\end{barticle}
\endbibitem

\bibitem[\protect\citeauthoryear{{Romano} et~al.}{2014b}]{Romano2014}
\begin{barticle}
\bauthor{\binits{P.} \bsnm{{Romano}}},
\bauthor{\binits{H.A.} \bsnm{{Krimm}}},
\bauthor{\binits{D.M.} \bsnm{{Palmer}}},
\bauthor{\binits{L.} \bsnm{{Ducci}}},
\bauthor{\binits{P.} \bsnm{{Esposito}}},
\bauthor{\binits{S.} \bsnm{{Vercellone}}},
\bauthor{\binits{P.A.} \bsnm{{Evans}}},
\bauthor{\binits{C.} \bsnm{{Guidorzi}}},
\bauthor{\binits{V.} \bsnm{{Mangano}}},
\bauthor{\binits{J.A.} \bsnm{{Kennea}}},
\bauthor{\binits{S.D.} \bsnm{{Barthelmy}}},
\bauthor{\binits{D.N.} \bsnm{{Burrows}}},
\bauthor{\binits{N.} \bsnm{{Gehrels}}},
\batitle{{The 100-month Swift catalogue of supergiant fast X-ray transients. I.
  BAT on-board and transient monitor flares}}.
\bjtitle{\aap}
\bvolume{562},
\bfpage{2}
(\byear{2014}b).
doi:\doiurl{10.1051/0004-6361/201322516}
\end{barticle}
\endbibitem

\bibitem[\protect\citeauthoryear{{Romano} et~al.}{2015}]{Romano2015}
\begin{barticle}
\bauthor{\binits{P.} \bsnm{{Romano}}},
\bauthor{\binits{E.} \bsnm{{Bozzo}}},
\bauthor{\binits{V.} \bsnm{{Mangano}}},
\bauthor{\binits{P.} \bsnm{{Esposito}}},
\bauthor{\binits{G.} \bsnm{{Israel}}},
\bauthor{\binits{A.} \bsnm{{Tiengo}}},
\bauthor{\binits{S.} \bsnm{{Campana}}},
\bauthor{\binits{L.} \bsnm{{Ducci}}},
\bauthor{\binits{C.} \bsnm{{Ferrigno}}},
\bauthor{\binits{J.A.} \bsnm{{Kennea}}},
\batitle{{Giant outburst from the supergiant fast X-ray transient IGR
  J17544-2619: accretion from a transient disc?}}
\bjtitle{\aap}
\bvolume{576},
\bfpage{4}
(\byear{2015}).
doi:\doiurl{10.1051/0004-6361/201525749}
\end{barticle}
\endbibitem

\bibitem[\protect\citeauthoryear{{Romanova} et~al.}{2003}]{Romanova2003}
\begin{barticle}
\bauthor{\binits{M.M.} \bsnm{{Romanova}}},
\bauthor{\binits{O.D.} \bsnm{{Toropina}}},
\bauthor{\binits{Y.M.} \bsnm{{Toropin}}},
\bauthor{\binits{R.V.E.} \bsnm{{Lovelace}}},
\batitle{{Magnetohydrodynamic Simulations of Accretion onto a Star in the
  ``Propeller'' Regime}}.
\bjtitle{\apj}
\bvolume{588},
\bfpage{400}--\blpage{407}
(\byear{2003}).
doi:\doiurl{10.1086/373990}
\end{barticle}
\endbibitem

\bibitem[\protect\citeauthoryear{{Romanova} et~al.}{2012}]{Romanova2012}
\begin{barticle}
\bauthor{\binits{M.M.} \bsnm{{Romanova}}},
\bauthor{\binits{G.V.} \bsnm{{Ustyugova}}},
\bauthor{\binits{A.V.} \bsnm{{Koldoba}}},
\bauthor{\binits{R.V.E.} \bsnm{{Lovelace}}},
\batitle{{MRI-driven accretion on to magnetized stars: global 3D MHD
  simulations of magnetospheric and boundary layer regimes}}.
\bjtitle{\mnras}
\bvolume{421},
\bfpage{63}--\blpage{77}
(\byear{2012}).
doi:\doiurl{10.1111/j.1365-2966.2011.20055.x}
\end{barticle}
\endbibitem

\bibitem[\protect\citeauthoryear{{Runacres} and {Owocki}}{2002}]{Runacres02}
\begin{barticle}
\bauthor{\binits{M.C.} \bsnm{{Runacres}}},
\bauthor{\binits{S.P.} \bsnm{{Owocki}}},
\batitle{{The outer evolution of instability-generated structure in radiatively
  driven stellar winds}}.
\bjtitle{\aap}
\bvolume{381},
\bfpage{1015}--\blpage{1025}
(\byear{2002}).
doi:\doiurl{10.1051/0004-6361:20011526}
\end{barticle}
\endbibitem

\bibitem[\protect\citeauthoryear{{Rybicki} et~al.}{1990}]{Rybicki90}
\begin{barticle}
\bauthor{\binits{G.B.} \bsnm{{Rybicki}}},
\bauthor{\binits{S.P.} \bsnm{{Owocki}}},
\bauthor{\binits{J.I.} \bsnm{{Castor}}},
\batitle{{Instabilities in line-driven stellar winds. IV - Linear perturbations
  in three dimensions}}.
\bjtitle{\apj}
\bvolume{349},
\bfpage{274}--\blpage{285}
(\byear{1990}).
doi:\doiurl{10.1086/168312}
\end{barticle}
\endbibitem

\bibitem[\protect\citeauthoryear{{Sako} et~al.}{2003}]{sako2003}
\begin{botherref}
\oauthor{\binits{M.} \bsnm{{Sako}}},
\oauthor{\binits{S.M.} \bsnm{{Kahn}}},
\oauthor{\binits{F.} \bsnm{{Paerels}}},
\oauthor{\binits{D.A.} \bsnm{{Liedahl}}},
\oauthor{\binits{S.} \bsnm{{Watanabe}}},
\oauthor{\binits{F.} \bsnm{{Nagase}}},
\oauthor{\binits{T.} \bsnm{{Takahashi}}},
{Structure and Dynamics of Stellar Winds in High-mass X-ray Binaries}.
ArXiv
\textbf{0309503}
(2003)
\end{botherref}
\endbibitem

\bibitem[\protect\citeauthoryear{{Sander} et~al.}{2012}]{SHT2012}
\begin{barticle}
\bauthor{\binits{A.} \bsnm{{Sander}}},
\bauthor{\binits{W.-R.} \bsnm{{Hamann}}},
\bauthor{\binits{H.} \bsnm{{Todt}}},
\batitle{{The Galactic WC stars. Stellar parameters from spectral analyses
  indicate a new evolutionary sequence}}.
\bjtitle{\aap}
\bvolume{540},
\bfpage{144}
(\byear{2012}).
doi:\doiurl{10.1051/0004-6361/201117830}
\end{barticle}
\endbibitem

\bibitem[\protect\citeauthoryear{{Sch{\"o}nherr}
  et~al.}{2014}]{schoenherr2014a}
\begin{barticle}
\bauthor{\binits{G.} \bsnm{{Sch{\"o}nherr}}},
\bauthor{\binits{F.-W.} \bsnm{{Schwarm}}},
\bauthor{\binits{S.} \bsnm{{Falkner}}},
\bauthor{\binits{T.} \bsnm{{Dauser}}},
\bauthor{\binits{C.} \bsnm{{Ferrigno}}},
\bauthor{\binits{M.} \bsnm{{K{\"u}hnel}}},
\bauthor{\binits{D.} \bsnm{{Klochkov}}},
\bauthor{\binits{P.} \bsnm{{Kretschmar}}},
\bauthor{\binits{P.A.} \bsnm{{Becker}}},
\bauthor{\binits{M.T.} \bsnm{{Wolff}}},
\bauthor{\binits{K.} \bsnm{{Pottschmidt}}},
\bauthor{\binits{M.} \bsnm{{Falanga}}},
\bauthor{\binits{I.} \bsnm{{Kreykenbohm}}},
\bauthor{\binits{F.} \bsnm{{F{\"u}rst}}},
\bauthor{\binits{R.} \bsnm{{Staubert}}},
\bauthor{\binits{J.} \bsnm{{Wilms}}},
\batitle{Formation of phase lags at the cyclotron energies in the pulse
  profiles of magnetized, accreting neutron stars}.
\bjtitle{A\&A}
\bvolume{564},
\bfpage{8}
(\byear{2014})
\end{barticle}
\endbibitem

\bibitem[\protect\citeauthoryear{{Sguera} et~al.}{2005}]{Sguera2005}
\begin{barticle}
\bauthor{\binits{V.} \bsnm{{Sguera}}},
\bauthor{\binits{E.J.} \bsnm{{Barlow}}},
\bauthor{\binits{A.J.} \bsnm{{Bird}}},
\bauthor{\binits{D.J.} \bsnm{{Clark}}},
\bauthor{\binits{A.J.} \bsnm{{Dean}}},
\bauthor{\binits{A.B.} \bsnm{{Hill}}},
\bauthor{\binits{L.} \bsnm{{Moran}}},
\bauthor{\binits{S.E.} \bsnm{{Shaw}}},
\bauthor{\binits{D.R.} \bsnm{{Willis}}},
\bauthor{\binits{A.} \bsnm{{Bazzano}}},
\bauthor{\binits{P.} \bsnm{{Ubertini}}},
\bauthor{\binits{A.} \bsnm{{Malizia}}},
\batitle{{INTEGRAL observations of recurrent fast X-ray transient sources}}.
\bjtitle{\aap}
\bvolume{444},
\bfpage{221}--\blpage{231}
(\byear{2005}).
doi:\doiurl{10.1051/0004-6361:20053103}
\end{barticle}
\endbibitem

\bibitem[\protect\citeauthoryear{{Sguera} et~al.}{2006}]{Sguera2006}
\begin{barticle}
\bauthor{\binits{V.} \bsnm{{Sguera}}},
\bauthor{\binits{A.} \bsnm{{Bazzano}}},
\bauthor{\binits{A.J.} \bsnm{{Bird}}},
\bauthor{\binits{A.J.} \bsnm{{Dean}}},
\bauthor{\binits{P.} \bsnm{{Ubertini}}},
\bauthor{\binits{E.J.} \bsnm{{Barlow}}},
\bauthor{\binits{L.} \bsnm{{Bassani}}},
\bauthor{\binits{D.J.} \bsnm{{Clark}}},
\bauthor{\binits{A.B.} \bsnm{{Hill}}},
\bauthor{\binits{A.} \bsnm{{Malizia}}},
\bauthor{\binits{M.} \bsnm{{Molina}}},
\bauthor{\binits{J.B.} \bsnm{{Stephen}}},
\batitle{{Unveiling Supergiant Fast X-Ray Transient Sources with INTEGRAL}}.
\bjtitle{\apj}
\bvolume{646},
\bfpage{452}--\blpage{463}
(\byear{2006}).
doi:\doiurl{10.1086/504827}
\end{barticle}
\endbibitem

\bibitem[\protect\citeauthoryear{{Sguera} et~al.}{2007}]{Sguera2007}
\begin{barticle}
\bauthor{\binits{V.} \bsnm{{Sguera}}},
\bauthor{\binits{A.B.} \bsnm{{Hill}}},
\bauthor{\binits{A.J.} \bsnm{{Bird}}},
\bauthor{\binits{A.J.} \bsnm{{Dean}}},
\bauthor{\binits{A.} \bsnm{{Bazzano}}},
\bauthor{\binits{P.} \bsnm{{Ubertini}}},
\bauthor{\binits{N.} \bsnm{{Masetti}}},
\bauthor{\binits{R.} \bsnm{{Landi}}},
\bauthor{\binits{A.} \bsnm{{Malizia}}},
\bauthor{\binits{D.J.} \bsnm{{Clark}}},
\bauthor{\binits{M.} \bsnm{{Molina}}},
\batitle{{IGR J18483-0311: an accreting X-ray pulsar observed by INTEGRAL}}.
\bjtitle{\aap}
\bvolume{467},
\bfpage{249}--\blpage{257}
(\byear{2007}).
doi:\doiurl{10.1051/0004-6361:20066762}
\end{barticle}
\endbibitem

\bibitem[\protect\citeauthoryear{{Sguera} et~al.}{2010}]{Sguera2010}
\begin{barticle}
\bauthor{\binits{V.} \bsnm{{Sguera}}},
\bauthor{\binits{L.} \bsnm{{Ducci}}},
\bauthor{\binits{L.} \bsnm{{Sidoli}}},
\bauthor{\binits{A.} \bsnm{{Bazzano}}},
\bauthor{\binits{L.} \bsnm{{Bassani}}},
\batitle{{XMM-Newton and INTEGRAL study of the SFXT IGR J18483-0311 in
  quiescence: hint of a cyclotron emission feature?}}
\bjtitle{\mnras}
\bvolume{402},
\bfpage{49}--\blpage{53}
(\byear{2010}).
doi:\doiurl{10.1111/j.1745-3933.2009.00798.x}
\end{barticle}
\endbibitem

\bibitem[\protect\citeauthoryear{{Sguera} et~al.}{2015}]{Sguera2015}
\begin{barticle}
\bauthor{\binits{V.} \bsnm{{Sguera}}},
\bauthor{\binits{L.} \bsnm{{Sidoli}}},
\bauthor{\binits{A.J.} \bsnm{{Bird}}},
\bauthor{\binits{A.} \bsnm{{Bazzano}}},
\batitle{{INTEGRAL discovery of unusually long broad-band X-ray activity from
  the Supergiant Fast X-ray Transient IGR J18483-0311}}.
\bjtitle{\mnras}
\bvolume{449},
\bfpage{1228}--\blpage{1237}
(\byear{2015}).
doi:\doiurl{10.1093/mnras/stv341}
\end{barticle}
\endbibitem

\bibitem[\protect\citeauthoryear{{Shakura} et~al.}{2013}]{Shakura2013}
\begin{barticle}
\bauthor{\binits{N.} \bsnm{{Shakura}}},
\bauthor{\binits{K.} \bsnm{{Postnov}}},
\bauthor{\binits{L.} \bsnm{{Hjalmarsdotter}}},
\batitle{{On the nature of `off' states in slowly rotating low-luminosity X-ray
  pulsars}}.
\bjtitle{\mnras}
\bvolume{428},
\bfpage{670}--\blpage{677}
(\byear{2013}).
doi:\doiurl{10.1093/mnras/sts062}
\end{barticle}
\endbibitem

\bibitem[\protect\citeauthoryear{{Shakura} et~al.}{2012}]{Shakura2012}
\begin{barticle}
\bauthor{\binits{N.} \bsnm{{Shakura}}},
\bauthor{\binits{K.} \bsnm{{Postnov}}},
\bauthor{\binits{A.} \bsnm{{Kochetkova}}},
\bauthor{\binits{L.} \bsnm{{Hjalmarsdotter}}},
\batitle{{Theory of quasi-spherical accretion in X-ray pulsars}}.
\bjtitle{\mnras}
\bvolume{420},
\bfpage{216}--\blpage{236}
(\byear{2012}).
doi:\doiurl{10.1111/j.1365-2966.2011.20026.x}
\end{barticle}
\endbibitem

\bibitem[\protect\citeauthoryear{{Shakura} et~al.}{2014}]{Shakura2014}
\begin{botherref}
\oauthor{\binits{N.} \bsnm{{Shakura}}},
\oauthor{\binits{k.} \bsnm{{Postnov}}},
\oauthor{\binits{L.} \bsnm{{Sidoli}}},
\oauthor{\binits{A.} \bsnm{{Paizis}}},
{Bright Flares in Supergiant Fast X-ray Transients}.
ArXiv e-prints, MNRAS in press
(2014)
\end{botherref}
\endbibitem

\bibitem[\protect\citeauthoryear{{Shaviv}}{1998}]{shaviv1998}
\begin{barticle}
\bauthor{\binits{N.J.} \bsnm{{Shaviv}}},
\batitle{{The Eddington Luminosity Limit for Multiphased Media}}.
\bjtitle{\apjl}
\bvolume{494},
\bfpage{193}--\blpage{197}
(\byear{1998}).
doi:\doiurl{10.1086/311182}
\end{barticle}
\endbibitem

\bibitem[\protect\citeauthoryear{{Shaviv}}{2000}]{shaviv2000}
\begin{barticle}
\bauthor{\binits{N.J.} \bsnm{{Shaviv}}},
\batitle{{The Porous Atmosphere of {$\eta$} Carinae}}.
\bjtitle{\apjl}
\bvolume{532},
\bfpage{137}--\blpage{140}
(\byear{2000}).
doi:\doiurl{10.1086/312585}
\end{barticle}
\endbibitem

\bibitem[\protect\citeauthoryear{{Shenar} et~al.}{2015}]{shenar2015}
\begin{botherref}
\oauthor{\binits{T.} \bsnm{{Shenar}}},
\oauthor{\binits{L.} \bsnm{{Oskinova}}},
\oauthor{\binits{W.-R.} \bsnm{{Hamann}}},
\oauthor{\binits{M.F.} \bsnm{{Corcoran}}},
\oauthor{\binits{A.F.J.} \bsnm{{Moffat}}},
\oauthor{\binits{H.} \bsnm{{Pablo}}},
\oauthor{\binits{N.D.} \bsnm{{Richardson}}},
\oauthor{\binits{W.L.} \bsnm{{Waldron}}},
\oauthor{\binits{D.P.} \bsnm{{Huenemoerder}}},
\oauthor{\binits{J.} \bsnm{{Ma{\'{\i}}z Apell{\'a}niz}}},
\oauthor{\binits{J.S.} \bsnm{{Nichols}}},
\oauthor{\binits{H.} \bsnm{{Todt}}},
\oauthor{\binits{Y.} \bsnm{{Naz{\'e}}}},
\oauthor{\binits{J.L.} \bsnm{{Hoffman}}},
\oauthor{\binits{A.M.T.} \bsnm{{Pollock}}},
\oauthor{\binits{I.} \bsnm{{Negueruela}}},
{A coordinated X-ray and Optical Campaign of the Nearest Massive Eclipsing
  Binary, \$$\backslash$delta\$ Orionis Aa: IV. A multiwavelength, non-LTE
  spectroscopic analysis}.
ArXiv e-prints
(2015)
\end{botherref}
\endbibitem

\bibitem[\protect\citeauthoryear{{Shull} and {Beckwith}}{1982}]{shull1982a}
\begin{barticle}
\bauthor{\binits{J.M.} \bsnm{{Shull}}},
\bauthor{\binits{S.} \bsnm{{Beckwith}}},
\batitle{Interstellar molecular hydrogen}.
\bjtitle{ARA\&A}
\bvolume{20},
\bfpage{163}--\blpage{190}
(\byear{1982})
\end{barticle}
\endbibitem

\bibitem[\protect\citeauthoryear{{Sidoli}}{2012}]{Sidoli2012}
\begin{bchapter}
\bauthor{\binits{L.} \bsnm{{Sidoli}}},
\bctitle{{Supergiant Fast X-ray Transients: a review}},
in \bbtitle{Proceedings of ''An INTEGRAL view of the high-energy sky (the first
  10 years)'' - 9th INTEGRAL Workshop and celebration of the 10th anniversary
  of the launch (INTEGRAL 2012). 15-19 October 2012. Bibliotheque Nationale de
  France, Paris, France. Published online at <A
  href=''http://pos.sissa.it/cgi-bin/reader/conf.cgi?confid=176''>http://pos.sissa.it/cgi-bin/reader/conf.cgi?confid=176</A>,
  id.11},
\byear{2012},
p. \bfpage{11}
\end{bchapter}
\endbibitem

\bibitem[\protect\citeauthoryear{{Sidoli} et~al.}{2010}]{Sidoli2010}
\begin{barticle}
\bauthor{\binits{L.} \bsnm{{Sidoli}}},
\bauthor{\binits{P.} \bsnm{{Esposito}}},
\bauthor{\binits{L.} \bsnm{{Ducci}}},
\batitle{{The longest observation of a low-intensity state from a supergiant
  fast X-ray transient: Suzaku observes IGRJ08408-4503}}.
\bjtitle{\mnras}
\bvolume{409},
\bfpage{611}--\blpage{618}
(\byear{2010}).
doi:\doiurl{10.1111/j.1365-2966.2010.17320.x}
\end{barticle}
\endbibitem

\bibitem[\protect\citeauthoryear{{Sidoli} et~al.}{2006}]{Sidoli2006}
\begin{barticle}
\bauthor{\binits{L.} \bsnm{{Sidoli}}},
\bauthor{\binits{A.} \bsnm{{Paizis}}},
\bauthor{\binits{S.} \bsnm{{Mereghetti}}},
\batitle{{IGR J11215-5952: a hard X-ray transient displaying recurrent
  outbursts}}.
\bjtitle{\aap}
\bvolume{450},
\bfpage{9}--\blpage{12}
(\byear{2006}).
doi:\doiurl{10.1051/0004-6361:20064940}
\end{barticle}
\endbibitem

\bibitem[\protect\citeauthoryear{{Sidoli} et~al.}{2007}]{Sidoli2007}
\begin{barticle}
\bauthor{\binits{L.} \bsnm{{Sidoli}}},
\bauthor{\binits{P.} \bsnm{{Romano}}},
\bauthor{\binits{S.} \bsnm{{Mereghetti}}},
\bauthor{\binits{A.} \bsnm{{Paizis}}},
\bauthor{\binits{S.} \bsnm{{Vercellone}}},
\bauthor{\binits{V.} \bsnm{{Mangano}}},
\bauthor{\binits{D.} \bsnm{{G{\"o}tz}}},
\batitle{{An alternative hypothesis for the outburst mechanism in supergiant
  fast X-ray transients: the case of IGR J11215-5952}}.
\bjtitle{\aap}
\bvolume{476},
\bfpage{1307}--\blpage{1315}
(\byear{2007}).
doi:\doiurl{10.1051/0004-6361:20078137}
\end{barticle}
\endbibitem

\bibitem[\protect\citeauthoryear{{Sidoli} et~al.}{2008}]{Sidoli2008}
\begin{barticle}
\bauthor{\binits{L.} \bsnm{{Sidoli}}},
\bauthor{\binits{P.} \bsnm{{Romano}}},
\bauthor{\binits{V.} \bsnm{{Mangano}}},
\bauthor{\binits{A.} \bsnm{{Pellizzoni}}},
\bauthor{\binits{J.A.} \bsnm{{Kennea}}},
\bauthor{\binits{G.} \bsnm{{Cusumano}}},
\bauthor{\binits{S.} \bsnm{{Vercellone}}},
\bauthor{\binits{A.} \bsnm{{Paizis}}},
\bauthor{\binits{D.N.} \bsnm{{Burrows}}},
\bauthor{\binits{N.} \bsnm{{Gehrels}}},
\batitle{{Monitoring Supergiant Fast X-ray Transients with Swift. I. Behavior
  outside outbursts}}.
\bjtitle{\apj}
\bvolume{687},
\bfpage{1230}--\blpage{1235}
(\byear{2008})
\end{barticle}
\endbibitem

\bibitem[\protect\citeauthoryear{{Sidoli} et~al.}{2012}]{Sidoli2012b}
\begin{barticle}
\bauthor{\binits{L.} \bsnm{{Sidoli}}},
\bauthor{\binits{S.} \bsnm{{Mereghetti}}},
\bauthor{\binits{V.} \bsnm{{Sguera}}},
\bauthor{\binits{F.} \bsnm{{Pizzolato}}},
\batitle{{The XMM-Newton view of supergiant fast X-ray transients: the case of
  IGR J16418-4532}}.
\bjtitle{\mnras}
\bvolume{420},
\bfpage{554}--\blpage{561}
(\byear{2012}).
doi:\doiurl{10.1111/j.1365-2966.2011.20063.x}
\end{barticle}
\endbibitem

\bibitem[\protect\citeauthoryear{{Sidoli} et~al.}{2013}]{Sidoli2013b}
\begin{barticle}
\bauthor{\binits{L.} \bsnm{{Sidoli}}},
\bauthor{\binits{P.} \bsnm{{Esposito}}},
\bauthor{\binits{V.} \bsnm{{Sguera}}},
\bauthor{\binits{A.} \bsnm{{Bodaghee}}},
\bauthor{\binits{J.A.} \bsnm{{Tomsick}}},
\bauthor{\binits{K.} \bsnm{{Pottschmidt}}},
\bauthor{\binits{J.} \bsnm{{Rodriguez}}},
\bauthor{\binits{P.} \bsnm{{Romano}}},
\bauthor{\binits{J.} \bsnm{{Wilms}}},
\batitle{{A Suzaku X-ray observation of one orbit of the supergiant fast X-ray
  transient IGR J16479-4514}}.
\bjtitle{\mnras}
\bvolume{429},
\bfpage{2763}--\blpage{2771}
(\byear{2013}).
doi:\doiurl{10.1093/mnras/sts559}
\end{barticle}
\endbibitem

\bibitem[\protect\citeauthoryear{{Sidoli} et~al.}{2015}]{Sidoli:2015}
\begin{barticle}
\bauthor{\binits{L.} \bsnm{{Sidoli}}},
\bauthor{\binits{A.} \bsnm{{Paizis}}},
\bauthor{\binits{F.} \bsnm{{F{\"u}rst}}},
\bauthor{\binits{J.M.} \bsnm{{Torrej{\'o}n}}},
\bauthor{\binits{P.} \bsnm{{Kretschmar}}},
\bauthor{\binits{E.} \bsnm{{Bozzo}}},
\bauthor{\binits{K.} \bsnm{{Pottschmidt}}},
\batitle{{Probing large-scale wind structures in Vela X-1 using off-states with
  INTEGRAL}}.
\bjtitle{\mnras}
\bvolume{447},
\bfpage{1299}--\blpage{1303}
(\byear{2015}).
doi:\doiurl{10.1093/mnras/stu2533}
\end{barticle}
\endbibitem

\bibitem[\protect\citeauthoryear{{Sim{\'o}n-D{\'{\i}}az} and
  {Stasi{\'n}ska}}{2008}]{SimonDiaz08}
\begin{barticle}
\bauthor{\binits{S.} \bsnm{{Sim{\'o}n-D{\'{\i}}az}}},
\bauthor{\binits{G.} \bsnm{{Stasi{\'n}ska}}},
\batitle{{The ionizing radiation from massive stars and its impact on HII
  regions: results from modern model atmospheres}}.
\bjtitle{\mnras}
\bvolume{389},
\bfpage{1009}--\blpage{1021}
(\byear{2008}).
doi:\doiurl{10.1111/j.1365-2966.2008.13444.x}
\end{barticle}
\endbibitem

\bibitem[\protect\citeauthoryear{{Sim{\'o}n-D{\'{\i}}az}
  et~al.}{2011}]{SimonDiaz11}
\begin{barticle}
\bauthor{\binits{S.} \bsnm{{Sim{\'o}n-D{\'{\i}}az}}},
\bauthor{\binits{N.} \bsnm{{Castro}}},
\bauthor{\binits{A.} \bsnm{{Herrero}}},
\bauthor{\binits{J.} \bsnm{{Puls}}},
\bauthor{\binits{M.} \bsnm{{Garcia}}},
\bauthor{\binits{C.} \bsnm{{Sab{\'{\i}}n-Sanjuli{\'a}n}}},
\batitle{{The IACOB project: A grid-based automatic tool for the quantitative
  spectroscopic analysis of O-stars}}.
\bjtitle{Journal of Physics Conference Series}
\bvolume{328}(\bissue{1}),
\bfpage{012021}
(\byear{2011}).
doi:\doiurl{10.1088/1742-6596/328/1/012021}
\end{barticle}
\endbibitem

\bibitem[\protect\citeauthoryear{{Smith} et~al.}{2012}]{Smith2012}
\begin{barticle}
\bauthor{\binits{D.M.} \bsnm{{Smith}}},
\bauthor{\binits{C.B.} \bsnm{{Markwardt}}},
\bauthor{\binits{J.H.} \bsnm{{Swank}}},
\bauthor{\binits{I.} \bsnm{{Negueruela}}},
\batitle{{Fast X-ray transients towards the Galactic bulge with the Rossi X-ray
  Timing Explorer}}.
\bjtitle{\mnras}
\bvolume{422},
\bfpage{2661}--\blpage{2674}
(\byear{2012}).
doi:\doiurl{10.1111/j.1365-2966.2012.20836.x}
\end{barticle}
\endbibitem

\bibitem[\protect\citeauthoryear{{Smith} and {Owocki}}{2006}]{so2006}
\begin{barticle}
\bauthor{\binits{N.} \bsnm{{Smith}}},
\bauthor{\binits{S.P.} \bsnm{{Owocki}}},
\batitle{{On the Role of Continuum-driven Eruptions in the Evolution of Very
  Massive Stars and Population III Stars}}.
\bjtitle{\apjl}
\bvolume{645},
\bfpage{45}--\blpage{48}
(\byear{2006}).
doi:\doiurl{10.1086/506523}
\end{barticle}
\endbibitem

\bibitem[\protect\citeauthoryear{{Sobolev}}{1960}]{Sobo60}
\begin{bbook}
\bauthor{\binits{V.V.} \bsnm{{Sobolev}}},
\bbtitle{{Moving envelopes of stars}}
(\bpublisher{Cambridge: Harvard University Press, 1960}, \blocation{???},
  \byear{1960})
\end{bbook}
\endbibitem

\bibitem[\protect\citeauthoryear{{Springmann} and {Pauldrach}}{1992}]{SP92}
\begin{barticle}
\bauthor{\binits{U.W.E.} \bsnm{{Springmann}}},
\bauthor{\binits{A.W.A.} \bsnm{{Pauldrach}}},
\batitle{{Radiation-driven winds of hot luminous stars. XI - Frictional heating
  in a multicomponent stellar wind plasma and decoupling of radiatively
  accelerated ions}}.
\bjtitle{\aap}
\bvolume{262},
\bfpage{515}--\blpage{522}
(\byear{1992})
\end{barticle}
\endbibitem

\bibitem[\protect\citeauthoryear{{Stevens}}{1991}]{stevens91}
\begin{barticle}
\bauthor{\binits{I.R.} \bsnm{{Stevens}}},
\batitle{{X-ray-illuminated stellar winds - Optically thick wind models for
  massive X-ray binaries}}.
\bjtitle{\apj}
\bvolume{379},
\bfpage{310}--\blpage{326}
(\byear{1991}).
doi:\doiurl{10.1086/170506}
\end{barticle}
\endbibitem

\bibitem[\protect\citeauthoryear{{Stewart} and
  {Fabian}}{1981}]{StewartFabian1981}
\begin{barticle}
\bauthor{\binits{G.C.} \bsnm{{Stewart}}},
\bauthor{\binits{A.C.} \bsnm{{Fabian}}},
\batitle{{The influence of mass loss on the observed X-ray spectra of
  early-type stars}}.
\bjtitle{\mnras}
\bvolume{197},
\bfpage{713}--\blpage{720}
(\byear{1981})
\end{barticle}
\endbibitem

\bibitem[\protect\citeauthoryear{{Sundqvist} and {Owocki}}{2013}]{Sundqvist13}
\begin{barticle}
\bauthor{\binits{J.O.} \bsnm{{Sundqvist}}},
\bauthor{\binits{S.P.} \bsnm{{Owocki}}},
\batitle{{Clumping in the inner winds of hot, massive stars from hydrodynamical
  line-driven instability simulations}}.
\bjtitle{\mnras}
\bvolume{428},
\bfpage{1837}--\blpage{1844}
(\byear{2013}).
doi:\doiurl{10.1093/mnras/sts165}
\end{barticle}
\endbibitem

\bibitem[\protect\citeauthoryear{{Sundqvist} et~al.}{2011}]{sund2011}
\begin{botherref}
\oauthor{\binits{J.O.} \bsnm{{Sundqvist}}},
\oauthor{\binits{S.P.} \bsnm{{Owocki}}},
\oauthor{\binits{J.} \bsnm{{Puls}}},
{The nature and consequences of clumping in hot, massive star winds}.
ArXiv
\textbf{1110.0485}
(2011)
\end{botherref}
\endbibitem

\bibitem[\protect\citeauthoryear{{Sundqvist} et~al.}{2010}]{Sundqv2010}
\begin{barticle}
\bauthor{\binits{J.O.} \bsnm{{Sundqvist}}},
\bauthor{\binits{J.} \bsnm{{Puls}}},
\bauthor{\binits{A.} \bsnm{{Feldmeier}}},
\batitle{{Mass loss from inhomogeneous hot star winds. I. Resonance line
  formation in 2D models}}.
\bjtitle{\aap}
\bvolume{510},
\bfpage{11}
(\byear{2010}).
doi:\doiurl{10.1051/0004-6361/200912842}
\end{barticle}
\endbibitem

\bibitem[\protect\citeauthoryear{{Sundqvist} et~al.}{2014}]{sund2014}
\begin{barticle}
\bauthor{\binits{J.O.} \bsnm{{Sundqvist}}},
\bauthor{\binits{J.} \bsnm{{Puls}}},
\bauthor{\binits{S.P.} \bsnm{{Owocki}}},
\batitle{{Mass loss from inhomogeneous hot star winds. III. An
  effective-opacity formalism for line radiative transfer in accelerating,
  clumped two-component media, and first results on theory and diagnostics}}.
\bjtitle{\aap}
\bvolume{568},
\bfpage{59}
(\byear{2014}).
doi:\doiurl{10.1051/0004-6361/201423570}
\end{barticle}
\endbibitem

\bibitem[\protect\citeauthoryear{{Sundqvist} et~al.}{2011}]{Sundqv2011}
\begin{barticle}
\bauthor{\binits{J.O.} \bsnm{{Sundqvist}}},
\bauthor{\binits{J.} \bsnm{{Puls}}},
\bauthor{\binits{A.} \bsnm{{Feldmeier}}},
\bauthor{\binits{S.P.} \bsnm{{Owocki}}},
\batitle{{Mass loss from inhomogeneous hot star winds. II. Constraints from a
  combined optical/UV study}}.
\bjtitle{\aap}
\bvolume{528},
\bfpage{64}
(\byear{2011}).
doi:\doiurl{10.1051/0004-6361/201015771}
\end{barticle}
\endbibitem

\bibitem[\protect\citeauthoryear{{Sundqvist} et~al.}{2012a}]{Sundqvist12}
\begin{barticle}
\bauthor{\binits{J.O.} \bsnm{{Sundqvist}}},
\bauthor{\binits{A.} \bsnm{{ud-Doula}}},
\bauthor{\binits{S.P.} \bsnm{{Owocki}}},
\bauthor{\binits{R.H.D.} \bsnm{{Townsend}}},
\bauthor{\binits{I.D.} \bsnm{{Howarth}}},
\bauthor{\binits{G.A.} \bsnm{{Wade}}},
\batitle{{A dynamical magnetosphere model for periodic H{$\alpha$} emission
  from the slowly rotating magnetic O star HD 191612}}.
\bjtitle{\mnras}
\bvolume{423},
\bfpage{21}--\blpage{25}
(\byear{2012}a).
doi:\doiurl{10.1111/j.1745-3933.2012.01248.x}
\end{barticle}
\endbibitem

\bibitem[\protect\citeauthoryear{{Sundqvist} et~al.}{2012b}]{sund2012}
\begin{barticle}
\bauthor{\binits{J.O.} \bsnm{{Sundqvist}}},
\bauthor{\binits{S.P.} \bsnm{{Owocki}}},
\bauthor{\binits{D.H.} \bsnm{{Cohen}}},
\bauthor{\binits{M.A.} \bsnm{{Leutenegger}}},
\bauthor{\binits{R.H.D.} \bsnm{{Townsend}}},
\batitle{{A generalized porosity formalism for isotropic and anisotropic
  effective opacity and its effects on X-ray line attenuation in clumped O star
  winds}}.
\bjtitle{\mnras}
\bvolume{420},
\bfpage{1553}--\blpage{1561}
(\byear{2012}b).
doi:\doiurl{10.1111/j.1365-2966.2011.20141.x}
\end{barticle}
\endbibitem

\bibitem[\protect\citeauthoryear{{Sundqvist} et~al.}{2013}]{Sundqvist13a}
\begin{barticle}
\bauthor{\binits{J.O.} \bsnm{{Sundqvist}}},
\bauthor{\binits{S.} \bsnm{{Sim{\'o}n-D{\'{\i}}az}}},
\bauthor{\binits{J.} \bsnm{{Puls}}},
\bauthor{\binits{N.} \bsnm{{Markova}}},
\batitle{{The rotation rates of massive stars. How slow are the slow ones?}}
\bjtitle{\aap}
\bvolume{559},
\bfpage{10}
(\byear{2013}).
doi:\doiurl{10.1051/0004-6361/201322761}
\end{barticle}
\endbibitem

\bibitem[\protect\citeauthoryear{{Sunyaev} et~al.}{2003}]{Sunyaev2003}
\begin{barticle}
\bauthor{\binits{R.A.} \bsnm{{Sunyaev}}},
\bauthor{\binits{S.A.} \bsnm{{Grebenev}}},
\bauthor{\binits{A.A.} \bsnm{{Lutovinov}}},
\bauthor{\binits{J.} \bsnm{{Rodriguez}}},
\bauthor{\binits{S.} \bsnm{{Mereghetti}}},
\bauthor{\binits{D.} \bsnm{{Gotz}}},
\bauthor{\binits{T.} \bsnm{{Courvoisier}}},
\batitle{{New source IGR J17544-2619 discovered with INTEGRAL}}.
\bjtitle{The Astronomer's Telegram}
\bvolume{190},
\bfpage{1}
(\byear{2003})
\end{barticle}
\endbibitem

\bibitem[\protect\citeauthoryear{{Swank} et~al.}{2007}]{Swank2007}
\begin{barticle}
\bauthor{\binits{J.H.} \bsnm{{Swank}}},
\bauthor{\binits{D.M.} \bsnm{{Smith}}},
\bauthor{\binits{C.B.} \bsnm{{Markwardt}}},
\batitle{{RXTE PCA Pointed Observations of IGR J11215-5952}}.
\bjtitle{The Astronomer's Telegram}
\bvolume{999},
\bfpage{1}
(\byear{2007})
\end{barticle}
\endbibitem

\bibitem[\protect\citeauthoryear{{Tarter} et~al.}{1969}]{Tarter1969}
\begin{barticle}
\bauthor{\binits{C.B.} \bsnm{{Tarter}}},
\bauthor{\binits{W.H.} \bsnm{{Tucker}}},
\bauthor{\binits{E.E.} \bsnm{{Salpeter}}},
\batitle{{The Interaction of X-Ray Sources with Optically Thin Environments}}.
\bjtitle{\apj}
\bvolume{156},
\bfpage{943}
(\byear{1969}).
doi:\doiurl{10.1086/150026}
\end{barticle}
\endbibitem

\bibitem[\protect\citeauthoryear{{Todt} et~al.}{2010}]{Todt2010}
\begin{barticle}
\bauthor{\binits{H.} \bsnm{{Todt}}},
\bauthor{\binits{M.} \bsnm{{Pe{\~n}a}}},
\bauthor{\binits{W.-R.} \bsnm{{Hamann}}},
\bauthor{\binits{G.} \bsnm{{Gr{\"a}fener}}},
\batitle{{The central star of the planetary nebula PB 8: a Wolf-Rayet-type wind
  of an unusual WN/WC chemical composition}}.
\bjtitle{\aap}
\bvolume{515},
\bfpage{83}
(\byear{2010}).
doi:\doiurl{10.1051/0004-6361/200912183}
\end{barticle}
\endbibitem

\bibitem[\protect\citeauthoryear{{Toropin} et~al.}{1999}]{Toropin1999}
\begin{barticle}
\bauthor{\binits{Y.M.} \bsnm{{Toropin}}},
\bauthor{\binits{O.D.} \bsnm{{Toropina}}},
\bauthor{\binits{V.V.} \bsnm{{Savelyev}}},
\bauthor{\binits{M.M.} \bsnm{{Romanova}}},
\bauthor{\binits{V.M.} \bsnm{{Chechetkin}}},
\bauthor{\binits{R.V.E.} \bsnm{{Lovelace}}},
\batitle{{Spherical Bondi Accretion onto a Magnetic Dipole}}.
\bjtitle{\apj}
\bvolume{517},
\bfpage{906}--\blpage{918}
(\byear{1999}).
doi:\doiurl{10.1086/307229}
\end{barticle}
\endbibitem

\bibitem[\protect\citeauthoryear{{Toropina} et~al.}{2006}]{toropina2006}
\begin{barticle}
\bauthor{\binits{O.D.} \bsnm{{Toropina}}},
\bauthor{\binits{M.M.} \bsnm{{Romanova}}},
\bauthor{\binits{R.V.E.} \bsnm{{Lovelace}}},
\batitle{{Spinning-down of moving magnetars in the propeller regime}}.
\bjtitle{\mnras}
\bvolume{371},
\bfpage{569}--\blpage{576}
(\byear{2006}).
doi:\doiurl{10.1111/j.1365-2966.2006.10667.x}
\end{barticle}
\endbibitem

\bibitem[\protect\citeauthoryear{{Toropina} et~al.}{2012}]{Toropina2012}
\begin{barticle}
\bauthor{\binits{O.D.} \bsnm{{Toropina}}},
\bauthor{\binits{M.M.} \bsnm{{Romanova}}},
\bauthor{\binits{R.V.E.} \bsnm{{Lovelace}}},
\batitle{{Bondi-Hoyle accretion on to a magnetized neutron star}}.
\bjtitle{\mnras}
\bvolume{420},
\bfpage{810}--\blpage{816}
(\byear{2012}).
doi:\doiurl{10.1111/j.1365-2966.2011.20093.x}
\end{barticle}
\endbibitem

\bibitem[\protect\citeauthoryear{{Torrej{\'o}n} et~al.}{2010a}]{Torrejon2010b}
\begin{barticle}
\bauthor{\binits{J.M.} \bsnm{{Torrej{\'o}n}}},
\bauthor{\binits{N.S.} \bsnm{{Schulz}}},
\bauthor{\binits{M.A.} \bsnm{{Nowak}}},
\bauthor{\binits{T.R.} \bsnm{{Kallman}}},
\batitle{{A Chandra Survey of Fluorescence Fe Lines in X-ray Binaries at High
  Resolution}}.
\bjtitle{ApJ}
\bvolume{715},
\bfpage{947}--\blpage{958}
(\byear{2010}a)
\end{barticle}
\endbibitem

\bibitem[\protect\citeauthoryear{{Torrej{\'o}n} et~al.}{2010b}]{Torrejon2010a}
\begin{barticle}
\bauthor{\binits{J.M.} \bsnm{{Torrej{\'o}n}}},
\bauthor{\binits{I.} \bsnm{{Negueruela}}},
\bauthor{\binits{D.M.} \bsnm{{Smith}}},
\bauthor{\binits{T.E.} \bsnm{{Harrison}}},
\batitle{{Near-infrared survey of high mass X-ray binary candidates}}.
\bjtitle{\aap}
\bvolume{510},
\bfpage{61}
(\byear{2010}b).
doi:\doiurl{10.1051/0004-6361/200912619}
\end{barticle}
\endbibitem

\bibitem[\protect\citeauthoryear{{Torrej{\'o}n} et~al.}{2015}]{Torrejon2015}
\begin{barticle}
\bauthor{\binits{J.M.} \bsnm{{Torrej{\'o}n}}},
\bauthor{\binits{N.S.} \bsnm{{Schulz}}},
\bauthor{\binits{M.A.} \bsnm{{Nowak}}},
\bauthor{\binits{L.} \bsnm{{Oskinova}}},
\bauthor{\binits{J.J.} \bsnm{{Rodes-Roca}}},
\bauthor{\binits{T.} \bsnm{{Shenar}}},
\bauthor{\binits{J.} \bsnm{{Wilms}}},
\batitle{{On the Radial Onset of Clumping in the Wind of the B0I Massive Star
  QV Nor}}.
\bjtitle{\apj}
\bvolume{810},
\bfpage{102}
(\byear{2015}).
doi:\doiurl{10.1088/0004-637X/810/2/102}
\end{barticle}
\endbibitem

\bibitem[\protect\citeauthoryear{{ud-Doula}}{2013}]{udDoula13}
\begin{bchapter}
\bauthor{\binits{A.} \bsnm{{ud-Doula}}},
\bctitle{{Stellar Winds, Magnetic Fields and Disks}},
in \bbtitle{Lecture Notes in Physics, Berlin Springer Verlag},
ed. by \beditor{\binits{J.-P.} \bsnm{{Rozelot}}},
\beditor{\binits{C..} \bsnm{{Neiner}}}
\bsertitle{Lecture Notes in Physics, Berlin Springer Verlag},
vol. \bseriesno{857},
\byear{2013},
p. \bfpage{207}.
doi:\doiurl{10.1007/978-3-642-30648-8\_8}
\end{bchapter}
\endbibitem

\bibitem[\protect\citeauthoryear{{ud-Doula} and {Owocki}}{2002}]{ud-Doula2002}
\begin{barticle}
\bauthor{\binits{A.} \bsnm{{ud-Doula}}},
\bauthor{\binits{S.P.} \bsnm{{Owocki}}},
\batitle{{Dynamical Simulations of Magnetically Channeled Line-driven Stellar
  Winds. I. Isothermal, Nonrotating, Radially Driven Flow}}.
\bjtitle{\apj}
\bvolume{576},
\bfpage{413}--\blpage{428}
(\byear{2002}).
doi:\doiurl{10.1086/341543}
\end{barticle}
\endbibitem

\bibitem[\protect\citeauthoryear{{{\v C}echura} and
  {Hadrava}}{2015}]{Cechura2015}
\begin{barticle}
\bauthor{\binits{J.} \bsnm{{{\v C}echura}}},
\bauthor{\binits{P.} \bsnm{{Hadrava}}},
\batitle{{Stellar wind in state transitions of high-mass X-ray binaries}}.
\bjtitle{\aap}
\bvolume{575},
\bfpage{5}
(\byear{2015}).
doi:\doiurl{10.1051/0004-6361/201424636}
\end{barticle}
\endbibitem

\bibitem[\protect\citeauthoryear{{{\v S}urlan} et~al.}{2012}]{sur2012}
\begin{barticle}
\bauthor{\binits{B.} \bsnm{{{\v S}urlan}}},
\bauthor{\binits{W.-R.} \bsnm{{Hamann}}},
\bauthor{\binits{J.} \bsnm{{Kub{\'a}t}}},
\bauthor{\binits{L.M.} \bsnm{{Oskinova}}},
\bauthor{\binits{A.} \bsnm{{Feldmeier}}},
\batitle{{Three-dimensional radiative transfer in clumped hot star winds. I.
  Influence of clumping on the resonance line formation}}.
\bjtitle{\aap}
\bvolume{541},
\bfpage{37}
(\byear{2012}).
doi:\doiurl{10.1051/0004-6361/201118590}
\end{barticle}
\endbibitem

\bibitem[\protect\citeauthoryear{{{\v S}urlan} et~al.}{2013}]{sur2013}
\begin{barticle}
\bauthor{\binits{B.} \bsnm{{{\v S}urlan}}},
\bauthor{\binits{W.-R.} \bsnm{{Hamann}}},
\bauthor{\binits{A.} \bsnm{{Aret}}},
\bauthor{\binits{J.} \bsnm{{Kub{\'a}t}}},
\bauthor{\binits{L.M.} \bsnm{{Oskinova}}},
\bauthor{\binits{A.F.} \bsnm{{Torres}}},
\batitle{{Macroclumping as solution of the discrepancy between H{$\alpha$} and
  P v mass loss diagnostics for O-type stars}}.
\bjtitle{\aap}
\bvolume{559},
\bfpage{130}
(\byear{2013}).
doi:\doiurl{10.1051/0004-6361/201322390}
\end{barticle}
\endbibitem

\bibitem[\protect\citeauthoryear{{van der Meer} et~al.}{2005}]{vdm2005}
\begin{barticle}
\bauthor{\binits{A.} \bsnm{{van der Meer}}},
\bauthor{\binits{L.} \bsnm{{Kaper}}},
\bauthor{\binits{T.} \bsnm{{di Salvo}}},
\bauthor{\binits{M.} \bsnm{{M{\'e}ndez}}},
\bauthor{\binits{M.} \bsnm{{van der Klis}}},
\bauthor{\binits{P.} \bsnm{{Barr}}},
\bauthor{\binits{N.R.} \bsnm{{Trams}}},
\batitle{{XMM-Newton X-ray spectroscopy of the high-mass X-ray binary 4U
  1700-37 at low flux}}.
\bjtitle{\aap}
\bvolume{432},
\bfpage{999}--\blpage{1012}
(\byear{2005}).
doi:\doiurl{10.1051/0004-6361:20041288}
\end{barticle}
\endbibitem

\bibitem[\protect\citeauthoryear{{van Loon} et~al.}{2001}]{vanloon2001}
\begin{barticle}
\bauthor{\binits{J.T.} \bsnm{{van Loon}}},
\bauthor{\binits{L.} \bsnm{{Kaper}}},
\bauthor{\binits{G.} \bsnm{{Hammerschlag-Hensberge}}},
\batitle{{Modelling the orbital modulation of ultraviolet resonance lines in
  high-mass X-ray binaries}}.
\bjtitle{\aap}
\bvolume{375},
\bfpage{498}--\blpage{526}
(\byear{2001}).
doi:\doiurl{10.1051/0004-6361:20010856}
\end{barticle}
\endbibitem

\bibitem[\protect\citeauthoryear{{Verner} et~al.}{1996}]{verner1996b}
\begin{barticle}
\bauthor{\binits{D.A.} \bsnm{{Verner}}},
\bauthor{\binits{G.J.} \bsnm{{Ferland}}},
\bauthor{\binits{K.T.} \bsnm{{Korista}}},
\bauthor{\binits{D.G.} \bsnm{{Yakovlev}}},
\batitle{Atomic data for astrophysics. ii. new analytic fits for
  photoionization cross sections of atoms and ions}.
\bjtitle{ApJ}
\bvolume{465},
\bfpage{487}
(\byear{1996})
\end{barticle}
\endbibitem

\bibitem[\protect\citeauthoryear{{Vink} et~al.}{2000}]{Vink00}
\begin{barticle}
\bauthor{\binits{J.S.} \bsnm{{Vink}}},
\bauthor{\binits{A.} \bsnm{{de Koter}}},
\bauthor{\binits{H.J.G.L.M.} \bsnm{{Lamers}}},
\batitle{{New theoretical mass-loss rates of O and B stars}}.
\bjtitle{\aap}
\bvolume{362},
\bfpage{295}--\blpage{309}
(\byear{2000})
\end{barticle}
\endbibitem

\bibitem[\protect\citeauthoryear{{Vink} et~al.}{2001}]{Vink01}
\begin{barticle}
\bauthor{\binits{J.S.} \bsnm{{Vink}}},
\bauthor{\binits{A.} \bsnm{{de Koter}}},
\bauthor{\binits{H.J.G.L.M.} \bsnm{{Lamers}}},
\batitle{{Mass-loss predictions for O and B stars as a function of
  metallicity}}.
\bjtitle{\aap}
\bvolume{369},
\bfpage{574}--\blpage{588}
(\byear{2001})
\end{barticle}
\endbibitem

\bibitem[\protect\citeauthoryear{{Vink} et~al.}{2010}]{Vink10}
\begin{barticle}
\bauthor{\binits{J.S.} \bsnm{{Vink}}},
\bauthor{\binits{I.} \bsnm{{Brott}}},
\bauthor{\binits{G.} \bsnm{{Gr{\"a}fener}}},
\bauthor{\binits{N.} \bsnm{{Langer}}},
\bauthor{\binits{A.} \bsnm{{de Koter}}},
\bauthor{\binits{D.J.} \bsnm{{Lennon}}},
\batitle{{The nature of B supergiants: clues from a steep drop in rotation
  rates at 22 000 K. The possibility of Bi-stability braking}}.
\bjtitle{\aap}
\bvolume{512},
\bfpage{7}
(\byear{2010}).
doi:\doiurl{10.1051/0004-6361/201014205}
\end{barticle}
\endbibitem

\bibitem[\protect\citeauthoryear{{Vink} et~al.}{2011}]{Vink11}
\begin{barticle}
\bauthor{\binits{J.S.} \bsnm{{Vink}}},
\bauthor{\binits{L.E.} \bsnm{{Muijres}}},
\bauthor{\binits{B.} \bsnm{{Anthonisse}}},
\bauthor{\binits{A.} \bsnm{{de Koter}}},
\bauthor{\binits{G.} \bsnm{{Gr{\"a}fener}}},
\bauthor{\binits{N.} \bsnm{{Langer}}},
\batitle{{Wind modelling of very massive stars up to 300 solar masses}}.
\bjtitle{\aap}
\bvolume{531},
\bfpage{132}
(\byear{2011}).
doi:\doiurl{10.1051/0004-6361/201116614}
\end{barticle}
\endbibitem

\bibitem[\protect\citeauthoryear{{Wade} et~al.}{2012}]{Wade12}
\begin{bchapter}
\bauthor{\binits{G.A.} \bsnm{{Wade}}},
\bauthor{\binits{J.H.} \bsnm{{Grunhut}}},
\bauthor{\bsnm{{MiMeS Collaboration}}},
\bctitle{{The MiMeS Survey of Magnetism in Massive Stars}},
in \bbtitle{Circumstellar Dynamics at High Resolution},
ed. by \beditor{\binits{A.C.} \bsnm{{Carciofi}}},
\beditor{\binits{T.} \bsnm{{Rivinius}}}
\bsertitle{Astronomical Society of the Pacific Conference Series},
vol. \bseriesno{464},
\byear{2012},
p. \bfpage{405}
\end{bchapter}
\endbibitem

\bibitem[\protect\citeauthoryear{{Waldron} and {Cassinelli}}{2001}]{wc2001}
\begin{barticle}
\bauthor{\binits{W.L.} \bsnm{{Waldron}}},
\bauthor{\binits{J.P.} \bsnm{{Cassinelli}}},
\batitle{{Chandra Discovers a Very High Density X-Ray Plasma on the O Star
  {$\zeta$} Orionis}}.
\bjtitle{\apjl}
\bvolume{548},
\bfpage{45}--\blpage{48}
(\byear{2001}).
doi:\doiurl{10.1086/318926}
\end{barticle}
\endbibitem

\bibitem[\protect\citeauthoryear{{Waldron} and {Cassinelli}}{2007}]{wc2007}
\begin{barticle}
\bauthor{\binits{W.L.} \bsnm{{Waldron}}},
\bauthor{\binits{J.P.} \bsnm{{Cassinelli}}},
\batitle{{An Extensive Collection of Stellar Wind X-Ray Source Region Emission
  Line Parameters, Temperatures, Velocities, and Their Radial Distributions as
  Obtained from Chandra Observations of 17 OB Stars}}.
\bjtitle{\apj}
\bvolume{668},
\bfpage{456}--\blpage{480}
(\byear{2007}).
doi:\doiurl{10.1086/520919}
\end{barticle}
\endbibitem

\bibitem[\protect\citeauthoryear{{Walter} and {Zurita
  Heras}}{2007}]{walter2007a}
\begin{barticle}
\bauthor{\binits{R.} \bsnm{{Walter}}},
\bauthor{\binits{J.} \bsnm{{Zurita Heras}}},
\batitle{Probing clumpy stellar winds with a neutron star}.
\bjtitle{A\&A}
\bvolume{476},
\bfpage{335}--\blpage{340}
(\byear{2007})
\end{barticle}
\endbibitem

\bibitem[\protect\citeauthoryear{{Walter} et~al.}{2006}]{Walter2006}
\begin{barticle}
\bauthor{\binits{R.} \bsnm{{Walter}}},
\bauthor{\binits{J.} \bsnm{{Zurita Heras}}},
\bauthor{\binits{L.} \bsnm{{Bassani}}},
\bauthor{\binits{A.} \bsnm{{Bazzano}}},
\bauthor{\binits{A.} \bsnm{{Bodaghee}}},
\bauthor{\binits{A.} \bsnm{{Dean}}},
\bauthor{\binits{P.} \bsnm{{Dubath}}},
\bauthor{\binits{A.N.} \bsnm{{Parmar}}},
\bauthor{\binits{M.} \bsnm{{Renaud}}},
\bauthor{\binits{P.} \bsnm{{Ubertini}}},
\batitle{{XMM-Newton and INTEGRAL observations of new absorbed supergiant
  high-mass X-ray binaries}}.
\bjtitle{\aap}
\bvolume{453},
\bfpage{133}--\blpage{143}
(\byear{2006}).
doi:\doiurl{10.1051/0004-6361:20053719}
\end{barticle}
\endbibitem

\bibitem[\protect\citeauthoryear{{Walter} et~al.}{2015}]{Walter15}
\begin{botherref}
\oauthor{\binits{R.} \bsnm{{Walter}}},
\oauthor{\binits{A.A.} \bsnm{{Lutovinov}}},
\oauthor{\binits{E.} \bsnm{{Bozzo}}},
\oauthor{\binits{S.S.} \bsnm{{Tsygankov}}},
{High-Mass X-ray Binaries in the Milky Way: A closer look with INTEGRAL}.
ArXiv e-prints
(2015)
\end{botherref}
\endbibitem

\bibitem[\protect\citeauthoryear{{Wang}}{2011}]{Wang2011}
\begin{barticle}
\bauthor{\binits{W.} \bsnm{{Wang}}},
\batitle{{Long-term hard X-ray monitoring of 2S 0114+65 with INTEGRAL/IBIS}}.
\bjtitle{\mnras}
\bvolume{413},
\bfpage{1083}--\blpage{1098}
(\byear{2011}).
doi:\doiurl{10.1111/j.1365-2966.2010.18192.x}
\end{barticle}
\endbibitem

\bibitem[\protect\citeauthoryear{{Watanabe} et~al.}{2003}]{Watanabe2003}
\begin{barticle}
\bauthor{\binits{S.} \bsnm{{Watanabe}}},
\bauthor{\binits{M.} \bsnm{{Sako}}},
\bauthor{\binits{M.} \bsnm{{Ishida}}},
\bauthor{\binits{Y.} \bsnm{{Ishisaki}}},
\bauthor{\binits{S.M.} \bsnm{{Kahn}}},
\bauthor{\binits{T.} \bsnm{{Kohmura}}},
\bauthor{\binits{U.} \bsnm{{Morita}}},
\bauthor{\binits{F.} \bsnm{{Nagase}}},
\bauthor{\binits{F.} \bsnm{{Paerels}}},
\bauthor{\binits{T.} \bsnm{{Takahashi}}},
\batitle{{Detection of a Fully Resolved Compton Shoulder of the Iron
  K{$\alpha$} Line in the Chandra X-Ray Spectrum of GX 301-2}}.
\bjtitle{\apjl}
\bvolume{597},
\bfpage{37}--\blpage{40}
(\byear{2003}).
doi:\doiurl{10.1086/379735}
\end{barticle}
\endbibitem

\bibitem[\protect\citeauthoryear{{Watanabe} et~al.}{2006}]{Watanabe2006}
\begin{barticle}
\bauthor{\binits{S.} \bsnm{{Watanabe}}},
\bauthor{\binits{M.} \bsnm{{Sako}}},
\bauthor{\binits{M.} \bsnm{{Ishida}}},
\bauthor{\binits{Y.} \bsnm{{Ishisaki}}},
\bauthor{\binits{S.M.} \bsnm{{Kahn}}},
\bauthor{\binits{T.} \bsnm{{Kohmura}}},
\bauthor{\binits{F.} \bsnm{{Nagase}}},
\bauthor{\binits{F.} \bsnm{{Paerels}}},
\bauthor{\binits{T.} \bsnm{{Takahashi}}},
\batitle{{X-Ray Spectral Study of the Photoionized Stellar Wind in Vela X-1}}.
\bjtitle{\apj}
\bvolume{651},
\bfpage{421}--\blpage{437}
(\byear{2006}).
doi:\doiurl{10.1086/507458}
\end{barticle}
\endbibitem

\bibitem[\protect\citeauthoryear{{White} and {Pravdo}}{1979}]{White1979}
\begin{barticle}
\bauthor{\binits{N.E.} \bsnm{{White}}},
\bauthor{\binits{S.H.} \bsnm{{Pravdo}}},
\batitle{{The discovery of 38.22 second X-ray pulsations from the vicinity of
  OAO 1653-40}}.
\bjtitle{\apjl}
\bvolume{233},
\bfpage{121}--\blpage{124}
(\byear{1979}).
doi:\doiurl{10.1086/183089}
\end{barticle}
\endbibitem

\bibitem[\protect\citeauthoryear{{White} et~al.}{1980}]{White1980}
\begin{barticle}
\bauthor{\binits{N.E.} \bsnm{{White}}},
\bauthor{\binits{R.H.} \bsnm{{Becker}}},
\bauthor{\binits{S.H.} \bsnm{{Pravdo}}},
\bauthor{\binits{E.A.} \bsnm{{Boldt}}},
\bauthor{\binits{S.S.} \bsnm{{Holt}}},
\bauthor{\binits{P.J.} \bsnm{{Serlemitsos}}},
\batitle{{The X-ray pulsars 4U 1145-61 and 1E 1145.1-6141}}.
\bjtitle{\apj}
\bvolume{239},
\bfpage{655}--\blpage{660}
(\byear{1980}).
doi:\doiurl{10.1086/158152}
\end{barticle}
\endbibitem

\bibitem[\protect\citeauthoryear{{Wilms} et~al.}{2000}]{wilms2000a}
\begin{barticle}
\bauthor{\binits{J.} \bsnm{{Wilms}}},
\bauthor{\binits{A.} \bsnm{{Allen}}},
\bauthor{\binits{R.} \bsnm{{McCray}}},
\batitle{On the absorption of x-rays in the interstellar medium}.
\bjtitle{ApJ}
\bvolume{542},
\bfpage{914}--\blpage{924}
(\byear{2000})
\end{barticle}
\endbibitem

\bibitem[\protect\citeauthoryear{{Wright} and
  {Barlow}}{1975}]{WrightBarlow1975}
\begin{barticle}
\bauthor{\binits{A.E.} \bsnm{{Wright}}},
\bauthor{\binits{M.J.} \bsnm{{Barlow}}},
\batitle{{The radio and infrared spectrum of early-type stars undergoing mass
  loss}}.
\bjtitle{\mnras}
\bvolume{170},
\bfpage{41}--\blpage{51}
(\byear{1975})
\end{barticle}
\endbibitem

\bibitem[\protect\citeauthoryear{{Yan} et~al.}{1998}]{Yan1998a}
\begin{barticle}
\bauthor{\binits{M.} \bsnm{{Yan}}},
\bauthor{\binits{H.R.} \bsnm{{Sadeghpour}}},
\bauthor{\binits{A.} \bsnm{{Dalgarno}}},
\batitle{Photoionization cross sections of he and h$_{2}$}.
\bjtitle{ApJ}
\bvolume{496},
\bfpage{1044}--\blpage{1050}
(\byear{1998})
\end{barticle}
\endbibitem

\bibitem[\protect\citeauthoryear{{Zurita Heras} and {Chaty}}{2009}]{Zurita2009}
\begin{barticle}
\bauthor{\binits{J.A.} \bsnm{{Zurita Heras}}},
\bauthor{\binits{S.} \bsnm{{Chaty}}},
\batitle{{Discovery of an eccentric 30 day period in the supergiant X-ray
  binary SAX J1818.6-1703 with INTEGRAL}}.
\bjtitle{\aap}
\bvolume{493},
\bfpage{1}--\blpage{4}
(\byear{2009}).
doi:\doiurl{10.1051/0004-6361:200811179}
\end{barticle}
\endbibitem

\end{thebibliography}

\end{document}